\documentclass[twoside,12pt]{report}
\usepackage{floatflt,bm,psfrag,layout}%,enumitem}
\usepackage{amsmath}
\usepackage{graphicx,amssymb,rotate}
\usepackage{mathrsfs}
\usepackage{color}
\usepackage{graphics} 
\usepackage{multirow}
\usepackage{float}
\usepackage{pstricks}
\usepackage[toc,page]{appendix}
\usepackage{pifont}
\usepackage[utf8]{inputenc}
\usepackage{tabls}
\usepackage{dashrule}
\usepackage{fancyhdr}

\usepackage{enumerate}

\usepackage{xcolor}
\usepackage{makeidx}
\usepackage{mathtools}
\usepackage{epsfig}
\usepackage{epstopdf}
\usepackage{soul}
\usepackage{array}
\usepackage{lscape}
\usepackage{rotating}
\usepackage{pdflscape}
\usepackage{booktabs}
\usepackage{tikz}
\usetikzlibrary{intersections}
\usepackage{xr}
\usepackage{soul}
\usepackage[numbers,sort&compress]{natbib}
\usepackage{notoccite}
\usepackage{ae,aecompl}
\usepackage{multicol}
\usepackage{tabularx}
\usepackage{newtxtext,newtxmath}
\usepackage[T1]{fontenc}

\DeclareUnicodeCharacter{FB01}{fi}
\DeclareUnicodeCharacter{2212}{\textendash}
\usepackage{comment}

\usepackage{caption}
\usepackage{subcaption}
\usepackage{twemojis}
\usepackage{cancel}
\usepackage[bottom]{footmisc}
\usepackage[bookmarks=true,bookmarksnumbered=true,%
pdftitle={Thesis},%
pdfauthor={Dhruv Ringe (IIT Indore)}, colorlinks = true,
            linkcolor = blue,
            urlcolor  = blue,
            citecolor = blue,
            anchorcolor = .]{hyperref}
\newcommand{\beq}{\begin{equation}}
\newcommand{\eeq}{\end{equation}}
\newcommand{\bea}{\begin{eqnarray}}
\newcommand{\eea}{\end{eqnarray}}

\def\nn{\nonumber}

\newcommand{\h}{\langle H \rangle}
\newcommand{\hu}{\langle \widetilde{H} \rangle}
\newcommand{\vs}{\langle S \rangle}
\newcommand{\fn}{\rm{FN}}

\newcommand{\cw}{\rm{CW}}
\newcommand{\eft}{\rm{EFT}}

\newcommand{\f}{\mathcal{F}}
\newcommand{\gsm}{\mathcal{G}_{SM}}
\newcommand{\ordone}{\mathcal{O}(1)}
\newcommand{\vev}{{\textit{vev}} }
\newcommand{\vevs}{{\textit{vev}}s }

\newcommand{\ms}{\overline{\rm{MS}}}
\newcommand{\Vcw}{V_{\rm{CW}}}

\newcommand{\tr}{{\rm{Tr}}}
\usepackage{aas_macros}

\newcommand{\gbl}{g_{BL}}
\def\a{\alpha}

\def\b{\beta}
\def\k{\kappa}

\def\l{\lambda}
\def\m{\mu}
\def\n{\nu}

\def\s{\sigma}

\usepackage[compact]{titlesec}
\usepackage{setspace}
\titlespacing*{\section}{0pt}{3ex}{0ex}
\titlespacing*{\subsection}{0pt}{3ex}{0ex}
\titlespacing*{\subsubsection}{0pt}{3ex}{0ex}

\begin{document}
\let\cleardoublepage\clearpage  %%% clears away extra blank pages after each chapter
%%%%%%%%%%%%%%%%%%Gap after equations%%%%%%%%%%%%%%%%%
\setlength{\abovedisplayskip}{5pt}
\setlength{\belowdisplayskip}{2pt}
\setlength{\abovedisplayshortskip}{2pt}
\setlength{\belowdisplayshortskip}{2pt} 
%%%%%%%%%%%%%%%%%%%%%%%%%%%%%%%%%%%%%%%%%%%%%%%%%%%%%%%%%%%
%\mainmatter
%%%%%%%%%%%%%%\pagestyle{fancy}%%%%%%%%%%%%%%%%%%%%%%%%
%
\graphicspath{{FirstPage/}}
\pagestyle{plain}
\pagenumbering{roman}\setcounter{page}{1}
\addcontentsline{toc}{section}{Title}

\begin{center}
\vspace*{-2 cm}
{\Large\bf{CONSTRAINING PARTICLE PHYSICS MODELS WITH GRAVITATIONAL WAVES FROM THE EARLY UNIVERSE}}\\
\vfill
{\large\textbf{Ph.D. Thesis}}\\
\vspace*{0cm}

\vfill\vspace*{0cm}
{\Large\textbf{by}}\\
\vspace*{0.5cm}
{\Large\textbf{DHRUV RINGE}}\\
\vspace*{1cm}

\includegraphics[width=5cm,clip]{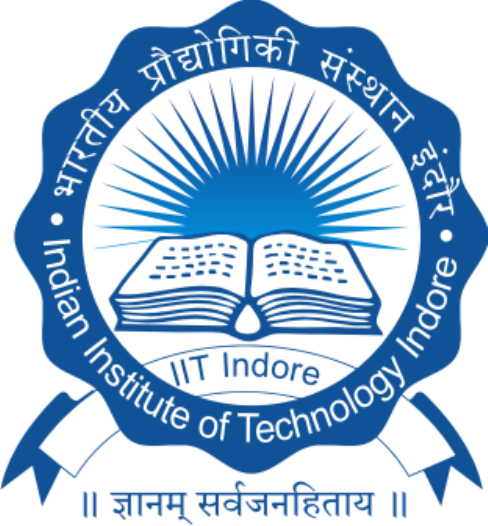}\\
\vspace*{1cm}

{\large \bf DEPARTMENT OF PHYSICS \\
INDIAN INSTITUTE OF TECHNOLOGY INDORE }\\ 
%Khandwa Road, Simrol, Indore - 453552, India}\\   
\textbf{SEPTEMBER 2024}
%\vspace*{1cm}

\end{center}

\newpage
\null
\newpage

\begin{center}
\vspace*{-2.0cm}
{\Large\bf{CONSTRAINING PARTICLE PHYSICS MODELS WITH GRAVITATIONAL WAVES FROM THE EARLY UNIVERSE}}\\

\vspace{5mm}
\vfill
{\large\textbf{A THESIS}}\\
\vspace*{0cm}
\vfill
\large{\it{Submitted in partial fulfillment of the \\requirements for the award of the degree}}\\
\vspace*{0.2cm}
{\Large\textbf{of}}\\
\vspace*{0.5cm}

{\large\textbf{DOCTOR OF PHILOSOPHY}}\\
\vspace*{0cm}

\vfill\vspace*{0cm}
{\Large\textbf{by}}\\
\vspace*{0.5cm}
{\Large\textbf{DHRUV RINGE}}\\
\vspace*{1cm}
\includegraphics[width=5cm,clip]{./Plots/IITimage.pdf}\\
\vspace*{1cm}

{\large \bf DEPARTMENT OF PHYSICS \\
INDIAN INSTITUTE OF TECHNOLOGY INDORE }\\ 
%Khandwa Road, Simrol, Indore - 453552, India}\\   
\textbf{SEPTEMBER 2024}
\vspace*{1cm}
\end{center}

\newpage
\null
\newpage

\noindent
\begin{tabular}{@{} l c @{}}
    \raisebox{-4.5 ex}{\includegraphics[width=1.75 cm]{./Plots/IITimage}} &  % Adjust width as needed
    {\large\textbf{INDIAN INSTITUTE OF TECHNOLOGY INDORE}}
\end{tabular}

\phantomsection
\addcontentsline{toc}{chapter}{Declaration}
\printindex
\vspace{.5 cm}
I hereby certify that the work which is being presented in the thesis entitled \textbf{CONSTRAINING PARTICLE PHYSICS MODELS WITH GRAVITATIONAL WAVES FROM THE EARLY UNIVERSE} in the partial fulfillment of the requirements for the award of the degree of \textbf{DOCTOR OF PHILOSOPHY} and submitted in the \textbf{DEPARTMENT OF PHYSICS}, \textbf{Indian Institute of Technology Indore}, is an authentic record of my own work carried out during the time period from July 2019 to September 2024 under the supervision of Subhendu Rakshit,  Professor, Department of Physics, IIT Indore.

The matter presented in this thesis has not been submitted by me for the award of any other degree of this or any other institute.

\begin{flushright}
Dhruv Ringe
\end{flushright}

\noindent\hdashrule[0 ex]{\textwidth}{.1mm}{1mm}

This is to certify that the above statement made by the candidate is correct to the best of my
knowledge.

\begin{flushright}
Prof. Subhendu Rakshit
\end{flushright}

\noindent\hdashrule[0 ex]{\textwidth}{.1mm}{1mm}
\textbf{DHRUV RINGE} has successfully given his/her Ph.D. Oral Examination held on \textbf{$4^{\rm{th}}$ April 2025}
\begin{flushright}
Prof. Subhendu Rakshit
\end{flushright}

\noindent\hdashrule[0 ex]{\textwidth}{.1mm}{1mm}

\newpage
\null
\newpage

\begin{center}
{\Huge\textbf{Acknowledgements}}\\
\phantomsection
%\addcontentsline{toc}{section}{Acknowledgements}
\printindex
\end{center}
\vspace{28pt}

\noindent 

I express my sincere gratitude to my supervisor, Prof. Subhendu Rakshit, for his support throughout my Ph.D. journey. I bow down to all my teachers who introduced me to the beautiful world of physics. I am grateful to my PSPC members, Dr. Manavendra Mahato and Dr. Bhargav Vaidya, for their suggestions over the years, and for keeping a keen eye on my progress. I thank Dr. Dipankar Das, Dr. Suman Majumdar, Dr. Amit Shukla, and Dr. Ranveer Singh for their friendly discussions. My seniors, Dr. Siddhartha Karmakar, Dr. Sujata Pandey, and Dr. Najimuddin Khan were just a phone call away whenever I needed guidance. Siddhartha's friendly mentorship and encouragement were especially invaluable as I navigated the inevitable ups and downs of Ph.D. life.

 Thanks to all my friends for our time together during this adventurous journey. Yoshini was my go-to for all the computer-related help. I thank Sanket for the late-night conversations about anything under the sun, and my wife Tanu for the joyful companionship. I will fondly remember Swapnesh, Souparna, Vasundhara, Omkar, Sudipta, Sayan, Jayesh, and `the gang': Manish, Avik, Pratik, Mrityunjay, Anugrah, Bhim, Ayushi, Ketan, Revanth, Abhishek, and Shreya, and all my friends over the years. This journey could not have been possible without the loving support of my family: Aai, Baba, Dhannu, Tanu, and Sheru. Mothe kaka, kaku, and Anna kaka were my local guardians and were a source of much joy.

I am thankful for the
support from DST, via SERB Grants no. MTR/2019/000997 and no. CRG/2019/002354. I also thank the Government of India UGC-JRF fellowship for funding me throughout my research journey. Finally, I thank the Physics department at IIT Indore, for providing the resources and facilities that made this research possible.

{\flushright{Dhruv Ringe}}
    
\newpage
\null
\newpage

\vspace*{5cm}
\begin{center}
    { \Large{Dedicated}}\\
    %\vspace{2mm}
    {\Large{to}} \\
    %\vspace{2mm}
    { \Large\bf{Aai and Baba}}

\end{center}
\newpage
\null
\newpage

\begin{center}
{\Huge\textbf{Abstract}}\\

\phantomsection
\addcontentsline{toc}{chapter}{Abstract}
\printindex
\end{center}
\vspace{28pt}
%abstract here
\noindent The standard model (SM) of particle physics faces several challenges today, including the existence of neutrino masses, the microscopic description of dark matter, non-zero baryon asymmetry, etc. The tests for most beyond SM (BSM) theories are limited by the accessible collider energy, which will increase to about $\lesssim100$ TeV in future colliders like FCC and ILC. To access physics at higher energies, we must look for alternatives to colliders. After the first detection of gravitational waves (GWs) by LIGO in 2015, GW astronomy has emerged as a promising alternative. In this thesis, we study the methods to constrain particle physics models using the stochastic GW imprints from cosmological phase transitions (PTs). Beginning with the theory and background, we describe how the GW background from first-order PTs (FOPTs) and topological defects such as domain walls (DWs) can constrain the model parameter space at upcoming GW observatories. The first BSM scenario involves a flavon FOPT in two ultraviolet-complete models of the Froggatt-Nielsen (FN) mechanism. In both models, for the FN symmetry-breaking scale $v_s = 10^{4-7}\,$GeV, the parameter space is constrained by GWs in upcoming observatories such as BBO, DECIGO, CE, and ET. However, the GW spectrum does not discriminate between the two models. Next, we consider FOPT in the doublet left-right symmetric model (DLRSM) during $SU(2)_R\times U(1)_{B-L}$ breaking. For the breaking scale $v_R=20,\,30,\,50\,$TeV, the parameter space can be constrained by GW observations at BBO, FP-DECIGO, and Ultimate DECIGO. A large number of points with detectable GW signals can be ruled out from the precise measurement of the trilinear Higgs coupling at future colliders. Finally, we discuss the GW spectrum generated by DWs formed after the spontaneous breaking of the discrete $\mathcal{P}$-symmetry imposed on DLRSM. Using Bayesian analysis, we fit the 15-year dataset from NANOGrav to the GW spectrum from DWs in DLRSM and find the best-fit values of the DW surface tension and the bias potential. The techniques of this thesis can be applied to other BSM scenarios in the future. 

\newpage
\null
\newpage

\begin{center}
{\Huge\bf List of Publications}\\

\phantomsection
\addcontentsline{toc}{chapter}{List of Publications}
\printindex
\end{center}
\vspace{28pt}
\begin{flushleft}
A. \underline{Published}:\\
\end{flushleft}

\begin{enumerate}

        \item \textbf{Probing intermediate scale Froggatt-Nielsen models at future gravitational wave observatories}\\ 
        Author: \textbf{Dhruv Ringe}\\
        Published in \emph{Phys.Rev.D} 107 (2023) 1, 015030~
        $\boldsymbol{\cdot}$~ \href{https://doi.org/10.1103/PhysRevD.107.015030}{\twemoji{link}}~DOI~$\boldsymbol{\cdot}$~\href{https://arxiv.org/abs/2208.07778}{\twemoji{link}}~e-Print

        \item \textbf{Gravitational wave imprints of the doublet left-right symmetric model}\\ 
        Authors: Siddhartha Karmakar, \textbf{Dhruv Ringe}\\
        Published in    \emph{Phys.Rev.D} 109 (2024) 7, 075034~
        $\boldsymbol{\cdot}$~ \href{https://doi.org/10.1103/PhysRevD.109.075034}{\twemoji{link}}~DOI~$\boldsymbol{\cdot}$~\href{https://arxiv.org/abs/2309.12023}{\twemoji{link}}~e-Print

        \item \textbf{Domain wall constraints on the doublet left-right symmetric model from pulsar timing array data}\\ 
        Authors: \textbf{Dhruv Ringe}\\
        Published in \emph{Phys.Rev.D}  111 (2025) 1, 015026~
        %https://doi.org/10.1103/PhysRevD.111.015026
        $\boldsymbol{\cdot}$~ \href{https://doi.org/10.1103/PhysRevD.111.015026}{\twemoji{link}}~DOI~$\boldsymbol{\cdot}$~\href{https://arxiv.org/abs/2407.14075}{\twemoji{link}}~e-Print~
\end{enumerate}

\begin{flushleft}
B. \underline{Submitted}:\\
\end{flushleft}

\begin{enumerate}
        \item \textbf{Stochastic gravitational wave background from Froggatt-Nielsen models}\\ 
        Authors: \textbf{Dhruv Ringe}\\
        Submitted to \emph{Bulletin of the Lebedev Physics Institute}~
        
\end{enumerate}

\begin{flushleft}
{\bf N.B}: Entries A1, A2, and A3 are part of my thesis.
\end{flushleft}

\newpage
\null
\newpage
%%%%%%%%%%%%%%%%%%%%%%%%%%%%%%%%%%%%%%%%%%%%%%%%%%%%%%%%%%%

\begingroup  % Keep spacing changes local
\singlespacing  % Alternative to \begin{singlespace}, use if loading setspace package

    {\hypersetup{linkbordercolor=white, linkcolor=black} \tableofcontents}

    \clearpage  % Ensure List of Figures starts on a new page
    \addcontentsline{toc}{chapter}{List of Figures}  % Add to ToC
    {\hypersetup{linkbordercolor=white} \listoffigures}
    \newpage
    \null 
    \clearpage  % Ensure List of Tables starts on a new page
    \addcontentsline{toc}{chapter}{List of Tables}  % Add to ToC
    {\hypersetup{linkbordercolor=white} \listoftables}

\endgroup
%%%%%%%%%%%%%%%%%%%%%%%%%%%%%%%%%%%%%%%%%%%%%%%%%%%%%%%%%%%
%include{ABb/ABb}
\hypersetup{citecolor=red}

% %%%%%%%%%%%%%%%%%%%%%%%%%%%%%%%%%%%%%%%%%%%%%%%%%%%%%%%%%%%
\newpage
\null
\newpage

\chapter{Introduction}
\label{chapter1}
\pagenumbering{arabic}
\markboth{\MakeUppercase{\chaptername\ \thechapter. 
Introduction}}{}

\linespread{0.1}
\graphicspath{{Chapter1/Figures}}
\pagestyle{headings}
\noindent\hrule height 1mm 
\vspace{4mm}
\noindent

The standard model (SM) of particle physics developed by Weinberg, Salam, and Glashow in the 60's, is one of the most successful theories in physics, with a long and eventful history of development, which culminated in the discovery of the Higgs boson at the LHC in 2012\,\citep{ATLAS:2012yve,CMS:2012qbp}. Despite the successes, the SM faces many theoretical and experimental challenges, including the existence of small neutrino masses, the particle nature of dark matter, the origin of a non-zero baryon asymmetry, etc. Many extensions beyond the SM (BSM) that aim to address these limitations are set up at high energy scales inaccessible to present-day colliders. Although the next generation of colliders such as the ILC and the FCC will be able to achieve relatively higher energies, they may take several decades to become operational, and will still be limited to testing BSM theories up to a scale of $\mathcal{O}(10)\,$TeV. 

The detection of gravitational waves (GWs) from a binary black hole merger in 2015\,\cite{Abbott_2016} by LIGO\,\cite{Harry:2010zz} was a historic event that kick-started the era of GW astronomy, and such detections have now become routine\,\citep{Acernese_2014,Abbott_2019,Abbott_2020}. In addition to the foreground signals which are localized in time and space, a time-independent stochastic GW background may also exist in the universe. There are good chances that upcoming GW observatories\,\cite{LISA:2017pwj,Corbin_2006,Musha:2017usi,Abbott_2017,Punturo:2010zz} and pulsar timing arrays (PTAs)\,\citep{NANOGrav:2020bcs,Manchester:2013ndt,Burke-Spolaor:2018bvk,Hobbs:2009yy,Lentati:2015qwp,Manchester:2012za} will be sensitive to such a GW spectrum in frequencies ranging from $\sim 10^{-9}$ Hz to $\sim 10^4$ Hz. The possible sources of a stochastic GW background include the density fluctuations from inflation \citep{Matarrese:1997ay,Mollerach:2003nq,Baumann:2007zm,Kohri:2018awv,DEramo:2019tit,Vagnozzi:2020gtf,Benetti:2021uea,PhysRevLett.122.201101}, topological defects in the vacuum \citep{Vilenkin:1984ib,Caldwell:1991jj,Hindmarsh:1994re,Preskill:1991kd,Gleiser:1998na,Hiramatsu:2013qaa}, and cosmological phase transitions (PTs)\,\citep{Caprini:2015zlo,Caprini:2018mtu,Friedrich:2022cak}. The GW imprint carries some information about the scale of the underlying microscopic physics. This thesis explores the methods to constrain models of particle physics, using GWs from cosmological PTs in the early universe. The upcoming GW observatories will be able to probe the PT scale from $\sim 100\,$MeV to $\sim 10^7\,$GeV, far beyond the reach of colliders. 

It is intriguing how elementary particles, the constituents that describe the functioning of nature at the microscopic level, are so intimately related to cosmology, which describes the universe at the largest length scales imaginable.
%As can be guessed, the study of cosmological phase transitions involves elements from diverse and seemingly unrelated topics, such as cosmology, particle physics, finite temperature field theory, and gravitational waves. 
This chapter discusses the basic theoretical background for the thesis. For each topic, I sketch the steps to arrive at important cornerstones by connecting the dots in between. The first section discusses basic concepts in cosmology. The next section is a brief review of the SM, its limitations, and its extensions relevant to this thesis. The next section describes how the finite temperature effective potential is used to study phase transitions. Next, I discuss the physics of gravitational waves, including their production and detection. Finally, I present a road map of the thesis.

%\vspace{4mm}
%\noindent\hrule height 0.5mm 
%\vspace{5mm}

\section{Cosmology}\label{chp1sec1}
\vspace{4mm}
Cosmology deals with length scales much larger than the typical size of galaxy clusters, where the universe appears to be homogeneous and isotropic to a very good approximation. This fact is known as the \textit{cosmological principle}. The universe is also expanding, with observations suggesting that in the past, the universe was much smaller and a lot hotter than it is today. In this section, I discuss some basic aspects of cosmology that are needed to understand cosmological phase transitions and the propagation of gravitational waves in an expanding universe. There are many books and reviews on the topic\,\cite{Baumann:2022mni,Kolb:1988aj, Breitbach:2018kma}.
\subsection{The FLRW metric}
The Friedmann-Lema\^{i}tre-Robertson-Walker (FLRW) metric, incorporates the cosmological principle and describes the expanding universe, in the language of general relativity. In 4d spacetime with coordinates $x^{\mu} = (t,{\bf{x}})$, the invariant interval for the FLRW metric is given by
\beq \label{eq: FLRW_k}
ds^2 = dt^2  - a(t)^2\left(\delta_{ij} + k\frac{x_ix_j}{1-k|{\bf{x}}|^2}\right)dx^idx^j\, ,
\eeq 
where $k$ is the curvature, which takes the value $0,+1,-1$, for flat, positive, and negative spatial curvatures respectively. From large-scale cosmological observations, the universe appears to be extremely flat, and we can set $k=0$, so that Eq.\,\eqref{eq: FLRW_k} becomes
\beq \label{eq: FLRW}
ds^2 = dt^2  - a(t)^2\delta_{ij}dx^idx^j\, .
\eeq 
The spatial coordinates $x_i$ are called comoving coordinates, $t$ is the cosmic time, and $a(t)$ is called the scale factor. The comoving coordinates $\bf{x}$, are related to physical coordinates ${\bf{x}}_{\rm{phys}}$ by, ${\bf{x}}_{\rm{phys}} = a(t) {\bf{x}}$, so that the physical velocity of an object is given by
\beq 
{\bf{v}}_{\rm{phys}} = \dot{\bf{x}}_{\rm{phys}}  = H{\bf{x}}_{\rm{phys}} + {\bf{v}}_{\rm{pec}}\,,
\eeq 
where a dot over a variable denotes the derivative with respect to time, $H\equiv \dot{a}/a$ is the \textit{Hubble parameter}, and  ${\bf{v}}_{\rm{pec}}$ is the \textit{peculiar velocity} with respect to a comoving observer.  

Due to expansion, a light signal emitted at time $t$ redshifts as it travels through space and reaches us at present time $t_0$. The fractional change in wavelength is 
\beq 
\frac{\delta\l}{\l}\equiv 1 + z = \frac{a(t_0)}{a(t)}\,,
\eeq 
where $z$ is called the \textit{redshift}, and is used as a measure of distance between the source and the Earth. Also, it is sometimes useful to work with the \textit{conformal time} $d\eta = dt / a(t)$, for which the FLRW invariant interval is
\beq 
\label{eq: FLRW_conf}
ds^2 = a^2(\eta) \, \left[ d\eta^2 - \delta_{ij}\, dx^i dx^j \right] \, .
\eeq 
The Einstein field equations dictate the evolution of the curvature of spacetime with the distribution of matter/energy, 
\beq \label{eq: einstein_equations}
G_{\mu\nu}(x) = 8\pi G T_{\mu\nu}(x)\, ,
\eeq 
where $G_{\mu\nu}$ is the Einstein tensor which takes into account the curvature of spacetime, and $T_{\mu\nu}$ is the energy-momentum tensor. The constant $G \approx 6.7\times 10^{-11}\,\rm{m^3/kg/s^2}$, is the Newton's gravitational constant. Sometimes another quantity called the reduced Planck mass, $M_p$, is used, which in natural units is given by $M_p = (8\pi G)^{-1/2} \approx 2.4 \times 10^{18}\,$GeV. 

For the FLRW metric, $T_{\mu\nu}$ is taken as the energy-momentum tensor for a perfect fluid, which is completely characterized by the energy density $\rho$ and pressure $P$ of the fluid
\beq 
T_{\mu\nu} = (\rho+P)\dot{x}_{\mu}\dot{x}_{\nu} - P g_{\mu\nu} \, ,
\eeq 
where $g_{\mu\nu}$ is the FLRW metric, and a dot over a variable denotes the derivative with respect to time. In the rest frame of the fluid, i.e., for a comoving observer, $\dot{x}_{\mu} = (1,0,0,0)$, so that
\beq \label{eq: Tmn FLRW}
T_{\mu\nu} = {\rm{diag}}(\rho,P,P,P) \, .
\eeq 
The energy-momentum tensor satisfies the continuity equation, $\nabla_{\mu}T_{\mu\nu} = 0$, where $\nabla_{\mu}$ is the covariant derivative. For the FLRW metric, the continuity equation gives
\beq \label{eq: cont}
\dot{\rho} + 3H(\rho+P) = 0\, ,
\eeq 
where $H=\dot{a}/a$, is called the Hubble parameter. Given a fluid equation of state of the form, $P = w\rho$, using Eq.\,\eqref{eq: cont}, we can determine how $a(t)$ evolves with time,   
\begin{itemize}
\setlength\itemsep{.1em}
\item Matter refers to a fluid of non-relativistic particles, which behave collectively as pressure-less dust, $w=0$, i.e. $P=0$. It includes ordinary matter, called `baryonic matter' comprising of nuclei and electrons, and `dark matter', which is only observed by its gravitational effects. Putting $w=0$ yields, 
$\rho(t)\propto 1/a(t)^3$. 
\item Radiation refers to a fluid of relativistic constituents, including a gas of photons, or any species of particles moving at relativistic speeds. For radiation, $w=1/3$, i.e. $P=\rho/3$, so we get, $\rho(t)\propto 1/a(t)^4$.
\item For a universe filled with vacuum energy, with negative pressure, $w=-1$. In this case, $\rho = \rm{const.}$\,. The vacuum energy is also referred to as `dark energy', and is related to the cosmological constant, $\Lambda$.
\end{itemize}

\begin{figure}[tbp]
\centering % \begin{center}/\end{center} takes some additional vertical space
\includegraphics[width=.75\textwidth]{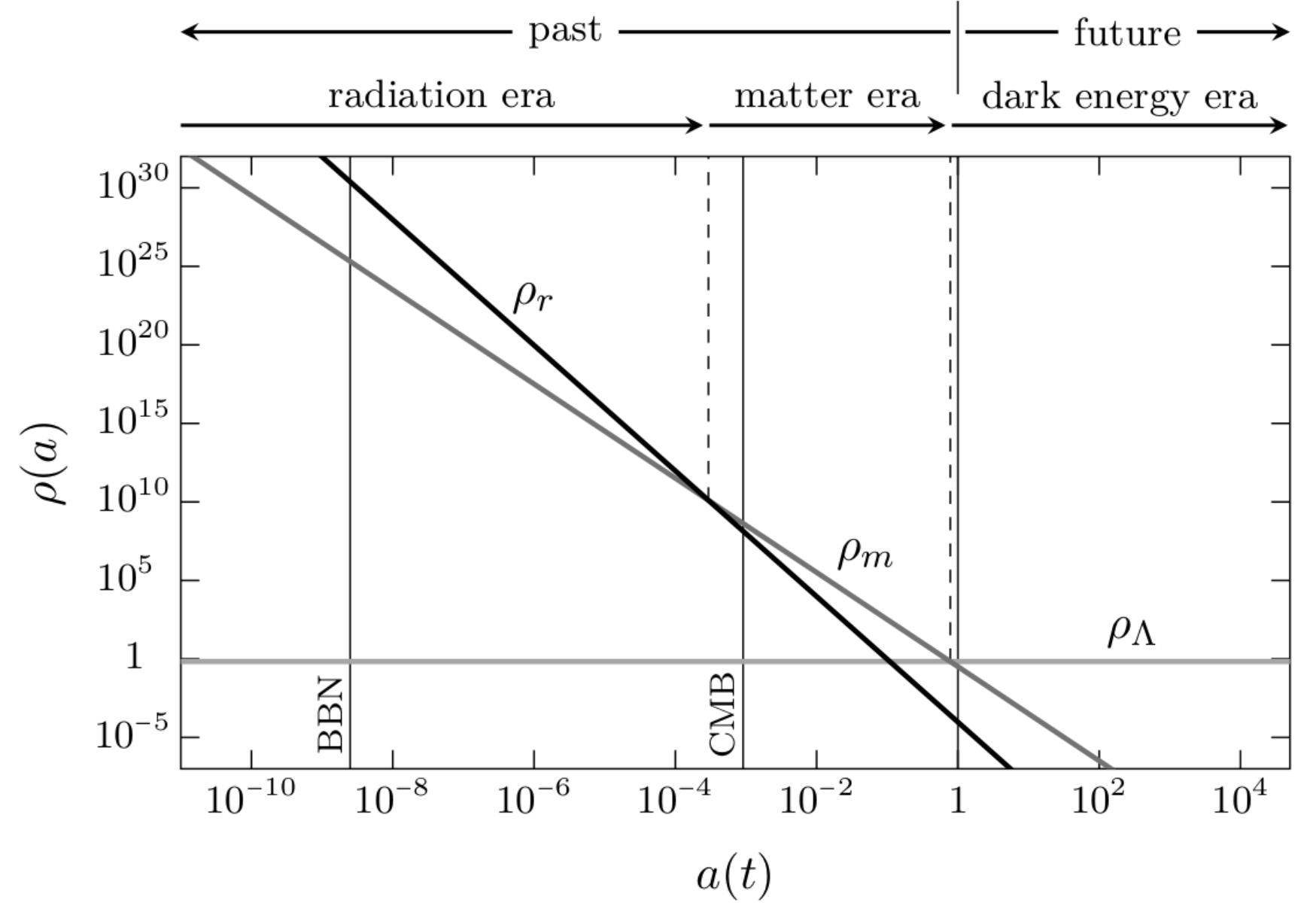}
\hfill
\caption{\label{fig: energy_density_eras} Energy density of the universe in different eras\,\cite{Baumann:2022mni}. On the $y$-axis, the numbers are taken as a fraction of the cosmological constant.}
\end{figure}

In the very early universe, all particles of the SM were relativistic, and the universe was radiation-dominated. As the universe cooled with expansion, some particle species became non-relativistic when the temperature became less than their respective masses and started behaving like matter. At present, only massless particles like photons contribute to the radiation. Neutrinos were relativistic for most of the history of the universe, but only recently became non-relativistic. Observations show that the expansion of the universe began accelerating in relatively recent history, signaling the beginning of the vacuum-dominated era. The evolution of the energy density of various components is shown in Fig.\,\ref{fig: energy_density_eras}.

Inserting the FLRW metric and $T_{\mu\nu}$ given in Eq.\,\eqref{eq: Tmn FLRW}, into the Einstein equations of Eq.\,\eqref{eq: einstein_equations}, we get the so-called \textit{Friedmann equations}, 
\begin{subequations}
    \beq \label{eq: friedmann1}
    H^2 = \left(\frac{\dot{a}}{a}\right)^2= \frac{8\pi G}{3}\rho\, ,
    \eeq 
    \beq \label{eq: friedmann2}
    \frac{\ddot{a}}{a} = -\frac{4\pi G}{3}(\rho+3P)\, ,
    \eeq 
\end{subequations}
where we have set the curvature $k=0$ for a flat universe in the first equation. Evaluating the first Friedmann equation for the present-day universe, we can define the \textit{critical energy density} of the universe
\beq \label{eq: critical density}
\rho_c = \frac{3 H_0^2}{8\pi G}\, ,
\eeq 
where, $H_0 = 100h~ {\rm{km/s/Mpc}}$, is the Hubble constant today, with the current value of $h=0.675\pm0.005$\footnote{There is a discrepancy in the measurement of $H_0$ from supernova data, and CMB observations, called the \textit{Hubble tension}. In this thesis, I ignore this discrepancy.}\,\cite{Planck:2018vyg}. The numerical value of the critical density is $\rho_c\approx 1.1h^2\times 10^{-5}$ protons$/\rm cm^3$. We can express the different forms of energy density measured today, as a fraction of the critical energy density
\beq 
\Omega_i = \frac{\rho_i}{\rho_c}\, ,
\eeq 
where $i=\{r,m,\Lambda, \cdots\}$ stands for radiation, matter, vacuum, etc.\,.
The present-day values indicated by the subscript `0' for $\Omega_i$, according to the $\Lambda$ Cold Dark Matter ($\Lambda$CDM) model are 
\bea
&\Omega_{r,0} \approx 10^{-4}\, ,& \qquad ({\rm{radiation}})\, , \nn\\
&\Omega_{m,0} \approx 0.05\, , & \qquad ({\rm{baryonic~matter}})\, , \nn\\
&\Omega_{dm,0} \approx 0.27\, , & \qquad ({\rm{dark~matter}})\, , \nn\\
&\Omega_{\Lambda,0} \approx 0.68\, , & \qquad ({\rm{dark~energy}})\, .\nn
\eea

The second Friedmann equation tells us how the scale factor evolves with time for the different equations of state $P= w\rho$. The result is,
\begin{subequations}
    \beq 
    \rho \propto a^{-3(1+w)}\, ,
    \eeq
    \beq 
    a(t) \propto t^{2/[3(1+w)]}\, ,\qquad {\rm{for}}~ w \neq -1\, .
    \eeq
\end{subequations}
In the radiation dominated era, $w=1/3$, so $a(t)\propto t^{1/2}$, while in the matter dominated era, $w=0$, so $a(t)\propto t^{2/3}$. In the vacuum-dominated era, the scale factor increases exponentially as, $a(t) \propto \exp(H_0\sqrt{\Omega_{\Lambda}} t)$. 

\subsection{Equilibrium thermodynamics}
So long as the plasma in the universe is in thermal equilibrium, we can assign a single temperature $T$ to the entire universe at a given time $t$. In equilibrium, the phase space distribution of particles in the plasma is given by the functions,
\beq 
f({\bf{p}}) = \frac{1}{e^{\b(E({\bf{p}})-\mu)}\mp 1}\,
\eeq 
where $-$ sign is for bosons, and $+$ sign is for fermions, and $\b = 1/T$. The chemical potential, $\mu$ is not important at early times so we neglect it. The energy of a particle is $E({\bf{p}}) = \sqrt{m^2+{\bf{p}}^2}$, where $m$ is the mass and ${\bf{p}}$ is the three-momentum. 

By integrating $f({\bf{p}})$ over the entire phase space, we can find the number density $n$, and the energy density $\rho$, for a particle species with $g$ internal degrees of freedom.
\begin{subequations}
\beq 
n(T) = \frac{g}{(2\pi)^3}\int d^3p ~f({\bf{p}}) = \frac{g}{2\pi^2}\int_{0}^{\infty} dp \frac{p^2}{e^{\b\sqrt{m^2+p^2}}\mp 1}\, ,
\eeq 
\beq
\rho(T) = \frac{g}{(2\pi)^3}\int d^3p ~f({\bf{p}})E({\bf{p}}) = \frac{g}{2\pi^2}\int_{0}^{\infty} dp \frac{p^2\sqrt{m^2+p^2}}{e^{\b\sqrt{m^2+p^2}}\mp 1}\, .
\eeq
\end{subequations}
While in general, these integrals need to be evaluated numerically, they have simple expressions in two limiting cases. In the relativistic limit where $T\gg m$, we get
\begin{subequations}
    \beq 
    n(T) \approx \frac{6}{5\pi^2}g T^3 \times\begin{cases} 
       1\,, & {\rm{bosons}} \\
       \frac{3}{4}\,, & {\rm{fermions}} 
   \end{cases}\, ,
    \eeq
    \beq 
    \rho(T) \approx \frac{\pi^2}{30}g T^4\times\begin{cases} 
       1\,, & {\rm{bosons}} \\
       \frac{7}{8}\,, & {\rm{fermions}}
   \end{cases}\, .
    \eeq 
\end{subequations}
Similarly, in the non-relativistic limit where $T\ll m$, we get
\begin{subequations}
    \beq 
    n(T) \approx g\left(\frac{mT}{2\pi}\right)^{3/2} e^{-m/T} \, ,
    \eeq
    \beq 
    \rho(T) \approx m\cdot n(T)\, .
    \eeq 
\end{subequations}
The factor $e^{-m/T}$ is called the Boltzmann suppression factor, which enables us to ignore the contribution of non-relativistic particles to the energy density to a very good approximation. 

In the relativistic case, the expressions for $n(T)$ and $\rho(T)$ are independent of $m$, so the contribution of each particle species can be added to get the corresponding total value
\beq \label{eq: rho_rad}
\rho_r = \frac{\pi^2}{30} T^4 g_*(T)\, , \qquad g_*(T) = \sum_b g_b(T) + \frac{7}{8}\sum_f g_f(T)\, ,
\eeq 
where the subscript $r$ stands for radiation, $b$ and $f$ stand for bosons and fermions respectively, and $g_*(T)$ denotes the effective number of relativistic degrees of freedom at temperature $T$. When all particles are in equilibrium and share the same temperature, finding $g_*(T)$ is an exercise in counting the degrees of freedom. For example, above the electroweak phase transition, for SM, the value is, $g_*(T) = 106.75$, which decreases at lower temperatures.

It is possible that the $i^{\rm{th}}$ species of particles has its own temperature $T_i\gg m_i$, then at temperature $T$ of the universe (i.e. photon temperature), 
\beq 
g_*(T) = \sum_{i\in b} g_i\left(\frac{T_i}{T}\right)^{4} + \frac{7}{8}\sum_{i\in f} g_i\left(\frac{T_i}{T}\right)^{4}\, .
\eeq 
For radiation, we can find the temperature dependence of the Hubble parameter using Eq.\,\eqref{eq: friedmann1}
\beq 
H = \frac{1}{M_p}\sqrt{\frac{\rho}{3}} = \frac{\pi T^2}{3M_p}\sqrt{\frac{g_*(T)}{10}}\, .
\eeq 

Using the first law of thermodynamics, $TdS = dU + PdV$, where $U$, $V$ and $S$ denote the internal energy, volume, and entropy of the system respectively, we can show that for relativistic species, the entropy density $s\equiv S/V$, is given by
\beq 
s = \frac{2\pi^2}{45}g_{*s}(T) T^3\, ,
\eeq 
where $g_{*s}(T)$ denotes the degrees of freedom which contribute to the entropy. For $T_i\gg m_i$, 
\beq 
g_{*s}(T) = \sum_{i\in b} g_i\left(\frac{T_i}{T}\right)^{3} + \frac{7}{8}\sum_{i\in f} g_i\left(\frac{T_i}{T}\right)^{3}\, .
\eeq 

If the universe expands adiabatically, the total entropy is conserved, i.e. $sa^3 = \rm{const}.\,$.

\subsection{Brief timeline of the universe}
According to the Big Bang model, the estimated age of the universe is $13.8$ Billion years. Close to the Big Bang, the universe was so hot that it was filled with a relativistic plasma composed of all kinds of particles. Just after the Big Bang, the universe expanded rapidly, while its temperature decreased. Whenever the temperature dropped below the mass of a particle species, that species \textit{decoupled}, i.e. went out of thermal equilibrium from the plasma, resulting in a change in the cosmological evolution. Table\,\ref{table: timeline} summarizes the timeline of the universe from the earliest moment to the epoch of photon decoupling. Among the different epochs, Big Bang nucleosynthesis (BBN), and the cosmic microwave background (CMB) have a great significance since they put stringent cosmological constraints, and restrict the possible modifications to the $\Lambda$CDM model.  

\begin{table}[tbp]
\begin{center}
{\setlength\tablinesep{3pt}
\setlength\tabcolsep{7pt}
\begin{tabular}{|c|c|c|c|}
\hline
Event & Time & Redshift & Temperature\\
\hline
Inflation & ? & - & -\\
Baryogenesis & ? & ? & ?\\
Dark-matter freeze-out & ? & ? & ?\\
EW phase transition & $20$\,ps & $10^{15}$ & $100$\,GeV\\
QCD phase transition & $20$\,$\mu$s & $10^{12}$ & $150$\,MeV\\
Neutrino decoupling & $1$\,s & $6\times 10^{9}$ & $1$\,MeV\\
Electron-positron annihilation & $6$\,s & $2\times 10^{9}$ & $500$\,keV\\
Big Bang nucleosynthesis & $3$\,min & $4\times 10^{8}$ & $100$\,keV\\
Radiation-matter equality & $50$\,kyr & $3400$ & $0.80$\,eV\\
Recombination & $290-370$\,kyr & $1090-1270$ & $0.25-0.29$\,eV\\
Photon decoupling & $370$\,kyr & $1090$ & $0.25$\,eV\\
\hline
\end{tabular}}
\end{center}
\caption{ Key epochs in the timeline of the universe. \label{table: timeline}}
\end{table}

Before the epoch of recombination, the universe was opaque to light, as photons were constantly scattered in the plasma. When the universe cooled sufficiently to allow the formation of neutral hydrogen, the photons could travel without being scattered. Redshifted to today, these photons are the source of the CMB. The CMB is therefore the earliest epoch that can be accessed by electromagnetic radiation. The CMB spectrum is the most precisely measured black body spectrum in nature, with a temperature of $\approx2.7$ K with tiny anisotropy in the temperature across the sky, $\delta T/T\sim 10^{-4}$. A series of satellite experiments have measured the anisotropy over the years, with the latest measurements made by the Planck satellite. The CMB observations put stringent constraints on cosmological models. 

BBN is one of the earliest epochs verified by astrophysical observations. The theory of BBN explains the formation of light nuclei such as deuterium, helium, and lithium in the early universe in terms of the baryon to photon ratio, $\eta\equiv n_B/n_{\gamma}$. It is in great agreement with the observed primordial abundances of these elements, as inferred from distant objects such as certain dwarf galaxies and quasars, where these elements could not have been formed via stellar nucleosynthesis. 

Currently, there are no direct probes to observe the history of the universe before BBN, and thus the early epochs in the universe model-dependent, as indicated by the question marks in Table\,\ref{table: timeline}. In addition to this list of events, there could be a series of phase transitions at high-temperatures, depending on the particle physics model. Since GWs are not screened unlike the CMB, the GW imprints of these phase transitions may be accessed by upcoming GW detectors. 

\section{The standard model: an overview}\label{chp1sec2}
\vspace{4mm}

The SM encapsulates our current understanding of all known elementary particles and their interactions. The elementary particles are called so because they appear to be point particles with no internal structure at the length scales currently accessible to colliders. The current list of elementary particles includes 
\begin{itemize}
\setlength\itemsep{.1em}
    \item Fermions: the quarks, namely up, down, charm, strange, top and bottom \,$\{u,d,c,s,t,b\}$ and the leptons, namely electron, muon, and tauon, $\{e,\mu,\tau\}$, and three corresponding neutrinos, $\{\nu_e,\nu_{\mu},\nu_{\tau}\}$.
    \item Gauge bosons; massive $W^{\pm}$ and $Z^0$ bosons, the massless photon $\gamma^0$, and gluons.
    \item A neutral scalar Higgs boson, $h$.
\end{itemize}  
The standard model is a mathematical description of these elementary particles, in the framework of quantum field theory. The gauge symmetry group of SM is, $$\mathcal{G_{\rm{SM}}} \equiv SU(3)_c\times SU(2)_L\times U(1)_{Y}\, ,$$ where the subscript $c$ stands for `color', $L$ stands for left-handed, and $Y$ stands for hypercharge. For the gauge groups $SU(3)_c$, $SU(2)_L$, and $U (1)_Y$, the corresponding vector fields are given by $V_{\mu} = G_{\mu}^{a=1,\cdots,8} , W_{\mu}^{a=1,2,3} , B_{\mu}$ respectively. The group generators in the fundamental representation are denoted by $t^a_V$, and the gauge couplings are denoted by $g_V$. For $SU(3)_c$, $g_V = g_s$, is the strong coupling constant, and $t^a_V = \frac{\l^a}{2}$, where $\l^a$ are the eight Gell-Mann matrices. For $SU(2)_L$, $g_V = g$, is the weak coupling constant, and $t^a_V = \frac{\tau^a}{2}$, where $\tau^a$ are the three Pauli matrices. For $U(1)_Y$, $g_V = g'$, and $t^a_V = \frac{\mathbb{I}}{2}$, proportional to the identity matrix.

\begin{table}[tbp]
\begin{center}
{\setlength\tablinesep{3pt}
\setlength\tabcolsep{7pt}
\begin{tabular}{|c|c|c|c|c|c|}
\hline
Field &~~~${ SU(3)_C}$ &~~ ${ SU(2)_L}$ ~& $T^3$ & $U(1)_Y$ & $Q=T^3+\frac{Y}{2}$ \\
\hline
$Q_L=\begin{pmatrix}
u_L \\[2pt]  d_L
\end{pmatrix}$ & {\bf 3} & {\bf 2} & $\begin{pmatrix}
 \frac{1}{2} \\[2pt] -\frac{1}{2} 
\end{pmatrix}$ & $\frac{1}{3}$ & $\begin{pmatrix}
\frac{2}{3} \\[2pt] -\frac{1}{3} 
\end{pmatrix}$ \\
$u_R$ & {\bf 3} & {\bf 1} & 0 & $\frac{4}{3}$ & $\frac{2}{3}$ \\
$d_R$ & {\bf 3} & {\bf 1}& 0 & $-\frac{2}{3}$ & $-\frac{1}{3}$ \\
$l_L=\begin{pmatrix}
\nu_L \\[2pt]  e_L
\end{pmatrix}$ & {\bf 1} & {\bf 2}& $\begin{pmatrix}
 \frac{1}{2} \\[2pt] -\frac{1}{2}
\end{pmatrix}$ & $-1$ & $\begin{pmatrix}
 0\\[2pt]-1  
\end{pmatrix}$ \\
$l_R$ & {\bf 1} & {\bf 1}& 0 & $-2$ & $-1$ \\ 
\hline
$H=\begin{pmatrix}
\phi^+ \\  \phi^0
\end{pmatrix}$  & {\bf 1} & {\bf 2} & $\begin{pmatrix}
\frac{1}{2} \\[2pt]  -\frac{1}{2}
\end{pmatrix}$ &  $1$ & $\begin{pmatrix}
1 \\[2pt] 0  
\end{pmatrix}$ \\
\hline
\end{tabular}}
\end{center}
\caption{ Representations of the SM fields under $\gsm$. $T^3$ is the third component of weak isospin of ${ SU(2)_L}$ group, $Y$ is the hypercharge quantum number, and $Q$ is the electric charge quantum number. \label{table: SMcharges}}
\end{table}

The matter fields are spin $1/2$ fermions that include leptons, which participate in electroweak interactions, and quarks, which participate in electroweak and strong interactions. Quarks are triplets of $SU(3)_c$ while leptons are $SU(3)_c$ singlets. We can split a four-component Dirac fermion, $\psi$, into its left-handed and right-handed components
\beq
\psi_L= \frac{1-\gamma_5}{2}\,\psi \qquad {\rm and} \qquad \psi_R= \frac{1+\gamma_5}{2}\,\psi\nn\, ,
\eeq
where $\gamma_5$ is the chirality matrix. The $SU(2)_L$ gauge group only acts on left-chiral fermions. The left and right-handed fermions have different representations of $SU(2)_L$, so SM is a \textit{chiral} theory. The left-handed quarks and leptons are $SU(2)_L$ doublets, and the right-handed components are $SU(2)_L$ singlets 
\beq
Q_L^i \equiv \begin{pmatrix}
u^i_{L} \\  d^i_{L}
\end{pmatrix}\,
,~
l^i_L \equiv \begin{pmatrix}
\nu^i_{L} \\  e^i_{L}
\end{pmatrix}\,,~u^i_R,~d^i_R, e^i_R
\label{lepquarlL}
\eeq
where $i=\{1,2,3\}$ represents the family index, $u$ represents the `up-type' quarks of the three generations $u,c,t$; and $d$ stands for the `down-type' quarks $d,s,b$. Similarly, $e$ denotes the three charged lepton generations, $e, \mu, \tau$, and $\nu$ are the corresponding neutrinos. There are no right-handed neutrinos in the SM. Thus, each generation\footnote{The different generations are sometimes also called flavors.} of fermions comes in five representations: $Q_L$, $l_L$, $u_R$, $d_R$, and $e_R$. 

The Higgs field, $H$, is an $SU(2)_L$ doublet 
\beq 
H = \begin{pmatrix}
\phi^+ \\  \phi^0
\end{pmatrix} ~=~ \frac{1}{\sqrt{2}}\begin{pmatrix}
 \rho_1+i\,\eta_1 \\
 \rho_2+i\,\eta_2
 \end{pmatrix}\nn\, .
\eeq
where $\phi^+$ is a charged complex scalar, and $\phi^0$ is a neutral complex scalar. The physical Higgs field, $h$, is obtained after spontaneous symmetry breaking, as discussed later. 

\subsection{The SM Lagrangian}
Given the field content and the symmetry group $\mathcal{G}_{SM}$, we can write down the SM lagrangian by listing all Lorentz-invariant, gauge-invariant, and renormalizable operators, with mass dimension $\leq 4$. The SM lagrangian can be written as a sum of four contributions
\beq\label{eq: L_SM}
\mathcal{L}_{\rm{SM}} = \mathcal{L}_{\rm{fermions}} + \mathcal{L}_{\rm{gauge}} + \mathcal{L}_{\rm{Higgs}} + \mathcal{L}_{\rm{Yukawa}}\, .
\eeq
The first piece contains the kinetic terms of the matter fields 
\bea
\mathcal{L}_{\rm{fermions}} &=& i\, \overline{Q}_L~\gamma^{\mu} D^L_\mu~ Q_L+i\, \overline{l}_L~\gamma^{\mu} D^L_\mu ~l_L \nn\\
&+&  i\,\overline{u}_R~\gamma^{\mu} D^R_\mu~ u_R + i\,\overline{d}_R~\gamma^{\mu} D^R_\mu~ d_R + i \,\overline{e}_R~\gamma^{\mu}D^R_\mu ~e_R \, .\label{eq: fermionew}
\eea
The covariant derivatives are of the form
\beq 
D_{\mu} = \partial_{\mu} - i\sum_V g_V t_V^a V_{\mu}^a \equiv \partial_{\mu} - ig_s\frac{\l^a}{2}G_{\mu}^a - ig\frac{\tau^a}{2}W_{\mu}^a - i{g'}\frac{Y}{2} B_{\mu}\, .
\eeq 
They act differently on different fields, depending upon the quantum numbers.
\bea  \label{eq: cov_der}
D_{\mu} Q_L &=&\left( \partial_{\mu} - ig_s\frac{\l^a}{2}G_{\mu}^a - ig\frac{\tau^a}{2}W_{\mu}^a - \frac{i}{2}g'\left(\frac{1}{3}\right)B_{\mu} \right)Q_L\, ,\nn\\
D_{\mu} l_L &=&\left( \partial_{\mu}  - ig\frac{\tau^a}{2}W_{\mu}^a - \frac{i}{2}g'\left(-1\right)B_{\mu} \right)l_L\, ,\nn\\
D_{\mu} u_R &=& \left( \partial_{\mu} - ig_s\frac{\l^a}{2}G_{\mu}^a - \frac{i}{2}g'\left(\frac{4}{3}\right)B_{\mu} \right)u_R\, ,\nn\\
D_{\mu} d_R &=& \left( \partial_{\mu} - ig_s\frac{\l^a}{2}G_{\mu}^a - \frac{i}{2}g'\left(-\frac{2}{3}\right)B_{\mu} \right)d_R\, ,\nn\\
D_{\mu} e_R &=&\left( \partial_{\mu}  - \frac{i}{2}g'\left(-2\right)B_{\mu} \right)e_R\, ,\nn\\
D_{\mu} H &=&\left( \partial_{\mu}  - ig\frac{\tau^a}{2}W_{\mu}^a - \frac{i}{2}g'\left(+1\right)B_{\mu} \right)H\, ,
\eea

The second piece in Eq.\,\eqref{eq: L_SM} is the gauge part of the SM lagrangian,
\begingroup
\allowdisplaybreaks
\beq
\mathcal{L}_{\rm{gauge}}= -\frac{1}{4} {G}_{\mu\nu}^a { G}^{a,\mu\nu}-\frac{1}{4} {W}_{\mu\nu}^i { W}^{i,\mu\nu}- \frac{1}{4} B_{\mu\nu} B^{\mu\nu}\,.\nn
\eeq
\endgroup\par
The field strength tensors are defined as, 
\begingroup
\allowdisplaybreaks
\bea
G_{\mu\nu}^a&=&\partial_\mu G_\nu^a - \partial_\nu G_\mu^a + g_s f^{abc} G_{\mu}^bG_{\nu}^c \, ,\nn\\
W_{\mu\nu}^i&=&\partial_\mu W_\nu^i - \partial_\nu W_\mu^i + g \epsilon^{ijk} W_{\mu}^jW_{\nu}^k \, ,\nn\\
B_{\mu\nu}&=&\partial_\mu B_\nu - \partial_\nu B_\mu \, ,\nn
\eea
\endgroup\par
where $f^{abc}$ and $\epsilon^{ijk}$ are the structure constants of $SU(3)_c$ and $SU(2)_L$ respectively.

The third piece of Eq.\,\eqref{eq: L_SM} involves the scalar Higgs doublet
\begingroup
\allowdisplaybreaks \bea
\mathcal{L}_{\rm{Higgs}} &=& (D^{\mu} H)^\dagger (D_{\mu} H) - V(H)\,.
\label{eq: LagHiggs}
\eea
\endgroup\par
%%%%%
where $V(H)$ is the SM Higgs potential and is given by,
\begingroup
\allowdisplaybreaks \bea
 V(H) &=& \m_H^2 H^\dagger H + \lambda (H^\dagger H)^2 \,,\label{SMSSBpot}
\eea
\endgroup\par
where $\m_H^2$ is a mass parameter, and $\l$ is the self-coupling of the Higgs field. 

The last piece of Eq\,\eqref{eq: L_SM} is the Yukawa lagrangian,
%%%%%
\begingroup
\bea
%%%%%
\mathcal{L}_{\rm{Yukawa}} &=&- Y^u_{ij}\overline{Q}^i_{L} \widetilde{H}~u^j_{R} -  Y^d_{ij}\overline{Q}^i_{L} H~ d^{~j}_{R} - Y^l_{ij}\overline{l}^i_{L} H~ e^{~j}_{R} + \rm{h.c.}
\label{eq: LagYuk}
\eea
\endgroup\par
where $\widetilde{H}=i \tau_2 H^*$, is the dual of the Higgs field $H$ and $H^*$ is the complex conjugation of $H$. The matrices $Y^d$, $Y^u$, $Y^e$  are the Yukawa couplings for the down-type quarks $(d,s,b)$, the up-type quarks $(u,c,t)$, and the charged leptons $(e,\mu,\tau)$ respectively. The indices $i,j$ are the generation indices, with $i,j\in\{1,2,3\}$.

Theoretically, a scalar doublet field is needed in the SM because we cannot write explicit mass terms for the gauge bosons of the form $M^2 A^{\mu}A_{\mu}$ in the lagrangian
since such terms are not gauge invariant. Similarly, explicit mass terms cannot be written for the quarks and leptons, since they are chiral under $SU(2)_L$. This prohibits the usual Dirac mass terms in the lagrangian of the form, 
$M\,\overline{\psi}\psi = M\,\overline{\psi}_L\psi_R + M\,\overline{\psi}_R\psi_L\, $. In the next section, I discuss how the gauge boson and fermion masses are dynamically generated by assigning a non-zero vacuum expectation value to the Higgs field. 

\subsection{Scalar sector: Spontaneous symmetry breaking}
\vspace{2 mm}
Spontaneous symmetry breaking (SSB) occurs when a theory is invariant under some symmetry, while its ground state (vacuum) does not exhibit this symmetry. This is a well-known phenomenon in statistical physics and also occurs in quantum field theory. 

The mechanism of SSB works with any symmetry group $G$ of the Lagrangian, which spontaneously breaks into its subgroup $H$ when a set of scalar fields $\{\phi^a\}$ acquires a non-zero \vev. According to the Goldstone theorem, if a continuous group $G$ is spontaneously broken, then massless scalars (called \textit{Goldstone bosons}) necessarily appear in the spectrum of possible excitations. The number of Goldstone bosons equals the number of broken symmetry generators. Moreover, if $G$ is a gauge symmetry, then these massless Goldstones are `eaten' by the gauge bosons, and an equal number of gauge bosons become massive. These would-be Goldstone bosons provide a non-zero longitudinal component to the polarization of the massive gauge bosons. For fermions, the mass terms can be generated by Yukawa interactions with the scalars responsible for SSB.

The generation of gauge boson and fermion masses via SSB in a toy example with $U(1)$ symmetry is discussed in Appendix\,\ref{app: ssb_toy}. In the case of SM, the mechanism is called the Brout-Englert-Higgs mechanism, or simply the Higgs mechanism, and is discussed below.

\subsubsection*{The Brout-Englert-Higgs mechanism}
The SSB mechanism in the SM was first proposed by Robert Brout,  Fran\c{c}ois Englert, and Peter Higgs. To give mass to the three massive gauge bosons of the SM, namely $W^{\pm}$ and $Z^0$, we need three broken generators, and an additional generator to account for the massless photon.  The symmetry-breaking pattern is:
$$SU(2)_L\times U(1)_Y\rightarrow U(1)_{\rm{em}}\, .$$

This is called electroweak symmetry breaking (EWSB). Introducing an $SU(2)_L$ doublet scalar field $H$ is the minimal way to achieve the desired breaking pattern. Recall that the Higgs potential is
\begingroup
\allowdisplaybreaks \bea
 V(H) &=& \mu_H^2 H^\dagger H + \lambda (H^\dagger H)^2 \,.
\eea
\endgroup\par
\begin{figure}[tbp]
\centering 
\includegraphics[width=.6\textwidth]{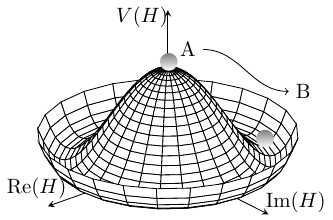}
\caption{\label{fig: higgs_potential} Schematic diagram illustrating the Higgs potential in two dimensions. Credit: tikz.net.}
\end{figure}

When $\mu_H^2$ is positive, the potential has only one global minimum at
$H = 0$. However, if $\mu_H^2$ becomes negative, the potential has non-trivial minima at $|H_0|\neq 0$, found by minimizing the potential:
\beq
\left .\frac{\partial V}{\partial H^{\dagger}}\right\vert_{H=H_0} = H_0(\mu_H^2 + 2\l|H_0|^2) = 0\,\implies |H_0|^2 = \sqrt{\frac{-\mu_H^2}{2\l}}\, .
\eeq
Fig.\,\ref{fig: higgs_potential} displays the shape of the Higgs potential in two dimensions. 

By convention, we define the Higgs \vev as $v_{\rm{EW}} = \sqrt{2}|H_0|$, so that
\beq 
v_{\rm{EW}} = \sqrt{\frac{-\mu_H^2}{\l}}\, .
\eeq 
After SSB, the chosen vacuum state must be neutral since $U(1)_{\rm{em}}$ is an unbroken symmetry of the vacuum. Thus we can write
\beq 
H = \frac{1}{\sqrt{2}}\begin{pmatrix}
 G_1 + iG_2 \\
 (h+v_{\rm{EW}}) + iG_0
\end{pmatrix}\, ,~~~~v\neq 0\, .
\eeq 
The fields $G_0, G_1, G_2$ are the would-be Goldstone bosons which are eaten by the massive gauge fields, and $h$ is the physical Higgs field. The would-be Goldstones are not physical and are absorbed as the longitudinal polarization of the gauge bosons. This is clearly seen in the unitary gauge, 
\beq 
H = e^{i\xi^i\tau^i}\frac{1}{\sqrt{2}}\begin{pmatrix}
 G_1 + iG_2 \\
 (h+v_{\rm{EW}}) + iG_0
\end{pmatrix} = \frac{1}{\sqrt{2}}\begin{pmatrix}
 0 \\
 h+v_{\rm{EW}}
\end{pmatrix}\label{eq: Higgs_unitary}
\eeq 
Substituting the above expression into the Higgs potential, we can read off the Higgs mass as $m_h^2 = 2\l v_{\rm{EW}}^2$. 

\subsection{Gauge sector}
The gauge boson mass matrix is derived from the Higgs kinetic term given in Eq.\,\eqref{eq: LagHiggs}. According to Eq.\,\eqref{eq: cov_der}, the gauge part of $D_{\mu}H$ reads 
\beq 
g\frac{\tau^i}{2} W_{\mu}^i + \frac{g'}{2}B_{\mu} = \frac{1}{2}\begin{pmatrix}
    g W_{\mu}^3 + g'B_{\mu} & \sqrt{2}gW^+\\
    \sqrt{2}gW^- & -g W_{\mu}^3 + g'B_{\mu}
\end{pmatrix}\, ,
\eeq 
where $W_{\mu}^{\pm} = \frac{1}{\sqrt{2}}(W^1_{\mu}\mp iW_{\mu}^2)$. Expanding $(D^{\mu}H)^{\dagger}D_{\mu}H$, we get the following mass terms for the gauge bosons 
\beq 
(D^{\mu}H)^{\dagger}D_{\mu}H \supset \frac{v_{\rm{EW}}^2}{2} \frac{1}{4}\left[(-gW^3_{\mu} + g'B_{\mu})^2 + 2g^2W^+W^-\right]\, .
\eeq 
We can read off the $W$ boson mass, $m^2_W = \frac{1}{4}g^2v_{\rm{EW}}^2$. For $W^3_{\mu}$ and $B_{\mu}$, we get the mass matrix:
\beq 
(D^{\mu}H)^{\dagger}D_{\mu}H \supset \frac{1}{2}\begin{pmatrix}
    B & W^3
\end{pmatrix}\begin{pmatrix}
    \frac{1}{4}g'^2v_{\rm{EW}}^2 & -\frac{1}{4}gg'v_{\rm{EW}}^2\\
    -\frac{1}{4}gg'v_{\rm{EW}}^2 & \frac{1}{4}g^2v_{\rm{EW}}^2
\end{pmatrix}\begin{pmatrix}
    B\\
    W^3
\end{pmatrix} \, .
\eeq 
This symmetric matrix can be diagonalized by the rotation matrix,
\beq 
R_{\theta_W} = \begin{pmatrix}
    \cos\theta_W & \sin\theta_W \\
    -\sin\theta_W & \cos\theta_W
\end{pmatrix}\, ,~~\begin{pmatrix}
    A \\
    Z^0
\end{pmatrix} = R_{\theta_W}\begin{pmatrix}
    B \\
    W^3
\end{pmatrix} \, .
\eeq 
The angle $\theta_W$ is called the Weinberg angle and is given by
\beq 
\cos\theta_W = \frac{g}{\sqrt{g^2+g'^2}}\, , ~~~\sin\theta_W = \frac{g'}{\sqrt{g^2+g'^2}}\, .
\eeq
After diagonalization, we get the massive $Z^0$ boson with $m_Z^2 = \frac{1}{4}(g^2+g'^2) v_{\rm{EW}}^2$, while the photon $A$ remains massless. Using the relations for $m_W$ and $m_Z$, we can define the $\rho$ parameter as, 
\beq 
\rho \equiv \frac{m_W^2}{m_Z^2\cos^2\theta_W} \, ,
\eeq 
whose value is 1 at the tree level in SM.

\subsection{Fermion sector}
The Yukawa lagrangian of Eq.\,\eqref{eq: LagYuk} generates the fermion masses via the Higgs mechanism. Substituting the expression of Eq.\,\eqref{eq: Higgs_unitary} in Eq.\,\eqref{eq: LagYuk}, the Yukawa Lagrangian can be written as
\beq
-\mathcal{L}_{\rm{Yukawa}} = \frac{(h+v_{\rm{EW}})}{\sqrt{2}}\left( Y^u_{ij}~\overline{u}^i_{L} ~u^j_{R} -  Y^d_{ij}~\overline{d}^i_{L}  d^{j}_{R} - Y^e_{ij}~\overline{e}^i_{L} e^{j}_{R} + \rm{h.c.}\right)\, .
\eeq

The fermion mass matrices in the flavor basis are given by,
\begingroup
\allowdisplaybreaks \beq
\mathcal{M}_{ij}^f = \frac{1}{\sqrt{2}} Y^f_{ij} v_{\rm{EW}}\,, ~~~ {\rm with},~~~ f=u,d~{\rm and }~e\,.
\eeq
\endgroup\par
These matrices can be diagonalized via bi-unitary transformations involving $3\times3$ unitary matrices $V_L^f$ and $V_R^f$
\beq
\left[\mathcal{M}^f_{\rm{diag}}\right]_{kk} = \frac{1}{\sqrt{2}} (V^f_L)_{ki} Y_{ij}^f (V^f_R)^\dagger_{jk} v_{\rm{EW}} = \frac{1}{\sqrt{2}} \left[Y_{\rm{diag}}^f\right]_{kk} v_{\rm{EW}} \label{femmas}\,,
\eeq

and the mass eigenstates of quarks and leptons are given by,
\begingroup
\allowdisplaybreaks \bea
u^\prime_{iL} &=& (V^u_L)_{ij} u_{jL}, ~~ u^\prime_{iR} = (V^u_R)_{ij} u_{jR}\,,\label{unit3}\\
d_{iL}^\prime &=& (V^d_L)_{ij} d_{jL},~ ~ d^\prime_{iR} = (V^d_R)_{ij} d_{jR}\,,\label{unit2}\\
e^\prime_{iL} &=& (V^e_L)_{ij} e_{jL}, ~~ e^\prime_{iR} = (V^e_R)_{ij} e_{jR}\label{unit1}\,.
\eea
\endgroup\par

In the SM, the neutrinos are massless due to the absence of right-handed neutrinos. As a result, the flavor and mass bases for the charged leptons coincide. However, for the quark sector, both bases are different. One can get from Eq.\,\eqref{unit3} and Eq.\,\eqref{unit2}, that the quarks of different flavors mix to form the mass eigenstates, which take part in flavor-changing charged current interactions. Using Eq.\,\eqref{eq: fermionew}, these interactions can be written as the following Lagrangian,
%%%%%%%%
\begingroup
\allowdisplaybreaks \bea
-\frac{g}{\sqrt{2}} (\bar{u^\prime}_L ~ \bar{c^\prime}_L~ \bar{t^\prime}_L)\gamma^\mu W^+_\mu V_{CKM}(d^\prime_L~ s^\prime_L~ b^\prime_L)^T + h.c.,\nn
\eea
\endgroup\par
%%%%%%%%
where $V_{CKM}=V^{u}_{L}V^{d\dagger}_{L}$, stands for the Cabibbo-Kobayashi-Maskawa matrix (CKM). It is parameterized by three mixing angles and a phase and is given by\,\cite{ParticleDataGroup:2014cgo},
%%%%%
%\vspace{0.2cm}
\begingroup
\allowdisplaybreaks \beq
V_{CKM}=\resizebox{0.7\textwidth}{!}{$\begin{pmatrix}
|V_{ud}|=  0.97425 \pm 0.00022 &  |V_{us}|= 0.2253 \pm 0.0008 & |V_{ub}|= 0.00413 \pm 0.00049 \\
|V_{cd}|=0.225 \pm 0.008 & |V_{cs}|= 0.986 \pm 0.016 & |V_{cb}|=0.0411 \pm 0.0013 \\
|V_{td}|=0.0084 \pm 0.0006 & |V_{ts}|=0.0400 \pm 0.0027 & |V_{tb}| =1.021 \pm 0.032
\end{pmatrix}$}.
\label{VCKM}
\eeq
\endgroup\par
%%%%%
%%%%%
However, the SM does not predict any flavor-changing neutral currents (FCNC) at the tree level.

\subsection{Limitations of the SM}
There are 19 input parameters to be experimentally determined in the SM: three gauge coupling constants $\{g,g',g_s\}$, the Higgs \vev $v_{\rm{EW}}$, the Higgs mass $m_h$, three charged-lepton masses $\{m_e,m_{\mu},m_{\tau}\}$, six quark masses $\{m_u,m_d,m_c,m_s,m_t,m_b\}$, three CKM mixing angles $\{\theta_{12},\theta_{23},\theta_{13}\}$, one weak $CP$-violating phase $\delta$, and one strong $CP$-violating parameter $\theta_{\rm{QCD}}$. Over the years, these parameters have been measured with ever-improving precision in collider experiments. Once these parameters are fixed, the SM predictions for the cross-sections for various processes, decay widths etc., match with the experimental data with astounding accuracy. Thus the SM has enjoyed enormous success and is one of the most predictive theories in modern physics.

Despite these successes, there are key challenges faced by the SM, as listed below
\begin{itemize}
\setlength\itemsep{.1em}
    \item Existence of tiny neutrino masses, as indicated by \textit{neutrino oscillations}. The minimal SM only describes massless left-handed neutrinos and does not explain neutrino masses. Moreover, for massive neutrinos, it is yet unknown if they are Dirac or Majorana fermions. To account for massive neutrinos, the SM has to be extended by adding new fields. 
    \item Origin of the \textit{baryon asymmetry} in the universe. The \textit{baryon to photon} ratio in the universe is, $n_{B}/n_{\gamma}\approx 10^{-9}$, where $n_B$ and $n_{\gamma}$ are the number densities for baryons and photons respectively. The SM does not explain how a non-zero baryon asymmetry can be generated if there is none in the initial stages of the universe. 
    \item The \textit{particle nature of dark matter}. We know about dark matter only through its gravitational effects. While neutrinos may partly contribute to dark matter, none of the SM particles can explain the entirety of dark matter.
    \item The \textit{hierarchy problem}: The existence of a light Higgs, $m_h = 125\,$GeV is theoretically puzzling since quantum corrections can drive the Higgs mass to the cutoff scale of the SM. 
    \item The \textit{SM flavor puzzle}: The Higgs mechanism generates fermion masses through Yukawa couplings. In the absence of an underlying mechanism, one would expect the fermion masses to be of the same order of magnitude. However, there is a huge hierarchy of the SM fermion masses, spanning over six orders of magnitude. This problem is discussed in more detail in Chapter\,\ref{chapter3}.
\end{itemize}
In addition to these limitations, there are other small discrepancies between the SM predictions and experimental data. 

\subsection{Extensions of the SM}
To address the shortcomings of the SM, different extensions have been proposed, with ideas such as \textit{supersymmetry}, \textit{grand-unified theories}(GUTs), \textit{composite Higgs models}, etc.\,. While the entire theory may be inaccessible to experimental scrutiny, such beyond SM (BSM) scenarios can often be simplified to make contact with phenomenology, leading to relatively simple extensions of the SM by adding new fields and/or new symmetries. Invariably, BSM scenarios lead to new kinds of phase transitions which may have interesting gravitational wave imprints. Therefore, below I briefly discuss some BSM scenarios relevant to the thesis.

\subsubsection{Scalar extensions}
In addition to the SM Higgs doublet, we can introduce new scalar multiplets in the theory in order to address one or more limitations of the SM. The extra scalar fields can lead to significant modifications of the Higgs couplings from the SM values, and can also alter the nature of EWSB. The representation of the scalar multiplets must not modify the $\rho$-parameter at tree level. 

The simplest scalar extension is obtained by adding a real scalar that is a singlet under $\mathcal{G}_{\rm{SM}}$, with the scalar potential given by
\beq 
V(\phi,H) = \frac{\mu_{\phi}^2}{2}\phi^2 + \frac{\l_3}{3}\phi^3+\frac{\l_{\phi}}{4}\phi^4 + \mu_1 \phi (H^{\dagger}H) + \frac{\l_{H\phi}}{2} \phi^2 (H^{\dagger}H) + V_{\rm{SM}}(H)\, ,  
\eeq 
where $V_{\rm{SM}}(H)$ is the SM Higgs potential. By imposing a symmetry on the potential, such as a $Z_2$ symmetry, some of the coupling constants can be set to zero. 

Similarly, we can add a complex singlet scalar, as discussed in Chapter\,\ref{chapter2}, allowing for more interesting possibilities. A very popular scalar extension is the two-Higgs doublet model (2HDM), which also arises in the scalar sector of the minimal supersymmetric SM (MSSM). 

\subsubsection{Gauge extensions}
In many BSM scenarios, the gauge group of SM is extended. For example, the SM has two accidental global symmetries, namely $U(1)_B$ and $U(1)_L$, corresponding to the baryon number and the lepton number. These symmetries are, however, anomalous, leading to $B$ and $L$ violating processes at the loop level. The combination $U(1)_{B-L}$ is anomaly-free, and can be promoted to a gauge symmetry. 

GUT theories aim to unify the strong and EW sectors as part of a single gauge group, eg. $SU(5)$, $SO(10)$, or $E(6)$. The fields are then arranged as different multiplets of the gauge group, and the SM is recovered after a series of SSBs, making these scenarios very interesting from the perspective of phase transitions. An intermediate step in many GUT theories leads to left-right symmetric models (LRSMs), with the gauge group $\mathcal{G_{\rm{LRSM}}} = SU(3)_c\times SU(2)_L\times SU(2)_R\times U(1)_{B-L}$. In Chapter\,\ref{chapter3} and Chapter\,\ref{chapter4}, I discuss phase transitions in a specific version of LRSM. 

\subsubsection{Flavor models}
In models aimed at addressing the SM flavor puzzle, an extra flavor symmetry is assumed, where the SM fermions belonging to different generations are assigned different charges. One popular flavor model is the Froggatt-Nielsen mechanism, which is discussed in detail for the quark sector in Chapter\,\ref{chapter2}. 

\subsubsection{Neutrino mass models}
The simplest way to generate Dirac neutrino masses is to add at least one right-handed neutrino $\nu_R$ to the minimal SM. However, the corresponding Yukawa couplings are extremely small owing to the tiny neutrino masses and are not natural. This problem is avoided by see-saw models, which introduce a Majorana mass for $\nu_R$ at a high scale, leading to a natural explanation for light left-handed neutrino masses. 
 
\section{The effective potential}\label{1_effective_potential}
\vspace{4mm}
The effective potential is an essential tool for studying cosmological phase transitions. It is a function of background scalar fields that enables us to compute higher-order quantum corrections to the vacuum states of the theory while retaining the semiclassical picture. The minima of the effective potential give all possible vacuum states.
%Sidney Coleman and Eric Weinberg calculated the first quantum corrections to the effective potential while Jackiw later expanded the result to two loops. 
At finite temperatures, the effective potential receives thermal corrections and helps track the evolution of vacuum states with temperature. For pedagogic reviews on the finite temperature effective potential, see \citep{Carrington:1991hz,Quiros:1999jp,Laine:2016hma,Breitbach:2018kma}. Below, I discuss some important theoretical and computational aspects of the effective potential. 

\subsection{Zero temperature effective potential}Consider a lagrangian density, $\mathcal{L}(\phi(x))$, for a scalar field $\phi$. 
The vacuum-to-vacuum transition amplitude is given by the path integral, 
\beq 
Z[J] \equiv \langle 0_{\rm{out}} | 0_{\rm{in}}\rangle_J = \int \mathcal{D}\phi~e^{i\int d^4x (\mathcal{L}(\phi(x)) + J(x)\phi(x))},
\eeq 
in presence of a source $J(x)$. The $n$-point Green's functions, $G^{(n)}(x_1,\cdots,x_n)$, can be computed by taking $n$ functional derivatives of $Z$ with respect to $J$
\beq
G^{(n)}(x_1,\cdots,x_n) = (-i)^n\frac{\delta^n Z}{\delta J(x_1)\cdots\delta J(x_n)}.
\eeq
Thus, $Z[J]$ is the generating functional of the $n$-point Green's functions
\beq 
Z[J] = \sum_{n=0}^{\infty} \frac{i^n}{n!} \int d^4x_1\cdots\int d^4x_n ~G^{(n)}(x_1,\cdots,x_n) J(x_1)\cdots J(x_n).
\eeq 
Another functional, $W[J]$, is defined using the relation
\beq 
Z[J] \equiv e^{iW[J]}\, .
\eeq 
It can be shown that $W[J]$ is the generating functional of all connected Green's functions, $G_c^{(n)}(x_1,\cdots,x_n)$. The functional derivative of $W[J]$ with respect to $J(x)$ 
\beq 
\frac{\delta W[J]}{\delta J(x)} = \frac{\int \mathcal{D}\phi~e^{i\int (\mathcal{L} + J\phi)}\phi(x)}{\int \mathcal{D}\phi~e^{i\int  (\mathcal{L} + J\phi)}} = \frac{\langle 0_{\rm{out}}|\phi|0_{\rm{in}}\rangle_J}{\langle 0_{\rm{out}}|0_{\rm{in}}\rangle_J}\, ,
\eeq 
is the vacuum expectation value of $\phi$ in the presence of a source $J(x)$. We define this as the classical field, $\phi_{\rm{cl}}$
\beq 
\phi_{\rm{cl}}(x) \equiv \frac{\delta W[J]}{\delta J(x)}= \frac{\langle 0_{\rm{out}}|\phi(x)|0_{\rm{in}}\rangle_J}{\langle 0_{\rm{out}}|0_{\rm{in}}\rangle_J}\, .
\eeq
The effective action, $\Gamma[\phi_{\rm{cl}}]$, is obtained by the Legendre transformation of $W[J]$,
\beq 
\Gamma[\phi_{\rm{cl}}] = W[J] - \int d^4x ~\phi_{\rm{cl}}(x) J(x)\, .
\eeq 
To see the relevance of the term `effective action' for $\Gamma$, let us take its functional derivative with respect to $\phi_{\rm{cl}}(x)$,
\begin{align}
\frac{\delta \Gamma [\phi_{\rm{cl}}]}{\delta \phi_{\rm{cl}}(x)} &= \int d^4y \frac{\delta W[J]}{\delta J(y)}\frac{\delta J(y)}{\delta \phi_{\rm{cl}}(x)} - \int d^4y ~\phi_{\rm{cl}}(y) \frac{\delta J(y)}{\delta \phi_{\rm{cl}}(x)}- J(x)\,,\nn\\
& = -J(x).
\end{align}
In particular, 
\beq\label{eq: classical gamma}
\left.\frac{\delta \Gamma [\phi_{\rm{cl}}]}{\delta \phi_{\rm{cl}}}\right|_{J=0} = 0\, ,
\eeq 
i.e., in the absence of external sources, $\phi_{\rm{cl}}$ extremizes the effective action, analogous to how the classical path extremizes the action in classical mechanics.

The effective action can be shown to be the generating functional of one-particle irreducible (1PI) Green's functions, $\Gamma^{(n)}(x_1,\cdots,x_n)$ 
\beq 
\Gamma[\phi_{\rm{cl}}] = \sum_{n=0}^{\infty} \frac{1}{n!} \int d^4x_1\cdots\int d^4x_n ~\Gamma^{(n)}(x_1,\cdots,x_n) \phi_{\rm{cl}}(x_1)\cdots \phi_{\rm{cl}}(x_n).
\eeq 
For translation-invariant vacuum states, $\phi_{\rm{cl}}(x)$ takes a constant value, $\phi_c$, independent of $x$. The effective action is then proportional to the spacetime volume, $\int d^4x$, and takes the form,
\beq\label{eq: v_eff0}
\Gamma[\phi_c] = \left(\int d^4x\right) \sum_{n=0}^{\infty} \frac{1}{n!}~\phi_c^n~ \Gamma^{(n)}(p=0)\, ,
\eeq 
where $\Gamma^{(n)}(p=0)$ are the 1PI Green's functions in Fourier space, evaluated at vanishing external momenta. The effective potential is an ordinary function of $\phi_c$, defined through the relation,
\beq\label{eq: gamma} 
\Gamma[\phi_c] = -\int d^4 x~ V_{\rm{eff}} (\phi_c)\, .
\eeq 

Comparing eq.\,\eqref{eq: gamma} with eq.\,\eqref{eq: v_eff0}, we get, 
\beq \label{eq: v_eff}
V_{\rm{eff}}(\phi_c) = -\sum_{n=0}^{\infty} \frac{1}{n!}~\phi_c^n ~\Gamma^{(n)}(p=0)\, .
\eeq 
The extremum condition of Eq.\eqref{eq: classical gamma} reduces to the condition,
\beq 
\frac{\partial}{\partial \phi_c} V_{\rm{eff}} = 0.
\eeq
The minima of the effective potential represent possible translation-invariant vacuum states, $\phi_c$, and $V_{\rm{eff}}(\phi_c)$ is the corresponding energy density. If there are more than one minima, the global minimum of $V_{\rm{eff}}$ is the true vacuum, while the other local minima represent metastable vacuum states which may decay by quantum mechanically tunneling to the true vacuum. 

\begin{figure}[tbp]
\centering 
\includegraphics[width=.8\textwidth]{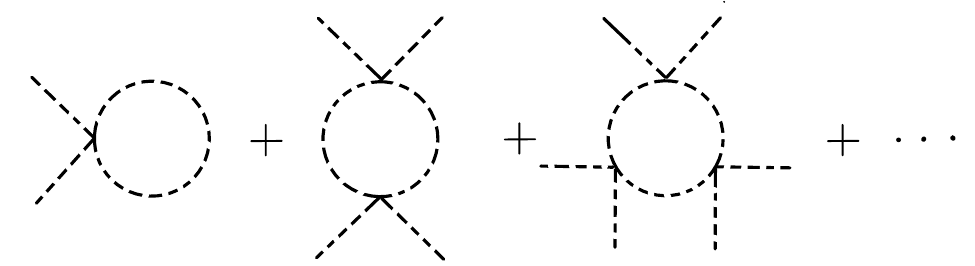}
\caption{\label{fig: V_scalar} Scalar one-loop contribution to the effective potential.}
\end{figure}

\begin{figure}[tbp]
\centering 
\includegraphics[width=.8\textwidth]{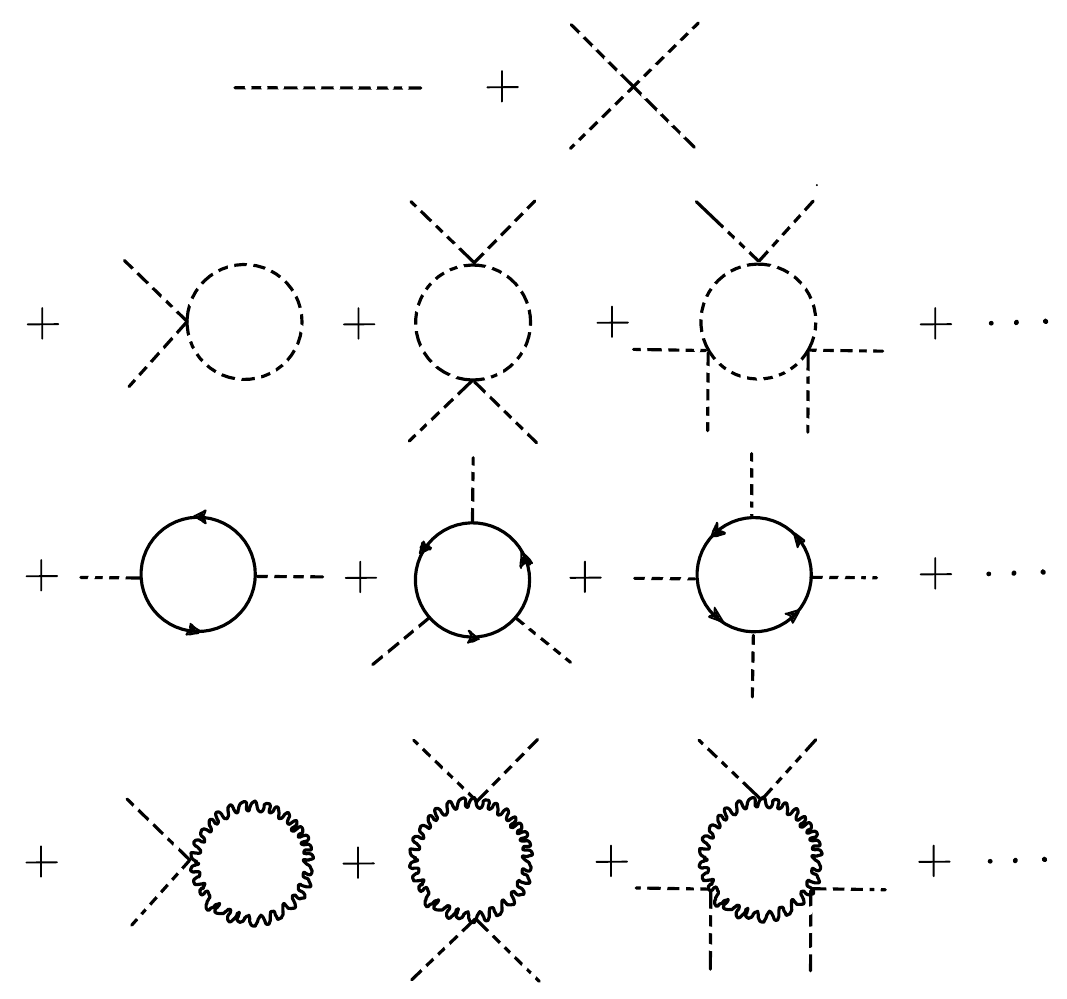}
\caption{\label{fig: V_eff} The effective potential to one-loop order, including the one-loop contributions from scalars (second row), fermions (third row), and gauge bosons (fourth row).}
\end{figure}

Eq.\,\eqref{eq: v_eff} implies that the $n^{\rm{th}}$ derivative of $V_{\rm{eff}}$ is the sum of all 1PI graphs with $n$ vanishing external momenta. We can decompose the effective potential as a loop expansion, with the expansion parameter, $\hbar$, (displayed explicitly here)
\beq 
V_{\rm{eff}} = V_0 + \hbar V_1 + \hbar^2 V_2 + \cdots
\eeq 
The tree-level contribution is just the ordinary potential entering $\mathcal{L}(\phi_c)$. The $l$-loop contribution can be obtained by evaluating the sum of all $l$-loop 1PI graphs. Let us evaluate the effective potential for a real scalar field, $\phi$, whose Lagrangian is,
\beq 
\mathcal{L}(\phi(x)) = \frac{1}{2}\partial_{\mu}\phi\partial^{\mu}\phi - \frac{1}{2}m^2\phi^2 - \frac{1}{4}\lambda \phi^4. 
\eeq 
The tree-level potential is given by, 
\beq 
V_0(\phi_c) = \frac{1}{2}m^2\phi_c^2 + \frac{1}{4}\lambda \phi_c^4\, , 
\eeq 
where $\phi_c$ is the classical field, also called the background field. For one-loop corrections, we need to evaluate the sum of all one-loop 1PI graphs shown in Fig.\,\ref{fig: V_scalar}. The $n^{\rm{th}}$ diagram has $2n$ external legs and contributes 
$$i\int \frac{d^4p}{(2\pi)^4}\frac{1}{2n}\left[\frac{3\lambda \phi_c^2}{p^2 - m^2 + i\epsilon}\right]^n\, .$$
The $n$ vertices contribute a factor $(-3i\lambda)^n$, with the factor of $3 = 4\times 3/4$ coming from the fact that for each vertex there are $4$ ways to attach to one of the external legs and 3 ways to attach to the other external leg.   The $n$ propagators contribute a factor, $i^n(p^2-m^2+i\epsilon)^{-n}$, and the $1/2n$ factor is a symmetry factor coming from overcounting due to the rotational and reflection symmetry of the diagram. The total contribution from all such graphs is, 
\bea\label{eq: Vloop}
V_1(\phi_c) &=& i\sum_{n=0}^{\infty}\int \frac{d^4p}{(2\pi)^4}\frac{1}{2n}\left[\frac{3 \lambda \phi_c^2}{p^2 - m^2 + i\epsilon}\right]^n\, \nn\\
&=& -\frac{i}{2}\int \frac{d^4p}{(2\pi)^4}\log\left[1 - \frac{3\lambda \phi_c^2}{p^2 - m^2 + i\epsilon}\right]\, .
\eea
Perform a Wick rotation, $p_E = (-ip^0,\vec{p})$, so that Eq.\eqref{eq: Vloop} becomes, 
\beq \label{eq: 1loop_scalar}
V_1(\phi_c) = \frac{1}{2}\int \frac{d^4p_E}{(2\pi)^4}\log(p_E^2 + m^2 + 3\lambda \phi_c^2)\, ,
\eeq 
where the field-independent term has been dropped. The quantity, $m^2 + 3\lambda \phi_c^2$, is called the field-dependent mass squared, $m^2(\phi_c)$, since
\beq 
m^2(\phi_c) = m^2 + 3\lambda \phi_c^2 = \frac{d^2 V_0(\phi_c)}{d \phi_c^2}.
\eeq 
The field-dependent mass is the mass of the quanta of $\phi$, in the presence of a background field $\phi_c$. The divergent expression on the right side of Eq.\eqref{eq: 1loop_scalar} can be regularized using dimensional regularization by introducing a small parameter $\epsilon = 4-d$, where $d$ is the dimension of spacetime. The regularized expression is
\beq
V_1(\phi_c) = \frac{m^4(\phi_c)}{64\pi^2}\left[\left\{\frac{2}{\epsilon} - \gamma_E + \log 4\pi \right\}  + \log\left(\frac{m^2(\phi_c)}{\mu^2}\right) - \frac{3}{2} + \mathcal{O}(\epsilon)\right]\, ,
\eeq
where $\gamma_E$ is the Euler-Mascheroni constant and $\mu$ is the renormalization scale. In the $\ms$ renormalization scheme, the counter-term cancels the contribution proportional to the quantity in curly brackets. The renormalized result is
\beq \label{eq: Vcw_scalar}
V_1(\phi_c) = \frac{m^4(\phi_c)}{64\pi^2}\left[ \log\left(\frac{m^2(\phi_c)}{\mu^2}\right) - \frac{3}{2} \right]\, .
\eeq 
This is the famous Coleman-Weinberg potential, often denoted by $\Vcw$. This formula is valid for the range of field values, for which the $\log$ term is $\mathcal{O}(1)$. Hence, the renormalization scale $\mu$ should be chosen wisely depending upon the scale of physics involved. Eq.\,\eqref{eq: Vcw_scalar} can be generalized for $N_s$ real scalars, $N_f$ fermions, and $N_g$ gauge bosons. Consider a general lagrangian, involving scalar fields, $\phi$,  fermions $\psi_i$, and gauge bosons, $A_{\mu}$,
\bea 
\mathcal{L} &=& -\frac{1}{4} \tr[F_{\mu\nu}F^{\mu\nu}] + \frac{1}{2} \tr[(D_{\mu}\phi)^{\dagger}D^{\mu}\phi] + i\overline{\psi}_i\cancel{D}\psi^i\nn\\
&& - \Gamma^i_{ja}\phi^a~ \overline{\psi}_i~\psi_j - V(\phi)\, .
\eea 
The scalars, fermions, and gauge bosons all contribute to the one-loop effective potential. For each species, the contribution comes from one-loop graphs where the internal line forming the loop comes from that species, as shown in Fig.\,\ref{fig: V_eff}. The procedure to obtain the final result is similar to that for scalars. For gauge bosons, the calculation is simplified in the Landau gauge in which there are no ghost contributions. In essence, Eq.\,\eqref{eq: 1loop_scalar} is generalizes to
\beq \label{eq: 1loop_all}
V_1(\phi_c) = \sum_a (-1)^{f_a}\frac{n_a}{2}\int \frac{d^4p_E}{(2\pi)^4}\log(p_E^2 + m_a^2(\phi_c))\, ,
\eeq
where $a$ runs over all scalar, fermion, and gauge boson degrees of freedom. The parameter $f_a$ is $0$ for bosons and $1$ for fermions, $n_a$ is the number of degrees of freedom, and $m_a$ is the field-dependent mass for each species. After renormalization, the Coleman-Weinberg potential in Landau Gauge, in $\ms$ scheme is given by
\beq \label{eq: 1_ColWein}
\Vcw(\phi_c) = \sum_{a} (-1)^{f_a}n_a\frac{m_a^4(\phi_c)}{64\pi^2}\left[ \log\left(\frac{m_a^2(\phi_c)}{\mu^2}\right) - c_a \right]\, ,
\eeq 
where $c_a = 3/2$ for scalars and fermions, and $5/6$ for gauge bosons. The procedure to evaluate field-dependent masses for all species is described in Appendix\,\ref{app: field}.
Note that we can always add a finite counter-term to the potential given in Eq.\,\eqref{eq: 1_ColWein}, $V_1 = \Vcw + V_{\rm{c.t.}}$\,, so that the one-loop contribution preserves the position of the minimum and the masses of the scalar fields at the respective tree-level values, i.e.,  
\beq
 \left.\frac{\partial V_1}{\partial \phi_c^a}\right|_{\phi_c = \phi_0} = 0\, ,~\left.\frac{\partial^2 V_1}{\partial \phi_c^a \partial \phi_c^b}\right|_{\phi_c = \phi_0} = 0\, . 
\eeq
This is known as the \textit{on-shell} renormalization scheme and is used extensively in this thesis. 

\subsection{Finite temperature effective potential}
The above computation of the one-loop effective potential is valid at zero temperature. However, the early universe was in a thermal bath of plasma comprising all sorts of particles. Particle interactions were screened by the plasma, and differed from those at zero temperature, thereby inducing thermal effects into the effective potential. The discussion here is broadly based on Refs.\,\cite{Laine:2016hma, Quiros:1999jp}. To obtain the effective potential at finite temperatures, we first take note of the striking resemblance between statistical mechanics and path integrals. 

Consider a quantum mechanical system in one dimension, described by the Hamiltonian
$$H = \frac{p^2}{2m} + V(q)\,.$$
At a temperature $T$, the equilibrium state of a canonical ensemble of such systems is described by the canonical density operator 
\beq \label{eq: density}
\hat{\rho} = \frac{e^{-\b \hat{H}}}{\tr \left[e^{-\beta \hat{H}}\right]}\,,
\eeq 
where $\beta = 1/T$ and the denominator is the partition function 
\beq \label{eq: partition}
Z = \tr \left[e^{-\beta \hat{H}}\right] = \sum_n  \langle n|e^{-\beta \hat{H}}|n\rangle \, ,
\eeq 
where $\{|n\rangle\}$ is a complete set of eigenstates. Clearly, $\tr[\hat{\rho}] = 1$. The canonical average of any operator $\hat{\mathcal{O}}$ is given by
\beq \label{eq: rhoA}
\langle \hat{\mathcal{O}} \rangle_{\b} = \tr[\hat{\mathcal{O}}\rho]\,.
\eeq 
The position and momentum eigenstates are given by, 
\bea 
&\hat{q}|q\rangle = q |q\rangle \, ,~\hat{p}|p\rangle = p |p\rangle \, ,&\nn\\
&\langle q|q'\rangle = \delta(q - q') \, ,~ \langle p|p'\rangle = 2\pi \delta(p - p') \, ,~\langle q|p\rangle = e^{iqp} \, ,&\nn
\eea  
and obey the completeness property 
$$\int dq |q\rangle \langle q| = \mathbb{I}\, , \qquad \int \frac{dp}{2\pi} |p\rangle \langle p| = \mathbb{I}\, .$$

Writing $\beta = N\epsilon$, we can write $Z$ as,
\bea \label{eq: Z1}
Z &=& \int dq\langle q|e^{-\beta \hat{H}}|q\rangle \nn\\
  &=& \int dq\langle q|e^{-\epsilon \hat{H}}\cdot e^{-\epsilon \hat{H}} \cdots e^{-\epsilon \hat{H}}|q\rangle \nn\\
  &=& \int\prod_{n=1}^N \frac{dp_ndq_n}{2\pi}\langle q_{n+1}|e^{-\epsilon \hat{H}}|p_n\rangle \langle p_n|q_n \rangle \, ,
\eea 
where in the second step, we introduced the identity matrix as $$\mathbb{I} = \int\frac{dp_ndq_n}{2\pi}|p_n\rangle \langle p_n|q_n \rangle \langle q_n | \, ,$$ between each pair of $e^{-\epsilon \hat{H}}$ factors, and identified $q_{N+1} = q_1 = q$. Performing the $dp$ integral, one can show that in the limit $N\rightarrow \infty$, Eq.\,\eqref{eq: Z1} yields
\beq 
Z = \tr [e^{-\beta \hat{H}}] = C\int \mathcal{D}q ~e^{-S_E}\, , 
\eeq 
with  
\beq 
C = \lim_{N\rightarrow\infty}\left(\frac{mN}{2\pi \b}\right)^{N/2}\,, ~~~ -S_E = -\int_0^{\b} d\tau \left[\frac{m}{2}\left(\frac{dq}{d\tau}\right)^2 + V(q)\right]\, .
\eeq 
Thus, we can identify the partition function $Z$ with a Euclidean path integral with the time variable $0\leq \tau \leq \b$, over periodic functions $q(\tau =0) = q(\tau = \b)$. The Minkowski path integral can be obtained by the identification, $\tau = i\, t$, giving $S_M=iS_E$.

We can extend the above procedure to quantum field theory, with the canonical density operator and the partition function defined by Eq.\,\eqref{eq: density} and Eq.\,\eqref{eq: partition} respectively, and the thermal average of an operator given by Eq.\,\eqref{eq: rhoA}. For a real scalar field, the partition function can be written as
\bea 
Z_{\b} &=& \tr [e^{-\beta \hat{H}}] = \int_{\phi(0,{\bf{x}})=\phi(\b,{\bf{x}})} \prod_{{\bf{x}}} [C\mathcal{D}\phi(\tau, {\bf{x}})] e^{-\left[\int_0^{\b}d\tau \int d^3x \mathcal{L}_E\right]} \, ,\\
\mathcal{L}_E &=& \frac{1}{2}\left(\frac{\partial \phi}{\partial \tau}\right)^2 + \frac{1}{2}\left(\frac{\partial \phi}{\partial x_i}\right)^2 + V(\phi)\, ,
\eea
where the path integral is taken over fields periodic in imaginary time, with period $\b$. 

Now consider arbitrary quantum fields $\phi(x)$, described by a Hamiltonian $\hat{H}$. At finite-$T$, the $n$-point Greens functions are defined by
\beq 
G_{\b}^{(n)}(x_1, \cdots,x_n) \equiv \left\langle \hat{T} \{\hat{\phi}(x_1) \cdots \hat{\phi}(x_n)\}\right\rangle_{\b}\, ,
\eeq 
where $\hat{T}$ stands for the time-ordering operator. The partition function can be shown to be the generating functional of the finite-$T$ Greens functions
\beq 
Z_{\b}[J] = \sum_{n=0}^{\infty} \frac{i^n}{n!} \int d^4x_1\cdots\int d^4x_n ~G_{\b}^{(n)}(x_1,\cdots,x_n) J(x_1)\cdots J(x_n).
\eeq 
As in the zero temperature case, we can define the generating functional for connected Greens functions $W_{\b}[J]$, and the finite-$T$ effective action by the Legendre transformation
\beq 
\Gamma_{\b}[\phi_{\rm{cl}}] = W_{\b}[J] - \int d^4x ~\phi_{\rm{cl}}(x) J(x)\, ,
\eeq 
where $\phi_{\rm{cl}}(x)\equiv \delta W_{\b}[J]/\delta J(x)$ is the classical field. For translationally invariant field states, we can again write $\phi_{\rm{cl}}(x) = \phi_c$, so that
\beq 
\Gamma_{\b}[\phi_{\rm{c}}] = -\int d^4x V^{\b}_{\rm{eff}}\, ,
\eeq 
where $V^{\b}_{\rm{eff}}$ is the finite-$T$ effective potential, which has the same properties as $V_{\rm{eff}}$: the minima represent the vacuum states of the theory.

Before discussing how $V^{\b}_{\rm{eff}}$ is computed, we note an interesting property of two-point correlation functions. In the Heisenberg picture, we have
\beq 
\hat{\phi}(x) = e^{it\hat{H}}\hat{\phi}(0,{\bf{x}})e^{-it\hat{H}}\,.
\eeq 
Setting $\tau = i t$ and using the cyclic property of the trace in Eq.\,\eqref{eq: rhoA}, we can derive the \textit{Kubo-Martin-Schwinger} (KMS) relation
\beq 
\langle\hat{\phi}(\tau,{\bf{x}})\hat{\phi}(0,{\bf{y}})\rangle_{\b} = \langle\hat{\phi}(\b,{\bf{y}})\hat{\phi}(\tau,{\bf{x}})\rangle_{\b}\, .
\eeq 
This implies $\hat{\phi}(0,{\bf{x}}) = \pm \hat{\phi}(\b,{\bf{x}})$, with the $+$ sign for commuting bosonic fields, and $-$ sign for anti-commuting fermionic fields. The KMS relation tells us that at finite temperature, the fields are periodic or anti-periodic in imaginary time, with period $\b$. As a result, we can decompose the field into Fourier components
\beq 
\phi(\tau,{\bf{x}}) = \sum_{n=-\infty}^{\infty} \phi_n({\bf{x}}) e ^{i\omega_n \tau}\, ,
\eeq 
where $\omega_n$ are called \textit{Matsubara frequencies}, given by
\beq 
\omega_n = \begin{cases} 
       \frac{2\pi n}{\b}\,, & {\rm{bosons}} \\
       \frac{(2n+1)\pi}{\b}\,, & {\rm{fermions}} 
   \end{cases}\, .
\eeq 
Due the KMS relation, the two-point Greens function $G_{\b}^{(2)}(\tau,{\bf{x}})$ can also be decomposed into Fourier modes
\beq 
\tilde{G}_{\b}^{(2)}(\omega_n,{\bf{p}}) = \int_{\a-\b}^{\a} d\tau\int d^3x\, e^{i\omega_n\tau-i{\bf{p}}\cdot{\bf{x}}}\,G_{\b}^{(2)}(\tau,{\bf{x}})  = \frac{1}{{\bf{p}}^2+\omega_n^2+i\epsilon}\, , 
\eeq 
where $0\leq \a \leq \b$. We can use this to derive Feynman rules for different fields\,\cite{Quiros:1999jp}. The finite temperature effective potential at one loop is now given by diagrams similar to those in Fig.\,\ref{fig: V_eff}, with the Matsubara modes of the scalars, fermions, and gauge bosons running in the loops. The resulting expression is
\beq 
\left[V^{\b}_{\rm{eff}}(\phi_c, T)\right]_{{\rm{1-loop}}} = \frac{T}{2}\sum_a(-1)^{f_a} \sum_n\int \frac{d^3p}{(2\pi)^3}\log\{|{\bf{p}}|^2+\omega_n^2+m_a^2(\phi_c)\}\, .
\eeq 
Skipping the intricacies of the calculation, this expression splits into a divergent temperature-independent contribution given in Eq.\,\eqref{eq: 1loop_all}, and a finite temperature-dependent contribution 
\bea
\left[V^{\b}_{\rm{eff}}(\phi_c, T)\right]_{{\rm{1-loop}}} =  V_1(\phi_c) + V_{1T}(\phi_c,T) \,.
\eea  
The temperature-dependent contribution is given by
\beq \label{eq: 1_V_1T}
V_{1T}(\phi,T) = \sum_{a} \frac{T^4}{2\pi^2} n_a J_{b/f}\bigg(\frac{m^2_a(\phi)}{T^2}\bigg)\, ,
\eeq 
where we have dropped the subscript `$c$', and the $J_{b/f}$ functions are for bosons and fermions respectively, given by
\beq 
J_{b/f}(x^2) = \pm \int_0^{\infty} dy~ y^2 \log\big(1\mp e^{-\sqrt{x^2+y^2}}\big)\, ,
\eeq 
where $x^2 = m^2(\phi)/T^2$. These functions have expansions for $x^2\ll 1$ and $x^2\gg 1$. For small $x^2$, i.e. high-$T$ limit \cite{Cline_1997}, 
\begin{comment}
\begin{eqnarray}\label{eq: 1_highT}
J_f(x^2,n) \approx &-&\frac{7\pi^4}{360} + \frac{\pi^2}{24} x^2 + \mathcal{O}(x^4)\,,\label{eq: highTf}\\
J_b(x^2,n) = &-&\frac{\pi^4}{45} + \frac{\pi^2}{12} x^2 - \frac{\pi}{6}\big(x^2\big)^{3/2}+ \mathcal{O}(x^4)\,.\label{eq: 1_highTb}
\end{eqnarray}
\end{comment}
\begin{subequations}\label{eq: 1_highT}
\beq
J_f(x^2,n) \approx -\frac{7\pi^4}{360} + \frac{\pi^2}{24} x^2 + \mathcal{O}(x^4)\,,
\eeq 
\beq
J_b(x^2,n) = -\frac{\pi^4}{45} + \frac{\pi^2}{12} x^2 - \frac{\pi}{6}\big(x^2\big)^{3/2}+ \mathcal{O}(x^4)\,.
\eeq 
\end{subequations}
For large $x^2$, i.e. low-$T$ limit, both fermions and bosons have the same expansion\,\cite{Cline_1997},
\begin{align}\label{eq: 1_lowT}
J_{b/f}(x^2,n) = &-\exp\bigg(-(x^2)^{1/2}\bigg)\bigg(\frac{\pi}{2}(x^2)^{3/2}\bigg)^{1/2}\sum_{l=0}^n \frac{1}{2^ll!}\frac{\Gamma(5/2+l)}{\Gamma(5/2-l)}(x^2)^{-l/2},
\end{align}
where $\Gamma(x)$ is the Euler Gamma function. In the low-$T$ limit, the thermal potential is Boltzmann-suppressed. 

\subsection{Symmetry restoration at high temperatures}
At high temperatures, the leading contribution to the effective potential comes from the tree-level piece and the thermal correction. Substituting Eq.\,\eqref{eq: 1_highT}  in Eq.\,\eqref{eq: 1_V_1T}, we get the high-$T$ thermal correction at leading order as
\beq \label{eq: 1_V1T_high}
    V_{1T}^{\rm{high}}(\phi,T) = \frac{T^2}{24}\left(\sum_b n_b m_b^2(\phi) + \frac{1}{2}\sum_f n_f m_f^2(\phi)\right) -\frac{\pi}{6} T\sum_b |m_b(\phi)|^3 + \cdots\, .
\eeq
Using this, we can write down a simplified formula for the effective potential. Consider a spontaneously broken theory with a real scalar field, described by the potential 
\beq
V(\phi) = -\frac{\mu_{\phi}^2}{2}\phi^2 + \frac{\l}{4}\phi^4\,,
\eeq
with $\mu_{\phi}^2>0$. There could also be other fermions and gauge bosons coupled to the scalar. The field-dependent masses are generally of the form $m^2(\phi) \sim \phi^2$. Thus
\bea
V_{\rm{eff}}(\phi, T) &\approx& V_0(\phi) + V_{1T}(\phi,T) \nn\\
&\approx& \left(-\frac{\mu_{\phi}^2}{2}+AT^2\right)\phi^2 -BT|\phi|^3 + \frac{\l}{4}\phi^4\, ,\label{eq: 1_V1T_high_approx}
\eea 
where $A$ and $B$ are model-dependent combinations of coupling constants and the particle degrees of freedom. The \vev is then a function of temperature, given by, $v(T) = \sqrt{(\mu_{\phi}^2 - 2AT^2)/\l}$. Thus at higher temperatures, the $\vev$ decreases and approaches zero. This phenomenon is called \textit{symmetry restoration} and was first discovered by Kirzhnits and Linde in Ref.\,\cite{kirzhnits1972} in the context of EW interactions, and was later confirmed by others. 

As the temperature of the universe decreases with expansion, the symmetry of the potential gets non-restored, eventually leading to a spontaneously broken theory at low temperatures. The coefficient $B$ of the cubic term is contributed solely by the bosonic degrees of freedom and is responsible for creating a barrier between the zero and non-zero minima of the effective potential. This barrier plays a major role in the occurrence of a first-order phase transition. When $B$ is vanishingly small, a second-order phase transition may occur.

\subsection{Thermal masses}
Particles in a plasma receive a temperature-dependent \textit{Debye mass} due to thermal self-energy corrections. 
Using the high-$T$ expansion of Eq.\,\eqref{eq: 1_V1T_high} without the cubic term, the thermal mass matrices can be written as, $\Pi_{ij} = c_{ij} T^2$, where $c_{ij}$ are, given by
\beq
    c_{ij} = \frac{1}{T^2}\left.\frac{\partial^2}{\partial\phi_i\partial\phi_j}V_{1T}^{\rm{high}}\right\vert_{\phi=\phi_c}\,.
\eeq
While all types of particles receive thermal mass corrections, only the thermal masses of bosons are important for the effective potential. In the case of gauge bosons, only the longitudinal polarization gets a Debye mass. The thermally corrected masses of each species with the above thermal masses are obtained as the eigenvalues of the matrix, $\mathcal{M}^2(\phi)+\Pi(T)$. 

\subsection{Daisy re-summation}\label{1_daisy}
The phenomenon of symmetry restoration at high temperatures, $T\gg \mu_{\phi}$, signals the breakdown of the perturbative loop expansion of the effective potential. This is because, at high-$T$, the thermal contribution of some higher-loop diagrams becomes as important as the one-loop contribution. The breakdown occurs due to infrared divergences in the loop contributions of bosons, for which the Matsubara frequencies $\omega_n$ vanish at $n = 0$\,\cite{Carrington:1991hz}. On the other hand, the infrared divergence is absent in the case of fermions. The most important higher-loop contribution comes from the so-called \textit{daisy diagrams} and can be re-summed. There are two alternative ways to perform the resummation, namely the \textit{Parwani method} \cite{Parwani:1991gq} and the \textit{Arnold-Espinosa} method \cite{Arnold-Espinosa}. 

In the Parwani method, the field-dependent mass is replaced with thermally corrected mass, i.e., $m_i^2(\phi)\rightarrow m_i^2(\phi) + \Pi_i(T)$, in the expressions of $V_{\cw}$ and $V_{1T}$. The daisy re-summed effective potential is given by
\beq \label{eq: parwani}
    V_{\rm{eff}} = V_0 + V_{\cw}(m_i^2(\phi) + \Pi_i(T)) + V_{\rm{c.t.}} + V_{1T}(m_i^2(\phi) + \Pi_i(T))\,.
\eeq

In the Arnold-Espinosa method, no such replacement for field-dependent mass is made, but an extra daisy term is added to the effective potential: 
\beq\label{eq: Vdaisy}
    V_{\rm{daisy}} = -\frac{T}{12\pi}\sum_i n_i \bigg((m_i^2(R) + \Pi_i(T))^{3/2} - (m_i^2(R))^{3/2}\bigg)\,.
\eeq
Thus, the effective potential is given by
\beq\label{eq: arnold_espinosa}
    V_{\rm{eff}} = V_0 + V_{\cw} + V_{\rm{c.t.}} + V_{1T} + V_{\rm{daisy}}.
\eeq
Both methods have some advantages and disadvantages. The Arnold-Espinosa method takes into account the daisy resummation consistently at the one-loop level, while the Parwani method mixes higher-order loop effects in the one-loop analysis.

\section{Cosmological phase transitions}\label{chp1sec4}
\vspace{4mm}
In the previous section, we saw how the effective potential evolves with temperature. In most cases involving a spontaneously broken symmetry at zero temperature, the symmetry of the potential gets restored at high temperatures. Therefore, at intermediate temperatures, a phase transition is possible as the \vev goes from zero to non-zero values. We can treat the \vev of the scalar field as the order parameter and analyze the nature of the phase transition. The \vev can evolve in different ways as the universe cools:
\begin{enumerate}
\setlength\itemsep{.1em}
    \item The \vev smoothly goes from zero to a non-zero value, without a phase transition. This scenario is called a smooth crossover. 
    \item The \vev continuously goes from zero to a non-zero value, but its first derivative is discontinuous at a particular temperature, indicating a second-order phase transition. 
    \item The \vev makes a discrete jump from zero to a non-zero value. In this case, there may be a range of intermediate temperatures in which the \vev can take two or more values, indicating the presence of multiple minima in the effective potential, separated by barriers. This is a first-order phase transition (FOPT). 
    \item Apart from the order of the phase transition, non-trivial topological defects may be formed during the phase transition. This depends on the topology of the vacuum manifold.  
\end{enumerate}
\begin{figure}[tbp]
\centering 
\includegraphics[width=.8\textwidth]{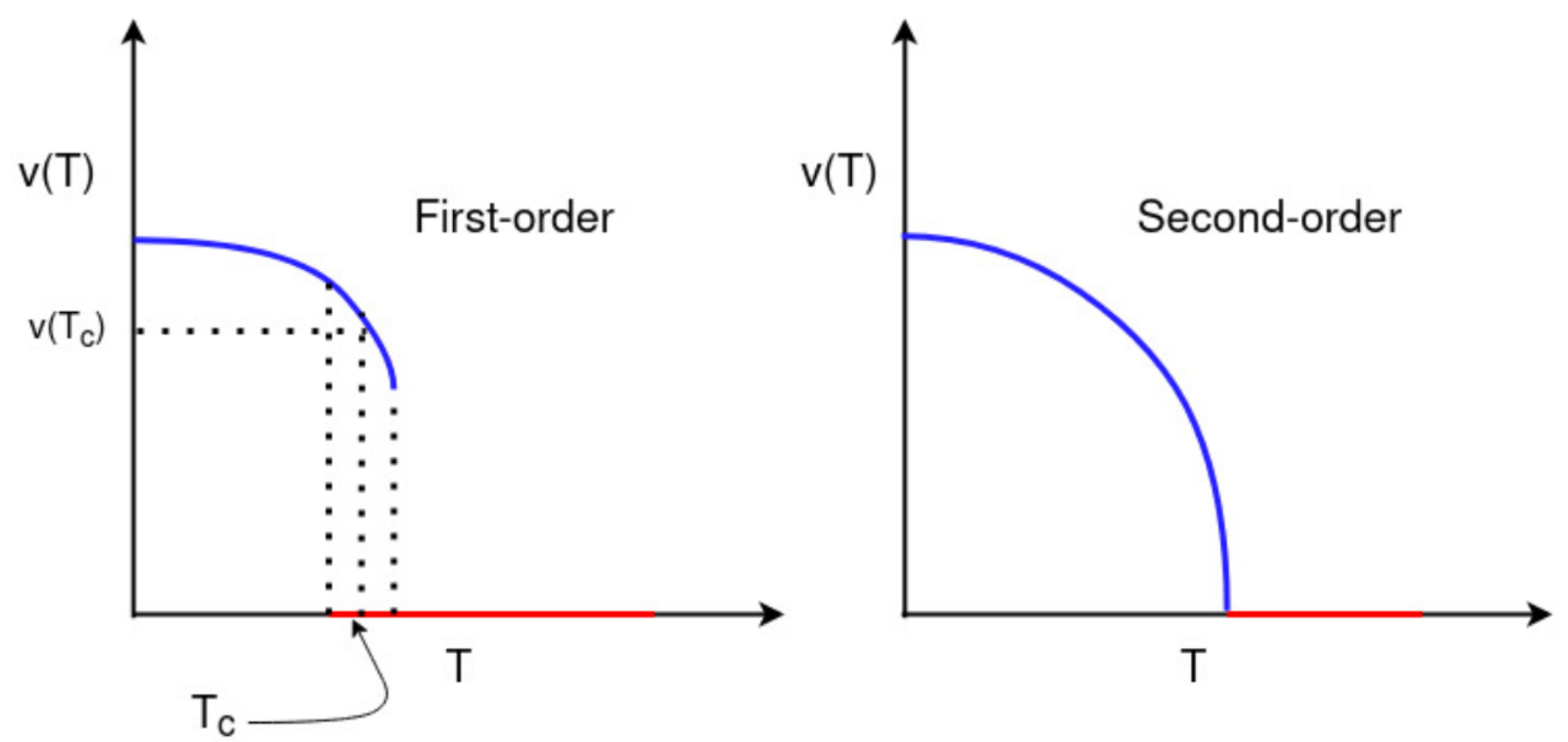}
\caption{\label{fig: vevs} Evolution of the \vev $v(T)$ in a first-order versus second-order PT.}
\end{figure}
Fig.\,\ref{fig: vevs} compares the evolution of the $\vev$ with temperature between a first-order and a second-order PT. Among the various possibilities for PT, FOPTs and the formation of certain topological defects are particularly interesting as these can produce a stochastic gravitational wave background. I discuss these in the sections below. 

\subsection{First-order phase transitions}
\begin{figure}[tbp]
\centering 
\includegraphics[width=.6\textwidth]{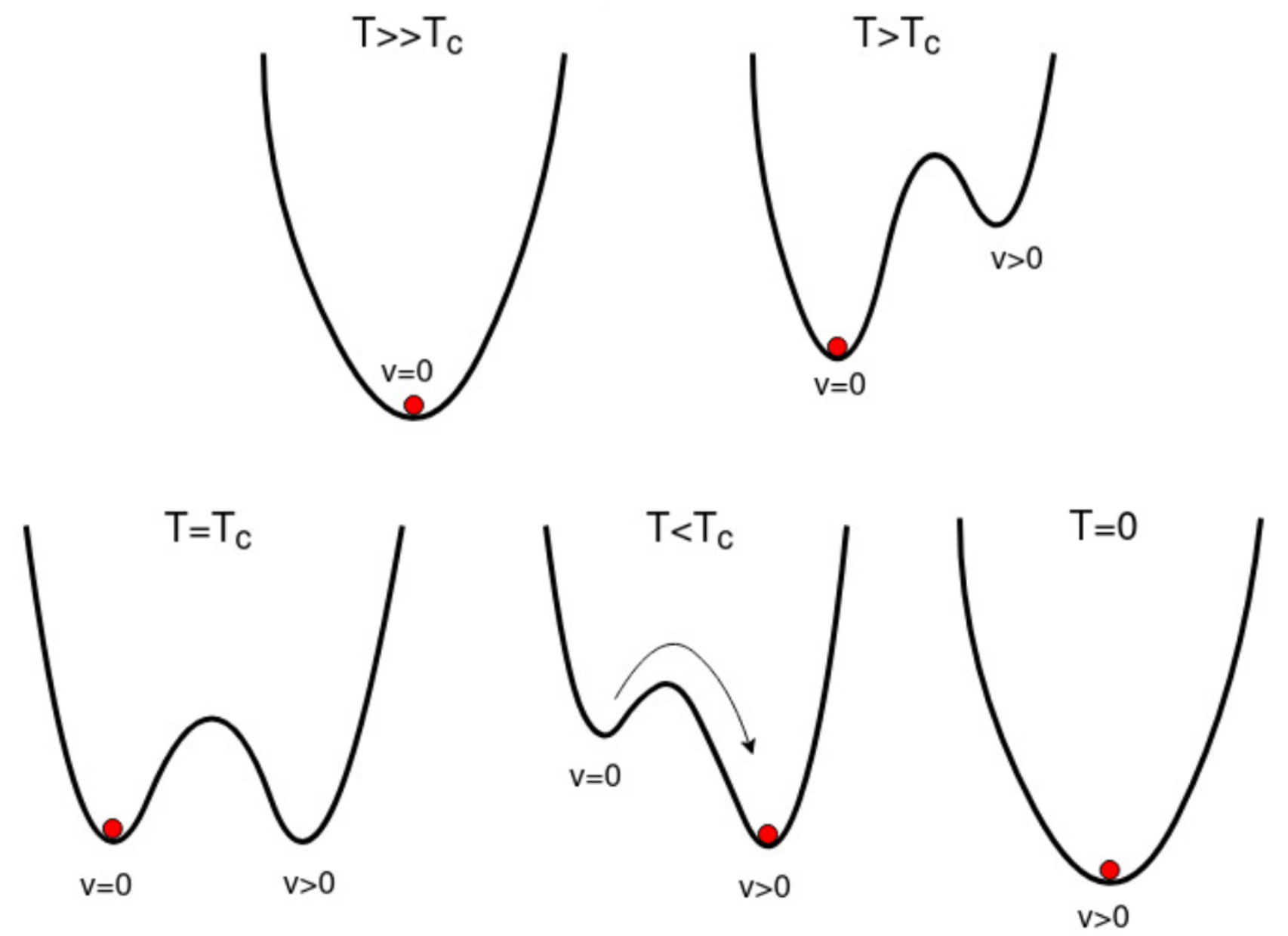}
\caption{\label{fig: FOPT_steps} Typical evolution of the effective potential in FOPT. The red dot represents the vacuum state at a given point in space.}
\end{figure}

A FOPT occurs when the effective potential features two (or more) minima separated by a barrier(s). The typical progression of events in a FOPT is shown in Fig.\,\ref{fig: FOPT_steps}. The effective potential has a single minimum at $\phi=0$ at very high temperatures, and the symmetry is restored.  As the universe expands and cools, a metastable minimum develops at some non-zero field value. During this time, the \vev remains zero since tunneling from the origin to the shallower minimum is energetically disallowed. The non-zero minimum deepens as the temperature drops and the two minima eventually become degenerate at the \textit{critical temperature}, $T_c$, where the vacuum can finally start tunneling via quantum mechanical or thermal effects. Below $T_c$, the $\phi=0$ minimum is metastable while the non-zero minimum becomes stable. The tunneling rate per unit volume increases with time and becomes comparable to the expansion rate, signaling the onset of phase transition. The tunneling rate can be found using semiclassical methods, first done for the zero temperature case by Callan and Coleman in Refs.\,\cite{PhysRevD.15.2929, Callan:1977pt}, and was extended to the finite temperature case by Linde in Ref.\,\cite{Linde:1980tt}. The discussion below on tunneling rates roughly follows Ref.\,\cite{Athron:2023xlk}.

\subsubsection{Tunneling rate in quantum mechanics}
First, let us consider a particle moving in a one-dimensional potential in quantum mechanics, with the Hamiltonian,
$$H(p,q) = \frac{p^2}{2m} + V(q)\,,$$
where $V(q)$ has a metastable minimum at $q=q_0$. This minimum is classically stable but is metastable in quantum mechanics due to the possibility of tunneling across the barrier. The amplitude to find the particle at the position $q=q_0$ at initial and final times $t_i$ and $t_f$, respectively is given as, 
\beq\label{eq: bounce1}
\langle q_0|e^{-\hat{H}(t_f-t_i)}| q_0\rangle = \int_{q(t_i) = q_0}^{q(t_f) = q_0} [dq]~ e^{iS[q]}\,,
\eeq 
where 
\beq 
S[q] = \int_{t_i}^{t_f} dt \left[\frac{1}{2}m\left(\frac{dq}{dt}\right)^2 - V(q)\right].
\eeq

\begin{figure}[tbp]
\centering 
\includegraphics[width=.8\textwidth]{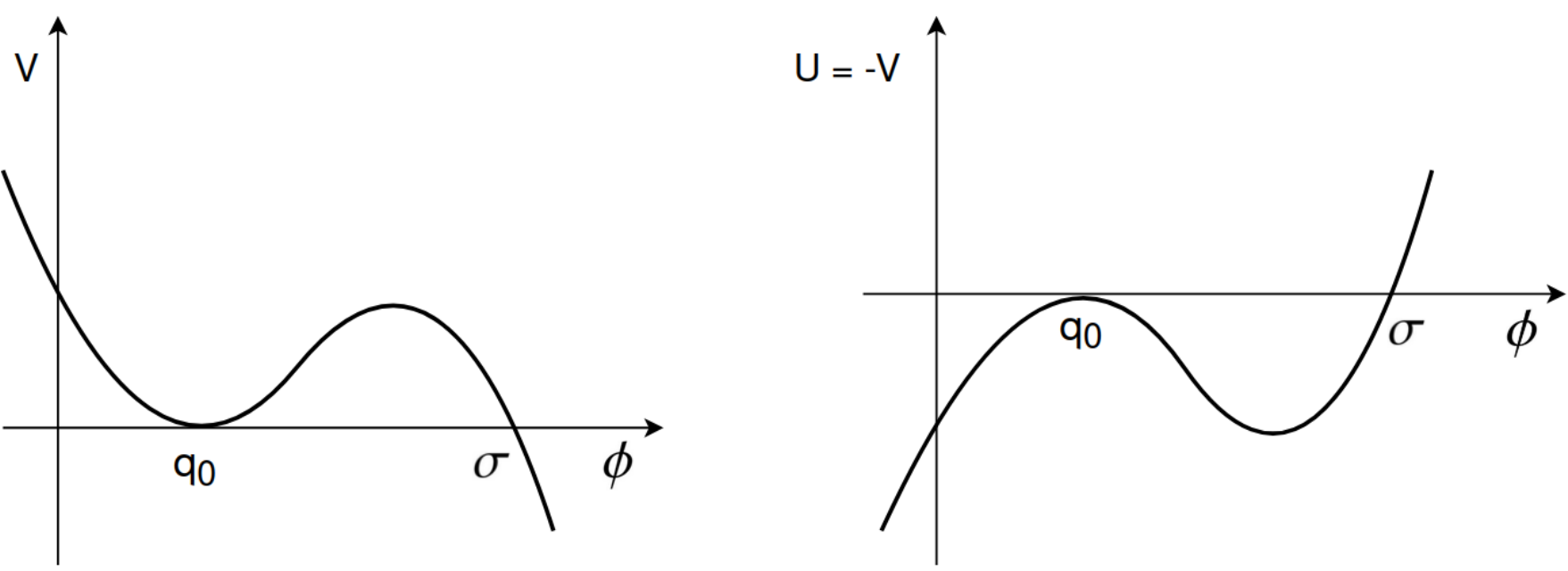}
\caption{\label{fig: tunnel} Left: Quantum mechanical potential with a metastable minimum at $q_0$. Right: The potential is inverted to solve the bounce equation of motion. }
\end{figure}

We want to eventually take the limit $t_i\rightarrow -\infty$ and $t_f\rightarrow +\infty$. After Wick rotation, $\tau = it$, expanding the Euclidean action $S_E$ around its minimum $S_E\simeq B + \delta S_E$, the RHS of Eq.\,\eqref{eq: bounce1} becomes, 
\beq\label{eq: bounce}
 \int_{q(\tau_i) = q_0}^{q(\tau_f) = q_0} [dq]~ e^{-S_E[q]} \approx \left[\int_{q(\tau_i) = q_0}^{q(\tau_f) = q_0} [dq]~ e^{-\delta S_E[q]}\right] e^{-B} \equiv C e^{-B}\,,
\eeq 
using the saddle point evaluation which assumes $S_E$ is large compared to $1$ (in natural units) i.e. semi-classical approximation. $B$ is the Euclidean action computed along the non-trivial path which extremizes $S_E$. This yields the equation of motion,
\beq \label{eq: eom_qm}
m\frac{d^2 q}{dt^2} = \frac{d V}{dq}\,,
\eeq 
with the boundary condition $q(t_i) = q(t_f) = q_0$. The above equation resembles the classical equation of motion of a particle moving in the inverted potential $V\rightarrow -V$, as shown in Fig.\,\ref{fig: tunnel}. The simplest non-trivial solution is the \textit{bounce} solution $q_b$ where the particle begins at $q_0$ in the distant past at $\tau=\tau_i$ with zero velocity, reaches the point $\sigma$ where $V(q_0) = V(\sigma)$ at $\tau = 0$, and returns to $q_0$ in the distant future at $\tau = \tau_f$. The prefactor $C$ depends on second-order variations of the action and is less important compared to the exponent $B$, which determines the order of magnitude for the tunneling amplitude.

In the limit $\mathcal{T}=\tau_f-\tau_i\rightarrow \infty$, the amplitude given in Eq.\,\eqref{eq: bounce1} depends only on the lowest energy state $E_0$ of the system 
\beq\label{eq: E0} 
\lim_{\mathcal{T}\rightarrow\infty}\langle q_0|e^{-\hat{H}\mathcal{T}}| q_0\rangle  = e^{-E_0\mathcal{T}} |\langle q_0|0\rangle|^2 = C e^{-B}
\eeq 
The decay rate $\Gamma$ is related to $E_0$ as, $\Gamma = -2 \,{\rm{Im}}\, E_0$, and has the form\,\cite{Andreassen:2016cvx} 
\beq \label{eq: decay}
\Gamma = |A|e^{-B}\,.
\eeq

\subsubsection{Tunneling rate in field theory}
In field theory, for a set of real scalar fields $\{\phi^a(\tau,\bf{x})\}$, the bounce equation of Eq.\,\eqref{eq: eom_qm} generalizes to a set of coupled partial differential equations
\beq \label{eq: eom_qft}
\left(\frac{\partial^2}{\partial\tau^2} + \nabla^2\right)\phi^a = \frac{\partial V}{\partial \phi^a}\,,
\eeq 
where $\nabla^2$ is the Laplacian over spatial coordinates. These equations are supplied with the boundary condition that the field starts in the false vacuum $(\phi=\phi_f)$ at $\tau\rightarrow -\infty$, turns around at $V=0$ at $\tau=0$, and returns to the false vacuum at $\tau\rightarrow +\infty$.

Assuming the bounce solution depends only on the combination $\rho=\sqrt{\tau^2+{\bf{x}}^2}$, Eq.\,\eqref{eq: eom_qft} becomes 
\beq \label{eq: bounce_rho}
\frac{\partial^2\phi^a}{\partial\rho^2} + \frac{3}{\rho}\frac{\partial\phi^a}{\partial\rho} = \frac{\partial V}{\partial \phi^a}\, ,
\eeq 
with the boundary conditions that the field starts at rest at $\rho=0$, and approaches $\phi_f$ as $\rho\rightarrow\infty$
$$\left.\frac{\partial{\phi^a}}{\partial\rho}\right\vert_{\rho = 0}\,,~~~\lim_{\rho\rightarrow\infty}\phi(\rho) = \phi_f\, .$$
Eq.\,\eqref{eq: bounce_rho} resembles a classical mechanics problem with $\rho$ as the time variable and $\phi$ as the position of a particle. It describes the motion of the particle moving in an inverted potential $-V$, with a damping term proportional to $1/\rho$. It was shown in Ref.\,\cite{PhysRevD.15.2929} that the existence of a bounce solution is guaranteed in the case of a single scalar field, using an overshoot/undershoot method. The $O(4)$ Euclidean bounce action can be written as
\beq 
S_4[\phi] = 2\pi^2\int d\rho~\rho^3\left[\frac{1}{2}\left(\frac{\partial\phi}{\partial\rho}\right)^2+V(\phi)\right]\, ,
\eeq 
where the factor $2\pi^2$ is the surface area of the unit 3-sphere. The bounce action can be analytically solved in the \textit{thin wall approximation}\,\cite{PhysRevD.15.2929}, where the potential is of the form
\beq 
V(\phi) = \frac{\l}{8}(\phi^2-a^2)^2 + \frac{\epsilon}{2a}(\phi-a)\,,
\eeq 
where $\epsilon$ is the potential difference between the two minima, assumed to be small. In this case, the bounce solution is
\beq \label{eq: thin}
\phi(\rho) \simeq \begin{cases}
    -a\,, & ~~\rho\gg R\,,\\
    \phi_1(\rho-R)\,, & ~~\rho\approx R\,,\\
    a\,, & ~~\rho\ll R\,.
\end{cases}
\eeq 
where $R$ is the value of $\rho$ at which the damping is small enough to get the field rolling, and $\phi_1$ is an odd function interpolating between the $\phi = \pm a$. The bounce action splits into a surface contribution and a bulk contribution. 

In physical space, the solution in Eq.\,\eqref{eq: thin} represents a bubble of radius $R$, with the true vacuum $\phi=a$ at its center, and the false vacuum $\phi=-a$ at infinity.  Analytically continuing the solution using $\tau = it$, the solution in real time is $\phi(t,{\bf{x}}) = \phi(\sqrt{t^2-|{\bf{x}}|^2})$. Since tunneling is a random process, when the space is filled with a metastable vacuum, the tunneling around a particular point in space manifests as the appearance of a small bubble containing the true vacuum. If the bubble is smaller than a critical radius $R_0$, the surface tension dominates and the bubble shrinks out of existence. If a bubble of size $\gtrsim R_0$ is formed at $t=0$, it expands rapidly according to the relation
\beq 
R^2(t) = R_0^2 + t^2\,.
\eeq 
Thus, at late times, the bubble wall moves close to the speed of light. 

Qualitatively similar behavior is shown by \textit{thick-walled} bubbles, which form when the potential difference between the two minima is significant. Ultimately, foregoing the details, the tunneling rate per unit volume in field theory at zero temperature is given by the expression\,\cite{Athron:2023xlk}
\beq 
\Gamma = R_0^{-4}\left(\frac{S_4[\phi]}{2\pi}\right)^2 e^{-S_4[\phi]}\,,
\eeq 
where $R_0$ is the size of the critical bubble.

At high temperature, the $O(4)$ symmetry of the bubble is lost, and instead, we have an $O(3)$ symmetric bubble, with the tunneling rate per unit volume given by \citep{Linde:1980tt,Kobakhidze_2017}
\beq \label{eq: tunnel_T}
\Gamma = \Gamma(T) =\left(\frac{S_3(T)}{2\pi T}\right)^{3/2} T^4 e^{-\frac{S_3(T)}{T}}\,,
\eeq 
where $S_3(T)$ is the $O(3)$ symmetric Euclidean bounce action. 

When analyzing FOPTs in different models, the tunneling rate is obtained by numerically solving the bounce equation and getting the Euclidean action, using software packages.  

If the friction due to the thermal plasma is negligible, the bubble walls can accelerate without any bounds and quickly approach the speed of light. This is called the \textit{runaway scenario}. On the other hand, if the scalar field is sufficiently coupled to the thermal plasma, it is called the \textit{non-runaway scenario}. In this case, the bubble walls encounter friction and reach a terminal wall velocity, $v_w$. The wall velocity could be sub-sonic, in which case the bubble expansion is called \textit{deflagration}, or supersonic, where it is called \textit{detonation}. These different cases have slightly different outcomes for the GW spectrum, as discussed in detail in Ref\,\cite{Caprini:2015zlo,Caprini:2018mtu,Caprini:2019egz}.

\subsubsection{Characterizing first-order phase transitions} \label{1_FOPT_params}
FOPTs produce a GW background, with the GW spectrum determined by the nature of FOPT. Therefore, we need to define quantities that characterize FOPTs. As mentioned earlier, at the critical temperature $T_c$ false vacuum and the true vacuum become degenerate. Below $T_c$, the tunneling rate per unit volume increases rapidly and may become comparable to the expansion rate of the universe at some point. In that case, the probability of nucleating a bubble in a Hubble volume becomes $\mathcal{O}(1)$. This temperature is called the \textit{nucleation temperature}, $T_n$, and is defined by the relation
\beq\label{eq: 1_nucl_criterion1}
\Gamma(T_n) \approx (H(T_n))^4\,, 
\eeq 
where $\Gamma$ is given by Eq.\,\eqref{eq: tunnel_T}. In the radiation-dominated era, for $g_*(T)\sim 100$, this gives\,\cite{Quiros:1999jp}
\begin{equation}\label{eq: 1_nucl_criterion2}
\frac{S_3(T_n)}{T_n} \simeq -4 \ln\left(\frac{\sqrt{8\pi}T_n}{M_p}\right),
\end{equation}
where $M_p$ is the reduced Planck mass. For the EWPT in SM, this condition simplifies to
\beq 
\frac{S_3(T_n)}{T_n} \approx 140\,.
\eeq 
Note that a solution of Eq.\,\eqref{eq: 1_nucl_criterion2} may not exist, or $T_n$ may turn out to be negative. In this case, the nucleation temperature is never achieved, indicating an incomplete FOPT. This situation generally occurs in \textit{supercooled phase transitions}, which involve many subtleties not relevant to this thesis. In the scenarios considered in this thesis, $T_n\simeq T_c$. 

An important physical quantity that characterizes the strength of FOPTs is the latent heat energy density, given by the energy density difference between the metastable and the stable phases\,\cite{Espinosa:2010hh}
\beq 
\rho_{\rm{vac}}(T) = \Delta \bigg(V-T\frac{\partial V}{\partial T}\bigg)\, ,
\eeq 
and is usually calculated at a temperature $T_*$ around which GW production is the most significant. In scenarios where $T_n\simeq T_c$, it is reasonable to take $T_*=T_n$. The latent heat density is usually normalized by the radiation energy density, $\rho_{\rm{rad}}(T) = \frac{\pi^2}{30}g_{*}T^4$, yielding a dimensionless parameter
\beq \label{eq: 1_alpha}
\a \equiv \left.\frac{\rho_{\rm{vac}}}{\rho_{\rm{rad}}}\right\vert_{T_*}\,.
\eeq 

Finally, we need the parameter $\beta$, which characterizes the rate at which the PT proceeds.
\beq \label{eq: 1_beta}
\b \equiv  -\left.\frac{dS}{dt}\right\vert_{t=t_*} = TH_*\left.\frac{dS}{dT}\right\vert_{T=T_*},
\eeq
where $S=S_3/T$, and $H_*$ is the Hubble parameter evaluated at $T=T_*$. The negative sign in the first equality is because the action decreases with time as $\Gamma(t)\sim e^{-\b t}$. The parameter $\b$ is usually normalized by $H_*$ to yield the dimensionless quantity $\b/H_*$.

\subsection{Topological defects}
Various types of topological defects may form during a PT, regardless of the order of the PT\,\cite{PhysRevD.26.435,Vachaspati:1997rr}. These include domain walls, cosmic strings, monopoles, and textures. Mathematically, the type of defect formed during SSB from $G\rightarrow H$, is governed by the topology of the vacuum manifold $G|H$. Unlike monopoles and textures, a network of domain walls or cosmic strings can produce a stochastic GW background. I briefly discuss these defects below.

\subsubsection{Domain walls}
Domain walls (DWs) are two-dimensional defects that form when the vacuum manifold is disconnected, as happens during the SSB of a discrete symmetry. 
\begin{figure}[tbp]
\centering 
\includegraphics[width=.8\textwidth]{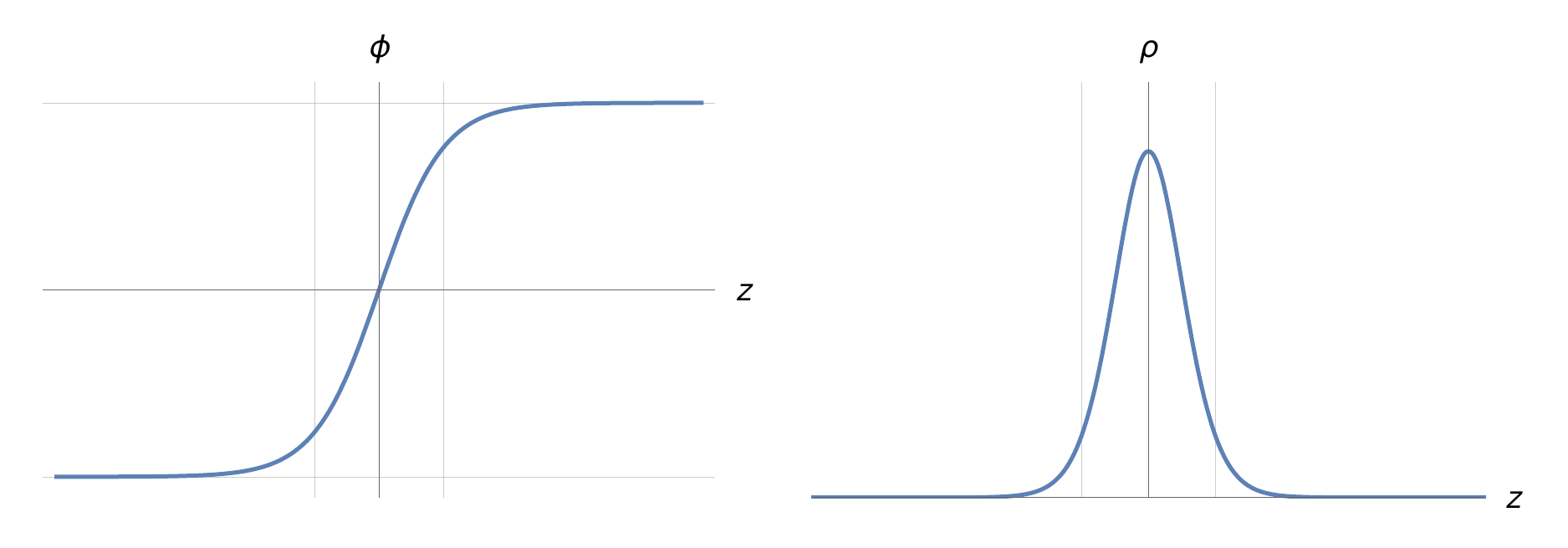}
\caption{\label{fig: dw} Left: The kink solution for the $Z_2$ DW. Right: Corresponding energy density. }
\end{figure}
The simplest example of DW formation is illustrated by a $Z_2$-symmetric potential for a real scalar field
\beq 
V(\phi) = \frac{\l}{4}(\phi^2-\eta^2)^2\,,
\eeq 
with $\l>0$. The potential has two degenerate minima at $\phi=\pm \eta$. Consider a field configuration with $\lim_{z\rightarrow -\infty}\phi(z) = -\eta$, and $\lim_{z\rightarrow +\infty}\phi(z) = +\eta$, then the classical equation of motion has a kink solution of the form 
\beq 
\phi(z) = \eta\, \tanh\left(\sqrt{\frac{\l}{2}}\eta \, z\right)\, ,
\eeq
as shown in the left panel of Fig.\,\ref{fig: dw}. The energy density is given by the `00' component of the energy-momentum tensor, $\rho(z) = \frac{1}{2}\left(\frac{d\phi}{dz}\right)^2 + V(\phi)$, is localized around $z=0$, as shown in the right panel of Fig.\,\ref{fig: dw}. 

At large distances, the kink solution appears as a membrane of energy, or a DW, located perpendicular to the $z$-axis at $z=0$, with a surface tension given by
\beq \label{eq: 1_sigmaDW}
\sigma = \int_{-\infty}^{\infty}\rho\,dz\propto \sqrt{\l}\,\eta^3\,.
\eeq
Akin to the $Z_2$ case, DWs can form in a PT if the potential has a set of disjoint, degenerate minima. When the universe cools from the symmetric phase at high temperature to the broken phase at low temperature, the field value at two distant points in space can fall into any of the minima with equal probability, leading to a network of DWs. However, DWs are severely constrained from cosmology and need to disappear before the onset of BBN. This can be done by introducing a bias in the potential, which breaks the degeneracy of the minima by a small amount. I discuss the formation and evolution of DWs in the doublet LRSM in Chapter\,\ref{chapter4}.

\subsubsection{Cosmic strings}
Cosmic strings (CSs) are one-dimensional defects formed when the vacuum manifold is not simply-connected, for example, during $U(1)$-breaking. The key parameter for CSs is the string tension $\mu$, or the energy per unit length. This is usually combined with Newton's constant to yield a dimensionless parameter $G\mu$. Unlike DWs, CSs are compatible with cosmology and can exist at late times. The fluctuations in CMB put an upper bound of $G\mu<10^{-6}$ on strings that can exist today. 

\section{Gravitational wave background}\label{chp1sec5}
Gravitational waves (GWs) are ripples in spacetime that are produced by time-dependent, non-spherical mass distributions. Like electromagnetic waves, GWs travel at the speed of light and can span a huge range of frequencies, depending on the production mechanism. While they were first predicted by Einstein over a century ago, the first direct detection of GWs from a binary black-hole merger by LIGO only happened in 2015\,\cite{Abbott_2016}. Unlike such mergers, which act as a foreground signal, stochastic GW backgrounds give rise to a time-independent spectrum, appearing as a noise in the detector that cannot be removed. In this section, I discuss the production and detection of such GW backgrounds, based on the detailed reviews\,\cite{Caprini:2018mtu,Athron:2023xlk}.  
\vspace{4mm}
%GWs are ripples in spacetime that move at the speed of light. They carry unique information about the processes that produced them, and therefore about the state of the universe at epochs and energy scales unreachable by any other means. The energy scales that GWs can probe extend far beyond the reach of presently available observational probes of the universe, mostly based on electromagnetic emission.
\subsection{Defining gravitational waves}
Consider a spacetime metric that adds a small perturbation to the Minkowski background metric, $\eta_{\mu\nu}\equiv\rm{diag}(+1,-1,-1,-1)$ 
\beq 
g_{\mu\nu}(x) = \eta_{\mu\nu} + h_{\mu\nu}(x)\,,~~~|h_{\mu\nu}(x)|\ll 1\, .
\eeq 
In other words, we are in the weak gravity limit, where terms of the order $h_{\mu\nu}^2$ or higher can be dropped, i.e. \textit{linearized gravity}.  Under infinitesimal general coordinate transformations, $x^{\mu}\rightarrow x'^{\mu} = x^{\mu}+ \xi^{\mu}(x)$, the perturbation $h_{\mu\nu}$ transforms as 
\begin{eqnarray}\label{eq:TensorChange}
h'_{\mu\nu}(x') = h_{\mu\nu}(x) - \partial_\mu\xi_\nu - \partial_\nu\xi_\mu\,.
\end{eqnarray}

We can write the affine connection $\Gamma^{\a}_{\mu\nu}$, the Riemann tensor $R^{\a}_{\mu\nu\b}$, the Ricci tensor $R_{\mu\nu}$, and the Ricci scalar $R$, to linear order in $h_{\mu\nu}$ as,
\bea
&&\hspace*{-1.2cm}\Gamma^\alpha_{~\,\mu\nu} \equiv {1\over2}g^{\alpha\beta}(\partial_\nu g_{\beta\mu} + \partial_\mu g_{\beta\nu} - \partial_\beta g_{\mu\nu}) \approx {1\over2}(\partial_\nu h^\alpha_{~\,\mu} + \partial_\mu h^\alpha_{~\,\nu} - \partial^\alpha h_{\mu\nu})\,,\\
&& \hspace*{-1cm} R^\alpha_{~\,\mu\nu\beta} = \partial_\nu\Gamma^\alpha_{~\,\mu\beta} - \partial_\beta\Gamma^\alpha_{~\,\mu\nu} \approx {1\over2}(\partial_\mu\partial_\nu h^\alpha_{~\,\beta} + \partial_\beta\partial^\alpha h_{\mu\nu} - \partial_\nu\partial^\alpha h_{\mu\beta} - \partial_\beta\partial_\mu h^\alpha_{~\,\nu})\,,\\
&& \hspace*{-1cm} R_{\mu\nu} \equiv -R^\alpha_{~\,\mu\nu\alpha} \approx {1\over2}(\partial_\nu \partial^\alpha h_{\alpha\mu} + \partial_\mu \partial^\alpha h_{\alpha\nu} - \partial_\mu\partial_\nu h - \Box h_{\mu\nu})\,,\\
&& \hspace*{-1cm} R = R^\mu_{~\,\mu} \approx (\partial^\alpha\partial^\beta h_{\alpha\beta}-\Box h)\,,
\eea
where $h \equiv h^{\a}_{~\a}$ is the trace of the metric perturbation, and the spacetime indices are raised and lowered using $\eta_{\mu\nu}$. Using these quantities, we can construct the Einstein tensor, to first order in $h_{\mu\nu}$
\bea\label{eq:EinsteinTensorLinTh}
\hspace*{-2cm} G_{\mu\nu} &\equiv& R_{\mu\nu} - {1\over2}\eta_{\mu\nu} R \nn\\
&=& {1\over2}(\partial_\nu \partial_\alpha h^\alpha_{\,~\mu} + \partial_\mu \partial^\alpha h_{\nu\alpha} - \partial_\mu\partial_\nu h - \Box h_{\mu\nu} - \eta_{\mu\nu}\partial^\alpha\partial_\beta h_\alpha^{~\beta}+\eta_{\mu\nu}\Box h) \nonumber\\
\hspace*{-2cm}  &=& \frac{1}{2}(\partial_\a \partial_\nu {\bar h}^\alpha_{~\,\mu} + \partial^\alpha \partial_\mu {\bar h}_{\nu\alpha} - \Box {\bar h}_{\mu\nu} - \eta_{\mu\nu}\partial_\alpha\partial^\beta {\bar h}^\alpha_{~\,\beta})\,,
\eea
where a new metric perturbation is introduced
\bea\label{eq:hTraceRev}
{\bar h}_{\mu\nu} \equiv h_{\mu\nu} - {1\over2}\eta_{\mu\nu}\,h\,.
\eea
Since ${\bar h} = -h$, the new metric perturbation is also called the trace-reversed perturbation. Under $x'^{\mu} \longrightarrow x^{\mu} + \xi^\mu$, $h_{\mu\nu}$ transforms as in Eq.~(\ref{eq:TensorChange}), whereas $\bar{h}_{\mu\nu}$ transforms as
\begin{eqnarray}\label{eq:TensorChangeTR}
{\bar h}'_{\mu\nu}(x') = {\bar h}_{\mu\nu}(x) + \xi_{\mu\nu}(x)\,,~~~ \xi_{\mu\nu}(x) \equiv \eta_{\mu\nu}\partial_\alpha\xi^\alpha - \partial_\mu\xi_\nu - \partial_\nu\xi_\mu\,.
\end{eqnarray}

We can always make an infinitesimal general coordinate transformation so that $\bar{h}_{\mu\nu}$ satisfies the \textit{Lorentz gauge} condition
\bea\label{eq:LorentzGauge}
\partial^\mu{\bar h}_{\mu\nu}(x) = 0\,.
\eea
This condition represents 4 constraints on ${\bar h}_{\mu\nu}$, reducing the number of independent degrees of freedom characterizing ${\bar h}_{\mu\nu}$ to $10-4=6$. However, there is still a remnant gauge freedom because one can make a transformation, ${\bar h}_{\mu\nu} \longrightarrow {\bar h}_{\mu\nu} + \xi_{\mu\nu}$, with $\xi_{\mu\nu} \equiv \eta_{\mu\nu}\partial^\alpha\xi_\alpha - \partial_\mu\xi_\nu - \partial_\nu\xi_\mu$. If the original ${\bar h}_{\mu\nu}$ satisfies Eq.\,\eqref{eq:LorentzGauge}, it will also be satisfied after the transformation provided $\Box\xi_{\mu\nu} = 0$. This happens if $\Box\xi_{\mu} = 0$. Using this, we can remove 4 more degrees of freedom. Thus $\bar{h}_{\mu\nu}$ has 2 independent degrees of freedom. 

Using Eq.\,\eqref{eq:EinsteinTensorLinTh} and Eq.\,\eqref{eq:LorentzGauge}, the linearized Einstein field equations in the Lorentz gauge, $G_{\mu\nu} = T_{\mu\nu}/M_p^2$, simplify to
\bea\label{eq:LinearizedEinsteinEqs}
\Box {\bar h}_{\mu\nu} = -{2\over M_p^2}T_{\mu\nu}\,,
\eea
where $M_p = (8\pi G)^{-1/2}$, is the reduced Planck's mass, and $T_{\mu\nu}$ is the energy-momentum tensor. The above equation is a wave equation with a source term, so the solutions are called gravitational waves. The general solution to the homogeneous wave equation can be written as the superposition of plane waves, 
\beq\label{eq:HomSolWaveEq}
{\bar h}_{\mu\nu}(x) = \int d^3k ~({\bar h}_{\mu\nu}({\bf k})e^{-ikx} + {\bar h}_{\mu\nu}^*({\bf k})e^{ikx})\,,
\eeq
where $kx \equiv k^\mu x_\mu = \omega(|{\bf k}|)t - {\bf k}\cdot {\bf x}$\,, $\omega(|{\bf k}|) = |{\bf k}|$\,, and ${\bar h}_{\mu\nu}({\bf k})$\,, are functions of the wave vector ${\bf k}$. The Lorentz gauge condition of Eq.\,\eqref{eq:LorentzGauge} implies that $k^\mu{\bar h}_{\mu\nu}({\bf k}) = 0$. 

\subsection{Gravitational waves in flat spacetime}
To study the propagation of GWs in empty space, let us restrict ourselves to a globally flat background, for which $T_{\mu\nu}=0$. The wave equation becomes
\beq 
\Box\bar{h}_{\mu\nu}(x) = 0\, .
\eeq 
The requirement of globally flat spacetime implies the condition: $\lim_{|\bf{x}|\rightarrow\infty}h_{\mu\nu}(x) = 0$.

Using Eq.~(\ref{eq:TensorChangeTR}), we can choose $\bar{h}_{\mu\nu}$ such that, ${\bar h} = {\bar h}_{0i} = 0$. By doing so, the normal and trace-reversed metric perturbations become equal ${\bar h}_{\mu\nu} = h_{\mu\nu}$. The Lorentz gauge condition implies ${\dot h}_{00} = -\partial_ih_{i0} = 0$\,, i.e. the `00' component which corresponds to the gravitational potential, is only a function of the spatial coordinates $h_{00} = V(\bf{x})$, so must vanish in flat spacetime. Thus we arrive at the \textit{tranverse traceless} (TT) gauge
\beq\label{eq:TTgauge}
h_{\mu 0} = 0\,,~~~~~~ h = h^i_{~\,i} = 0\,,~~~~~~\partial_ih_{ij} = 0\,,
\eeq
Consider a plane wave travelling in the $\hat{\eta} = \bf{k}/|\bf{k}|$ direction. We can fix $\hat{z} = \hat{\eta}$ so that in the TT gauge, only $h_{11}, h_{12}, h_{21}$ and $h_{22}$ are non-zero, with $h_{12} = h_{21}$, and $h_{11} = -h_{22}$. Thus we are left only with 2 independent components, which are denoted by $h_{\times} \equiv h_{12} = h_{21}$ and $h_+ \equiv h_{11} = - h_{22}$. These two components, $h_{\times}$ and $h_{+}$, correspond to the two physical polarization states of the GW. The perturbed line element, due to the passing of a GW, is given by
\beq
ds^2 = dt^2 - dz^2 - (1+h_{+})dx^2 - (1-h_{+})dy^2 - 2h_{\times}dxdy\,.
\eeq

The retarded wave solution to the linearized Einstein equations with the source term given in Eq.\,\eqref{eq:LinearizedEinsteinEqs} is given by 
\beq 
\bar{h}_{\mu\nu} = 4G\int d^3x' \frac{T_{\mu\nu}(t-|\bf{x} - \bf{x'}|,\bf{x'})}{|\bf{x} - \bf {x'}|}\, .
\eeq 
Far away from the source, $|\bf{x} - \bf {x'}|\approx |\bf{x}| = r$. Then $\bar{h}_{ij}$ is given by
\beq 
\bar{h}_{ij}(t,r) = \frac{2G\ddot{Q}_{ij}(t-r)}{r}\, ,
\eeq 
where $\ddot{Q}_{ij}$ is the second time derivative of the quadrupole moment $Q_{ij}$, given by
\beq 
Q_{ij} = \int d^3x'~ T_{00}(t,{\bf{x'}})~ x'_ix'_j \, .
\eeq 
GWs are therefore contributed only by second or higher-order multipoles in the multipole expansion, and need a time-dependent, non-spherical distribution of mass to be produced. The GWs carry energy, with the GW energy-momentum tensor in the TT gauge given by 
\beq\label{eq:GWenmontens}
	T^{\rm GW}_{\mu\nu}=\frac{\langle \nabla_\mu h_{\alpha \beta}\nabla_\nu h^{\alpha \beta} \rangle}{32\pi \,G}\,,
\eeq
where $\langle ...\rangle$ denotes the time average. From Eq.~\eqref{eq:GWenmontens}, the energy density of GWs is
\begin{equation}\label{eq:GWendensTT}
	\rho_{\rm GW}= T^{\rm GW}_{00}=\frac{\langle {\dot{h}}_{ij} \,{\dot{h}^{ij}} \rangle}{32\pi \,G}\,,
\end{equation}
For an asymptotically flat spacetime, $\dot{h}_{ij}$ denotes the derivative with respect to the time variable of the Minkowski metric. We can compute the luminosity $L$ by integrating $T^{\rm GW}_{00}$ over a closed surface enclosing the GW source\,\cite{Breitbach:2018kma}
\beq \label{eq: GW_luminosity}
L = \frac{G}{5}\langle {\dddot{Q}}^{\rm{TT}}_{ij} \,{\dddot{Q}_{\rm{TT}}^{ij}} \rangle\, .
\eeq 

\subsection{Gravitational waves in an expanding universe}

We can repeat the above analysis for GWs in a curved background. Write the metric as
\beq 
g_{\mu\nu} = \bar{g}_{\mu\nu} + \delta g_{\mu\nu}\, ,
\eeq 
where $\bar g_{\mu\nu}$ is the background metric, and the metric perturbation is $h_{\mu\nu} =\delta g_{\mu\nu}$. The trace-reversed metric now becomes $\bar{h}_{\mu\nu} =h_{\mu\nu} - \frac{1}{2}\bar g^{\alpha\beta}h_{\alpha\beta}$, and the Lorentz gauge condition becomes $\nabla^\mu \bar{h}_{\mu\nu} = 0$, where $\nabla$ is the covariant derivative in presence of $\bar{g}_{\mu\nu}$. The GW energy-momentum tensor is given by Eq.\,\eqref{eq:GWenmontens}. In the case of the FLRW metric, the calculation is significantly simplified due to its symmetries. For GWs in the FLRW background, the invariant interval can be written as
\beq
\label{GWcosmo}
ds^2 = dt^2 - a^2(t) \, (\delta_{ij} + h_{ij}) \, dx^i dx^j \, ,
\eeq  
where $h_{ij}$ are the tensor spatial perturbations, with
\beq
\label{TT}
\partial_i h_{ij} = h_{ii} = 0 \, .
\eeq

The TT conditions (\ref{TT}), leave two independent degrees of freedom, corresponding to the two GW polarization states. Introducing the conformal time $d\eta = dt / a(t)$, Eq.\eqref{GWcosmo} becomes
\beq 
\label{GWconf}
ds^2 = a^2(\eta) \, \left[ d\eta^2 - (\delta_{ij} + h_{ij}) \, dx^i dx^j \right] \, .
\eeq

The GW equation of motion is found using the linearized Einstein equations over the FLRW background, which lead to 
\beq
\label{gweqx}
\ddot{h}_{ij}(\mathbf{x}, t) + 3 \, H \, \dot{h}_{ij}(\mathbf{x}, t) -  
\frac{\mathbf{\nabla}^2}{a^2} \, h_{ij}(\mathbf{x}, t) = 16\pi G \, \Pi_{ij}^{TT}(\mathbf{x}, t)\,,
\eeq
where $\mathbf{\nabla}^2 = \partial_i \, \partial_i$, a dot denotes derivative with respect to $t$, $H = \dot{a} / a$ is the Hubble rate, and $\Pi_{ij}^{TT}$ is the TT part of the anisotropic stress. The anisotropic stress is given by
\beq
\label{Piij}
a^2 \, \Pi_{ij} = T_{ij} - P \, a^2 \, (\delta_{ij} + h_{ij})\,,
\eeq
where $T_{ij}$ denotes the spatial components of the energy-momentum tensor of the source, and $P$ is the background pressure. In terms of the conformal time, $\eta$, we can define
\beq 
H_{ij}(\mathbf{k}, \eta) = a\,h_{ij}(\mathbf{k}, \eta) \,,
\eeq 
then, in Fourier space, Eq.~(\ref{gweqx})  becomes
\beq 
\label{gweq2}
H_{ij}''(\mathbf{k}, \eta) + \left(k^2 - \frac{a''}{a}\right)\,H_{ij}(\mathbf{k}, \eta) = 
16\pi G \, a^3\,\Pi_{ij}^{TT}(\mathbf{k}, \eta) \,,
\eeq 
where primes denote derivatives with respect to $\eta$, and $k = |\mathbf{k}|$ is the comoving wave-number. The TT perturbation $h_{ij}$ can be decomposed into the two polarizations $r = +, \times$, as
\beq
\label{hrketa}
h_{ij}(\mathbf{x}, t) = \sum_{r = + , \times} \, \int \frac{d^3 \mathbf{k}}{(2 \pi)^{3}} \, h_r(\mathbf{k},t) \,
e^{- i \mathbf{k} \cdot \mathbf{x}} \, e_{ij}^{r}(\mathbf{\hat{k}})\, ,
\eeq
where the polarization tensors $e_{ij}^{r}(\mathbf{\hat{k}})$ are taken to be real and satisfy 
$e_{ij}^{r}(\mathbf{- \hat{k}}) = e_{ij}^{r}(\mathbf{\hat{k}})$. 

In the case where there is no source, $\Pi_{ij}^{TT}(\mathbf{x}, t)=0$, the solution for a generic scale factor obeying a power law $a(\eta)=a_n\eta^n$, eg. in the case of radiation ($n=1$) and matter ($n=2$) domination, can be obtained as
\begin{equation}
	h_r({\bf k},\eta)=\frac{A_r(\mathbf{k})}{a(\eta)}\eta\, j_{n-1}(k\eta)+\frac{B_r(\mathbf{k})}{a(\eta)}\eta\,y_{n-1}(k\eta)\,,
	\label{hsol_for_n}
\end{equation}
where $j_n(x), y_n(x)$ are the spherical Bessel functions, and $A_r(\mathbf{k})$ and $B_r(\mathbf{k})$ are dimensional constants, to be determined from the initial conditions. 

For a homogeneous and isotropic, unpolarized and Gaussian GW background, the Fourier amplitudes $h_r(\mathbf{k}, \eta)$ are considered to be random variables. The power spectrum can be written as
\beq
\label{powerspec}
\langle h_r(\mathbf{k}, \eta) \, h^*_{p}(\mathbf{q}, \eta) \rangle = \frac{8\pi^5}{k^3} \,
\delta^{(3)}(\mathbf{k} - \mathbf{q}) \, \delta_{r p} \, h_c^2(k, \eta)\,,
\eeq
where $h_c$ is dimensionless, real and depends only on the time $\eta$ and the comoving wave-number $k = |\mathbf{k}|$. From this we can get
\beq
\label{hcketa}
\langle h_{ij}(\mathbf{x}, \eta) \, h_{ij}(\mathbf{x}, \eta) \rangle = 2 \, \int_0^{+\infty} \frac{dk}{k} \, h_c^2(k, \eta)\,,
\eeq

The energy density in GWs is given by the 00-component of the energy-momentum tensor as seen in Eq.~\eqref{eq:GWendensTT}
\beq
\label{rhogw}
\rho_{\rm GW} \, = \, \frac{\langle \dot{h}_{ij}(\mathbf{x}, t) \, \dot{h}_{ij}(\mathbf{x}, t) \rangle}{32 \pi G} \,= \, 
\int_0^{+\infty} \frac{dk}{k}\,\frac{d \rho_{\rm GW}}{d \mathrm{log} k}\,,
\eeq

To connect with present-day GW observations, we must evaluate $\rho_{\rm{GW}}$ today, by accounting for the expansion of the universe. When redshifted to today, the present-day frequency corresponding to the comoving wave-number $k$, is $f = k / (2 \pi \, a_0)$, where the subscript `$0$' denotes the value at present time. The characteristic GW amplitude per logarithmic frequency interval today is defined as,
\beq
\label{hcf}
h_c(f) = h_c(k, \eta_0)\,.
\eeq
A stochastic background is often characterized by the spectral density
\beq
\label{Shf}
S_h(f) = \frac{h_c^2(f)}{2 f} \, ,
\eeq
with units of $\mathrm{Hz}^{-1}$. This quantity can be directly compared to the noise in a GW detector, as discussed in Sec.\,\ref{sec: interferometers} on interferometers. 

The GW spectrum is defined in terms of the GW energy density $\rho_{\rm{GW}}$, as \cite{Caprini:2015zlo},
\beq\label{eq: GW_spectrum}
\Omega_{\rm{GW}}(f)\equiv \frac{1}{\rho_c}\frac{d\rho_{\rm {GW}}}{d\ln f},
\eeq
where $f$ is the frequency and $\rho_c$ is the critical energy density of the universe, defined in Eq.\,\eqref{eq: critical density}.

\subsection{GWs from FOPTs and topological defects}
During a FOPT, when two bubbles collide and coalesce, the energy distribution around them is time-dependent and non-spherical. As a result, each coalescence generates tiny amounts of GWs according to Eq.\,\eqref{eq: GW_luminosity}. Since bubble nucleation is a stochastic process, the GW production is also stochastic and lasts for the entire duration of the PT. The latent heat of the colliding bubbles gets distributed into GWs via three main processes: (i) bubble wall collisions ($\Omega_{\rm{col}}$), (ii) sound waves produced in the thermal plasma ($\Omega_{\rm{sw}}$) following the collision, and (iii) the resulting turbulence ($\Omega_{\rm{turb}}$) generated in the plasma. The contributions can be added in the linear approximation:
\beq
h^2 \Omega_{\rm{GW}} \simeq h^2 \Omega_{\rm{col}}+h^2 \Omega_{\rm{sw}}+h^2 \Omega_{\rm{turb}}.
\eeq
Depending on the details of the PT, the contribution of each process can be important or unimportant. The full formulae are given in the subsequent chapters wherever needed. Each contribution is of the form 
\beq 
h^2\Omega_i(f) \propto v_w  \left(\frac{\k_i\a}{1+\a}\right)^{x_i} \times S_i(f)
\eeq 
where the efficiency factor $\k_i$ is the fraction of the latent heat going into the process $i\in\{\rm{col,~sw,~turb}\}$, $x_i$ is a positive exponent, and $S_i(f)$ is the spectral shape modeled as a broken power law with a peak frequency (redshifted to today) of the form
\beq 
f_i \propto \frac{1}{v_w}\left(\frac{\b}{H_*}\right) T_*\,.
\eeq 
The spectral indices for each process are obtained by numerical simulation of bubble collisions. 

In DWs and CSs, GWs are produced due to the vibrational modes of these defects. The spectrum from DWs is a broken power-law, with peak amplitude $\propto \sigma^2$, and peak frequency $\propto T_{\rm{ann}}$, while for CSs, the amplitude is $\propto (G\mu)^2$, and the spectrum is much flatter. We state the full GW spectrum wherever needed in the subsequent chapters.

\subsection{Detection}
The principle of detecting GWs is based on the property that they induce a quadrupolar deformation of spacetime. There are two broad categories of GW detectors; the man-made detectors are laser interferometers, while extremely low-frequency GWs can be inferred using pulsar timing arrays.

\subsubsection{Interferometers}\label{sec: interferometers}
Due to the quadrupolar nature of GWs, we can efficiently detect them using Michelson interferometers with freely suspended masses. When a linearly polarized GW passes through an interferometer perpendicular to its plane, with the polarization axes aligned with the arms of the detector, one arm contracts while the other arm expands. This creates a phase difference between the laser beams reflected from each arm, which changes with time when the wave passes, changing the interferometric pattern.

Consider an interferometer with two arms of equal length, $L$, lying along the $x$ and $y$ axes. The passing of a GW at time $t$ creates a strain $h(t)=(\Delta L_x(t)-\Delta L_y(t))/L$. In the TT gauge, this strain is related to the GW as $h(t)=F_+ h_+(t)+F_\times h_\times(t)$, where $F_{+,\times}(\hat\Omega, \psi)$ are the detector pattern functions which depend on the geometry of the system, and the relative orientation of GWs with the detector. $\hat\Omega$ is the direction of the incident GW, and $\psi$ is the angle of polarization with respect to the $xy$ axes in a plane perpendicular to $\hat{\Omega}$. Then the Fourier transform of the signal is \cite{Maggiore:1999vm}
\beq    
\tilde h(f)=\int d\hat\Omega \, [F_+(\hat\Omega, \psi) h_+(f,\hat\Omega)+F_\times (\hat\Omega, \psi) h_\times(f,\hat\Omega)]\,.
\eeq
For an unpolarized GW, the $\psi$ dependence cancels, and since the time average $\langle h(f)\rangle$
is zero, we can define the second moment of signal distribution as
\beq
\langle h^2(t)\rangle = F \int_0^\infty df \, S_h(f)\,, ~~~~~~~~~ F=\int \frac{d\hat\Omega}{4\pi}\,[F_+^2(\hat\Omega, \psi)+F_\times^2(\hat\Omega, \psi)] \,,
\eeq 
where $S_h(f)$ is the spectral density, defined earlier, and $F$ is the detector pattern function averaged over all possible directions of the GW. If the angle between the two arms is $\a$, then \cite{Maggiore:1999vm} $F=2/5\sin^2\a$. 

\begin{figure}[tbp]
\centering 
\includegraphics[width=.8\textwidth]{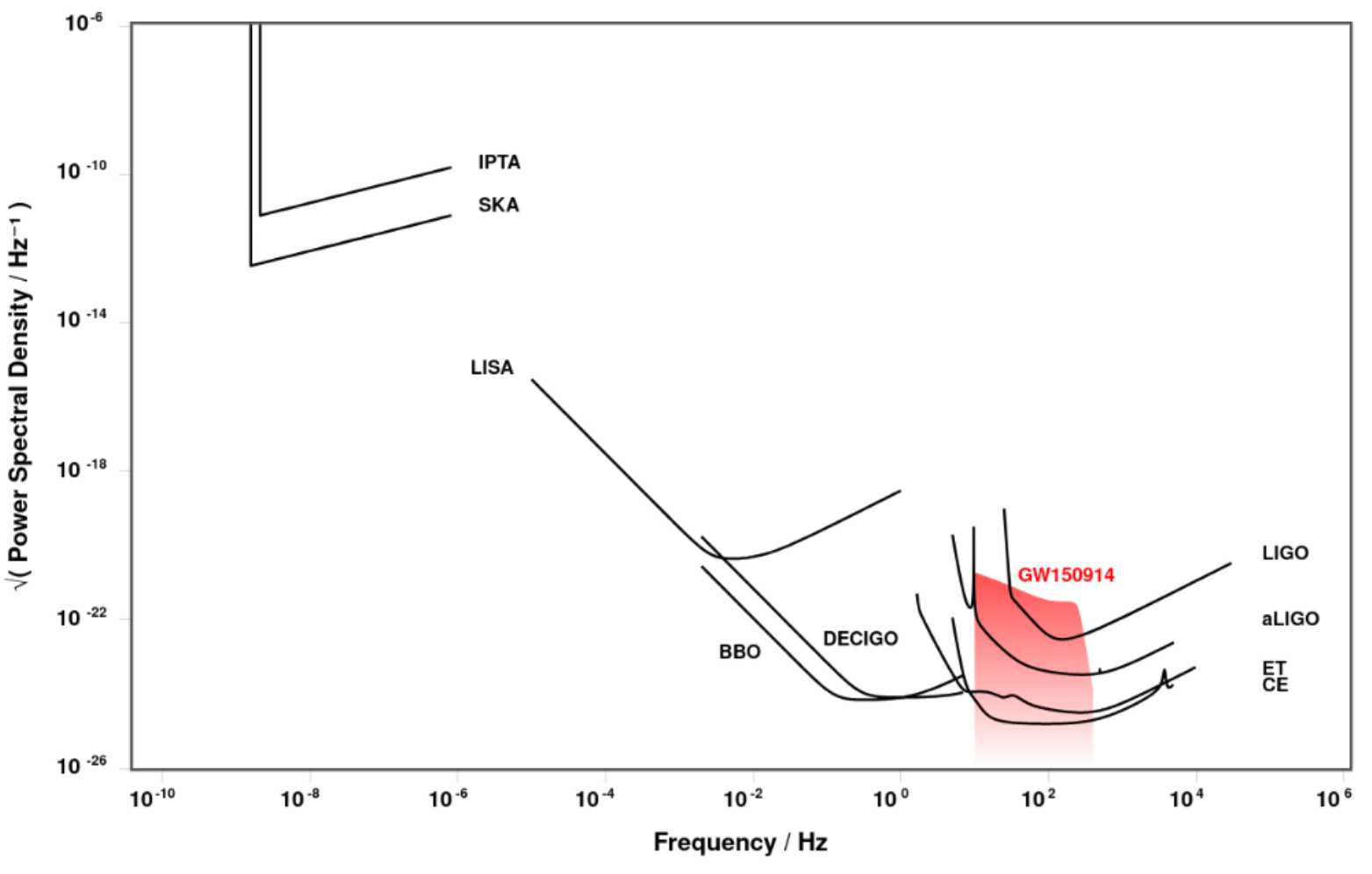}
\caption{\label{fig: gwplotter} Noise power spectral density for various GW detectors. The first GW detection by LIGO is shown in red. Credit: gwplotter.com }
\end{figure}

The total output of a detector is the GW signal plus the noise, $S(t)= h(t)+n(t)$. Analogous to the signal $h(t)$, we can define the noise spectral density $S_n(f)$ as, 
\begin{equation}
\langle \tilde n^*(f) \tilde n(f') \rangle =\delta(f-f') \frac{S_n(f)}{2}\,, ~~~~~~~~~ \langle n^2(t)\rangle =  \int_0^\infty df \, S_n(f)\,. 
\end{equation}
The level of noise in the detector is given by the strain sensitivity $h_n(f)=\sqrt{S_n(f)}$. Fig.\,\ref{fig: gwplotter} shows $h_n(f)$ for a set of GW detectors.

The prospect of detecting a GW signal in a given GW observatory can be quantified using the \textit{signal-to-noise ratio} (SNR), defined as \cite{Athron:2023xlk,Thrane:2013oya}
\beq\label{eq: 1_SNR}
    {\rm{SNR}} = \sqrt{n_{\rm{det}}\tau\int_{f_{\rm{min}}}^{f_{\rm{max}}} df \left[\frac{\Omega_{\rm{GW}}(f) h^2}{\Omega_{\rm{sens}}(f) h^2}\right]^2} ,
\eeq
where $\tau$ is the time period (in seconds) over which the detector is active, and the integration is carried out over the entire frequency range $[f_{\rm{min}},f_{\rm{max}}]$ of the detector. For calculations, we take $\tau = 1$ year. The factor $n_{\rm{det}}$ is two for experiments aimed at GW detection via cross-correlation measurement, or one for experiments aimed at detection via auto-correlation measurement (eg, LISA). $\Omega_{\rm{sens}}(f)$ is the noise energy density power spectrum for the chosen detector, related to $S_n(f)$ by\footnote{I would like to thank the reviewer of our paper\,\cite{Karmakar:2023ixo} for helping us understand the differences between the various sensitivity curves.}:
\beq 
h^2\Omega_{\rm{sens}}(f) = \frac{2\pi^2}{3 H_0^2} f^3 S_n\,.
\eeq 
A signal is detectable if the observed SNR value exceeds a threshold SNR, denoted as $\rm{SNR_{thres}}$. The value of $\rm{SNR_{thres}}$ varies from detector to detector. We take $\rm{SNR_{thres}} = 1$ for the purpose of discussion. Using $\rm{SNR_{thres}}$ and $\tau$, we can obtain \textit{power-law integrated sensitivity curves} (PLISCs) introduced in Ref.\,\cite{schmitz_2020_3689582}, which have a nice graphical interpretation: any power-law signal crossing the PLISC of a detector has an SNR$>\rm{SNR_{thres}}$. 

\subsubsection{Pulsar timing arrays}
A pulsar is a rotating neutron star with a strong magnetic field that emits two collimated beams of electromagnetic radiation along the direction of its magnetic axis. Since the magnetic axis does not necessarily coincide with the rotation axis, the two beams of electromagnetic radiation rotate like light beams from a lighthouse. Whenever a beam of a pulsar points toward the Earth, radio telescopes observe a tiny blip in the radio signal, which repeats at regular intervals coinciding with the period of the pulsar. The arrival times of the pulses can be predicted very accurately over long times. Especially millisecond pulsars, which have rotation periods of the order of a millisecond, have very small timing irregularities. 

Using pulsars as clocks is called \textit{pulsar timing}. In this process, first, a \textit{time of arrival} (TOA) is assigned to each pulse of a pulsar. The measured TOAs are compared with the predictions of a theoretical model that incorporates the various conditions that may affect the signal, such as the relative motion between the pulsar and Earth, the evolution of the pulsar’s rotation, the propagation of the pulses through
the interstellar medium, etc.\,. Next, the \textit{timing residuals}, or the differences between the best-fit model predictions and observed TOAs are obtained. There could be various factors that cause timing residuals. It was realized that the effect of a GW background on the timing residuals of various pulsars could be distinguished and isolated by looking at correlations among the timing residuals of different pulsars. An isotropic GW background causes different frequency shifts $\frac{\Delta\nu}{\nu}$, along different directions in the sky, giving rise to a unique spatial correlation, called the Hellings-Downs (HD) correlation 
\beq 
\left\langle\frac{\Delta\nu}{\nu}(\hat{\zeta})\frac{\Delta\nu}{\nu}(\hat{\xi})\right\rangle \propto x \, \mathrm{log} x - \frac{x}{6} + \frac{1}{3}
\hspace*{0.5cm} \mbox{with} \hspace*{0.5cm} 
\,x = \frac{1 - \mathrm{cos} \theta}{2}\,,
\eeq 
where $\hat{\zeta}$ and $\hat{\xi}$ are two unit vectors, separated by an angle $\theta$ in the sky. This particular dependence on $\theta$ is a unique signature of a stochastic GW background. Various collaborations of radio telescopes, called \textit{pulsar timing arrays} (PTAs) perform timing observations for several pulsars across the sky, for detecting a stochastic GW background. Some major PTAs are, the North American Nanohertz Observatory for Gravitational Waves (NANOGrav),\,\cite{NANOGrav:2023gor} the European Pulsar Timing Array (EPTA) \,\cite{EPTA:2023fyk}, Parkes Pulsar Timing Array (PPTA)\,\cite{Reardon:2023gzh}, and the Indian Pulsar Timing Array (InPTA), which are all part of the International Pulsar Timing Array (IPTA) consortium.

The GW background is characterized by its amplitude $h_c(f)$, given by
\beq   
h_c(f) = A_{\a} \, \left(\frac{f}{\mathrm{year}^{-1}}\right)^{\a} \,.
\eeq
The value of $\a$ depends on the specific process that produces the GW background. For example, in the case of SMBHBs, $\a=2/3$. 

\section{Thesis roadmap}
\vspace{4mm}
With the theoretical background in place, we can put together a plan to study the nature of PTs in different scenarios.  The first step is to choose a BSM model with scalar fields, that incorporates spontaneous breaking of the symmetry group in one or more steps. The model could have a plethora of motivations depending on the observations it aims to address. Next, we choose a point in the model parameter space allowed by theoretical and experimental constraints. For this point, the one-loop finite temperature effective potential $V_{\rm{eff}}(\phi,T)$ is constructed by employing the steps described in  Sec.\,\ref{1_effective_potential} and using the master equations Eq.\,\eqref{eq: parwani} and Eq.\,\eqref{eq: arnold_espinosa}. 

To locate the possible PTs in the theory, the phases i.e. the minima of $V_{\rm{eff}}$ are tracked as a function of the temperature. An abrupt jump in the phase is used to tag the PT as first-order. This is done using the Python-based software package \texttt{CosmoTransitions}\,\cite{Wainwright:2011kj}, which outputs a list of all possible phases, the order of each PT, and the energy density of each phase at PT. In the case of FOPT, the package computes $T_c$ and the tunneling action $S_3/T$ for the transition between two phases. This information is useful to compute the FOPT parameters $T_n$, $\a$, and $\frac{\b}{H_*}$, given in Eq.\,\eqref{eq: 1_nucl_criterion2}, Eq.\,\eqref{eq: 1_alpha}, and Eq.\,\eqref{eq: 1_beta} respectively. For DWs, the computation involves numerically solving the kink equations to obtain the DW profiles and the surface tension given in Eq.\,\eqref{eq: 1_sigmaDW}. 

\begin{figure}[tbp]
\centering 
\includegraphics[width=.9\textwidth]{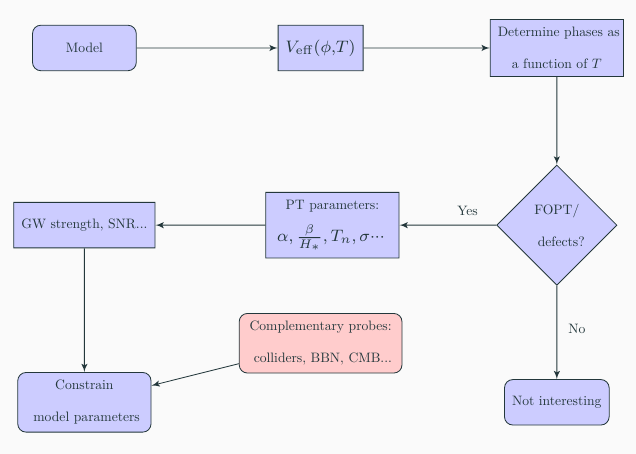}
\hfill
\caption{\label{fig: roadmap} Flowchart of the methodology adopted in this thesis.}
\end{figure}

Using the PT parameters, we can finally compute the GW spectrum $\Omega_{\rm{GW}}(f)$, and use the sensitivities of various GW detectors to find the SNR using Eq.\,\eqref{eq: 1_SNR}. By doing so, obtain constraints on the model parameter space from GWs. These constraints can be combined with other constraints from colliders and cosmology. This procedure is summarized in Fig.\,\ref{fig: roadmap}.

 %Introduction
\newpage
\null
\newpage
\chapter{First-order flavon phase transition}
\label{chapter2}
\linespread{0.1}
\graphicspath{{Chapter2/Figures/}}
\pagestyle{headings}
\hrule height 1mm
\vspace{4mm}
\begin{comment}
A major aspect of this thesis is analyzing pp collision data collected by the ALICE detector during Run-2 of the LHC. The ALICE detector is specially designed to analyze and study the properties of QGP. This chapter describes the ALICE apparatus, the tools and methodology for vertex and track reconstruction, and the performance of the sub-detectors relevant to the analyses presented in this thesis.

\vspace{4mm}
\noindent\hrule height 0.5mm 
\vspace{5mm}
\end{comment}
\section{Introduction}\label{chp2sec1}
\vspace{4mm}

The SM does not explain the hierarchy of fermion masses, or equivalently the hierarchy of Yukawa couplings, which range from $Y\sim 1$ for the top-quark, to $Y\sim 10^{-6}$ for the electron. Accounting for neutrino masses further enhances this hierarchy, making the problem much worse. Equally puzzling is the hierarchy of quark mixing angles, and the anarchy of leptonic mixing angles. This problem of Yukawa hierarchies has come to be known as the SM flavor puzzle \citep{Babu:2009fd,Grossman:2017thq,Nir:2020jtr}.

This chapter is based on Ref.\,\cite{Ringe:2022rjx}. In this chapter, we revisit the elegant solution to the SM flavor puzzle, originally proposed by Froggatt and Nielsen \cite{Froggatt:1978nt}, and focus on the quark sector. The observed hierarchy of quark masses and mixings  is generated by non-renormalizable operators involving  one or more complex scalar fields (flavons). The flavons spontaneously break the flavor symmetry imposed on the SM fields by acquiring a non-zero vacuum expectation value (\vev). A crucial parameter in FN models is the ratio $\epsilon = \frac{\langle S \rangle}{\Lambda}$, where $\langle S \rangle$ is the \vev of the flavon, and $\Lambda$ is the cutoff scale of the effective theory, associated with the mass of the messenger vector-like fermions (VLFs) in the UV completion. 

The UV completion of FN is a tedious task due to the involvement of several VLFs and is further complicated for realizations of FN with large $U(1)_{\fn}$ charge assignments. Some explicit UV constructions are found for example in \citep{Nir-Seiberg1,Nir-Seiberg2,Calibbi2012yj,delaVega:2021ugs}. While the method to determine the minimum number of vector-like fermions was first given in \cite{Nir-Seiberg1}, a simple procedure was given in \cite{Calibbi2012yj} to find the complete set of charge assignments. Using the procedure of \cite{Calibbi2012yj}, we construct minimal, non-supersymmetric, UV complete models for two simple realizations of the FN mechanism with one flavon. It should be noted however that adding several heavy fermions threatens the vacuum stability of the flavon potential, which demands the inclusion of additional bosons to restore the vacuum stability. 

Given an FN model, the ratio $\epsilon$ is fixed to some extent by the observed pattern of quark masses and mixing angles, whereas the scale $\Lambda$ itself is undetermined. The phenomenology of neutral meson mixing puts a lower bound of a few TeV on $\Lambda$ \cite{huntingflavon}. On the other hand, searches for vector-like fermions based on collider results \citep{Ellis:2014dza,Buchkremer:2012dn} and electroweak vacuum stability \citep{Blum:2015rpa,Gopalakrishna:2018uxn,Borah:2020nsz} put a bound of $\gtrsim1$ TeV. There are also bounds coming from processes such as the Higgs pair production at the LHC via flavons \cite{Arroyo-Urena:2022oft}. Planned future colliders such as the International Linear Collider (ILC) \citep{Behnke:2013xla,Baer:2013cma} and the Future Circular Collider (FCC) \citep{FCC:2018byv,FCC:2018evy,FCC:2018vvp} may push the lower bound on $\Lambda$ to $\sim 50$ TeV. Beyond that, the cutoff is practically unconstrained and could lie anywhere between $\sim 10$ TeV and the Planck scale $(\sim 10^{19}$ GeV). However, if the flavon undergoes a SFOPT, then the corresponding stochastic background may be detected by upcoming GW detectors, for symmetry-breaking scales as high as $10^4-10^7$ GeV. The corresponding GW signature has a peak frequency of $10^{-1}-10^2$ Hz, making it an ideal energy scale to be probed at observatories like BBO, DECIGO, ET, and CE. While several authors have discussed the GW background arising at such intermediate scales in different contexts \citep{Dev:2016feu,Addazi:2018nzm,Croon:2018kqn,Baldes:2018nel,Dev:2019njv,Greljo:2019xan,VonHarling:2019rgb,Huang:2020bbe,Chu:2022mjw,Ghoshal:2020vud,Dasgupta:2022isg,Ahmadvand:2021vxs,DiBari:2021dri,DiBari:2020bvn}, as far as we know, this is the first attempt to probe FN models incorporating UV completions, with GWs.

This chapter is organized as follows: in section \ref{sec:FN}, we review the FN mechanism and discuss two realizations involving a single flavon. In section \ref{sec: UV}, we construct the UV theory for the two FN models by giving the particle content involving heavy vector-like quarks. In section \ref{sec: potential}, we construct the one-loop finite temperature effective potential needed to analyze the flavon phase transition (PT). The issue of vacuum stability of FN UV completions is discussed in section \ref{sec: stability}. In section \ref{sec: 2_FOPT} we define the parameters characterizing a first-order phase transition (FOPT), and give the parameter scans showing regions of SFOPT in the two models considered. In section \ref{sec:GW}, we discuss the prospects of detecting the GW signature at future GW observatories. Finally, we give our concluding remarks in section \ref{sec: conclusion}.

\section{Froggatt-Nielsen mechanism}
\label{sec:FN}
In its simplest form, the FN mechanism is achieved by imposing a global horizontal $U(1)_{\fn}$ symmetry on the SM fields. This symmetry prohibits the inclusion of the usual Yukawa terms in the SM Lagrangian. Instead, we can write $U(1)_{\fn}$ preserving effective operators of the form:
\begin{equation}
  - \mathcal{L}_{\fn} = y_{ij}\left(\frac{S}{\Lambda}\right)^{n_{ij}} \overline{f}_{Li} H f_{Rj} + \rm{h.c.} 
\end{equation}
where $y_{ij}$ are $\mathcal{O}(1)$ couplings, $f_L$, $f_R$ respectively represent left-handed fermion doublets and right-handed fermion singlets of the SM, and $i,j$ denote the family index. $\Lambda$ is the cutoff scale, related to the mass scale of particles of the UV theory. $S$ is a complex scalar field called the flavon, and $H$ is the SM Higgs doublet. The FN mechanism works both in the lepton sector as well as the quark sector. Here we focus exclusively on the quark sector,
\begin{equation}\label{eq:FN}
-\mathcal{L}_{\fn} = y^u_{ij}\left(\frac{S}{\Lambda}\right)^{n^u_{ij}} \overline{Q}^i_{L} \widetilde{H}~u^j_{R} + y^d_{ij}\left(\frac{S}{\Lambda}\right)^{n^d_{ij}} \overline{Q}^i_{L} H~ d^{~j}_{R} + \rm{h.c.}
\end{equation} 
where $\widetilde{H} = i \sigma^2 H^*$. Without loss of generality, we can choose the FN charges of the scalars as: $Q_{\fn}(S) = -1$ and $Q_{\fn}(H) = 0$. The power of the $\frac{S}{\Lambda}$ term is: $$n^{u/d}_{ij} = Q_{\fn}(\overline{Q}_L^i)+Q_{\fn}(u_R^{~j}/d_R^{~j}).$$

When $U(1)_{\fn}$ is spontaneously broken by the non-zero \vev of the flavon, the FN operators give rise to the familiar SM Yukawa terms:
\begin{equation} \label{eq: yukSM}
-\mathcal{L}_{\rm{Y}} = Y^u_{ij}\overline{Q}^i_{L} \widetilde{H}~u^j_{R} + Y^d_{ij} \overline{Q}^i_{L} H~ d^{~j}_{R} + \rm{h.c.},
\end{equation}
where,
\begin{equation}\label{eq: yij}
Y^{u/d}_{ij} = y^{u/d}_{ij} \epsilon^{n^{u/d}_{ij}},
\end{equation}
and $\epsilon$ is defined as the ratio,
\begin{equation}
\epsilon \equiv \frac{\langle S \rangle}{\Lambda} = \frac{v_s}{\sqrt{2}\Lambda}
\end{equation}
For $\epsilon<1$, the desired hierarchy of quark masses is generated with appropriate charge assignments. Corresponding quark mass matrices are: $${\bf{\mathcal{M}^{u/d}}} = \frac{v_{\rm{EW}}}{\sqrt{2}}{\bf{Y^{u/d}}},$$ where  $v_{\rm{EW}} = 246.2$\,GeV, is the electroweak \vev.

It is useful to parameterize the hierarchy in terms of the Cabibbo angle: $\lambda=\sin\theta_{C}\sim 0.23$. The effective couplings $y^{u/d}_{ij}$ are all required to be $\mathcal{O}(1)$, but their exact value is unimportant, allowing some freedom in assigning charges to the quarks.

Below, we consider two simple examples that implement the FN mechanism with a single flavon. 

\begin{itemize}
\item \textbf{Model 1}: $\epsilon = \lambda = 0.23$ \cite{Binetruy:1996xk}
\begin{equation}\label{eq: model1}
Q_{\fn}(\overline{Q}_L)=(3,2,0),~Q_{\fn}(u_R)=(5,2,0),~Q_{\fn}(d_R)=(4,3,3)\,,
\end{equation}
where the FN charges of the three generations of $\overline{Q}_L,~u_L,$ and $u_R$ are given in parentheses.  The Yukawa matrices are:
\begin{gather}\label{eq: yuk_model1}
{\bf{Y^u}} \sim \begin{pmatrix}
\epsilon^8 & \epsilon^5 & \epsilon^3\\
\epsilon^7 & \epsilon^4 & \epsilon^2\\
\epsilon^5 & \epsilon^2 & 1
\end{pmatrix},~~~
{\bf{Y^d}} \sim \begin{pmatrix}
\epsilon^7 & \epsilon^6 & \epsilon^6\\
\epsilon^6 & \epsilon^5 & \epsilon^5\\
\epsilon^4 & \epsilon^3 & \epsilon^3
\end{pmatrix}, 
\end{gather}

where each entry of the matrices is multiplied by an $\mathcal{O}(1)$ factor $y^{u/d}_{ij}$. It is useful to note the $\epsilon$-dependence of the determinants:

\begin{equation}\label{det1}
\det {\bf{Y^u}} \propto \epsilon^{12},~~~\det {\bf{Y^d}} \propto \epsilon^{15}.
\end{equation}

\item \textbf{Model 2}: This is a non-supersymmetric variation of the charge assignment considered in \cite{Berkooz:2004kx}. $\epsilon = \lambda^2 = 0.05$, 
\begin{equation}\label{eq: model2}
Q_{\fn}(\overline{Q}_L)= Q_{\fn}(u_R)=(2,1,0),~Q_{\fn}(d_R)=(1,1,1).
\end{equation}
The Yukawa matrices are \footnote{Note that the matrix $\bf{Y^d}$ differs from the matrix in \cite{Berkooz:2004kx} by a factor of $\epsilon$. This is because of the presence of two Higgs doublets in supersymmetric theories, which can take care of the additional hierarchy.}:
\begin{gather}\label{eq: yuk_model2}
{\bf{Y^u}} \sim \begin{pmatrix}
\epsilon^4 & \epsilon^3 & \epsilon^2\\
\epsilon^3& \epsilon^2 & \epsilon\\
\epsilon^2 & \epsilon & 1
\end{pmatrix},~~~
{\bf{Y^d}} \sim \begin{pmatrix}
\epsilon^3 & \epsilon^3 & \epsilon^3\\
\epsilon^2 & \epsilon^2 & \epsilon^2\\
\epsilon & \epsilon & \epsilon
\end{pmatrix}.
\end{gather}
The determinants are:
\begin{equation}\label{det2}
\det {\bf{Y^u}} \propto \epsilon^6,~~~\det {\bf{Y^d}} \propto \epsilon^6.
\end{equation}
\end{itemize}

The FN mechanism can also be realized with more complicated abelian/non-abelian groups instead of $U(1)_{\fn}$, that may or may not be gauged. As a result, there can be models with more than one flavon field. See Appendix \ref{app: model3} for an example of a two-flavon model. As we'll see in the next section, lower powers of $\epsilon$ imply the need for fewer heavy quarks, and hence a simpler UV theory. 

\section{UV completion of FN models}\label{sec: UV}

\begin{figure}[tbp]
\centering % \begin{center}/\end{center} takes some additional vertical space
\includegraphics[width=.95\textwidth]{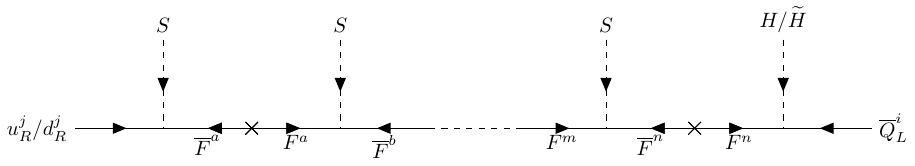}
\hfill
\caption{\label{fig: chain} A chain diagram of the UV theory, contributing to an FN operator corresponding to $Y^{u/d}_{ij}$. The number of insertions of the $S$ field is equal to the power of $\epsilon$ in $Y^{u/d}_{ij}$. Here $F$s represent the heavy VLQs sitting at the scale $\Lambda$. Yukawa couplings at the vertices are all $\mathcal{O}(1)$. Arrows indicate the flow of $U(1)_{\fn}$ charge, which is conserved at each vertex.}
\end{figure}

In the high-energy theory, the effective operators are listed in \eqref{eq:FN}. are generated from chain diagrams of the kind shown in Fig.\,\ref{fig: chain}. These diagrams are mediated by a set of heavy vector-like quarks (VLQs) with mass $M\sim \Lambda$. The heavy VLQs can be $SU(2)_L$ singlets or doublets. For simplicity, we take all the VLQs to be $SU(2)_L$ singlets, as it does not impact our ultimate goal of analyzing the flavon PT. We represent the chain diagram of Fig.\,\ref{fig: chain} using the shorthand notation,
\begin{equation*}
u^j/d^j-F^a-F^b-\cdots -F^m-F^n-\overline{Q}^i
\end{equation*}
The scalar field at each vertex is suppressed, and only the SM fermions and the mediating VLQs are shown. For $SU(2)_L$ singlet VLQs, it is assumed that an $H/\widetilde{H}$ field is attached to $\overline{Q}$, while an $S$ field is attached to each of the other vertices.

The singlet VLQs come in `up-type' or `down-type' representations of the SM gauge group, $\mathcal{G}_{\rm{SM}}=SU(3)_c\times SU(2)_L\times U_1(Y),$ depending upon the type of light quark they interact with. The notation used here is: 
$$(T_L, T_R)\sim ({\bf{3}, \bf{1}}, 4/3)~~({\rm{up\, type}}),$$
$$(B_L, B_R)\sim ({\bf{3}, \bf{1}},-2/3)~~({\rm{down\, type}}).$$ 

The UV Lagrangian contains fermion bilinears and Yukawa terms of the form:
\begin{equation}\label{eq: UV_yuk}
-\mathcal{L}_{\rm{UV}}\supset M\overline{T}^i_L T^i_R + \sum_{i,j}\kappa^{u}_{ij} S \overline{T}^i_L T^j_R  + \sum_{i,j}\gamma^u_{ij} \overline{Q}^i_L \widetilde{H} T^j_R + \sum_{i,j}\omega^d_{ij} S \overline{T}^i_L u^j_R + \rm{h.c.},
\end{equation}
where $\kappa$'s, $\gamma$'s, and $\omega$'s are $\mathcal{O}(1)$ Yukawa couplings, and we have taken the same bare mass $M$ for all VLQs. There are similar terms for the down-type quarks. The indices $\{i,j\}$ in each sum run over the respective number of quarks for each species. In addition to the above terms, we can also write down a Yukawa term for the top quark: $Y^u_{33}\overline{u}^3_L u^3_R + \rm{h.c.}$, since it is allowed by the $U(1)_{\fn}$ symmetry. After SSB, the fermionic mass Lagrangian takes the form,
\begin{equation}
-\mathcal{L}_{\rm{UV}} \supset \overline{\f^u}_L \mathcal{M}^u \f^u_R + \overline{\f^d}_L \mathcal{M}^d \f^d_R + \rm{h.c.},
\end{equation}
where $\f^u_L = \big(Q^1,Q^2,Q^3, T^1,T^2\cdots\big)_L^T$, 
$\f^u_R = \big(u,c,t, T^1,T^2\cdots\big)_R^T$, 
$\f^d_L = \big(Q^1,Q^2,Q^3, B^1,B^2\cdots\big)_L^T$, $\f^d_R = \big(d,s,b, B^1,B^2\cdots\big)_R^T$. 

The mass matrices $\mathcal{M}^{u/d}$ depend on the specific model. Physical fermion masses are given by the singular values of $\mathcal{M}^{u/d}$, i.e. eigenvalues of $\big(\mathcal{M}^{\dagger}\mathcal{M}\big)^{1/2}$. In \citep{Nir-Seiberg1,Nir-Seiberg2}, it was shown that for a Yukawa matrix $\textbf{Y}$ of the SM fermions with $\det\textbf{Y}\propto \epsilon^n$, at least $n$ VLQs are required in the UV theory. We now follow the procedure elucidated in \cite{Calibbi2012yj} to assign charges to the VLQs and build the particle content for our UV models.  

\subsection{Model 1}
According to \eqref{det1}, at least $12$ u-type VLQs and $15$ d-type VLQs are needed. For the up sector, first we select three elements from the $\textbf{Y}^u$ matrix in \eqref{eq: yuk_model1}, which contribute to the determinant \footnote{Every term in the determinant comes with the same power of $\epsilon$. This is a general feature of choosing a $U(1)$ symmetry.}. For each element, we form a chain diagram with the number of VLQs equal to the corresponding power of $\epsilon$. Here we choose, $Y^u_{11}\propto \epsilon^8$, $Y^u_{22}\propto \epsilon^4$, and $Y^u_{33}\propto \epsilon^0$. Next, we use a separate set of VLQs to construct each chain. The $U(1)_{\fn}$ charge assignment is determined by conserving the charge at every vertex.
\begin{eqnarray}
&u_R^1-T^8-T^7-T^6-T^5-T^4-T^3-T^2-T^1-\overline{Q}_L^1& \nonumber \\
&u_R^2-T'^4-T'^3-T'^2-T'^1-\overline{Q}_L^2& \nonumber
\end{eqnarray}

\noindent Similarly, for the down sector, we have: $Y^d_{11}\propto \epsilon^7$, $Y^d_{22}\propto \epsilon^5$, and $Y^d_{33}\propto \epsilon^3$. The chains are, 
\begin{eqnarray}
&d_R^1-B^7-B^6-B^5-B^4-B^3-B^2-B^1-\overline{Q}_L^1& \nonumber \\
&d_R^2-B'^5-B'^4-B'^3-B'^2-B'^1-\overline{Q}_L^2& \nonumber \\
&d_R^3-B''^3-B''^2-B''^1-\overline{Q}_L^3& \nonumber
\end{eqnarray}

\begin{table}[tbp]
\small
\begin{center}
\vspace{0.5 cm}
\begin{tabular}{|c|c c c c c c c c|}
\hline
 $Q_{\fn}$ & -3    & -2     & -1      & 0      & 1      & 2      & 3       & 4    \\ 
\hline
up-type    & $T^1$ & $T^2$  & $T^3$   & $T^4$  & $T^5$  & $T^6$  & $T^7$   & $T^8$\\  
           &       & $T'^1$ & $T'^2$  & $T'^3$ & $T'^4$ &        &         &      \\    
\hline
down-type  & $B^1$ & $B^2$  & $B^3$   & $B^4$  & $B^5$  & $B^6$  & $B^7$   &      \\  
           &       & $B'^1$ & $B'^2$  & $B'^3$ & $B'^4$ & $B'^5$ &         &      \\
           &       &        &         & $B''^1$& $B''^2$& $B''^3$&         &      \\
\hline
\end{tabular}
\end{center}
\caption{$U(1)_{\fn}$ charges for VLQs in model 1. A total $12+15=27$ heavy quarks are needed in this model. \label{table:table1}
}
\end{table}

\noindent Chains for the rest of the terms can now be written using the VLQs found above. The $U(1)_{\fn}$ charge assignments for the VLQs are displayed in table \ref{table:table1}. The resulting mass matrices are,

\begin{subequations}\label{subeq:model1}
\begin{equation}
\mathcal{M}^u = \begin{psmallmatrix}
0& 0& 0& \hu& 0& 0& 0& 0& 0 & 0 & 0 & 0 & 0 & 0 & 0\\
0& 0& 0& 0& \hu& \hu& 0& 0& 0 & 0 & 0 & 0 & 0 & 0 & 0\\
0& 0& \hu& 0& 0& 0& 0& 0& \hu & \hu & 0 & 0 & 0 & 0 & 0\\
0& 0& 0& M& \vs& \vs& 0& 0& 0 & 0 & 0 & 0 & 0 & 0 & 0\\
0& 0& 0& 0& M& 0& \vs& \vs& 0& 0 & 0 & 0 & 0 & 0 & 0\\
0& 0& 0& 0& 0& M& \vs& \vs& 0& 0 & 0 & 0 & 0 & 0 & 0\\
0& 0& \vs& 0& 0& 0& M& 0& \vs& \vs & 0 & 0 & 0 & 0 & 0\\
0& 0& \vs& 0& 0& 0& 0& M& \vs& \vs & 0 & 0 & 0 & 0 & 0\\
0& 0& 0& 0& 0& 0& 0& 0& M& 0 & \vs & \vs & 0 & 0 & 0\\
0& 0& 0& 0& 0& 0& 0& 0& 0& M & \vs & \vs & 0 & 0 & 0\\
0& \vs& 0& 0& 0& 0& 0& 0& 0& 0 & M & 0 & \vs & 0 & 0\\
0& \vs& 0& 0& 0& 0& 0& 0& 0& 0 & 0 & M & \vs & 0 & 0\\
0& 0& 0& 0& 0& 0& 0& 0& 0& 0 & 0 & 0 & M & \vs & 0\\
0& 0& 0& 0& 0& 0& 0& 0& 0& 0 & 0 & 0 & 0 & M & \vs\\
\vs& 0& 0& 0& 0& 0& 0& 0& 0& 0 & 0 & 0 & 0 & 0 & M\\
\end{psmallmatrix},
\end{equation}
\begin{equation}
\mathcal{M}^d = \begin{psmallmatrix}
0& 0& 0& \h& 0& 0& 0& 0& 0& 0& 0& 0& 0& 0& 0& 0& 0& 0\\
0& 0& 0& 0& \h& \h& 0& 0& 0& 0& 0& 0& 0& 0& 0& 0& 0& 0\\
0& 0& 0& 0& 0& 0& 0& 0& \h& \h& \h& 0& 0& 0& 0& 0& 0& 0\\
0& 0& 0& M& \vs& \vs& 0& 0& 0& 0& 0& 0& 0& 0& 0& 0& 0& 0\\
0& 0& 0& 0& M& 0& \vs& \vs& 0& 0& 0& 0& 0& 0& 0& 0& 0& 0\\
0& 0& 0& 0& 0& M& \vs& \vs& 0& 0& 0& 0& 0& 0& 0& 0& 0& 0\\
0& 0& 0& 0& 0& 0& M& 0& \vs& \vs& \vs& 0& 0& 0& 0& 0& 0& 0\\
0& 0& 0& 0& 0& 0& 0& M& \vs& \vs& \vs& 0& 0& 0& 0& 0& 0& 0\\
0& 0& 0& 0& 0& 0& 0& 0& M& 0& 0& \vs& \vs& \vs& 0& 0& 0& 0\\
0& 0& 0& 0& 0& 0& 0& 0& 0& M& 0& \vs& \vs& \vs& 0& 0& 0& 0\\
0& 0& 0& 0& 0& 0& 0& 0& 0& 0& M& \vs& \vs& \vs& 0& 0& 0& 0\\
0& 0& 0& 0& 0& 0& 0& 0& 0& 0& 0& M& 0& 0& \vs& \vs& \vs& 0\\
0& 0& 0& 0& 0& 0& 0& 0& 0& 0& 0& 0& M& 0& \vs& \vs& \vs& 0\\
0& 0& 0& 0& 0& 0& 0& 0& 0& 0& 0& 0& 0& M& \vs& \vs& \vs& 0\\
0& \vs& \vs& 0& 0& 0& 0& 0& 0& 0& 0& 0& 0& 0& M& 0& 0& \vs\\
0& \vs& \vs& 0& 0& 0& 0& 0& 0& 0& 0& 0& 0& 0& 0& M& 0& \vs\\
0& \vs& \vs& 0& 0& 0& 0& 0& 0& 0& 0& 0& 0& 0& 0& 0& M& \vs\\
\vs& 0& 0& 0& 0& 0& 0& 0& 0& 0& 0& 0& 0& 0& 0& 0& 0& M
\end{psmallmatrix}.
\end{equation}
\end{subequations}
The first three rows and columns of each matrix correspond to the light quarks in UV theory, and the VLQs are arranged in order of increasing $Q_{\fn}$. We have suppressed the $\mathcal{O}(1)$ Yukawa couplings given in \eqref{eq: UV_yuk}, which accompany the elements with \vevs.

\subsection{Model 2}
According to \eqref{det2}, at least $6$ u-type VLQs and $6$ d-type VLQs are needed. Repeating the procedure described above, for the up sector we have: $Y^u_{11}\propto \epsilon^4$, $Y^u_{22}\propto \epsilon^2$, and $Y^u_{33}\propto\epsilon^0$. Therefore we have the chains,
\begin{eqnarray}
&u_R^1-T^4-T^3-T^2-T^1-\overline{Q}_L^1& \nonumber \\
&u_R^2-T'^2-T'^1-\overline{Q}_L^2& \nonumber
\end{eqnarray}

\noindent For the down sector, we have: $Y^d_{11}\propto \epsilon^3$, $Y^d_{22}\propto\epsilon^2$, and $Y^d_{33}\propto\epsilon^1$. This gives,
\begin{eqnarray}
&d_R^1-B^3-B^2-B^1-\overline{Q}_L^1& \nonumber \\
&d_R^2-B'^2-B'^1-\overline{Q}_L^2& \nonumber \\
&d_R^3-B''^1-\overline{Q}_L^3& \nonumber
\end{eqnarray}

\begin{table}[tbp]
\small
\begin{center}
\vspace{.5 cm}
\begin{tabular}{|c|c c c c|}
\hline
 $Q_{\fn}$ & -2    & -1     & 0       & 1      \\ 
\hline
up-type    & $T^1$ & $T^2$  & $T^3$   & $T^4$  \\  
           &       & $T'^1$ & $T'^2$  &        \\    
\hline
down-type  & $B^1$ & $B^2$  & $B^3$   &        \\  
           &       & $B'^1$ & $B'^2$  &        \\
           &       &        & $B''^1$ &        \\
\hline
\end{tabular}
\end{center}
\caption{$U(1)_{\fn}$ charges for VLQs in model 2. A total $6+6=27$ heavy quarks are needed in this model. \label{table:table2}}
\end{table}

\noindent The $U(1)_{\fn}$ charge assignments are listed in table \ref{table:table2}. Clearly, model 2 has a simpler particle content than model 1, although this advantage comes at the expense of a larger scale separation between $v_s$ and $M$. 
The mass matrices are,
\begin{subequations}\label{subeq:model2}
\begin{equation}
\mathcal{M}^u = 
\begin{psmallmatrix}
0& 0& 0& \hu& 0& 0& 0& 0& 0\\
0& 0& 0& 0& \hu& \hu& 0& 0& 0\\
0& 0& \hu& 0& 0& 0& \hu& \hu& 0\\
0& 0& 0& M& \vs& \vs& 0& 0& 0\\
0& 0& \vs& 0& M& 0& \vs& \vs& 0\\
0& 0& \vs& 0& 0& M& \vs& \vs& 0\\
0& \vs& 0& 0& 0& 0& M& 0& \vs\\
0& \vs& 0& 0& 0& 0& 0& M& \vs\\
\vs& 0& 0& 0& 0& 0& 0& 0& M\\
\end{psmallmatrix},
\end{equation}

\begin{equation}
\mathcal{M}^d = 
\begin{psmallmatrix}
0& 0& 0& \h& 0& 0& 0& 0& 0\\
0& 0& 0& 0& \h& \h& 0& 0& 0\\
0& 0& 0& 0& 0& 0& \h& \h& \h\\
0& 0& 0& M& \vs& \vs& 0& 0& 0\\
0& 0& 0& 0& M& 0& \vs& \vs& \vs\\
0& 0& 0& 0& 0& M& \vs& \vs& \vs\\
\vs& \vs& \vs& 0& 0& 0& M& 0& 0\\
\vs& \vs& \vs& 0& 0& 0& 0& M& 0\\
\vs& \vs& \vs& 0& 0& 0& 0& 0& M\\
\end{psmallmatrix}
\end{equation}
\end{subequations}
 As for model 1, the Yukawa couplings have been suppressed. 
 
\subsection{Couplings of the UV theory}
Let us denote the Yukawa couplings of the UV theory given in \eqref{eq: UV_yuk} collectively by $\eta^{u/d}_{ij}$. These are related to the $\mathcal{O}(1)$ couplings $y^{u/d}_{ij}$ described in \eqref{eq: yij} by relations of the form,
\begin{equation}\label{eq: yuk_highlow}
y^{u/d}_{ij} \sim \prod\eta^{u/d}_{kl},
\end{equation}
where $\eta^{u/d}_{kl}$ are the relevant couplings appearing in the chain diagrams corresponding to $y^{u/d}_{ij}$. Although these couplings are $\mathcal{O}(1)$, their exact value is unimportant for our purpose. The simplest possible choice is to keep all the Yukawas unity. Another way is to choose them from a uniform distribution between, say, $[1/3,3]$. Here we take all Yukawas to be equal and parameterize them as,
\begin{equation} \label{eq:scaling}
\eta^{u/d}_{ij} = y,
\end{equation}
where $y$, chosen to be a real number, is a scaling factor, which we treat as a free parameter characterizing the overall Yukawa strength. This simplifying assumption is reasonable, and the factor $y$ will be useful for bookkeeping. In the limit $y=0$, the VLQs completely decouple from the scalar sector, while the perturbativity of Yukawas requires that $|y| \leq \sqrt{4\pi}$. We make a further simplification by choosing $y\geq 0$. 

\section{Effective potential}\label{sec: potential}
To study the flavon PT, we need to construct the finite-temperature effective potential described in Sec.\,\ref{1_effective_potential}. 
\begin{comment}
At zero temperature, the effective potential for a scalar field $\phi$ can be written in a loop expansion,
\begin{equation}
V(\phi)|_{T=0} = V_0(\phi) + V_1(\phi) + V_2(\phi) + \cdots ,
\end{equation} 
where the subscript indicates the loop order. The first term on the R.H.S., $V_0$ is the `classical' or tree-level contribution, while the first quantum corrections begin at the one-loop level. The temperature dependence, on the other hand, is induced at the one-loop level. Hence the one-loop temperature-dependent effective potential can be written schematically as:
\begin{equation}
V(\phi,T) = V_0(\phi) + V_1(\phi) + V_{1T}(\phi,T)
\end{equation}
\end{comment}
Below, we compute each contribution to the effective potential separately.

\subsection{Tree-level potential}
The Higgs-flavon scalar potential is essentially an extension of the SM by a complex scalar (CxSM) \cite{Gonderinger:2012rd}, with a global $U(1)_{\fn}$ symmetry,
\begin{eqnarray}
    V_0(S,H) &=& -\mu_H^2 |H|^2  + \lambda_H |H|^4 + \lambda_{HS}|S|^2|H|^2 \nonumber\\
              &-& \mu_S^2 |S|^2 + \lambda_S|S|^4,
\end{eqnarray}

\noindent where all parameters are assumed to be real ($CP$ conserving potential). Tree-level vacuum stability requires,

\begin{equation}
\lambda_{H}\geq 0,~\lambda_{S} \geq 0.
\end{equation}

\noindent Further, we take $\lambda_{HS} \geq 0.$
\noindent Write the fields as: $$ H =  \begin{pmatrix}
G_{+}\\
\frac{h+i G_0}{\sqrt{2}}\\
\end{pmatrix}, ~~  S = \frac{1}{\sqrt{2}}s ~e^{i\rho},
$$ 
where $h,~s$ are real scalar fields, $G_0,~G_{\pm}$ are SM goldstone bosons, and $\rho$ is a pseudocalar. 

In recent years, it has become popular to associate the pseudoscalar $\rho$ with an Axion-like Particle (ALP), to additionally address the strong $CP$ problem of SM. The pseudoscalar field, in this case, has been variously called the `flavorful axion' \citep{Wilczek:1982rv,Davidson:1981zd,PhysRevD.29.1504}, the `flaxion' \cite{Ema:2016ops} or the `axiflavon' \cite{Calibbi:2016hwq}. The ALP can acquire a tiny mass via the QCD anomaly, making it a pseudo-Nambu Goldstone (pNG) boson. Alternately, an explicit $U(1)_{\fn}$ breaking term such as $V_0(S,H)\supset \lambda_1(S^2+S^{*2})$ can generate a small mass term for $\rho$, with the mass scale constrained by low energy phenomenology \cite{huntingflavon}. The pseudoscalar plays no role in the dynamics of the flavon PT, which is governed by the real scalar field $s$, so we will not comment on it further.

When written in terms of field components, the potential also has a remnant  $Z_2$ symmetry where $s\rightarrow-s$. Spontaneous breaking of such a discrete symmetry by the \vev of $s$ can lead to the formation of domain walls, which are disallowed by cosmology as they would eventually dominate the energy density of the universe, and directly contradict the cosmological principle. An additional $Z_2$-breaking term $V_0(S,H)\supset \lambda_2(S+S^*)$,
ensures that domain walls disintegrate fast enough to respect cosmological constraints. The contribution of these explicit symmetry-breaking terms is, however small and insignificant for analyzing the nature of the flavon PT.

In terms of components, the potential is, 
\begin{eqnarray}
V_0(h,s,G_0,G_{\pm}) &=& -\frac{\mu^2_H}{2} \big(h^2+G_0^2+ 2G_+G_-\big) + \frac{\lambda_H}{4} \big(h^2+G_0^2+ 2G_+G_-\big)^2\nonumber\\
&+& \frac{\lambda_{HS}}{4}s^2\big(h^2+G_0^2+ 2G_+G_-\big)
- \frac{\mu^2_S}{2} s^2 + \frac{\lambda_S}{4} s^4.
\end{eqnarray}
Setting the goldstones to zero,
\begin{equation}
V_0(h,s) = -\frac{\mu^2_H}{2} h^2 + \frac{\lambda_H}{4} h^4 +\frac{\lambda_{HS}}{4}h^2s^2- \frac{\mu^2_S}{2} s^2 + \frac{\lambda_S}{4} s^4.
\end{equation}
Suppose at zero temperature the fields get a \vev: $$\langle h \rangle = v_h,~~\langle s \rangle = v_s.$$
We take $\mu^2_H$, $\lambda_{HS}$, $\lambda_S$ and $v_s$ as free parameters. 
 Minimization conditions are,
\begin{subequations}
\begin{equation}
\left.\frac{\partial V_0}{\partial h}\right\vert_{(v_h,v_s)} = 0,
\end{equation}
\begin{equation}
\left.\frac{\partial V_0}{\partial s}\right\vert_{(v_h,v_s)} = 0.
\end{equation}
\end{subequations}
These imply,
\begin{subequations}
\begin{equation}\label{min1}
\big(-\mu^2_H + \lambda_H v_h^2 +\frac{\lambda_{HS}}{2} v_s^2\big)v_h = 0,
\end{equation}
\begin{equation}
\big(-\mu^2_S + \lambda_S v_s^2 +\frac{\lambda_{HS}}{2}v_h^2\big)v_s = 0.
\end{equation}
\end{subequations}
Since $v_s\neq 0$, the second equation gives, 
\begin{eqnarray}\label{min2}
\mu_S^2 = \lambda_S v_s^2 + \frac{\lambda_{HS}}{2}v_h^2,
\end{eqnarray}
while in the first equation, we set $v_h = 0$, so that $\mu_H^2$ is an undetermined parameter. 
The scalar mass matrix is computed from the Hessian of the potential. Write $(h,s) = (\phi_1,\phi_2)$, then,
$$m^2_{ij} = \left.\frac{\partial^2 V_0}{\partial \phi_i\partial\phi_j}\right\vert_{(v_h,v_s)}$$
Setting $v_h=0$, 
\begin{gather}
\mathcal{M}^2_{\rm{scalar}} = \begin{pmatrix}
-\mu^2_H+\frac{\lambda_{HS}}{2} v_s^2 & 0 \\
0 & 2\lambda_S v_s^2 \\
\end{pmatrix}.
\end{gather}

The Hessian must be positive-definite, which gives us an additional constraint,
\begin{equation}
\mu^2_H\leq \frac{\lambda_{HS}}{2} v_s^2.
\end{equation}

We make the choice,
\begin{equation}
\mu^2_H = 0.01 ~\lambda_{HS} v_s^2,
\end{equation}
which numerically ensures that if a PT is possible, it occurs along the $s$ direction at one-loop level. This is, however, not the only possible choice, and it was verified that changing the multiplying coefficient up to an order of magnitude doesn't affect our results significantly. With the above choice, we can set $h=0$ in the potential in the rest of the analysis.
\begin{equation}
V_0(s) = -\frac{\mu_S^2}{2} ~s^2 + \frac{\lambda_S}{4}~s^4.
\end{equation}

The minimization condition \eqref{min2} now becomes,
\begin{equation}
\mu^2_S = \lambda_S v_s^2.
\end{equation}

\subsection{One-loop correction}
One-loop correction to the effective potential is given by the Coleman-Weinberg (CW) formula \cite{ColWein}. Following Eq.\,\eqref{eq: 1_ColWein}, in the $\ms$ scheme, in Landau gauge, we have 
\begin{equation}\label{eq:ColWein}
V_{\rm{CW}}(h,s) = \frac{1}{64\pi^2}\sum_{i} (-1)^{f_i}n_i m^4_i(h,s)\bigg[\log\bigg(\frac{m^2_i(h,s)}{\mu^2}\bigg)- c_i\bigg],
\end{equation}
where the sum runs over all particles that couple to the flavon\footnote{We retain the field $h$ for the sake of completeness, although finally, $h$ is to be set to zero.}. These include the SM Higgs ($h$), the SM  goldstones ($G_0,G_{\pm}$), the flavon ($s$), and the SM quarks and VLQs ($F$). As the SM gauge bosons do not directly couple to $s$, they do not contribute to the one-loop effective potential. In the above equation, $\mu$ is the renormalization scale, which we set to $\mu^2 = v_s^2$. The factor $f_i$ is $+1$ $(0)$ for fermions (bosons). $n_i$ is the number of degrees of freedom: (1,1,3,12) for ($h,~s,~G_{0,{\pm}},~F$), and $c_i = 5/6$ for gauge bosons and $c_i = 3/2$ for others.

The field-dependent masses, $m_i^2(h,s)$, are obtained by expanding the tree-level Lagrangian around a constant field value $(h,s)$, and reading off the respective mass terms for the respective species. For the scalars, we get,
\begin{subequations}
\begin{equation}\label{eq: m2hs}
\mathcal{M}^2(h,s) = \begin{pmatrix}
-\mu^2_H + 3\lambda_H h^2 + \frac{\lambda_{HS}}{2}s^2 & \lambda_{HS} hs\\
\lambda_{HS} hs & \lambda_S (3s^2-v_s^2) + \frac{\lambda_{HS}}{2}h^2\\
\end{pmatrix},
\end{equation}
\begin{equation}\label{eq: m2gg}
m^2_{G_{0,\pm}}(s) = -\mu^2_H + \lambda_H h^2 + \frac{\lambda_{HS}}{2} s^2.
\end{equation}
\end{subequations}
Note that $m^2_i(s)<0$ for the above scalars near the origin, implying an imaginary field-dependent mass. This well-known problem occurs for scalars associated with a spontaneously broken symmetry, leading to a complex potential. The imaginary part of the potential is related to particle decay \cite{Weinberg:1987vp}. However, the dynamics of PT  is governed only by the real part of the potential, and thus we always take the real part of the potential in our analysis.

\begin{figure}[tbp]
\centering % \begin{center}/\end{center} takes some additional vertical space
\includegraphics[width=.55\textwidth]{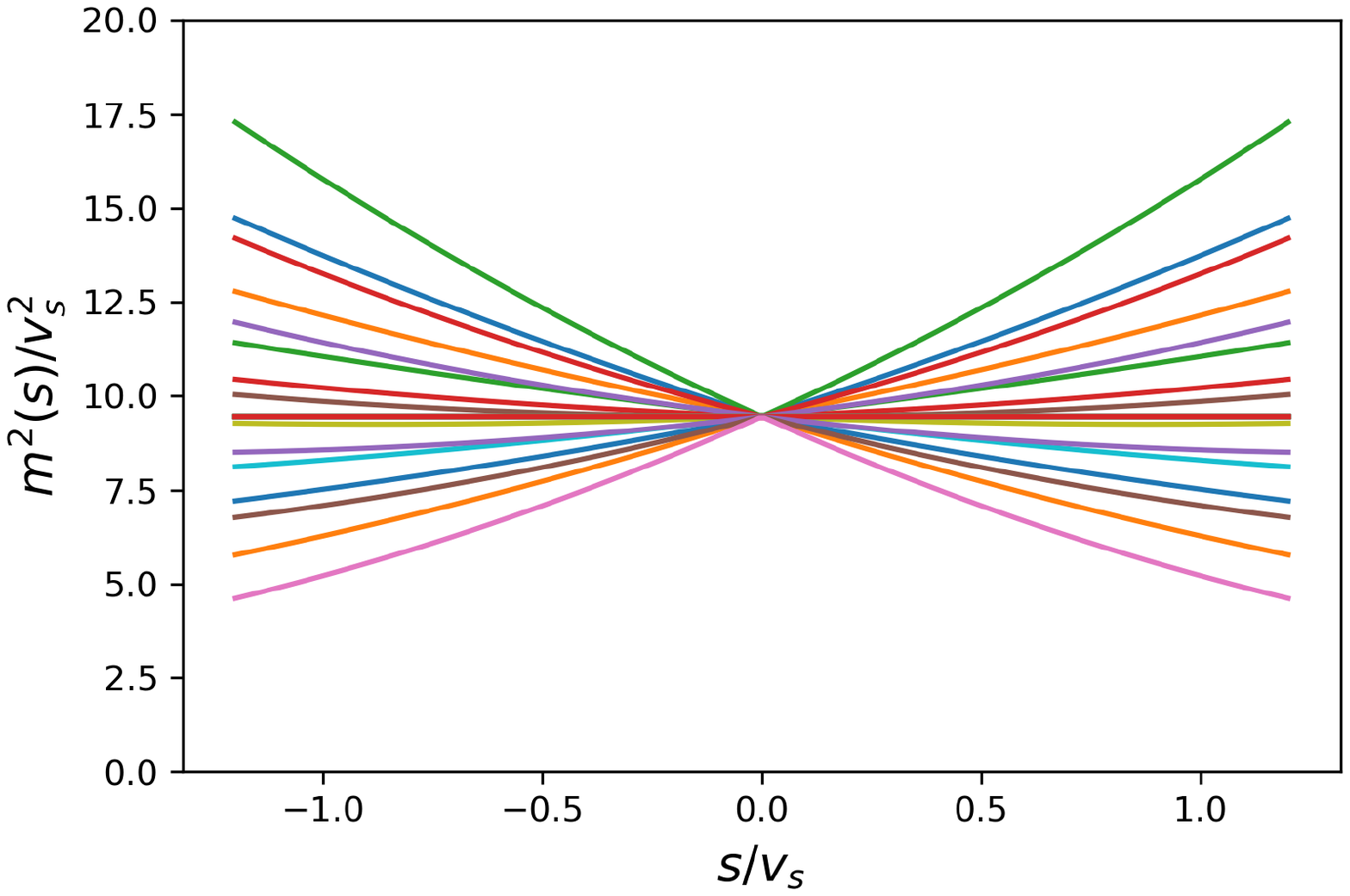}
\hfill
\caption{\label{fig: m_vlq} Field-dependent mass squared for the VLQs of model 1. Here $y=0.5$ and $v_s = 10^5$ GeV. Physical VLQ masses are defined at $\frac{s}{v_s}=1.$}
\end{figure}

For fermions, the field-dependent mass matrices are obtained from the physical mass matrices $\mathcal{M}^{u/d}$ given in \eqref{subeq:model1} and \eqref{subeq:model2}, by the substitution:
$$ \hu\rightarrow \frac{1}{\sqrt{2}}\begin{pmatrix}
h\\
0\\
\end{pmatrix},~\h\rightarrow \frac{1}{\sqrt{2}}\begin{pmatrix}
0\\
h\\
\end{pmatrix},~\s\rightarrow \frac{s}{\sqrt{2}}.$$ 
$m_i^2(h,s)$ are then given by singular values of the field-dependent $\mathcal{M}^{u/d}$. For matrices of dimension larger than 5, the eigenvalues need to be evaluated numerically. The lightest three eigenvalues correspond to the SM quarks, while the rest are the VLQs. The contribution of the light quarks to the potential is negligible and is therefore ignored. In Fig.\,\ref{fig: m_vlq}, we plot the squared field-dependent masses for the VLQs of model 1. In the region $0\le s\lesssim M$, with $h=0$, the VLQ masses are well fitted by quadratic polynomials,
\begin{equation}\label{eq:VLQ param}
m^2_i(s) = M^2 + a_i~yMs + b_i~y^2s^2,
\end{equation}

where $a_i$ and $b_i$ are dimensionless fitting parameters, and $y$ is the scaling factor described in \eqref{eq:scaling}. For the largest and smallest VLQ masses, the parameters were found to be $|a_i|,~|b_i|\sim 1$, while they are smaller for others.

The $\overline{\rm{MS}}$ renormalization takes care of the UV divergent terms arising from integrals over momenta running in the loop.  However, additional finite counter-terms can be added to $V_{\rm{CW}}$ so that the \vev and flavon mass are unchanged from their tree-level values. Write,
\begin{subequations}
\begin{equation}\label{eq: ct1}
V_1(s) = V_{\cw}(s) + V_{\rm{c.t.}}(s),
\end{equation}
\begin{equation}\label{eq: ct2}
V_{\rm{c.t.}}(s) = -\frac{\delta\mu^2_s}{2} s^2 + \frac{\delta\lambda_s}{4} s^4.
\end{equation}
\end{subequations}
This is the on-shell renormalization scheme. From \eqref{eq: ct1} and \eqref{eq: ct2}, the finite renormalization conditions are: 
\begin{subequations}
\begin{equation}
\left.\frac{\partial V_1}{\partial s}\right\vert_{v_s} = 0,
\end{equation}
\begin{equation}
\left.\frac{\partial^2 V_1}{\partial s^2}\right\vert_{v_s} = 0.
\end{equation}
\end{subequations}

These determine the parameters $\delta\mu^2_s$ and $\delta\lambda_s$,
\begin{subequations}
\begin{equation}
\delta\mu^2_s = \frac{1}{2 v_s}\left(3\left.\frac{\partial V_{\cw}}{\partial s}\right\vert_{v_s}-v_s \left.\frac{\partial^2 V_{\cw}}{\partial s^2}\right\vert_{v_s}\right),
\end{equation}
\begin{equation}
\delta\lambda_s = \frac{1}{2 v_s^3}\left(\left.\frac{\partial V_{\cw}}{\partial s}\right\vert_{v_s}-v_s\left.\frac{\partial^2 V_{\cw}}{\partial s^2}\right\vert_{v_s}\right).
\end{equation}
\end{subequations}

The scale-independence of the full effective potential \cite{ColWein} is spoiled by the appearance of $\mu$ in the CW formula. This is a result of truncating the potential at one-loop level. Approximate scale-independence is recovered when the Logs are $\mathcal{O}(1)$. Hence the renormalization scale needs to be chosen wisely, depending upon the regime of interest. In the scenario considered here, there are two scales: $v_s$ and the VLQ mass scale $M$, with the scale separation quantified by the ratio,
\begin{equation}\label{eq: epsilon}
\epsilon \sim \frac{1}{\sqrt{2}}\frac{v_s}{M}.
\end{equation}
For model 1, $\epsilon = 0.23$, while for model 2, $\epsilon = 0.05$, implying a larger scale separation for model 2. To study the PT, we are interested in the behavior of the potential around $s\sim v_s$. For model 1, the largest Log is $\sim 2.5$, while for model 2, the largest Log is $\sim 5.5$ at $\mu = v_s$. Hence the Logs are $\mathcal{O}(1)$. We treat these Logs as small and analyze the one-loop potential at the fixed scale $\mu = v_s$, as the detailed RGE analysis is beyond the scope of this work.

\subsection{Finite temperature contribution}
 The one-loop finite temperature correction is the free energy associated with the scalar field, and is given by $\phi$ Eq.\,\eqref{eq: 1_V_1T}, 
\begin{equation}
V_{1T}(\phi,T) = \sum_i n_i\frac{T^4}{2\pi^2} J_{b/f}\bigg(\frac{m^2_i(\phi)}{T^2}\bigg)\,,\nn
\end{equation}
where the function $J_b$ ($J_f$) is defined for bosons (fermions). Write $x^2 \equiv \frac{m^2_i}{T^2}$, then $J_{b/f}$ is given by,
\begin{equation}
J_{b/f}(x^2) = \pm \int_0^{\infty} dy~ y^2 \log\big(1\mp e^{-\sqrt{x^2+y^2}}\big)\,. \nn
\end{equation} 
\begin{comment}
These functions have well known  expansions for $x^2<<1$, and $x^2>>1$ . For small $x^2$ \cite{Cline_1997}, 
\begin{eqnarray}%\label{eq: highT}
J_f(x^2,n) = &-&\frac{7\pi^4}{360} + \frac{\pi^2}{24} x^2 + \frac{1}{32}x^4\big(\log x^2- c_f\big)\nonumber\\
&-&\pi^2x^2\sum_{l=2}^n\left(-\frac{1}{4\pi^2} x^2\right)^l\frac{(2l-3)!!\zeta(2l-1)}{(2l)!!(l+1)}\left(2^{2l-1}-1\right),\label{eq: highTf}\\
J_b(x^2,n) = &-&\frac{\pi^4}{45} + \frac{\pi^2}{12} x^2 - \frac{\pi}{6}\big(x^2\big)^{3/2}-\frac{1}{32}x^4\big(\log x^2- c_b\big)\nonumber\\
&+&\pi^2x^2\sum_{l=2}^n\left(-\frac{1}{4\pi^2} x^2\right)^l\frac{(2l-3)!!\zeta(2l-1)}{(2l)!!(l+1)},\label{eq: highTb}
\end{eqnarray}
where, $c_+=3/2+2\log\pi-2\gamma_E$, and $c_-=c_++2\log 4$. Here $\gamma_E$ is the Euler-Mascheroni constant, and $\zeta(x)$ is the Riemann-$\zeta$ function. $n$ is the order till we want to expand the sum. Usually taking $n=3,4$ gives good results.

For large $x^2$, both fermions and bosons have the same expansion at order $n$ \cite{Cline_1997},
\begin{align}\label{eq:lowT}
J_{b/f}(x^2,n) = &-\exp\bigg(-(x^2)^{1/2}\bigg)\bigg(\frac{\pi}{2}(x^2)^{3/2}\bigg)^{1/2}\sum_{l=0}^n \frac{1}{2^ll!}\frac{\Gamma(5/2+l)}{\Gamma(5/2-l)}(x^2)^{-l/2},
\end{align}
where $\Gamma(x)$ is the Euler Gamma function. For intermediate values of $x^2$, the functions $J_{b/f}(x^2)$ can be obtained by interpolating between the large $x^2$ and small $x^2$ approximation. 
\end{comment}
The finite temperature effects have been computed in this work using the \texttt{CosmoTransitions} package \cite{Wainwright:2011kj}.

\subsection{Daisy resummation}
As discussed in Sec.\,\ref{1_daisy}, the perturbative loop expansion of the effective potential breaks down near the critical temperature $T_c$ due to infrared divergences from the bosonic Matsubara zero-modes \cite{Carrington:1991hz}. The leading contribution to the divergences comes from multiloop `daisy diagrams', which need to be resummed. We use the Parwani method for resummation discussed in Eq.\,\eqref{eq: parwani}. Thermal masses are obtained by substituting \eqref{eq: m2hs}, \eqref{eq: m2gg} in the high-$T$ expansion \eqref{eq: 1_V1T_high} to leading order in $T$. These are:
\begin{eqnarray}\label{eq: daisy}
\Pi_{h,G_{0,\pm}}(T) &=& T^2\bigg(\frac{g_1^2}{16}+\frac{3 g_2^2}{16}+\frac{\lambda_{HS}}{12} + \frac{\lambda_H}{2} + \frac{y_t^2}{4}\bigg),\\
\Pi_{s}(T) &=& T^2\bigg(\frac{\lambda_{HS}}{6} + \frac{\lambda_S}{3}\bigg),
\end{eqnarray}
where $g_1,~g_2$ are the usual SM gauge couplings, and $y_t$ is the top Yukawa coupling. All SM couplings are taken at the scale $\mu = v_s$, and are found by evolving the SM RGEs \cite{Buttazzo:2013uya}. 

Now that all the contributions to the effective potential are defined, we finally have:
\begin{equation}
V_{\rm{eff}}(s,T) = V_0(s) + V_{\cw}(s,T) + V_{{\rm{c.t.}}}(s) + V_{1T}(s,T),
\end{equation}
where the Daisy resummation is incorporated by adding the thermal masses to the field-dependent masses in $V_{\cw}$ and $V_{1T}$.

\section{Vacuum stability}\label{sec: stability}
Fermions contribute with a negative sign to the CW formula \eqref{eq:ColWein}, and tend to destabilize the potential at large field values. In the context of FN, vacuum stability has been addressed earlier \citep{Giese:2019khs}. Similarly, Higgs vacuum stability of the SM in the presence of vector-like fermions has also been discussed \citep{Blum:2015rpa,Gopalakrishna:2018uxn,Borah:2020nsz,Higaki:2019ojq}.  Typically, UV completions of FN contain several ($\mathcal{O}(10)$) heavy fermions with $\mathcal{O}(1)$ Yukawa couplings to the flavon, destabilizing the effective potential. Here we discuss the vacuum stability along the flavon direction. 

The negative contribution of VLQs can be seen from the effective quartic coupling at one-loop level. Setting the renormalization scale $\mu^2 = v_s^2$ and using \eqref{eq: ct1},
\begin{eqnarray}\label{eq: instability}
\frac{\lambda_{s,\rm{eff}}}{4} &=& \frac{1}{4!} \left.\frac{\partial^4}{\partial s^4}\big(V_0+V_1\big)\right\vert_{s=0}\nonumber\\
&\approx& \frac{\lambda_s+\delta\lambda_s}{4} + \frac{9}{64\pi^2}\log\big(\lambda_s\big)+\frac{\lambda^2_{HS}}{64\pi^2}\log\left(\frac{\lambda_{HS}}{100}\right)\nonumber\\
&+& \sum_{i\in {\rm{VLQ}}} \frac{y^4}{64\pi^2}
\bigg\{a_i^4-12 a^2_ib_i - 12b_i^2 \log \bigg(\frac{M^2}{v_s^2}\bigg)\bigg\},
\end{eqnarray}
where the second term on the RHS is the flavon contribution, the third term is contributed by $h,~G_{0,\pm}$, and we have used the fitting polynomials \eqref{eq:VLQ param} in the last term for the VLQ contribution. Since the coefficients $a_i$ and $b_i$ are all $\mathcal{O}(1)$, the VLQ term is dominated by the factors of 12, and hence the fermionic contribution to the quartic coupling is negative. Assuming for simplicity that each term in the VLQ sum contributes roughly equally, the total contribution also increases with the number of VLQs. Lastly, the VLQ contribution scales as $y^4$, therefore, increasing the Yukawa factor by a small amount dramatically increases the negative contribution.

The instability scale, $\Lambda_{\rm{inst}}$ is estimated from the field value at which the potential falls below the height of the local minimum at $s=v_s$. 
For $\lambda_S\sim\mathcal{O}(0.1)$, $\lambda_{HS}\sim\mathcal{O}(1)$ and $y=1$, $\Lambda_{\rm{inst}}$ was found around the VLQ mass scale $M$. Increasing the Yukawa scaling factor further pushes the instability scale lower. Naively, the potential can be stabilized by increasing the couplings ($\lambda_S$, $\lambda_{HS}$), to counteract the fermionic contribution. However, this procedure is limited by the perturbativity bounds on the quartic couplings: $\lambda_S, \lambda_{HS}\leq 4\pi$. For large couplings, the one-loop results also become unreliable, and any FOPT observed from the one-loop analysis can't be trusted. As is evident from \eqref{eq: instability}, the fermion contribution to the quartic coupling is highly sensitive to the scaling factor $y$. Hence, we can push $\Lambda_{\rm{inst}}$ to at least a few times of $M$ by decreasing $y$. We found that by keeping $y = 0.5$, $\Lambda_{\rm{inst}}$ is atleast $~3M$ for the smallest value of the couplings $(\lambda_S,\lambda_{HS})$  in the parameter region of interest \footnote{Keeping $y=0.5$ for all couplings uniformly can contribute slightly to the hierarchy of fermion masses as some combinations $y_{ij}$ may be smaller than $\mathcal{O}(1)$ according to \eqref{eq: yuk_highlow}, but this effect is tolerable.}. Beyond the instability scale, the potential can be stabilized by adding new heavy bosonic degrees of freedom which couple to the flavon as illustrated in Appendix \ref{app: bosons}. If the mass scale of the new bosons is high compared to $M$, we can write down an Effective Field Theory (EFT), by introducing higher-dimensional operators in the flavon potential in a model-independent way (see for example \cite{Blum:2015rpa}),
\begin{equation}
V_{1} \rightarrow V_{1} + \frac{1}{\Lambda_b^2} |S|^6,
\end{equation}
where $\Lambda_b\sim \Lambda_{\rm{inst}}$ is the cut-off scale associated with the new bosons and the Wilson coefficient is taken to be unity. For sufficiently heavy bosons, the thermal contribution is suppressed by the Boltzmann factor, as can be seen from the small $T$ expansion \eqref{eq: 1_lowT}. Hence, the heavy bosonic fields play an insignificant role in determining the nature and strength of the PT. We assume that vacuum stability at field values beyond $\Lambda_{\rm{inst}}$ is taken care of by such bosons, without bothering about any particular construction.

\section{First order flavon phase transition}
\label{sec: 2_FOPT}

\subsection{A case for flavon FOPT}
To generate an FOPT, a barrier must be induced between the two degenerate minima at $T_c$. Although the conventional wisdom is that only bosons contribute to the barrier to generate a SFOPT, it has been shown in the context of electroweak baryogenesis that a SFOPT is also possible in purely fermionic extensions of the SM \citep{Carena_2005,Fairbairn_2013,
Davoudiasl_2013,Egana_Ugrinovic_2017,Angelescu_2019,Cao:2021yau}. It is therefore interesting to see the effect of the heavy VLQs on the strength of FOPT in a flavon PT. Let us parametrize the finite temperature effective potential, $V_{\rm{eff}}$ as,
\begin{equation}
V_{\rm{eff}}(s,T) = \frac{\mu^2(T)}{2} s^2 + \frac{\lambda_3(T)}{3} s^3 + \frac{\lambda_4(T)}{4} s^4 + \frac{\lambda_6(T)}{6} s^6.
\end{equation}  
The following two cases may arise:
\begin{itemize}
\item \textbf{Case 1}: When the dimension 6 term is negligible, and 
$$\mu^2(T_c)>0,~\lambda_3(T_c)<0,~\lambda_4(T_c)>0.$$
The cubic term is primarily responsible for inducing the barrier. Such a cubic term is generated, for example, by bosons in the high-$T$ expansion as in \eqref{eq: 1_highT}, while fermions do not induce any such contribution. However, both fermions and bosons have the same expansion in the low-$T$ limit \eqref{eq: 1_lowT}, and hence can contribute equally to the formation of a barrier \cite{Egana_Ugrinovic_2017}.

\item \textbf{Case 2}: If the cubic term is small compared to the others, then a barrier can also be formed when \cite{Grojean:2004xa},
$$\mu^2(T_c)>0,~\lambda_4(T_c)<0,~\lambda_6(T_c)>0.$$
Although the $|S|^6$ operator may be generated by integrating out the stabilizing heavy bosonic fields, we checked that it plays a negligible role in forming a barrier in our parameter space of interest.
\end{itemize}

\begin{figure}[tbp]
\centering % \begin{center}/\end{center} takes some additional vertical space
\includegraphics[width=.55\textwidth]{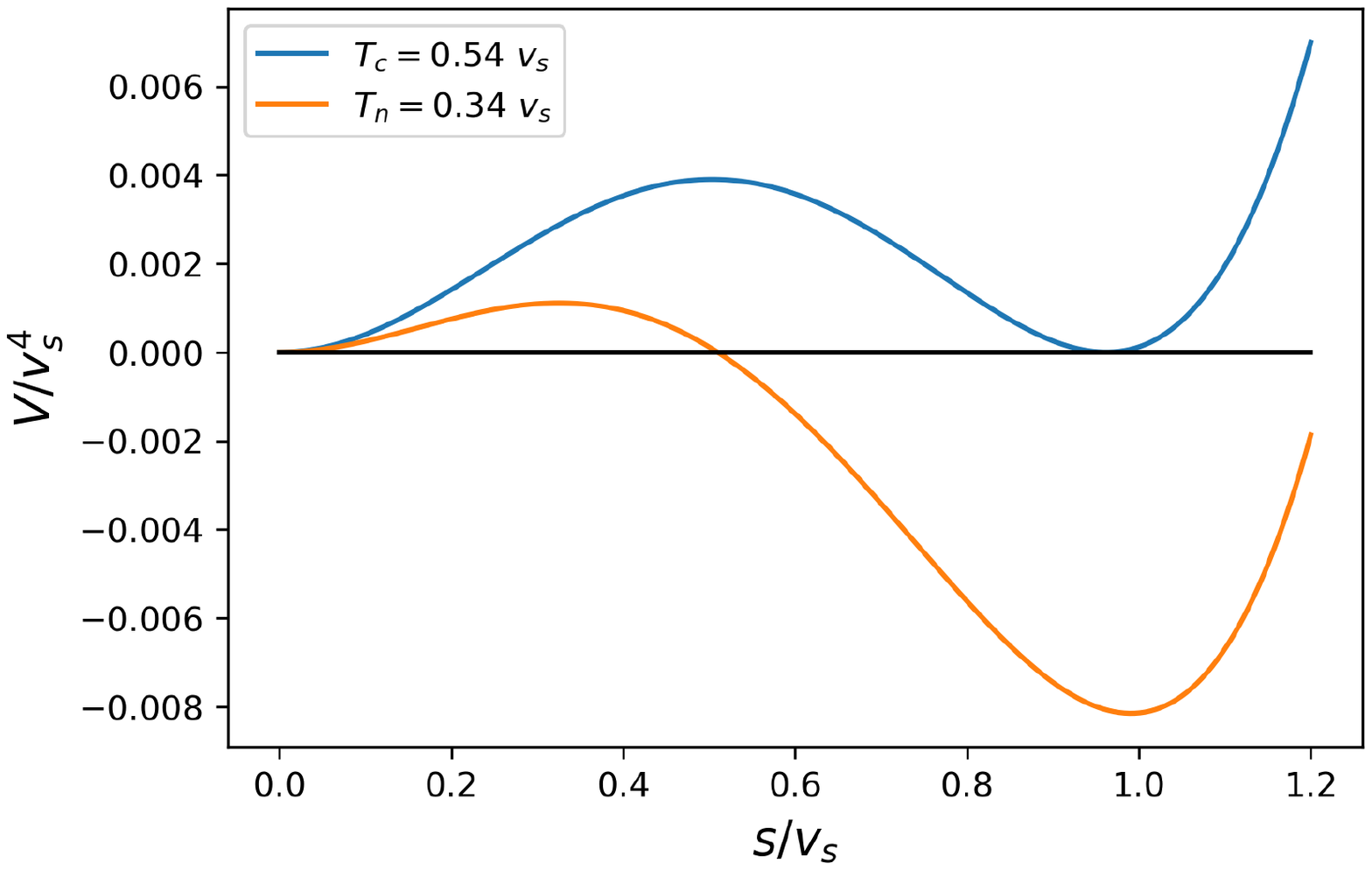}
\hfill
\caption{\label{fig: nearPT} Shape of the potential of model 1 near the FOPT, for $v_s=10^5$ GeV, $y=0.5$, $\lambda_S = 0.105$, $\lambda_{HS} = 4.253$.}
\end{figure}

The first case is relevant to our scenario, where the effect of a possible $|S|^6$ operator is taken to be small, as discussed in the previous section. In Fig.\,\ref{fig: nearPT}, we depict the barrier formed at $T=T_c$ (blue), and at $T=T_n$ (orange), for a benchmark point in model 1. The ratio, $\frac{v_s(T_n)}{T_n}\gtrsim 2.7$, indicates a SFOPT. At $T=T_n$, the bosons $\{h,s,G_{0,\pm}\}$ have masses $\lesssim T_n$, while the VLQ masses are larger than $T_n$.  As a result, the bosonic contribution to the effective potential is governed by the high-$T$ expansion \eqref{eq: 1_highT}, which incidentally justifies the need for taking daisy resummation into account. On the other hand, the fermionic contribution is governed by the low-$T$ expansion \eqref{eq: 1_lowT}. This behavior is observed in the entire region of parameter space considered in the next section. Therefore, barrier formation happens primarily due to bosons via the cubic term in the high-$T$ expansion, and the fermionic contribution is comparatively small due to the Boltzmann suppression factor in \eqref{eq: 1_lowT}. In the analysis below, the VLQs are observed to reduce the strength of FOPT slightly when compared to the no-VLQ case.

\subsection{Parameter space}
Now we discuss the prospects of achieving a SFOPT in the two FN models \eqref{eq: model1} and \eqref{eq: model2}. The parameters considered are, $\{y,v_s,\lambda_S,\lambda_{HS}\}$. The VLQ mass scale $M$ is directly related to $v_s$ via \eqref{eq: epsilon}, and is therefore not an independent parameter. 

In what follows, we keep $v_s = 10^5$ GeV, $y\in\{0,0.5\}$, and perform a parameter scan in the $\lambda_S-\lambda_{HS}$ plane, with $\lambda_S\in [0, 0.3]$, and $\lambda_{HS} \in [2.5,6]$. By restricting ourselves to this particular region of the plane, we ensure that perturbativity is obeyed and that the one-loop calculations are reliable. For the two models, we choose $y=0.5$, along with the described lower bounds on $\lambda_S$ and $\lambda_{HS}$ so that the instability scale $\lambda_{\rm{inst}}$ is at least $\gtrsim 3M$. For each point on the plane, we explore the phase structure of the finite temperature potential using the `findallTransitions' and `calcTctrans' methods of the Python-based package \texttt{CosmoTransitions} \cite{Wainwright:2011kj}. In case of an FOPT, the tunneling action $S_3$ is computed using the class `pathDeformation', and $T_n$ is numerically obtained using the condition \eqref{eq: 1_nucl_criterion2}. Next, the other two PT parameters, namely $\alpha$ and $\beta/H_*$ are computed using Eq.\,\eqref{eq: 1_alpha} and Eq.\,\eqref{eq: 1_beta} respectively. The value of $g_*$ used in the calculation is $g_* = 106.75+1 = 107.75$ since the flavon contributes an extra degree for freedom.  To compute $\beta$, we plot the action around $T=T_n$ and calculate the first derivative using the fourth-order finite difference formula. 

\begin{figure}[tbp]
\centering % \begin{center}/\end{center} takes some additional vertical space
\includegraphics[width=.32\textwidth]{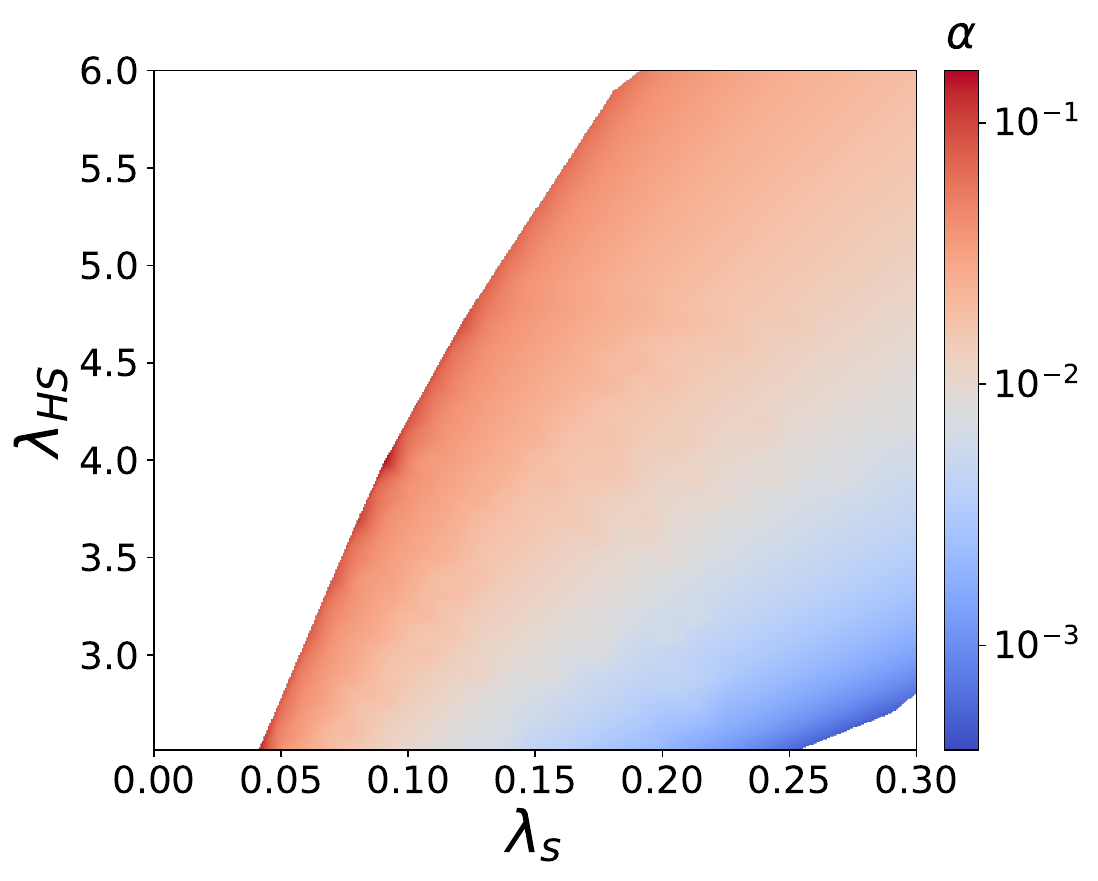}
\hfill
\includegraphics[width=.32\textwidth]{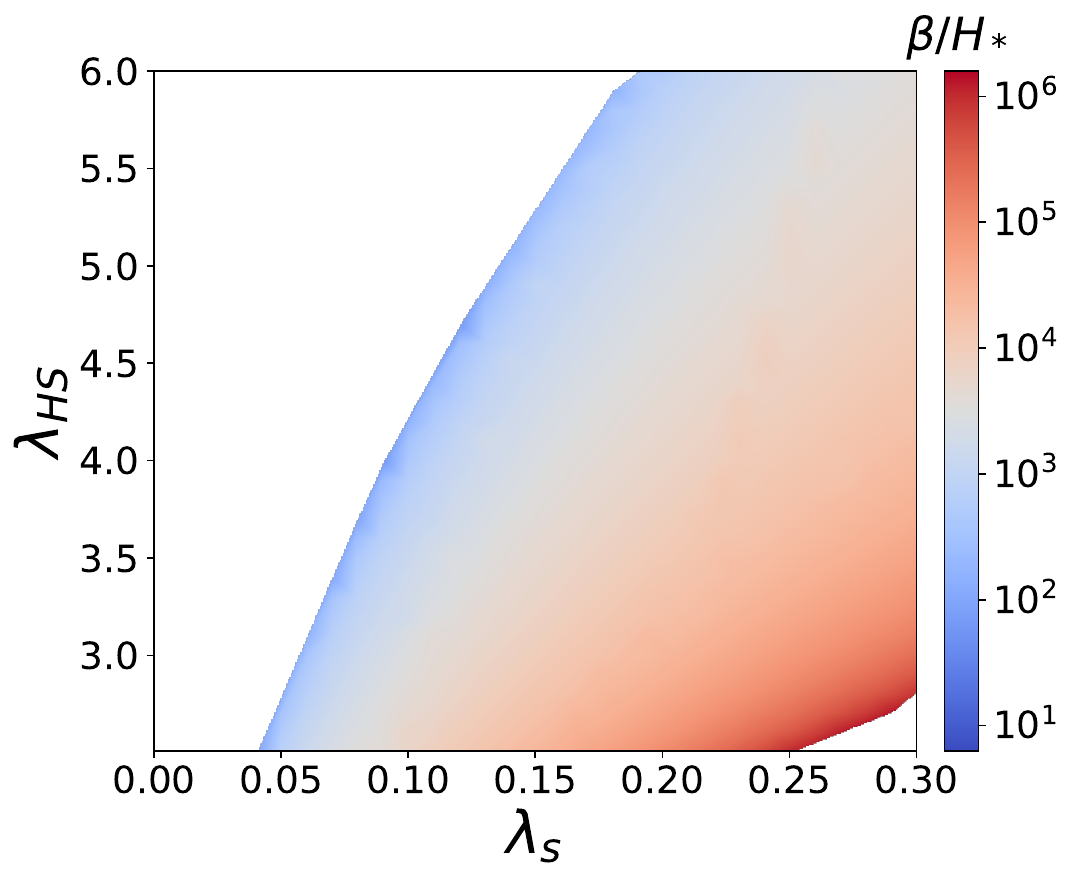}
\hfill
\includegraphics[width=.32\textwidth]{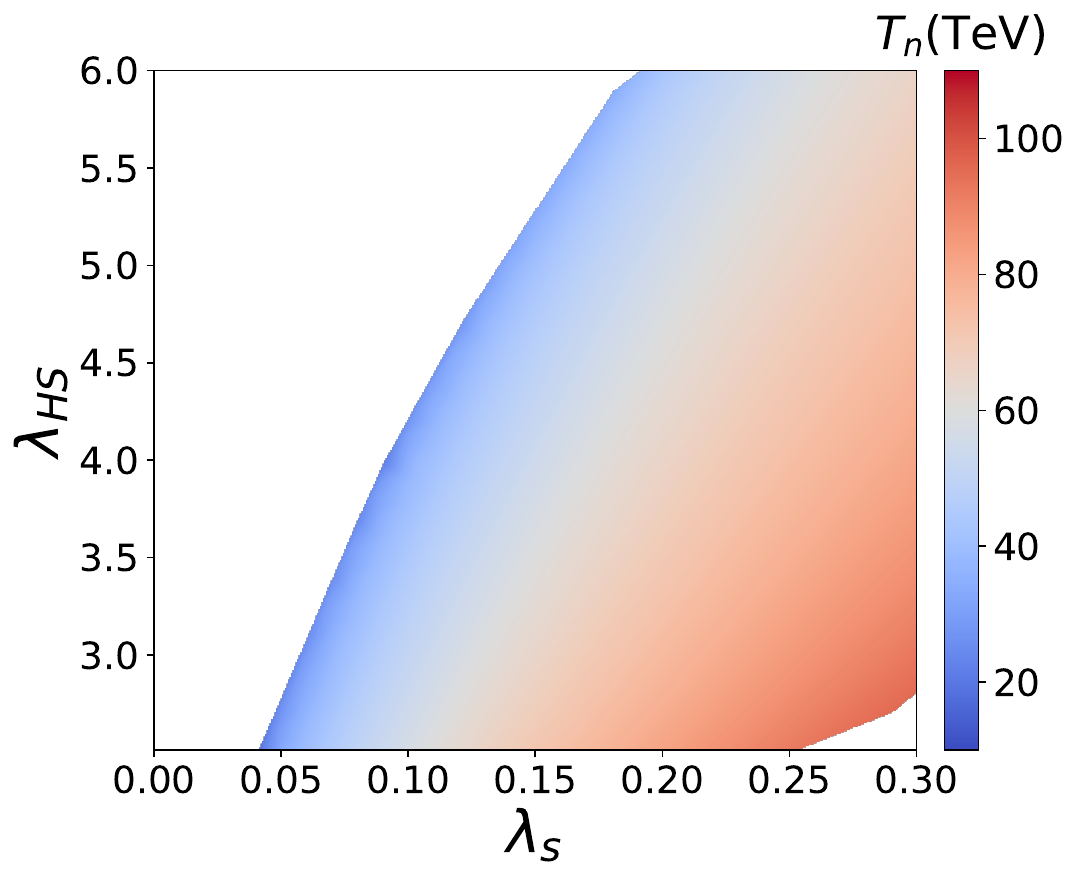}

\caption{\label{fig: noVLQ} FOPT parameter space when the influence of VLQs is switched off. Colors show PT parameters: $\alpha$ (left), $\beta/H_*$ (middle), $T_n$ (right). Here, $y=0$, and $v_s = 10^5$ GeV.}
\end{figure}

First, let us examine the region of FOPT without taking the VLQs into consideration, which corresponds to setting $y=0$ in either model 1 or model 2. Let us call it the `reference model'. The region of FOPT is shown in Fig.\,\ref{fig: noVLQ}, where $v_s$ is set to $10^5$ GeV, as a function of $\alpha$ (left panel), $\beta/H_*$ (middle panel), and $T_n$ (right panel). The observed range of the PT parameters is:  $\alpha\in [10^{-3},10^{-1}]$, $\beta/H_* \in [10^1,10^6]$, and $T_n \in [10,100]$ TeV. A SFOPT occurs when $\alpha$ is large, and $\beta/H_*$ is small. The region of FOPT is defined by a sharp boundary on the left-hand side of the $\lambda_S-\lambda_{HS}$ plane. In the white region beyond the left boundary, the metastable minimum at $s=0$, is deep enough so that the tunneling to the $s=v_s$ minimum never achieves completion, as the condition \eqref{eq: 1_nucl_criterion2} is never met. For a given $\lambda_{HS}$, the boundary occurs at the lowest value of $\lambda_S$ where the FOPT can reach completion, and this is where it is the strongest. As we move to the right along the $\lambda_S$-axis, the FOPT becomes progressively weaker, as shown by the plots in Fig.\,\ref{fig: noVLQ}. Beyond the boundary on the right-hand side, the FOPT is either extremely weak or the PT is second order. Notice that the nucleation temperature $T_n$ increases with increasing couplings $\lambda_S$ and $\lambda_{HS}$.

\begin{figure}[tbp]
\centering % \begin{center}/\end{center} takes some additional vertical space
\includegraphics[width=.32\textwidth]{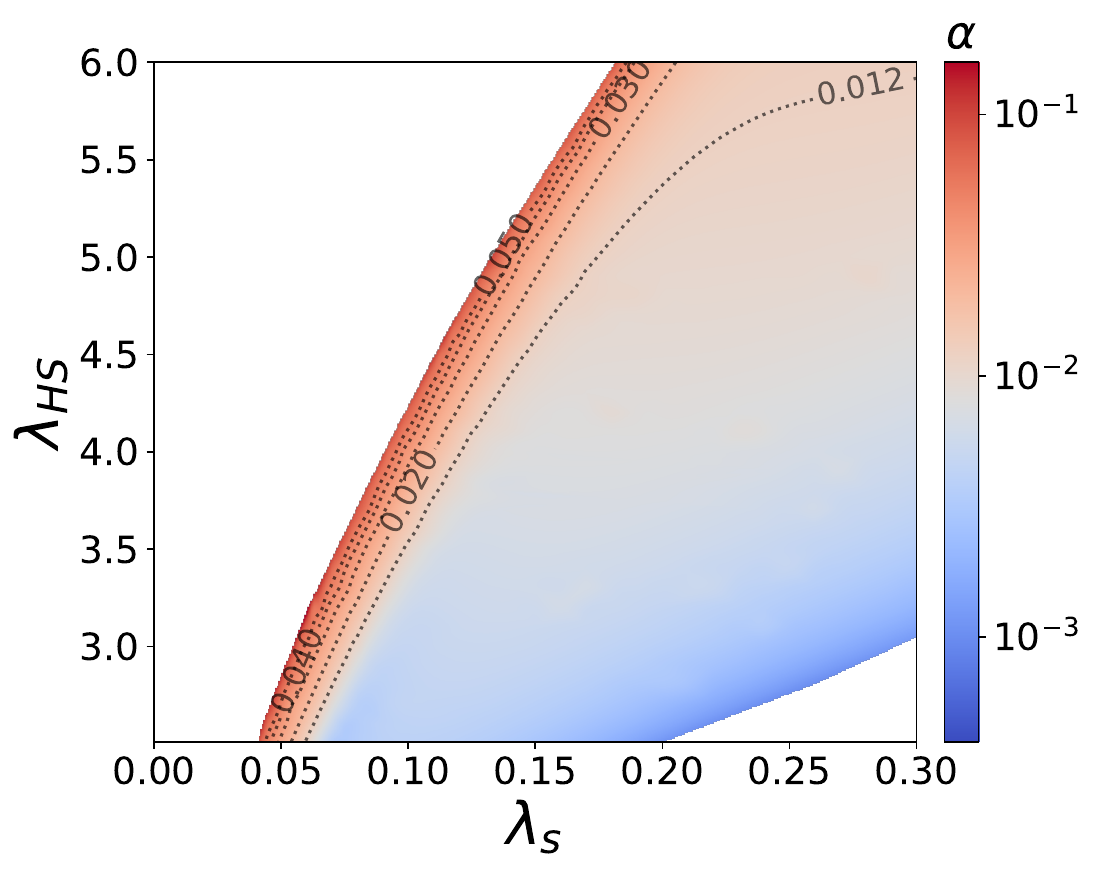}
\hfill
\includegraphics[width=.32\textwidth]{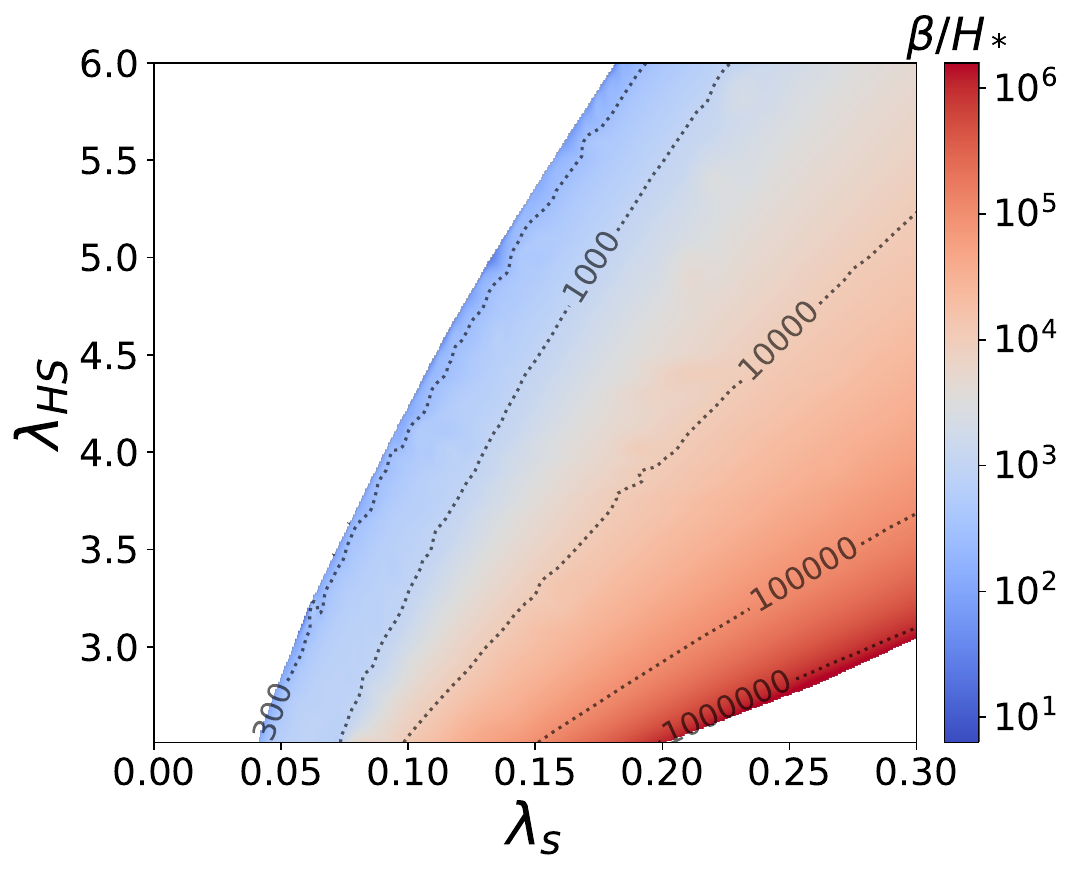}
\hfill
\includegraphics[width=.32\textwidth]{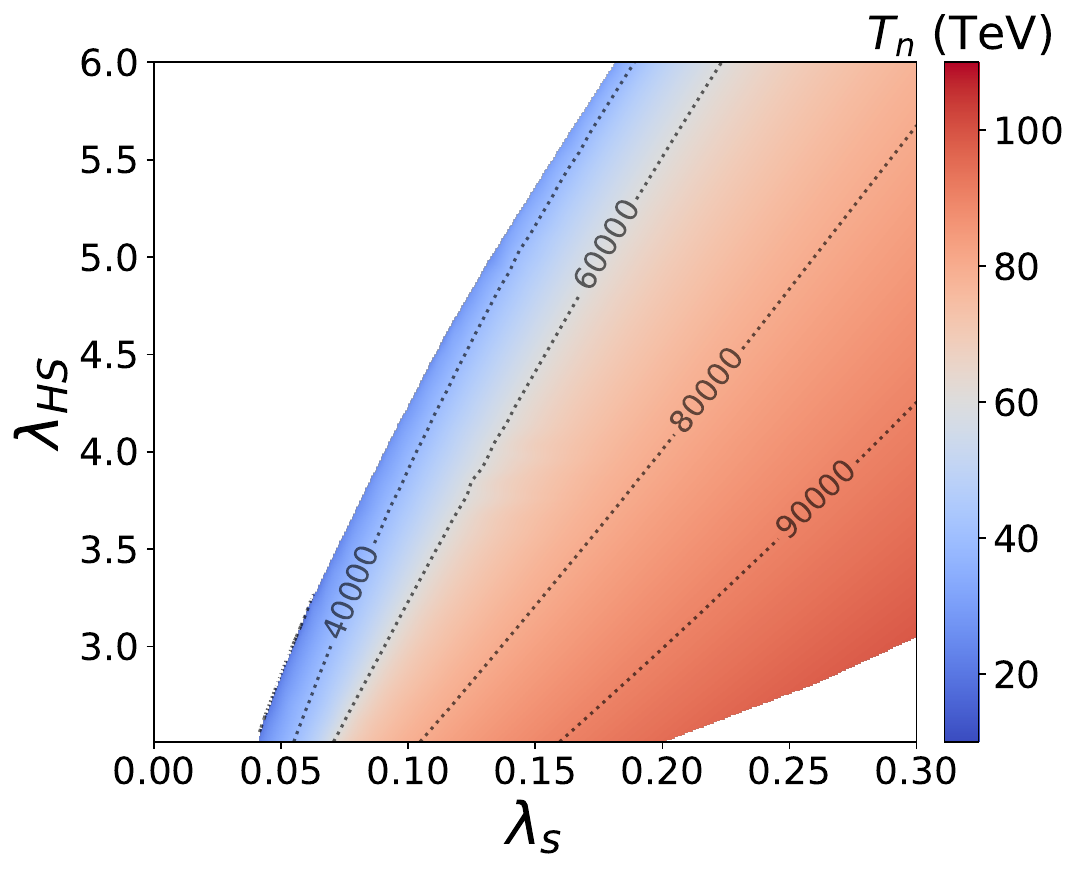}
\hfill
\includegraphics[width=.32\textwidth]{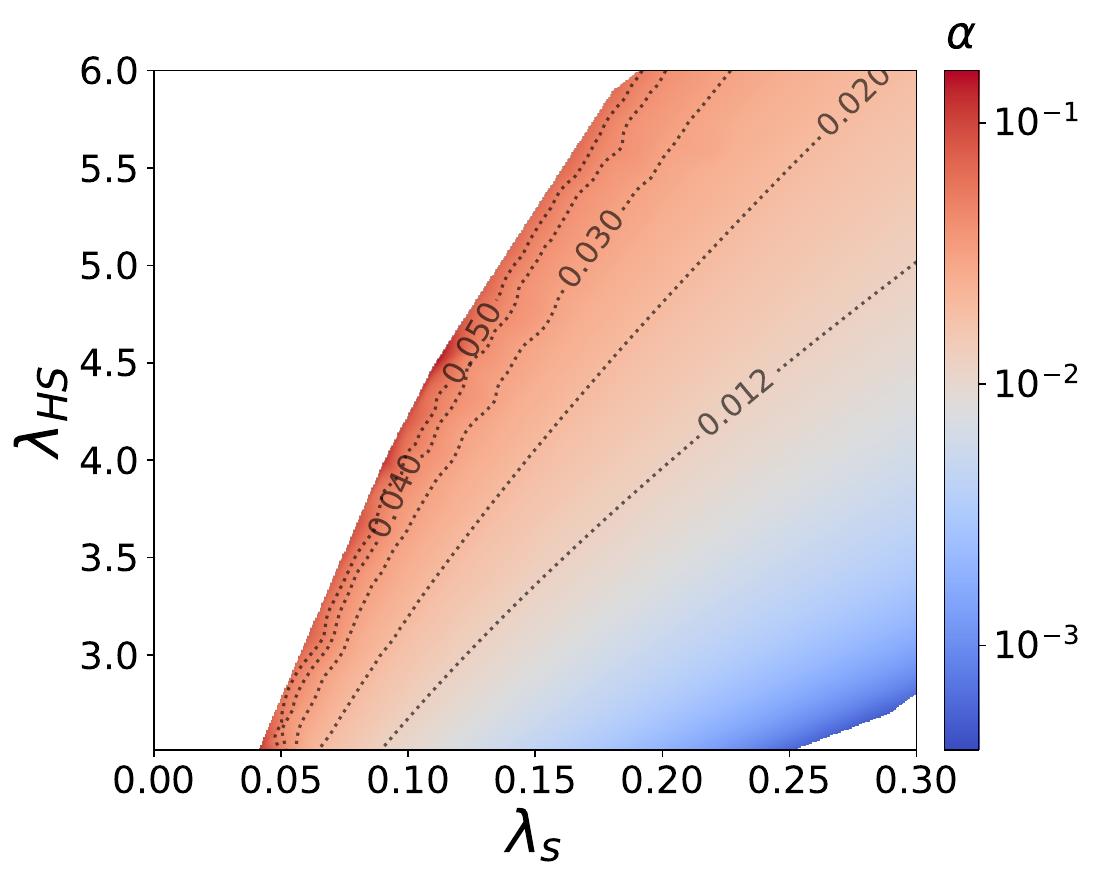}
\hfill
\includegraphics[width=.32\textwidth]{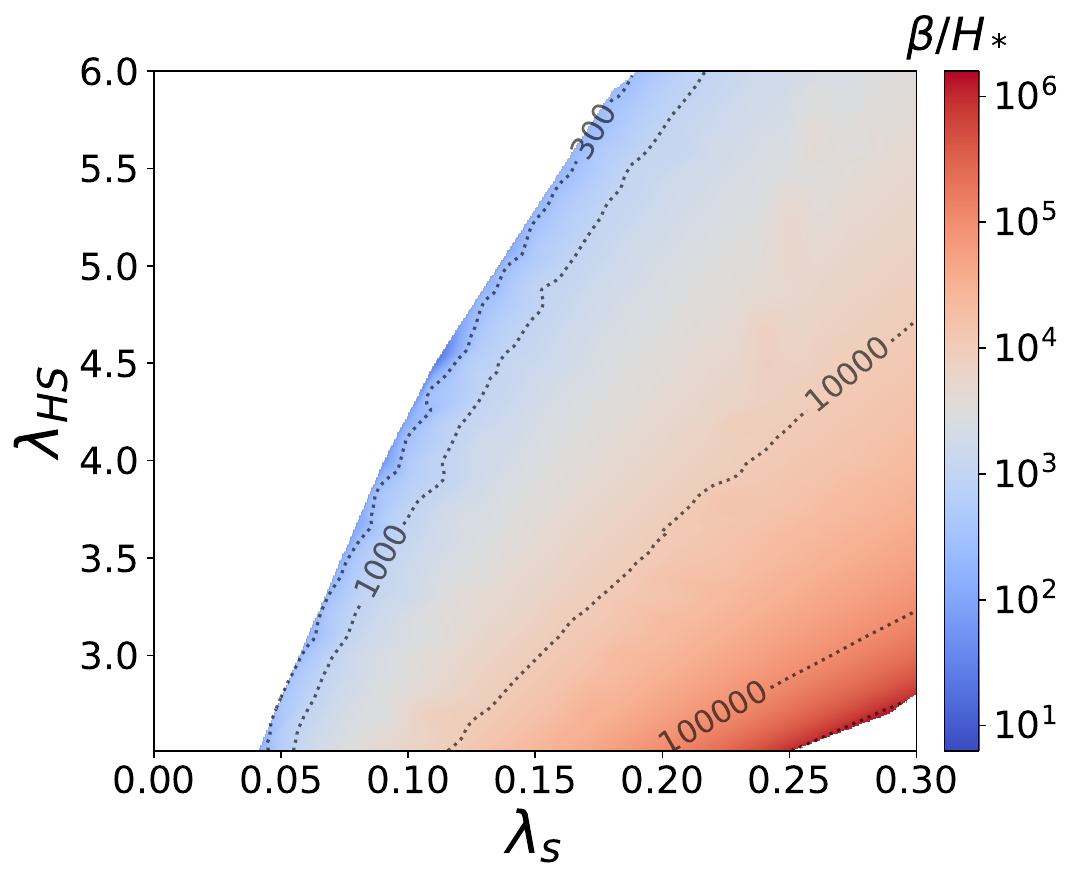}
\hfill
\includegraphics[width=.32\textwidth]{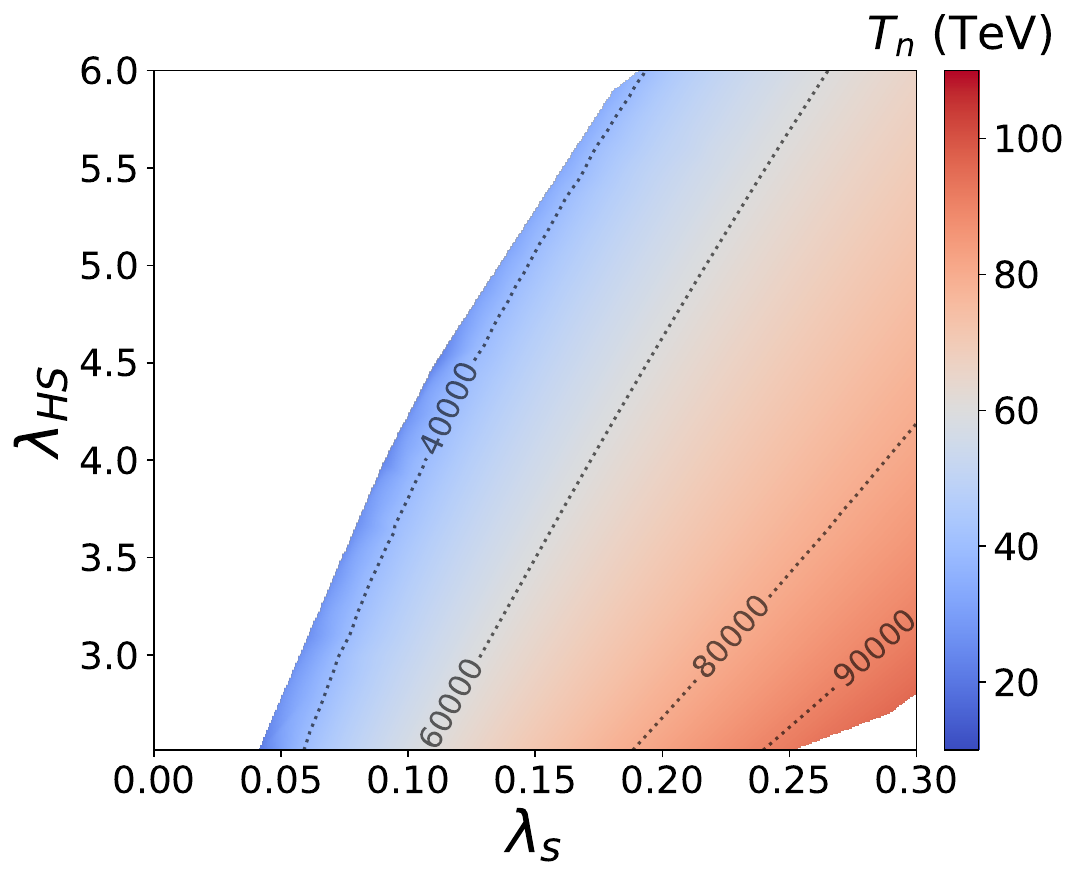}
\caption{\label{fig: param_scan} PT parameters for model 1 (top row) and model 2 (bottom row) with flavon \vev, $v_s=10^5$ GeV, and $y=0.5$. In each row, the left panel indicates the variation of $\alpha$, with contours $\alpha=0.012,~0.02,~0.03,~0.04,~0.05$. In the middle panel, $\beta/H_*$ is depicted with contours $\beta/H_*= 3\times10^2,~ 10^3,10^4,~10^5,~10^6$. In the right panel, $T_n$ is depicted with contours $T_n=4,~6,~8,~9$ $\times 10^4$ GeV.}
\end{figure}

Scans showing the allowed region of FOPT for model 1 are given in the top row of Fig.\,\ref{fig: param_scan}. We see the same qualitative trends as in the case of the reference model, as we move along the $\lambda_S$-axis in the plots of $\alpha$, $\beta/H_*$, and $T_n$. The biggest difference is seen in the left panel depicting $\alpha$, where the change near the left boundary is more dramatic compared to the reference model. The effect of the VLQs of model 1 is therefore to reduce the region of SFOPT slightly. The color gradient in the plots of $\beta/H_*$ and $T_n$ is similarly more dramatic compared to that of the reference model. To note the differences between model 1 and model 2, we overlay contours for a few chosen values of $\alpha$, $\beta/H_*$, and $T_n$.

In the bottom row of Fig.\,\ref{fig: param_scan}, we show the parameter scans for model 2, with the same choice of contours as model 1. In the left panel depicting $\alpha$, we see that the contours for $\alpha=0.012,~0.02$ appear more spread out when compared to the respective contours for model 1, indicating a slower fall in the $\alpha$ values in model 2 as we increase $\lambda_S$. Close to the left boundary, however, the contours of the two models are comparable. We find that the plots resemble those of the reference model more closely as compared to model 1. This is because of the larger scale separation between $v_s$ and the VLQ mass scale $M$ for model 2, compared to model 1, due to which the VLQ contribution is more suppressed in model 2. As a result, the influence of VLQs is less pronounced in model 2 as compared to model 1. In the next section, we describe the GW signature arising from a SFOPT.

\section{Gravitational wave background}
\label{sec:GW}
The GW spectrum is defined in terms of the GW energy density $\rho_{\rm{GW}}$, as \cite{Caprini:2015zlo},
\begin{equation}
\Omega_{\rm{GW}}(f)\equiv \frac{1}{\rho_c}\frac{d\rho_{\rm {GW}}}{d\ln f},
\end{equation}
where $f$ is the frequency and $\rho_c$ is the critical energy density of the universe, given by,
\begin{equation}
\rho_c = \frac{3H_0^2}{8\pi G}.
\end{equation}
\noindent Here, $H_0 = 100h~ {\rm{km/s/Mpc}}$, is the Hubble constant today, with the current value of $h=0.675\pm0.005$, and $G$ is Newton's gravitational constant.

% If there is sufficient friction in the plasma, the bubble walls end up reaching a terminal velocity $v_w$. This is called the non-runaway scenario. On the on the other hand, if the friction provided by the plasma is not sufficient, the bubble walls may continue to accelerate forever in the plasma rest frame, which is referred to as the runaway case. 

Given a strong FOPT, bubbles of the stable phase larger than a critical size expand rapidly.  If there is sufficient friction in the plasma, the bubble walls may end up reaching a terminal velocity $v_w$. GWs are produced when the bubbles collide and coalesce with each other. The latent heat of the colliding bubbles gets distributed into GWs via three main processes: bubble wall collisions ($\Omega_{\rm{col}}$), sound waves produced in the thermal plasma ($\Omega_{\rm{sw}}$), and the resulting MHD turbulence ($\Omega_{\rm{turb}}$). The contributions can be added in the linear approximation:
\begin{equation}
h^2 \Omega_{\rm{GW}} \simeq h^2 \Omega_{\rm{col}}+h^2 \Omega_{\rm{sw}}+h^2 \Omega_{\rm{turb}}.
\end{equation}

In this chapter, we have considered the spontaneous breaking of a global $U(1)_{\rm{FN}}$ symmetry, where no gauge bosons acquire a mass when they cross the bubble walls from the symmetric phase to the broken phase.  Also, there are no other light particles in the symmetric phase that become heavy in the broken phase. As a result, the friction exerted by the plasma on the bubble walls is expected to be negligible, which means that the walls can accelerate without any bounds. The runaway scenario \citep{Espinosa:2010hh,Bodeker:2017cim} is therefore relevant, and we take $v_w=1$. The contribution from bubble walls is expected to be the most significant while the effect of sound waves and MHD turbulence is subdominant. For the sake of completeness, though, we list all three contributions to $h^2\Omega_{\rm{GW}}$ \cite{Caprini:2015zlo} in the runaway scenario 
\begin{eqnarray}
h^2\Omega_{\rm{col}}(f) &=& 1.67\times 10^{-5} \left(\frac{H_*}{\beta}\right)^2 \left(\frac{\kappa_{\phi}\alpha}{1+\alpha}\right)^2\left(\frac{100}{g_*}\right)^{1/3}\left(\frac{0.11v_w^3}{0.42+v_w^2}\right)S_{\rm{col}}(f),\label{eq: 2_collisions}\\
h^2\Omega_{\rm{sw}}(f) &=& 2.65\times 10^{-6} \left(\frac{H_*}{\beta}\right)^2 \left(\frac{\kappa_{v}\alpha}{1+\alpha}\right)^2\left(\frac{100}{g_*}\right)^{1/3}v_w ~S_{\rm{sw}}(f),\label{eq: 2_soundwaves}\\
h^2\Omega_{\rm{turb}}(f) &= &3.35\times 10^{-4} \left(\frac{H_*}{\beta}\right)^2 \left(\frac{\kappa_{\rm{turb}}\alpha}{1+\alpha}\right)^{3/2}\left(\frac{100}{g_*}\right)^{1/3}v_w ~S_{\rm{turb}}(f). \label{eq: 2_turbulence}
\end{eqnarray}

\noindent The spectral shape functions, $S_{\rm{col}}$, $S_{\rm{sw}}$ and $S_{\rm{turb}}$ determine the power-law behavior of each contribution at low and high frequencies. These are, 
\begin{eqnarray}
S_{\rm{col}}(f) &=& \left(\frac{f}{f_{\rm{col}}}\right)^{2.8}\frac{3.8}{1+2.8(f/f_{\rm{col}})^{3.8}},\\
S_{\rm{sw}}(f) &=& \left(\frac{f}{f_{\rm{sw}}}\right)^3 \left(\frac{7}{4+3(f/f_{\rm{sw}})^2}\right)^{7/2},\\
S_{\rm{turb}}(f) &=& \left(\frac{f}{f_{\rm{turb}}}\right)^3\frac{1}{[1+(f/f_{\rm{turb}})]^{11/3}(1+8\pi f/h_*)}.
\end{eqnarray}

Here, $h_*$ is the Hubble rate at $T=T_*$, 
\begin{equation}
h_* = 1.65\times 10^{-7}~{\rm{Hz}}\left(\frac{T_*}{100~{\rm{GeV}}}\right)\left(\frac{g_*}{100}\right)^{1/6}.
\end{equation} 
Taking into account the expansion of the universe from $T=T_*$ to the present day, the red-shifted peak frequencies are,
\begin{eqnarray}
f_{\rm{col}} &=& 1.65\times 10^{-5}~{\rm{Hz}}
\left(\frac{\beta}{H_*}\right)\left(\frac{T_*}{100~{\rm{GeV}}}\right)\left(\frac{g_*}{100}\right)^{1/6}\left(\frac{0.62}{1.8-0.1v_w+v_w^2}\right),\label{eq: 2_fcol}\\
f_{\rm{sw}} &=& 1.9\times 10^{-5}~{\rm{Hz}}~\frac{1}{v_w}\left(\frac{\beta}{H_*}\right)\left(\frac{T_*}{100~{\rm{GeV}}}\right)\left(\frac{g_*}{100}\right)^{1/6},\label{eq: 2_fsw}\\
f_{\rm{turb}} &=& 2.7\times 10^{-5}~{\rm{Hz}}~\frac{1}{v_w}\left(\frac{\beta}{H_*}\right)\left(\frac{T_*}{100~{\rm{GeV}}}\right)\left(\frac{g_*}{100}\right)^{1/6}.\label{eq: 2_fturb}
\end{eqnarray}

We also need the efficiency factors: $\kappa_{\phi}$, $\kappa_{\rm{sw}}$, $\kappa_{\rm{turb}}$ representing the fraction of the latent heat which is shared by the three different processes. In the runaway scenario, these are,
\begin{equation}
\kappa_{\phi} = \frac{\alpha-\alpha_{\infty}}{\alpha}, ~~\kappa_v = \frac{\alpha_{\infty}}{\alpha}\kappa_{\infty}, ~~\kappa_{\rm{turb}} = \epsilon_{\rm{MHD}}\kappa_v.
\end{equation}
Here, $\epsilon_{\rm{MHD}}$ is the turbulent fraction of bulk motion in plasma, which is at most $5\%$-$10\%$ \cite{Hindmarsh:2015qta}. Also, $\kappa_{\infty}$ is given by,
\begin{eqnarray}
\kappa_{\infty} &\equiv& \frac{\alpha_{\infty}}{0.73+0.083\sqrt{\alpha_{\infty}}+\alpha_{\infty}}, \\
\alpha_{\infty} &\simeq& \frac{5}{4\pi^2}\frac{\sum_i c_i\Delta m_i^2}{g_*T_*^2},
\end{eqnarray}
where the sum over $i$ runs over all particles that are light in the symmetric phase and acquire mass in the broken phase, with $\Delta m_i^2$ as the mass squared difference in the two phases. Also,
$c_i = n_i (n_i/2)$ for bosons (fermions) with $n_i$ the number of degrees of freedom of the particle. Since in our case there are no such light particles that become heavy as they cross the bubble wall, we expect $\alpha_{\infty}$ to be negligible. As a result $\alpha\gg\alpha_{\infty}$, and only the contribution from bubble collisions is significant,
\begin{equation}
h^2\Omega_{\rm{GW}} \simeq h^2\Omega_{\rm{col}}.
\end{equation}

\subsection{Detection prospects}
To get a detectable GW signal resulting from FOPT, we need large $\alpha$, and small $\beta/H_*$, as is evident from the dependence of expressions \eqref{eq: 2_collisions}, \eqref{eq: 2_soundwaves}, and \eqref{eq: 2_turbulence} on these quantities. In the parameter scans considered in section \ref{sec: 2_FOPT}, we saw that the strongest FOPT occurs near the left boundary, as shown in the plots of Fig.\,\ref{fig: noVLQ} and Fig.\,\ref{fig: param_scan}.

\begin{figure}[tbp]
\centering % \begin{center}/\end{center} takes some additional vertical space
\includegraphics[width=.49\textwidth]{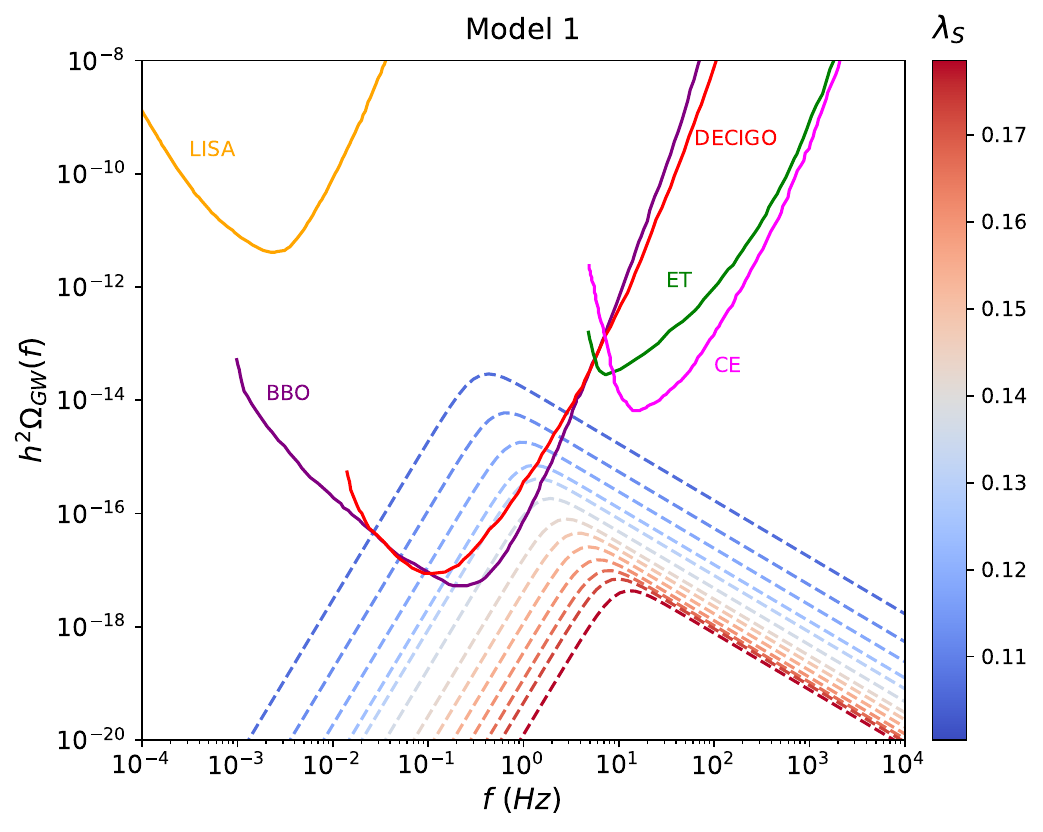}
\hfill
\includegraphics[width=.49\textwidth]{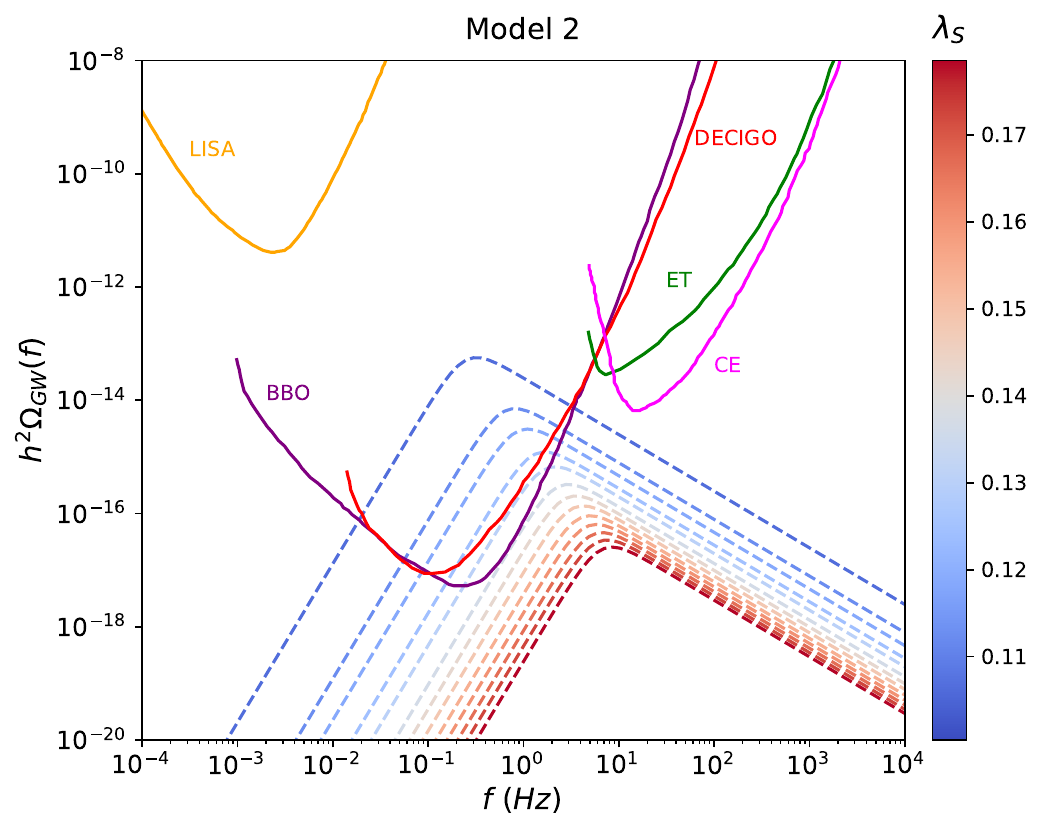}
\caption{\label{fig: gw} GW spectra for model 1 and model 2 for $\lambda_s\in[0.1,0.18]$. Here $\lambda_{HS} = 4.253$, $v_s=10^5$ GeV, and $y=0.5$.}
\end{figure}

\begin{figure}[tbp]
\centering % \begin{center}/\end{center} takes some additional vertical space
\includegraphics[width=.47\textwidth]{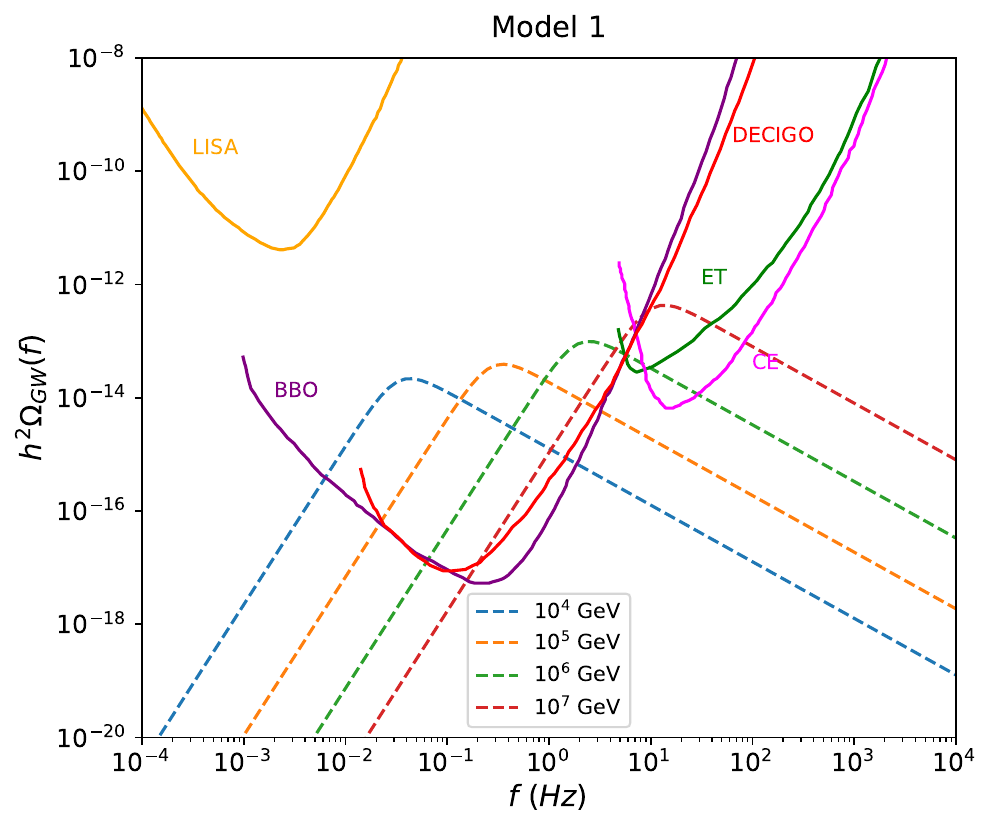}
\hfill
\includegraphics[width=.47\textwidth]{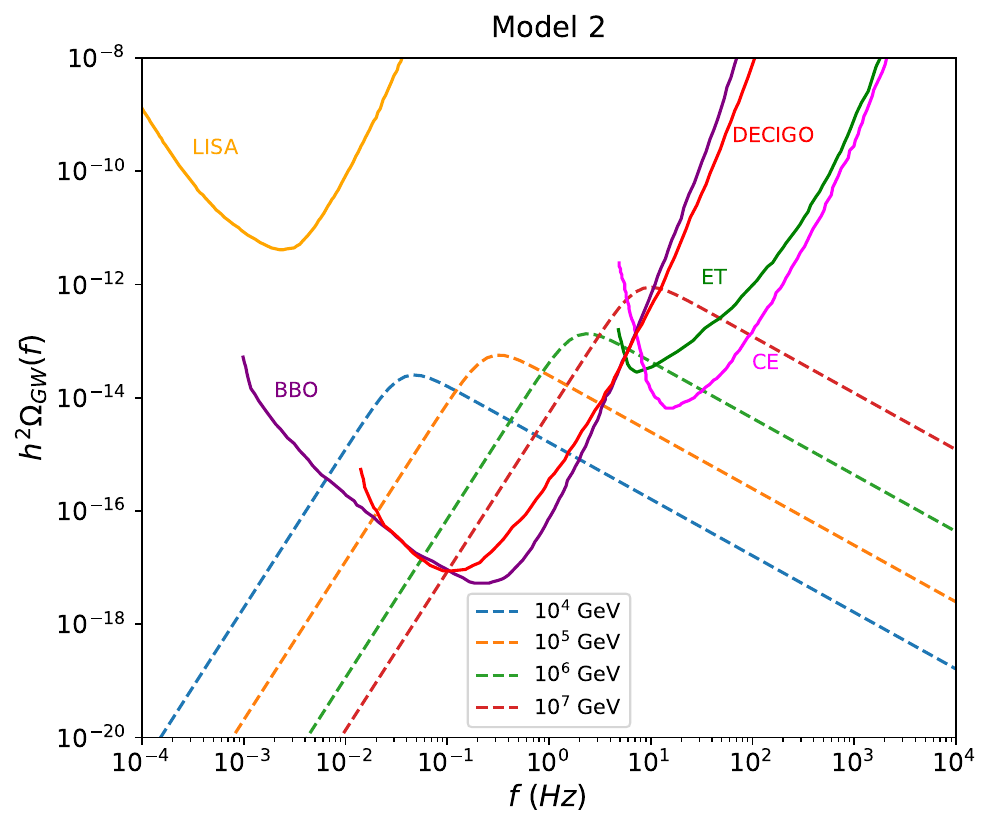}
\caption{\label{fig: gw_vs} GW spectra for model 1 and model 2 for $v_s = 10^{4,5,6,7}~{\rm{GeV}}$. Here $\lambda_s =0.105$, $\lambda_{HS} = 4.253$, and $y=0.5$. The peak frequency is seen to shift from left to right as $v_s$ increases.}
\end{figure} 

An example of GW spectra highlighting the dependence on model parameters is shown in Fig.\,\ref{fig: gw}. For model 1 and model 2, the dashed curves show the GW spectrum, when $\lambda_{HS} = 4.253$, and $\lambda_S$ is varied from $\lambda_S=0.1$ to $\lambda_S$ =0.18. The other parameters are kept fixed at $v_s=10^5$ GeV, and $y=0.5$. Solid curves are the power law integrated sensitivity curves (PLISC) \cite{Thrane:2013oya} for the GW observatories LISA \cite{LISA:2017pwj}, BBO \cite{Corbin_2006}, DECIGO \cite{Musha:2017usi}, ET \cite{Punturo:2010zz}, and CE \cite{Abbott_2017}. The strength of the GW spectrum decreases as $\lambda_S$ increases. This corresponds to moving horizontally from left to right, starting close to the left boundary in the $\lambda_S$-$\lambda_{HS}$ plane in the plots of Fig.\,\ref{fig: param_scan}. As the GW strength decreases, it eventually goes below the noise floor of DECIGO and BBO, becoming undetectable.

In Fig.\,\ref{fig: gw_vs}, we show the effect of varying the scale $v_s$  on the GW spectrum for a benchmark value of $(\lambda_S,\lambda_{HS})$. Taking $v_s = 10^4,~10^5, ~10^6,~10^7~{\rm{GeV}}$, we observe that the peak frequency shifts from left to right as $v_s$ increases. This can be explained from Eq.\,\eqref{eq: 2_fcol}, Eq.\,\eqref{eq: 2_fsw} and Eq.\,\eqref{eq: 2_fturb}, where the peak frequency is proportional to $T_*\sim v_s$. The height of the peak amplitude is seen to increase with $v_s$ due to the temperature dependence of the RHS of Eq.\,\eqref{eq: 1_nucl_criterion2}, which leads to a larger amount of supercooling at higher $T$. Different detectors are suitable for different scales $v_s$. For $v_s =10^{4,5}$ GeV, the chosen benchmark point gives a strong GW signal at DECIGO and BBO for both model 1 and model 2. While for $v_s=10^6$ GeV, the signal would also be detectable at CE, in addition to DECIGO and BBO. For $v_s=10^7$ GeV, the signal can be detected at BBO, DECIGO, CE, and ET. For the entire viable region of FOPT, the GW signal is not observable at LISA, since the peak amplitude is almost always found to be smaller than $10^{-12}$.

\begin{figure}[tbp]
\centering % \begin{center}/\end{center} takes some additional vertical space
\includegraphics[width=.47\textwidth]{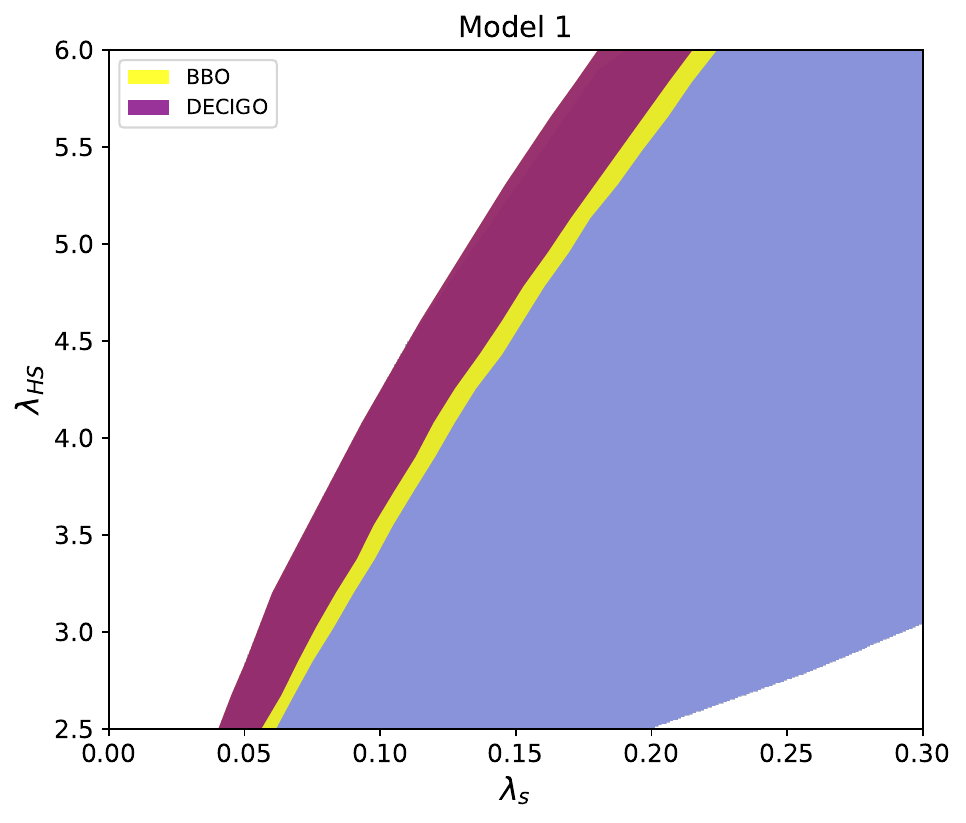}
\hfill
\includegraphics[width=.47\textwidth]{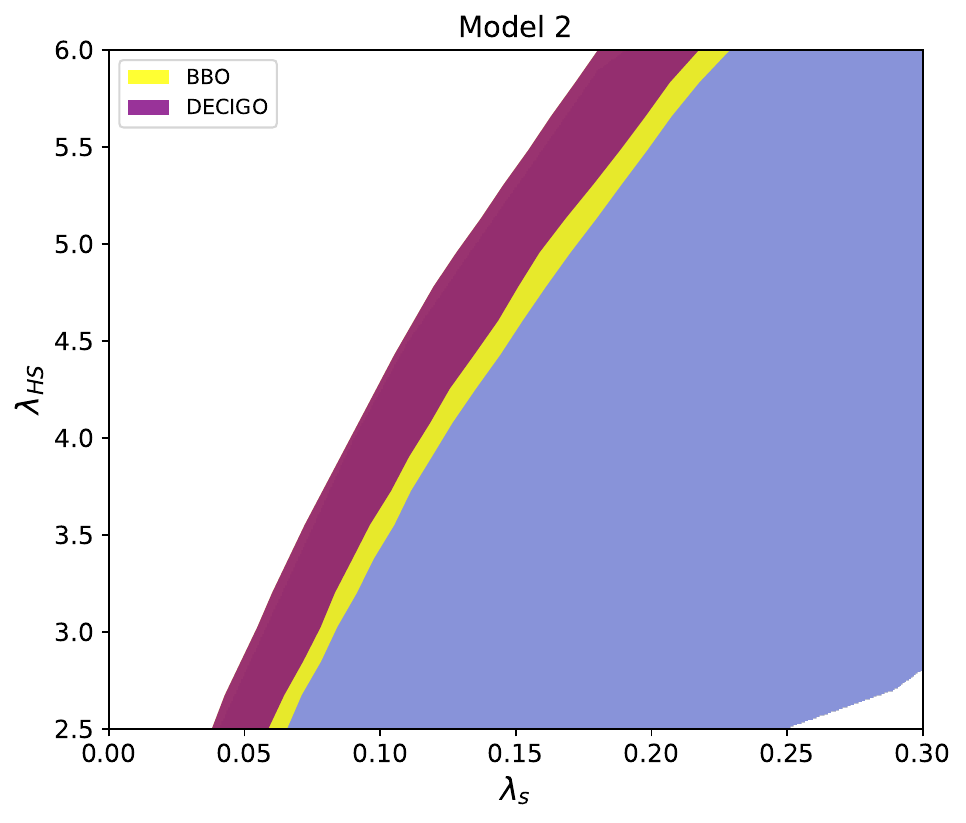}
\caption{\label{fig: gw_detection} Regions of GW detection for model 1 and model 2, with $v_s=10^5$ GeV, and $y=0.5$. The region coloured in purple overlaps the yellow region, since BBO is more sensitive compared to DECIGO.} 
\end{figure}

\begin{figure}[tbp]
\centering % \begin{center}/\end{center} takes some additional vertical space
\includegraphics[width=.47\textwidth]{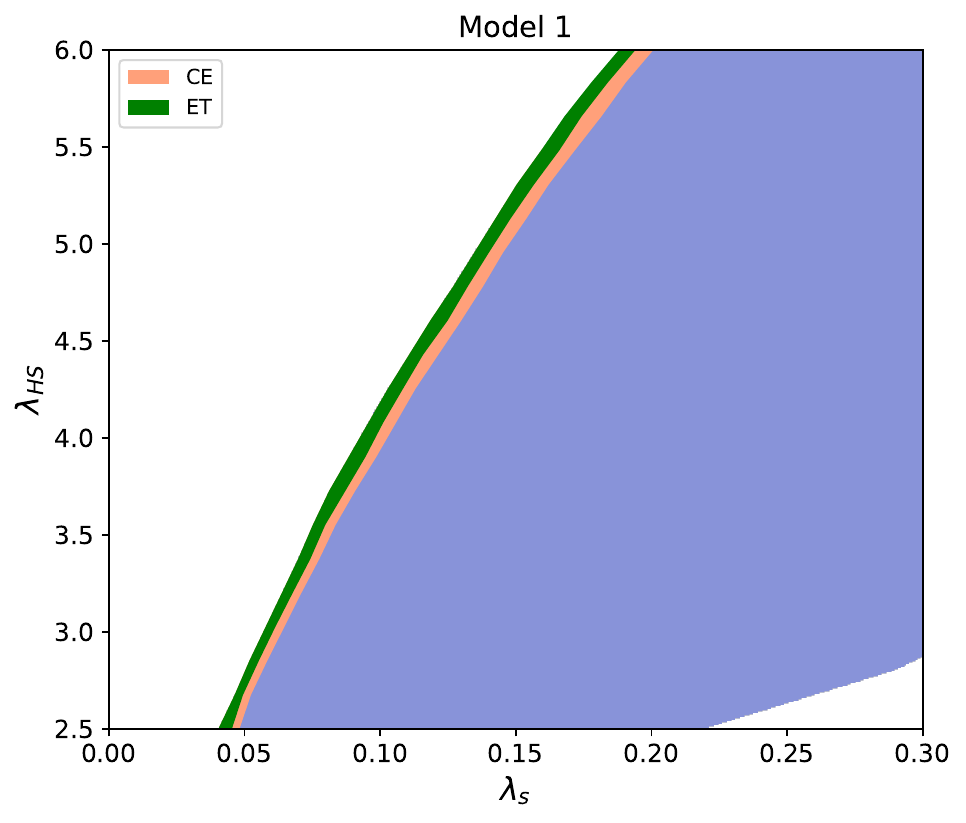}
\hfill
\includegraphics[width=.47\textwidth]{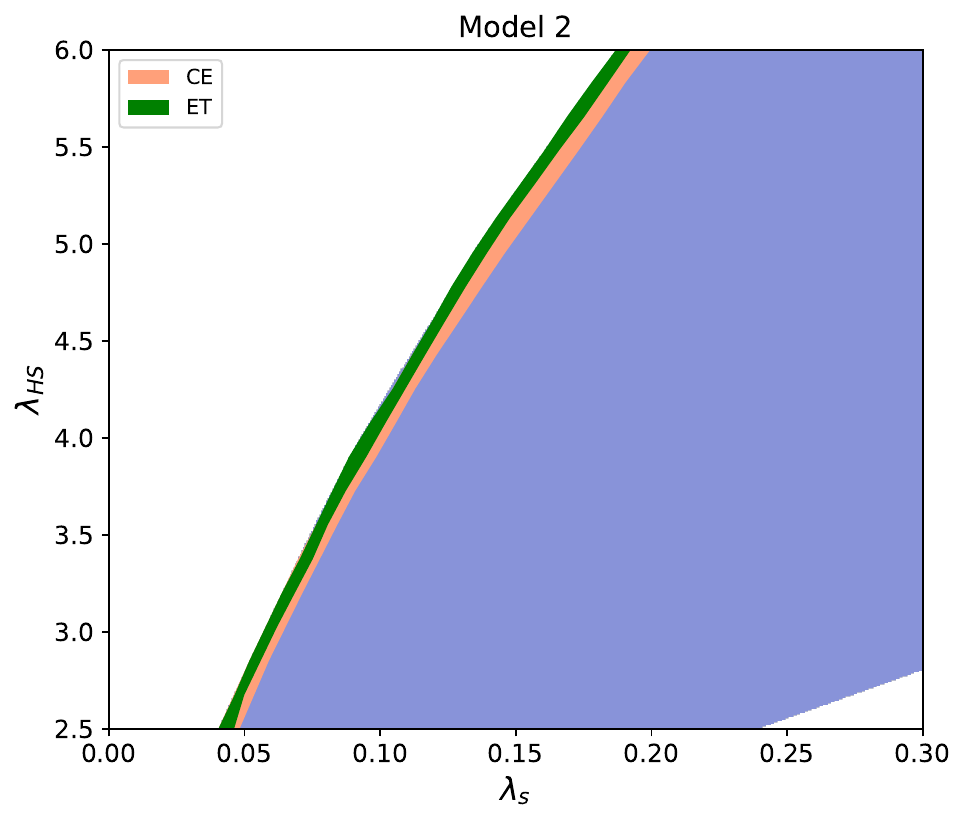}
\caption{\label{fig: gw_detection_vs1.e7} Regions of GW detection for model 1 and model 2, with $v_s=10^7$ GeV, and $y=0.5$. The region colored in green overlaps the peach region since CE is more sensitive compared to ET.} 
\end{figure}

Next, in Fig.\,\ref{fig: gw_detection}, we showcase the detectable regions of the parameter space for model 1 (left panel) and model 2 (right panel), when $v_s=10^5$ GeV, and $y=0.5$. The colored region corresponds to the region sensitive to DECIGO (purple), and BBO (yellow). The two regions overlap with each other; the region corresponding to BBO is broader than that of DECIGO since the former is more sensitive. In the blue region, the GW signal is too weak to be detected. In Fig.\,\ref{fig: gw_detection_vs1.e7}, we show the detectable parameter space for both models with $v_s=10^7$ GeV, and $y=0.5$. Now the parameter space is accessible to ET and CE in addition to DECIGO and BBO. For convenience, we show only regions for ET (green) and CE (peach), as the regions for BBO and DECIGO are much smaller. The detectable region is narrower compared to the region at $10^5$ GeV.  The best detection prospects for $v_s=10^7$ GeV are at CE.

Previously in Fig.\,\ref{fig: param_scan}, it was found that the constant $\alpha$ contours for chosen values lie closer to the boundary for model 1 as compared to model 2. However, it is interesting to note that apart from minor qualitative differences, the GW signal strength is comparable for the two models in the detectable region, both at $v_s=10^5$ GeV and $v_s=10^7$ GeV. This is because the regions of detectability require $\alpha\gtrsim 0.03$, where the contours in the two models are similar. This implies that a detectable GW background cannot discriminate between model 1 and model 2. The GW signal is therefore not sensitive to the specific $U(1)_{\fn}$ charge assignment, as long as there is only one flavon that breaks the symmetry. 

\subsection{Influence of $\lambda_{HS}$ on Higgs-flavon mixing}
Now we discuss the implications of having $\mathcal{O}(1)$ coupling $\lambda_{HS}$ as seen in Fig.\,\ref{fig: noVLQ} and Fig.\,\ref{fig: param_scan}, on the Higgs mass and mixing, for relatively low flavon masses ($\sim$ few TeV). Recall that the parameters discussed in this chapter are evaluated at the scale $\mu^2=v_s^2$, i.e., the scale at which GW production occurs. Below this scale, the flavon field can be integrated out, and we are left with the EFT Lagrangian given by \cite{Gorbahn:2015gxa}:
\begin{equation}
-\mathcal{L}_{\eft} \supset -\widetilde{\mu}^2_H(\mu) |H|^2 + \widetilde{\lambda}_H(\mu) |H|^4 + \frac{\overline{c}_H}{2v^2} \partial_{\mu}|H|^2\partial^{\mu}|H|^2,
\end{equation}
where $\widetilde{\mu}^2_H$ and $\widetilde{\lambda}_H$ are low-energy effective couplings. The coefficient $\overline{c}_H$ is induced by Higgs-flavon mixing, $\overline{c}_H = \frac{\lambda_{HS}^2 v^2 v_s^2}{m_S^4} \equiv x^2$, with $m^2_S=2\lambda_S v_s^2$, and $v=246.2$ GeV. 

In principle, a $|H|^6$ term is also generated, but its coefficient vanishes unless the flavon symmetry is explicitly broken. Regardless of the $|H|^6$ operator, the $\overline{c}_H$ operator affects the normalization of the Higgs kinetic term after EWSB. A rescaling of the Higgs field is therefore required,
\begin{equation}
h\rightarrow h'=\sqrt{1+\overline{c}_H} ~ h \approx h\bigg(1+\frac{x^2}{2}\bigg).
\end{equation}
The normalization rescales all couplings of $h$ by different amounts. For example, the coupling between the Higgs and the weak bosons is modified by \cite{Gorbahn:2015gxa}, 
\begin{equation}
g_{hVV} = \frac{g_V v}{\sqrt{1+x^2}}\approx g_V v \bigg(1-\frac{x^2}{2}\bigg).
\end{equation}
The Higgs mass is given by \cite{Dawson:2021jcl},
\begin{equation}
m_H^2 = 2\widetilde{\lambda}_Hv^2(1+2x^2).
\end{equation}

Defining $\lambda\equiv \widetilde{\lambda}_H(1+2x^2)$, gives us back the SM relation,
\begin{equation}
\lambda = \frac{m_H^2}{2v^2}.
\end{equation}
Taking an extreme benchmark point from our parameter space, $\lambda_{HS}=5,~\lambda_S=0.2$, with the lowest value of $v_s=10^4~\rm{GeV}$, we get, $x^2\approx 0.0946$. Hence, even for this extreme case with $v_s=10^4~\rm{GeV}$, the normalization effects are small \cite{Gorbahn:2015gxa}. 

At the EW scale, the heavy flavon also modifies the triple Higgs coupling \cite{Curtin:2014jma} according to,
\begin{equation}
\lambda_3 \simeq \frac{m_H^2}{2v} + \frac{\lambda_{HS}^3 v^3}{24\pi^2 m_S^2},
\end{equation}
where the first term is the SM contribution: $\lambda_3^{\rm{SM}}\equiv \frac{m_H^2}{2v}$, and the second term is induced by the flavon via loop effects. The current bound on $\lambda_3$ from Higgs pair production data at the LHC is $-9\lesssim \lambda_3/\lambda_3^{\rm{SM}} \lesssim 15$ \citep{CMS:2017ihs,CMS:2017orf,ATLAS:2018rnh}, and is therefore too weak to exclude any parameter space of our model.

\section{Conclusions}\label{sec: conclusion}
In this chapter, we have explored the possibility of probing FN models with GW detectors, if the flavon undergoes a SFOPT. To do so, we have constructed two minimal, non-supersymmetric UV completions of the FN mechanism involving a single flavon.  To avoid any possible complications arising from gauging it, we have chosen the family symmetry group as global $U(1)$. However, the two models considered here are UV complete only in the sense that they generate the effective FN operators which give rise to the hierarchy of quark Yukawas at low scale. We find that it is essential to add heavy bosons couple to the flavon, to counter the destabilizing effect of adding several fermions to the theory. We have chosen a region of parameter space where the effect of the heavy bosons on the nature and strength of FOPT is insignificant, to make predictions independent of any specific realization of heavy bosons. 

To avoid stringent constraints from collider experiments, we have focused on the intermediate energy scale, $10^4$-$10^7$ GeV for the scale of the FN mechanism. We find that in the $\lambda_{S}$-$\lambda_{HS}$ plane, a strong FOPT is observed near the boundary separating the FOPT region from the region where the PT never completes. The region for observing a GW signal therefore lies within a thin strip close to this boundary. Detection prospects are the best when $v_s \sim 10^4$-$10^5$ GeV, in which case the peak of the GW signal lies in the sensitivity range of BBO and DECIGO. Higher scales of $v_s \sim 10^6$-$10^7$ GeV can be probed by BBO, DECIGO, as well as CE and ET.

Detecting a stochastic GW background can become a powerful tool to probe the flavor symmetry-breaking scale in the coming years, along with complementary collider signatures. For the two models considered, the GW signatures are of comparable strength when the same set of parameters is used. Therefore, we conclude that it is not possible to discriminate between them based solely on the GW signature. We take the liberty to extrapolate this finding to all models which involve a single flavon, when the family symmetry group is $U(1)_{\fn}$. As such, models involving more than one flavon may be discriminated by GWs since they may feature multiple peaks, owing to different scales of the \vevs, but this possibility remains to be seen. It would be interesting to see if the GW signature can shed some light on the FN symmetry group. 

 %Flavon PT
\newpage
\null
\newpage

\chapter{FOPT in DLRSM}
\label{chapter3}
\linespread{0.1}
\graphicspath{{Chapter3/Figures}}
\pagestyle{headings}
%{\bf{\Huge Chapter 5 }}\\
%\noindent\rule{5.5cm}{1.5pt} \\
%\vspace*{0.8cm}\\
%{\bf{\Huge Metastability in Standard Model}}\\
\noindent\hrule height 1mm 
\vspace{4mm}
%\noindent\rule{14cm}{2.8pt} 
% %%%%%%%%%%%%

\section{Introduction}
Left-right symmetric models\,(LRSMs)\,\cite{Pati:1974yy,Mohapatra:1974gc,PhysRevD.11.566,Senjanovic:1975rk,Senjanovic:1978ev} provide an attractive scenario for addressing several limitations of the standard model (SM). In LRSM, the SM gauge group is extended from $\mathcal{G}_{\rm{SM}} = SU(3)_c\times SU(2)_L\times U(1)_{Y}$ to $\mathcal{G}_{\rm{LRSM}} = SU(3)_c\times SU(2)_L\times SU(2)_R\times U(1)_{B-L}$, and the right-chiral fermions transform as doublets under the $SU(2)_R$ subgroup. The non-observation of a right-handed charged current at colliders puts a lower bound on the $SU(2)_R\times U(1)_{B-L}$-breaking scale, $v_R\gtrsim \mathcal{O}(10)$\,TeV, while the upper bound remains unconstrained. 

There are different realizations of LRSM, depending on the scalars involved in the spontaneous breaking of $\mathcal{G}_{\textrm{LRSM}}$ to $\mathcal{G}_{\rm{SM}}$. These also differ in the mechanism for generating fermion masses. The triplet left-right symmetric model\,(TLRSM)\,\cite{Maiezza:2016ybz,PhysRevD.44.837,Senjanovic:2016bya}, contains a scalar bi-doublet, and two $SU(2)$ triplets. The charged fermion masses are then generated by the bi-doublet, whereas the neutrino masses are generated by the type-II seesaw mechanism\,\cite{Schechter:1981cv}. On the other hand, the scalar sector of a doublet left-right symmetric model\,(DLRSM) consists of a scalar bi-doublet and a pair of $SU(2)$ doublets\,\cite{Senjanovic:1978ev,Mohapatra:1977be}. In DLRSM, neutrino masses can be incorporated by extending the model with an additional charged singlet scalar\,\cite{Babu:1988qv, FileviezPerez:2016erl, FileviezPerez:2017zwm, Babu:2020bgz}. 
Contrary to TLRSM, where the \vev of the triplets is constrained to be small, there are no sources of custodial symmetry breaking in DLRSM at the tree level. Apart from TLRSM and DLRSM, other variations have also been discussed in the literature and have different experimental consequences\,\cite{Ma:2012gb,Frank:2020odd,Graf:2021xku,Akhmedov:1995vm}. 

\begin{comment}
The era of gravitational wave\,(GW) astronomy was kickstarted by the observation of GW from a binary black hole merger by the aLIGO collaboration in 2015\,\cite{Abbott_2016}. Several ground-based and space-based observatories such as LISA\,\cite{LISA:2017pwj}, DECIGO\,\cite{Seto:2001qf}, BBO\,\cite{Corbin:2005ny}, ET\,\cite{Punturo:2010zz}, and CE\,\cite{LIGOScientific:2016wof} are planned and will be functional in the coming decades. Various phenomena in the early universe, such as inflation, cosmic strings, domain walls, and strong first-order phase transitions\,(SFOPT) can lead to a stochastic GW background\,\cite{Caprini:2015zlo,Athron:2023xlk,LISACosmologyWorkingGroup:2022jok,Caprini:2018mtu}. The
upcoming GW observatories will be capable of detecting the GW background from SFOPT upto symmetry-breaking scales as high as $10^6- 10^7$\,GeV\,\cite{Dev:2016feu,Addazi:2018nzm,Ringe:2022rjx}. 
\end{comment}

In the context of LRSM, GW astronomy presents a novel approach to probe the scale $v_R$ by studying the possibility of an observable GW background from SFOPT within the LRSM.  Different realizations of LRSM have been explored in the literature for GW imprints: through SFOPT in TRLSM\,\cite{Brdar:2019fur, Li:2020eun}, and in an LR model with seesaw-like fermion mass generation\,\cite{Graf:2021xku}, and also from domain walls arising out of the breaking of the discrete parity symmetry\,\cite{Borah:2022wdy}. However, GW imprints of DLRSM have not yet been explored in the literature. Recently, it was shown that the pattern of electroweak symmetry breaking (EWSB) in DLRSM can be vastly different from the other versions of LRSM, with interesting consequences from precision observables\,\cite{Bernard:2020cyi} and Higgs data\,\cite{Karmakar:2022iip}. It is therefore interesting to study the GW signature in DLRSM.

EWSB in DLRSM happens via three \vevs: $\k_1,~\k_2$, coming from the bi-doublet, and $v_L$ coming from the $SU(2)_L$ doublet. These are constrained by the relation, $\k_1^2+\k_2^2+v_L^2=v^2$, where $v = 246$\,GeV. It is useful to define the \vev ratios: $r = \k_2/\k_1,~w = v_L/\k_1$. As the custodial symmetry is preserved at the tree level, the \vevs $\k_1,~ \k_2$, and $v_L$ can all be sizable, i.e. $r$ and $w$ can be $\mathcal{O}(0.1)$ and $\ordone$ respectively. In Ref.\,\cite{Bernard:2020cyi}, it was shown that the EW precision data prefers a large value of $w$. Further, in Ref.\,\cite{Karmakar:2022iip} it was shown that the measurement of the Higgs signal strength and meson mixing bounds prefer large values of $r$ and $w$. It is therefore interesting to note that, unlike TLRSM, EWSB in DLRSM can be considerably different from that in SM, even though the $SU(2)_R\times U(1)_{B-L}$-breaking dynamics is decoupled from the EW scale. In this chapter, we ask \textbf{(i)} whether DLRSM can lead to a detectable GW background in some region of the parameter space, and \textbf{(ii)} whether this region of the parameter space prefers a special pattern of EWSB.

The rest of the chapter is organized as follows. In Sec.\,\ref{Sec: 3_Model}, we give a brief review of DLRSM: field content, symmetry breaking, and mass generation in the gauge, fermion, and gauge sectors. In Sec.\,\ref{subsec: theory} and Sec.\,\ref{subsec: higgs}, we discuss the theoretical bounds and the constraints from the Higgs data, respectively. In Sec.\,\ref{sec: 3_effective potential}, we construct the one-loop finite temperature effective potential required to study the phase transition (PT) associated with $SU(2)_R\times U(1)_{B-L}$-breaking. We then describe our procedure for scanning the parameter space in Sec.\,\ref{sec: paramscan}. In Sec.\,\ref{sec: 3_GW}, we discuss the GW background obtained for points with SFOPT. In Sec.\,\ref{sec: detection prospects}, we compute the signal-to-noise ratio (SNR) for six benchmark points, at various planned GW detectors such as FP-DECIGO, BBO, and Ultimate-DECIGO. In Sec.\,\ref{Sec: collider}, we discuss future collider probes that can complement the GW signal. Finally, in Sec.\,\ref{sec: summary}, we summarize our key findings and present concluding remarks.

\section{The model}
\label{Sec: 3_Model}
We follow the notation of Refs.\,\cite{Bernard:2020cyi,Karmakar:2022iip} for the scalar potential and \vev structure of the scalar multiplets. The fermion content of the model has the following charges under the LRSM gauge group, $\mathcal{G}_{\rm{LRSM}} = SU(3)_c \times SU(2)_L\times SU(2)_R \times U(1)_{B-L}$,
\bea
Q_L &=& \begin{pmatrix}
 u_L \\
 d_L 
\end{pmatrix} \sim (3,2,1,1/3), \hspace{10mm}
Q_R = \begin{pmatrix}
 u_R \\ 
 d_R 
\end{pmatrix} \sim (3,1,2,1/3), \nn \\
L_L &=& \begin{pmatrix}
 \nu_L \\
 e_L 
\end{pmatrix} \sim (1,2,1,-1), \hspace{12mm}
L_R = \begin{pmatrix}
 \nu_R \\
 e_R 
\end{pmatrix} \sim (1,1,2,-1),
\label{eq:fermion}
\eea
where the quantum numbers of the  multiplets under the sub-groups of $\mathcal{G}_{\rm{LRSM}}$ are indicated in brackets. We have suppressed the family index $i\in\{1,2,3\}$ for three generations of quarks and three generations of leptons. The right-handed neutrino $\nu_R$ is needed to complete the $SU(2)_R$ lepton doublet. This choice of fermions is required for the cancellation of the $U(1)_{B-L}$ gauge anomaly and ensures that the model is manifestly symmetric under the transformations: $Q_L \leftrightarrow Q_R$, $L_L \leftrightarrow L_R$. 

\subsection{Scalar sector}
\label{sec:ScalarSector}
The scalar sector of DLRSM includes a complex bi-doublet $\Phi$ needed to generate charged fermion masses, and two doublets $\chi_L$ and $\chi_R$, which participate in the EW- and LR-symmetry breaking, respectively. These scalar multiplets and their charges under $\mathcal{G}_\text{LRSM}$ are:
\bea
\Phi = \begin{pmatrix}
    \phi_1^0 & \phi_2^+ \\
    \phi_1^- & \phi_2^0
\end{pmatrix} \sim (1,2,2,0), ~ \chi_L = \begin{pmatrix}
    \chi_L^+ \\
    \chi_L^0
\end{pmatrix} \sim (1,2,1,1), ~ \text{and}~\chi_R = \begin{pmatrix}
    \chi_R^+ \\
    \chi_R^0
\end{pmatrix} \sim (1,1,2,1).\nn\\
\eea

We take the potential to be parity-symmetric, i.e., the couplings of `L' and `R' fields are equal. This imposes an additional discrete symmetry $\mathcal{P}: L\leftrightarrow R$ on the Lagrangian. The most general, CP-conserving, renormalizable scalar potential is then given by,
\bea\label{eq: 3_potential}
V  &=& V_2 + V_3 + V_4,\nn\\
V_2  &=& -\m_1^2\tr(\Phi^{\dagger}\Phi) - \m_2^2\ [\tr(\tilde{\Phi}\Phi^{\dagger})+ \tr(\tilde{\Phi}^{\dagger} \Phi)] - \m_3^2\ [\chi_L^{\dagger} \chi_L + \chi_R^{\dagger} \chi_R], \nn\\
V_3 &=& \m_4\  [\chi_L^{\dagger} \Phi \chi_R + \chi_R^{\dagger} \Phi^{\dagger} \chi_L] + \m_5\  [\chi_L^{\dagger} \tilde{\Phi} \chi_R + \chi_R^{\dagger}\tilde{\Phi}^{\dagger}\chi_L ], \nn\\
V_4 &=& \l_1[\tr(\Phi^{\dagger}\Phi)]^2 + \l_2\ [ [\tr(\tilde{\Phi} \Phi^{\dagger})]^2
 + [\tr(\tilde{\Phi}^{\dagger} \Phi)]^2 ]  + \l_3\text{Tr}(\tilde{\Phi} \Phi^{\dagger}) \, \tr(\tilde{\Phi}^{\dagger} \Phi)  \nn\\
 &&+ \l_4\tr(\Phi^{\dagger}\Phi) \, [\tr(\tilde{\Phi}\Phi^{\dagger})+ \tr(\tilde{\Phi}^{\dagger}\Phi)]  + \rho_1\  [(\chi_L^{\dagger} \chi_L )^2 + (\chi_R^{\dagger} \chi_R )^2]
 + \rho_2\  \chi_L^{\dagger} \chi_L \chi_R^{\dagger}\chi_R \nn\\
 &&+ \alpha_1\tr(\Phi^{\dagger} \Phi ) [\chi_L^{\dagger}\chi_L + \chi_R^{\dagger}\chi_R ]
 + \Big\{\frac{\alpha_2}{2} \ [\chi_L^{\dagger} \chi_L  \tr(\tilde{\Phi} \Phi^{\dagger} ) + \chi_R^{\dagger} \chi_R  \tr(\tilde{\Phi}^{\dagger} \Phi )] + {\rm h.c.} \Big\} \nn\\
 &&+ \alpha_3\ [\chi_L^{\dagger}\
 \Phi \Phi^{\dagger}\chi_L + \chi_R^{\dagger} \Phi^{\dagger} \Phi  \chi_R  ] 
 + \alpha_4\ [\chi_L^{\dagger}\
 \tilde{\Phi} \tilde{\Phi}^{\dagger}\chi_L + \chi_R^{\dagger} \tilde{\Phi}^{\dagger} \tilde{\Phi}  \chi_R  ],
 \label{eq:scalarpotential}
\eea
with $\tilde{\Phi}\equiv \sigma_2\Phi^*\sigma_2$. The potential has mass parameters: $\{\m_{1,2,3,4,5}\}$, and quartic couplings: $\{\l_{1,2,3,4},\a_{1,2,3,4},\rho_{1,2}\}$. We assume all parameters to be real for simplicity. 

The neutral scalars can be written in terms of real and imaginary components,
\bea
\phi_1^0 &=& \frac{1}{\sqrt{2}} (\phi_{1r}^0 + i \phi_{1i}^0), \,\,\,\,
\phi_2^0 = \frac{1}{\sqrt{2}} (\phi_{2r}^0 + i \phi_{2i}^0), \nn\\
\chi_L^0 &=& \frac{1}{\sqrt{2}} (\chi_{Lr}^0 + i \chi_{Li}^0), \,\,\,\,
\chi_R^0 = \frac{1}{\sqrt{2}} (\chi_{Rr}^0 + i \chi_{Ri}^0). 
\eea
We assign non-zero \vevs only to the real components of the neutral scalars and do not consider CP- or charge-breaking minima. The \vev structure is denoted by
\beq
     \langle\Phi \rangle = \frac{1}{\sqrt{2}}\begin{pmatrix}
     \k_1 & 0\\
     0 & \k_2
 \end{pmatrix}, ~ \langle\chi_L\rangle = \frac{1}{\sqrt{2}} \begin{pmatrix}
     0\\
     v_L
 \end{pmatrix}, ~ \langle\chi_R \rangle = \frac{1}{\sqrt{2}}\begin{pmatrix}
     0\\
     v_R
 \end{pmatrix}\,.
 \eeq

The pattern of symmetry breaking is as follows:
$$SU(2)_L\times SU(2)_R \times U(1)_{B-L}\xrightarrow{\mathit{~~v_R~~}} SU(2)_L\times U(1)_{Y} \xrightarrow{\mathit{\k_1,\k_2,v_L}} U(1)_Y . $$
The \vev $v_R$ of the doublet $\chi_R$ breaks $SU(2)_R\times U(1)_{B-L}$, while the three \vevs $\k_1,~\k_2$, and $v_L$ trigger EWSB. Note that the discrete LR symmetry $\mathcal{P}$, is also broken by $v_R$, which leads to the formation of domain walls \cite{Borah:2022wdy,Chakrabortty:2019fov,Mishra:2009mk,Borah:2011qq,Banerjee:2020zxw,Borboruah:2022eex}. Such a domain wall network can dominate the energy density of the universe at late times. To avoid domination, a small bias term can be introduced via explicit LR-breaking operators, so that the domain walls become unstable and decay before the epoch of big bang nucleosynthesis. For example, the bias was generated by Planck-suppressed higher-dimensional operators in Ref.\,\cite{Borah:2022wdy}. Due to large suppression, these do not affect the nature and strength of the $SU(2)_R \times U(1)_{B-L}$ breaking PT.

As mentioned earlier, the EW \vevs can be conveniently expressed in terms of the \vev ratios $r$ and $w$ as, $\k_2 = r \k_1$ and $v_L = w \k_1$.
Then, $\k_1^2 (1 + r^2 + w^2) = v^2$, i.e., the value of $\k_1$ is fixed for a given $r$ and $w$. The absence of a right-handed charged current in collider experiments implies a hierarchy of scales $v_R\gg v$. 

In terms of the \vevs $\k_1, \k_2, v_L$, and $v_R$, the minimization conditions are,
\beq
\frac{\partial V}{\partial \k_1} = \frac{\partial V}{\partial \k_2} = \frac{\partial V}{\partial v_L} = \frac{\partial V}{\partial v_R} = 0.
\eeq
Using the minimization conditions, we trade $\m_1^2$, $\m_2^2$, $\m_3^2$, and $\m_5$ for the \vevs, $\m_4$, and quartic couplings (see Appendix \ref{appendix: min} for full expressions). Thus, the parameters of the DLRSM scalar sector reduce to
\bea 
\{\l_{1,2,3,4}, \a_{1,2,3,4}, \rho_{1,2}, \m_4, r, w, v_R \}\,\, .
\label{eq: scalar_sector_params}
\eea

The CP-even, CP-odd, and charged scalar mass matrices are obtained using 
\beq
m^2_{ij} = \left.\frac{\partial^2 V}{\partial \varphi_i \,\partial\varphi_j}\right\vert_{\langle \varphi \rangle},
\eeq
where
\beq\label{eq: varphi}
\varphi \equiv \{\phi^0_{1r},\phi^0_{2r},\chi^0_{Lr},\chi^0_{Rr},\phi^0_{1i},\phi^0_{2i},\chi^0_{Li},\chi^0_{Ri},\phi^{\pm}_{1},\phi^{\pm}_{2},\chi^{\pm}_{L},\chi^{\pm}_{R}\}\,\,.
\eeq
Physical scalar masses and mixing angles are obtained by diagonalizing these matrices. We denote the physical spectrum of scalars by:
 CP-even scalars, $h,~ H_1,~ H_2,~ H_3$,
 CP-odd scalars, $A_1,~A_2$, and the 
charged scalars, $H_1^{\pm},~H_2^{\pm}$. 

The lightest CP-even scalar, $h$ has a mass of the order $v$, and is identified with the SM-like Higgs with mass $\sim 125$ GeV. Using non-degenerate perturbation theory, $m_h$ is estimated as\,\cite{Bernard:2020cyi,Karmakar:2022iip}
\bea 
m_{h, \text{analytic}}^2 &=& \frac{\k_1^2}{2 (1+r^2 +w^2)}\times\nn\\
&&\Bigg( 4 \Big(\l_1 (r^2+1)^2 + 4 r (\l_4(r^2+1)+r \l_{23}) + w^2 (\a_{124} + r^2 (\a_1+\a_3) \nn\\ &&+\a_2 r) + \rho_1 w^4 \Big) 
 -\frac{1}{\rho_1}(\a_{124} + r^2 (\a_1+\a_3) + \a_2 r + 2 \rho_1 w^2)^2 \Bigg)\,,
\label{eq:mh_analytic}
\eea
where, $\a_{124} \equiv \a_1+r\a_2+\a_4$, and $\l_{23} = 2\l_2+\l_3$. In the limit $r,w\rightarrow 0$, the above expression simplifies to
\beq\label{eq: mH_approx}
m_{h, \text{analytic}}^2 = v^2\left(2\l_1 - \frac{(\a_1+\a_4)^2}{2\rho_1}\right).
\eeq
However, it was pointed out in Ref.\,\cite{Karmakar:2022iip} that for certain values of the quartic parameters, the analytical estimate for $m_h$ may not suffice. 

The other scalars have masses of the order $v_R$. To $\mathcal{O}\big(\k_1/v_R\big)$, these masses are related to each other as
\bea
&m_{H_1}^2\simeq m_{A_1}^2\simeq m_{H^{\pm}_1}^2 \approx \frac{1}{2}(\a_3-\a_4) v_R^2\,, &\nn\\
&m_{H_2}^2\simeq m_{A_2}^2\simeq m_{H^{\pm}_2}^2 \approx \frac{1}{2}(\rho_2 - 2 \rho_1)v_R^2\,, &\nn \\
&m_{H_3}^2 = 2\rho_1 v_R^2\,, &\nn \\
&m_{H_2}^2>m_{H_1}^2\,.&\nn 
\eea
The first two mass expressions are valid in the limit $r,w\rightarrow 0$. Positive-definite nature of the CP-even mass matrix leads to two approximate criteria: $\rho_2 > 2 \rho_1$ and $\a_3 > \a_4$. In our analysis, we calculate the scalar masses and mixing numerically. The full analytic expressions at the leading order can be found in the Appendix of Ref.\,\cite{Karmakar:2022iip}. 

For the CP-even scalars, the mass-squared matrix is diagonalized by the orthogonal matrix $O$, 
\bea 
O^T \mathcal{M}^2_\text{CPE} O = \big(\mathcal{M}^{\text{diag}}_\text{CPE}\big)^2,\,\,\,\,\,\, X_\text{physical} = O^T X\, ,
\label{eq:ortho}
\eea
where $X = (\phi_{1r}^0, \phi^0_{2r}, \chi^0_{Lr}, \chi_{Rr}^0)^T$, $X_\text{physical} = (h, H_1, H_2, H_3)^T$. The scalars $H_1$ and $A_1$ can contribute to the mixing of $K_0 - \bar{K}_0$ system, leading to the constraint, $m_{H_1, A_1} > 15$\,TeV \,\cite{Zhang:2007da}. The scalar $H_3$ predominantly originates from the doublet $\chi_R$ and its coupling to the SM particles is suppressed by the element $O_{41} \sim v^2/v_R^2$. So, collider searches allow it to be much lighter than $H_1$. 

The triple Higgs coupling $(c_{h^3})$ in DLRSM is given by\,\cite{Karmakar:2022iip}
\bea\label{eq: chhh} 
 c_{h^3} &=& \frac{\k_1}{2} \Big(2 (\l_1 + r \l_4) O_{11}^3 + 2(r\l_1 + \l_4) O_{21}^3 + 2 w \rho_1 O_{31}^3 + 2(r(\l_1+4\l_2 + 2\l_3)+3\l_4) O_{11}^2 O_{21} \nn \\
 && \hspace{20pt} 
 +  2(\l_1+4\l_2 + 2\l_3 +3 \l_4 r) O_{11} O_{21}^2 + w (\a_1 + \a_4) O_{11}^2 O_{31}  + (\a_1 + r \a_2 + \a_4) O_{11} O_{31}^2 \nn\\
 &&  \hspace{20pt} 
 + w (\a_1 + \a_3) O_{21}^2 O_{31} + (\a_2 + r(\a_1 + \a_3)) O_{21} O_{31}^2   \Big) \,\,\, ,
 \label{kappah}
\eea
with the corresponding coupling multiplier $\kappa_h = c_{h^3}/c_{h^3}^\textrm{SM}$, where $c_{h^3}^\textrm{SM} = m_h^2/2 v = \l^{\rm{SM}} v$.

\subsection{Fermion sector}
The fermion multiplets couple to the bi-doublet $\Phi$ via Yukawa terms:
\bea 
\mathcal{L}_{\rm{Y}} \supset - \bar{Q}_{Li} (y_{ij} \Phi + \tilde{y}_{ij} \tilde{\Phi}) Q_{Rj} + \rm{h.c.} \,\, , 
\label{eq:uvyukawaterms}
\eea
which leads to the mass matrices for the quarks:
\bea 
M_U = \frac{1}{\sqrt{2}}(\k_1 y + \k_2 \tilde{y}), \,\,\,\,
M_D = \frac{1}{\sqrt{2}}(\k_2 y + \k_1 \tilde{y})\,\,,   \nn
\eea 
where $M_U$ and $M_D$ stand for up-type and down-type mass matrices in the flavor basis, respectively. To obtain the physical basis of fermions, these mass matrices need to be diagonalized through unitary transformations described by the left- and right-handed CKM matrices ($V_{L,R}^\textrm{CKM}$). Manifest left-right symmetry implies $V_{R}^\textrm{CKM} = V_{L}^\textrm{CKM}$.
For the calculation of the effective potential in the next section, it is enough to take $y\approx\text{diag}(0,0,y_{33})$ and $\tilde{y}\approx\text{diag}(0,0,\tilde{y}_{33})$. In the limit $V^\textrm{CKM}_{33}\approx 1$, 
\bea\label{eq: y_33}
y_{33} &=& \frac{\sqrt{2}(1+r^2+w^2)^{1/2}}{v(1-r^2)}(m_t-r m_b), \nn\\
\tilde{y}_{33} &=& \frac{\sqrt{2}(1+r^2+w^2)^{1/2}}{v(1-r^2)}(m_b-r m_t),
\eea
where the top and bottom quark masses are  $m_t = 173.5$ GeV, and $m_b \approx 5$ GeV. In the limit $r,w\rightarrow 0 $, $y_{33}$ and $\tilde{y}_{33}$ reduce to the SM Yukawa couplings $y_t$ and $y_b$ respectively. However, we do not make any such assumption and use Eq.\,\eqref{eq: y_33}, allowing $r,w$ to be arbitrary. 

The couplings of the SM-like Higgs with the third-generation quarks are given by: 
\bea \label{eq: hff}
c_{htt\,(hbb)} = \frac{\k_1}{\k_{-}^2}\,\Big((O_{11}-r O_{21})m_{t(b)} + (O_{21}-r O_{11}) (V_{L}^{\text{CKM}}\,\hat{M}_{D(U)}\,V_{R}^{\text{CKM} \dagger})_{33} \Big), 
\label{eq:htthbb}
\eea 
where $\k^2_- = \k_1^2-\k_2^2 = \k_1^2(1-r^2)$ and $\hat{M}_{U(D)}$ denotes the diagonal up\,(down)-type quark mass matrix. Here $O_{ij}$ are the elements of the orthogonal transformation matrix appearing in Eq.\,\eqref{eq:ortho}. Then the coupling multipliers, $\k_b$ and $\k_t$ are: $\k_f = c_{hff}/c^{\rm{SM}}_{hff}$, where $c_{hff}^{\rm{SM}} = m_f/v$ and $f = t, b$.

Since $V_{L,R}^{\rm{CKM}}\approx \mathbf{1}$, Eq.\,\eqref{eq: hff}, becomes,
\bea
c_{htt} &\approx & \frac{\k_1}{\k^2_-}\big(O_{11}(m_t-r m_b) + O_{21}(m_b-r m_t)\big),\nn\\
c_{hbb} &\approx & \frac{\k_1}{\k^2_-}\big(O_{11}(m_b - r m_t) + O_{21}(m_t - r m_b)\big)\nn
\eea
Note that there is a hierarchy, $O_{21}\ll O_{11}\sim 1$, $m_t\gg m_b$, and $r\ll 1$. The SM couplings are recovered by setting $O_{11}=1, O_{21}=0, r=0$, in the above expressions. For a large $\phi_{1r}^0- \phi_{2r}^0$ mixing, i.e. $O_{21} \gtrsim \mathcal{O}(10^{-2})$ or large $\k_2$, i.e. $r \sim \mathcal{O}(10^{-1})$, the deviation of $hb\bar{b}$ coupling from the SM value can be quite large due to the multiplicative factors proportional to $O_{21}m_t$, and $rO_{11}m_t$. On the other hand, the deviation of $ht\bar{t}$ coupling is proportional to $O_{21}m_b$ and $r O_{21} m_t$, and is therefore rather small for the current precision of $\k_t$ measurement.

The Yukawa term for leptons is similar to that of quarks given in Eq. (\ref{eq:uvyukawaterms}). However, since neutrino masses are tiny, generating them in DLRSM would lead to a large hierarchy among lepton Yukawa couplings. Moreover, the neutrinos could be Majorana, in which case DLRSM cannot account for them. In Refs.  \cite{FileviezPerez:2016erl, FileviezPerez:2017zwm, Babu:1988qv, Babu:2020bgz}, neutrino masses were explained by adding a singlet charged scalar to DLRSM. In Appendix\,\ref{sec: neutrino mass}, we show that this extra field does not modify the strength of FOPT.

\subsection{Gauge sector}
\label{sec:gaugesector}
In this chapter, we work under the assumption of manifest left-right symmetry of the UV-Lagrangian, i.e., $g_R = g_L = g$. Here, $g_{L(R)}$ are the gauge couplings of $SU(2)_{L(R)}$, and $g$ is the $SU(2)_L$ gauge coupling of SM.
The mass matrix for charged gauge bosons is
\bea 
\mathcal{L}_{\text{mass}} \supset 
\frac{g^2}{8} \begin{pmatrix}
W_L^+ & W_R^+
\end{pmatrix}
\begin{pmatrix}
v^2 & -2 \k_1 \k_2 \\
-2  \k_1 \k_2  & V^2
\end{pmatrix}
\begin{pmatrix}
 W_L^- \\
 W_R^- \\
\end{pmatrix}   \,\,,
\eea
where, $v^2 = \k_1^2 + \k_2^2 + v_L^2$ and $V^2 = \k_1^2 + \k_2^2 + v_R^2$. The physical charged gauge bosons have masses,
\bea 
m^2_{W_{1,2}} = \frac{g^2}{4} \Big(v^2 +  V^2   \mp \sqrt{(v^2 -  V^2)^2 + 16  \k_1^2 \k_2^2 } \Big) \,\,\,,
\label{Wmasses}
\eea
$W_1^{\pm}$ is identified as the SM $W^{\pm}$ boson and $W_2^{\pm}$ is the new charged gauge boson with mass $\sim \mathcal{O}(v_R)$. The mixing matrix is characterized by an orthogonal rotation with angle $\xi  \simeq   2\k_1\k_2/v_R^2$. 

Similarly, the neutral gauge boson mass matrix is,
\bea 
\mathcal{L}_{\text{mass}} \supset \frac{1}{8}\begin{pmatrix}
W^{3\m}_{L} & W^{3\m}_{R} & B^{\m}
\end{pmatrix} 
\begin{pmatrix}
g^2 v^2 & -g^2 \k_+^2 & -g g_{BL} v_L^2 \\
 & g^2 V^2 & -g g_{BL} v_R^2 \\
 & & g_{BL}^2 (v_L^2 + v_R^2) 
\end{pmatrix}
\begin{pmatrix}
W^3_{L\m} \\ W^3_{R\m} \\ B_{\m} 
\end{pmatrix} \,\,, \nn\\
\eea
where $\k_+^2 = \k_1^2 + \k_2^2$, $g_{BL}$ is the gauge coupling of $U(1)_{B-L}$, and here some of the elements have been suppressed since the matrix is symmetric. The lightest eigenstate is massless and identified as the photon, while the other two states have masses
\bea
m^2_{Z_1,Z_2} &=& \frac{1}{8}\Big(g^2 v^2 + g^2 V^2 + g_{BL}^2 (v_L^2 + v_R^2) \nn\\
&&\mp \sqrt{(g^2 v^2 + g^2 V^2 + g_{BL}^2 (v_L^2 + v_R^2))^2 + 4 (g^4 + 2 g^2 g_{BL}^2)(\k_+^4 - v^2 V^2) } \Big) \,\,.\nn\\
\label{Zmasses}
\eea
The lighter mass eigenstate $Z_1$ corresponds to the SM $Z$ boson, while $Z_2$ has a mass  $\sim\mathcal{O}(v_R)$. 

In the limit $\k_1, \k_2, v_L \ll v_R$  the mixing matrix is~\cite{Dev:2016dja} 
\bea 
\begin{pmatrix} A_{\m} \\ Z_{1\mu}  \\ Z_{2\m} \end{pmatrix} =
\begin{pmatrix} 
s_W & c_W s_Y & c_W c_Y \\  -c_W & s_W s_Y & s_W c_Y \\ 0 & c_Y & s_Y
\end{pmatrix}
  \begin{pmatrix} W^3_{L\m} \\ W^3_{R\m} \\ B_{\m}  \end{pmatrix} ,
\eea 
where
\bea 
s_W &\equiv & \sin \theta_W = \frac{\gbl}{\sqrt{g^2 + 2 \gbl^2 }}\,\,,\,\,\,\,\,
c_W \equiv  \cos \theta_W = \sqrt{\frac{g^2 + \gbl^2}{g^2 + 2 \gbl^2}}\,\,,\nn\\
s_Y &\equiv &  \sin \theta_Y = \frac{\gbl}{\sqrt{g^2 + \gbl^2}}\,\,,\,\,\,\,\,
c_Y \equiv  \cos \theta_Y = \frac{g}{\sqrt{g^2 + \gbl^2}}\,\,. 
\eea
We fix $g_{BL} = g g'/(g^2 -g'^2)^{1/2}$, where $g'$ is the gauge coupling for $U(1)_Y$ of SM. Direct searches for spin-1 resonances have put a lower limit on the masses of the new charged and neutral gauge bosons. In DLRSM, the masses of such new gauge bosons are $m_{W_2} \sim g v_R/2 = $ and $m_{Z_2} \sim m_{W_2}/\cos \theta_Y$. Recently, the  lower limit on the mass of $W_2$ in DLRSM has been estimated to be, $m_{W_2} > 5.1$\,TeV\,\cite{Solera:2023kwt}, which leads to a lower bound on $v_R$, $v_R > 2 m_{W_2}/g = 15.7$\,TeV. The constraint on $m_{Z_2}$ is comparatively weaker, $m_{Z_2}>4.3$ TeV. Therefore, the lowest value of $v_R$ we use in our benchmark scenarios is $v_R = 20$\,TeV.

\subsection{Theoretical bounds}\label{subsec: theory}
We incorporate the following theoretical constraints:
\begin{itemize}
\setlength\itemsep{0.1 em}
    \item {\it Perturbativity:} The quartic couplings of the scalar potential, $\{\l_{1,2,3,4},~\a_{1,2,3,4},~\rho_{1,2}\}$, are subjected to the upper limit of $4\pi$ from perturbativity. Moreover, the Yukawa couplings of the DLRSM Lagrangian must satisfy the perturbativity bound $y_{33}, \tilde{y}_{33} < \sqrt{4\pi}$, with $y_{33}, \tilde{y}_{33}$  defined in Eq.\,\eqref{eq: y_33},
    These constrain the value of \vev ratios roughly to $r < 0.8$ and $w < 3.5$ \cite{Karmakar:2022iip}. 
    
    \item {\it Unitarity:} The scattering amplitudes of $2 \rightarrow 2$ processes involving scalars and gauge bosons must satisfy perturbative unitarity. To $\mathcal{O}(\k_1/v_R)$, these constraints can be expressed in terms of the masses of the new scalars in DLRSM \cite{Bernard:2020cyi},
    \bea
&0 <   \rho_1  <  \frac{8 \pi}{3}\, , \,\,\text{or,}\,\, \frac{m_{H_3}^2}{v_R^2}  < \frac{16 \pi}{3} \,\,\,,&  \nn\\ 
&\frac{(c_{H_3})^2}{k^4} \, \frac{m^2_{H_3}}{v_R^2} < \frac{16 \pi}{3} \,\,\, ,& \nn\\
&2 \frac{w^2}{k^2} \sum_{i=1,2} F_i^2 \frac{m^2_{H^\pm_i}}{v_R^2} + \frac{c_{H_3}}{k^2} \frac{m^2_{H_3}}{v_R^2}  <  16 \pi\,\,\,,& \nn\\
&2 \frac{w^2}{k^2} \sum_{i=1,2} S_i^2 \frac{m^2_{H^\pm_i}}{v_R^2} + \frac{c_{H_3}}{k^2} \frac{m^2_{H_3}}{v_R^2}  <  16 \pi\,\,\, ,&
\label{eq:unitaritycondition}
\eea
    where $k^2= 1+r^2+w^2$ and $F_i, S_i,$ and $c_{H_3}$ are defined in terms of the parameters of the potential \cite{Bernard:2020cyi}.
\item {\it Boundedness from below:} The scalar potential must be bounded from below\,(BFB) along all directions in field space. This leads to additional constraints on the quartic couplings of the model. The full set of such constraints was derived in Ref.\,\cite{Karmakar:2022iip}, which we have implemented in our numerical analysis.   
\end{itemize}

\subsection{Constraints from $h(125)$ data}\label{subsec: higgs}
\label{sec: hdata_constraints}

In the following, we qualitatively describe the constraints on DLRSM from Higgs-related measurements at the LHC. 

\begin{itemize}
    \item  The key constraint comes from the measurement of the mass of SM-like Higgs, $m_h = 125.38 \pm 0.14\,$GeV \cite{CMS:2012qbp}. If the theoretical bounds of perturbativity and boundedness from below are taken into account together with $m_{h, \text{analytic}} \simeq 125$\,GeV, it leads to an upper bound on the \vev ratio, $w \lesssim 2.93 + 4.35 r - 0.48 r^2$. 

    \item One of the most stringent constraints on the DLRSM parameter space comes from the measurement of $hb\bar{b}$ coupling, $\k_b = 0.98^{+0.14}_{-0.13}$\,\cite{ATLAS:2020qdt}. If the mixing between $\phi^0_{1r}$ and $\phi^0_{2r}$ takes large values, $\k_b$ can deviate from unity, thereby ruling out a large region of parameter space allowed by theoretical bounds and the measurement of $m_h$. However, $ht\bar{t}$ coupling is not significantly modified and does not result in any new constraints. 
    
    \item As discussed in Sec.~\ref{sec:gaugesector}, a large value of $v_R$ ensures that the mixings between the SM-like and heavier gauge bosons are rather small, $\xi \sim \mathcal{O}(v^2/v_R^2)$. Therefore, the $h W_1 W_1$ and $h Z_1 Z_1$ couplings are quite close to their SM values and do not lead to any additional constraints on the DLRSM parameter space.
        
    \item The trilinear coupling of the SM-like Higgs given in Eq.\,\eqref{eq: chhh}, does not necessarily align with the SM value. As seen in Eq.\,\eqref{eq: chhh}, for non-zero mixings, particularly, $O_{21}\neq 0$, the parameters appearing in the parenthesis can individually take a wide range of values, leading to a potentially significant deviation of $c_{h^3}$ from $c^{\rm{SM}}_{h^3}$. In our analysis, we impose the ATLAS bound of $\k_h = [-2.3, 10.3]$ at 95$\%$ CL\,\cite{ATLAS:2019pbo}. 
\end{itemize}

\section{Effective Potential}\label{sec: 3_effective potential}
In this section, we construct the full one-loop finite temperature effective potential\,\cite{Quiros:1999jp,Laine:2016hma} required to study the nature of the PT associated with the breaking of $SU(2)_R \times U(1)_{B-L}$. Below we describe the procedure step by step. 

The tree-level effective potential is obtained by setting all the fields to their respective background field value in the potential given in Eq.\,\eqref{eq: 3_potential}. The CP-even neutral component of $\chi_{\rm{R}}$ is responsible for breaking the $SU(2)_{\rm{R}}\times U(1)_{B-L}$ gauge group, whose background value we denote by $R$. Since $v_R\gg v$, all other field values can be set to zero. 
Hence, in the notation of Eq.\,\eqref{eq: varphi}, the background fields are
$$\langle\varphi\rangle = \{0,0,0,R,0,0,0,0,0,0,0,0\}.$$
The tree-level effective potential is then given by
\beq
    V_0(R) = -\frac{\mu_3^2}{2}R^2 + \frac{\rho_1}{4}R^4.
\eeq
At the one-loop level, the zero-temperature correction to the effective potential is given by the Coleman-Weinberg (CW) formula \cite{ColWein}. In the Landau gauge, with $\overline{\rm{MS}}$ renormalization scheme, the CW potential given in Eq.\,\eqref{eq: 1_ColWein} becomes
\beq
    V_{\cw}(R) = \frac{1}{64\pi^2}\sum_i (-1)^{f_i} n_i m_i^4(R)\left[\log\bigg(\frac{m_i^2(R)}{\mu^2}\bigg) - c_i \right],
\eeq
where $i$ runs over all species coupling to the $SU(2)_R\times U(1)_{B-L}$-breaking field $\chi^0_{Rr}$. The field-dependent mass, $m_i(R)$ is the mass of the species $i$ in the presence of the background field $R$. When there is mixing between the different species, the masses are extracted as the eigenvalues of the corresponding mass matrices. The expressions for the field-dependent masses can be found in Appendix\,\ref{appendix: field}. In Appendix\,\ref{sec: neutrino mass} we take the minimal mechanism of neutrino mass generation of refs.\,\cite{Babu:1988qv, FileviezPerez:2016erl} and show that the right-handed neutrino $\n_R$ and the extra charged scalar do not contribute to the effective potential. Therefore the contributions only come from the CP-even scalars: $\{\phi_{1r}^0, \phi_{2r}^0, \chi_{Lr}^0, \chi_{Rr}^0\}$, CP-odd scalars: $\{\phi_{1i}^0, \phi_{2i}^0, \chi_{Li}^0, \chi_{Ri}^0\}$, charged scalars: $\{\phi_1^{\pm}, \phi_2^{\pm}, \chi_L^{\pm}, \chi_R^{\pm}\}$, and gauge bosons $W_{L,R}^{\pm}$, $Z_{L,R}$ and $B$. The factor $f_i$ is 0 (1) for bosons\,(fermions), and the number of degrees of freedom $n_i$ are,
\bea
&n_{\phi^0_{1r}}=n_{\phi^0_{2r}}=n_{\chi^0_{Lr}}=n_{\chi^0_{Rr}}= 1,& \nn\\
&n_{\phi^0_{1i}}=n_{\phi^0_{2i}}=n_{\chi^0_{Li}}=n_{\chi^0_{Ri}}=1,& \nn\\
&n_{\phi^{\pm}_{1}}=n_{\phi^{\pm}_{2}}=n_{\chi^{\pm}_{L}}=n_{\chi^{\pm}_{R}}=2,&\nn\\
&n_{W^{\pm}_{Lt}}=n_{W^{\pm}_{Rt}}=4,&\nn\\
&n_{W^{\pm}_{Ll}}=n_{W^{\pm}_{Rl}}=2,&\nn\\
&n_{Z_{Lt}}=n_{Z_{Rt}}=n_{B_{t}}=2,&\nn\\
&n_{Z_{Ll}}=n_{Z_{Rl}}=n_{B_{l}}=1.&\nn
\eea
The subscripts $t$ and $l$ stand for transverse and longitudinal polarizations of the gauge bosons. The constant $c_i = 5/6$ for gauge bosons, and $3/2$ for all other fields. We set the renormalization scale $\mu = v_R$ to ensure the validity of the CW formula by having $\ordone$ logs.

We impose the on-shell renormalization condition so that the position of the minimum and the mass of the CP-even scalar $\chi^0_{Rr}$ calculated from the one-loop potential coincides with the corresponding tree-level value. This is achieved by introducing a counter-term potential
\beq
    V_{\rm{c.t.}}(R) = -\frac{\delta\mu_3^2}{2} R^2 + \frac{\delta\rho_1}{4}R^4\,,
\eeq
where the unknown coefficients $\delta\mu_3^2$ and $\delta\rho_1$ are fixed by demanding
\begin{subequations}
\beq
     \left. \frac{\partial}{\partial R}(V_{\cw} + V_{\rm{c.t.}})\right\vert_{R=v_R} = 0\,,
\eeq
\beq
     \left. \frac{\partial^2}{\partial R^2}(V_{\cw} + V_{\rm{c.t.}})\right\vert_{R=v_R} = 0\,.
\eeq
\end{subequations}
This leads to
\begin{subequations}
\beq
     \delta\mu_3^2 = \frac{3}{2 v_R}\left.\frac{\partial V_{\cw}}{\partial R}\right\vert_{R=v_R} - \frac{1}{2}\left.\frac{\partial^2 V_{\cw}}{\partial R^2}\right\vert_{R=v_R},
\eeq
\beq
     \delta\rho_1 = \frac{1}{2 v_R^3}\left.\frac{\partial V_{\cw}}{\partial R}\right\vert_{R=v_R} - \frac{1}{2 v_R^2}\left.\frac{\partial^2 V_{\cw}}{\partial R^2}\right\vert_{R=v_R}.
\eeq
\end{subequations}
Then the one-loop contribution to the effective potential is
\begin{equation}
    V_1 = V_{\cw}+V_{\rm{c.t.}}.
\end{equation}

Next, we include the one-loop finite temperature correction given in Eq.\,\eqref{eq: 1_V_1T} using \texttt{CosmoTransitions}. 
We add the daisy corrections via the Arnold-Espinosa method given in Eq.\,\eqref{eq: arnold_espinosa} so that the effective potential is given by
\beq
    V_{\rm{eff}} = V_0 + V_{\cw} + V_{\rm{c.t.}} + V_{1T} + V_{\rm{D}}\,.\nn
\eeq
In our analysis, we use the Arnold-Espinosa method, as it takes into account the daisy resummation consistently at the one-loop level, while the Parwani method mixes higher-order loop effects in the one-loop analysis. 

\begin{figure}[tbp] 
\centering 
\includegraphics[width=.49\textwidth]{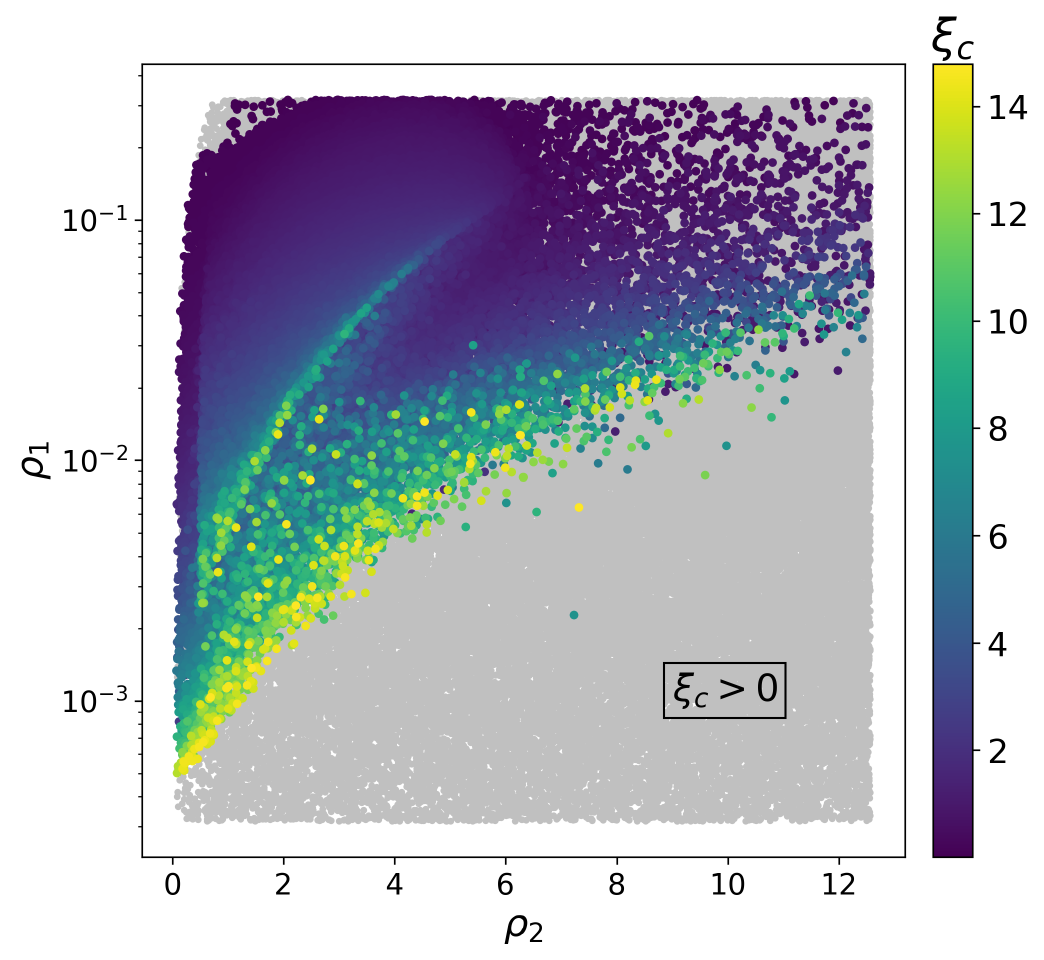}
\includegraphics[width=.49\textwidth]{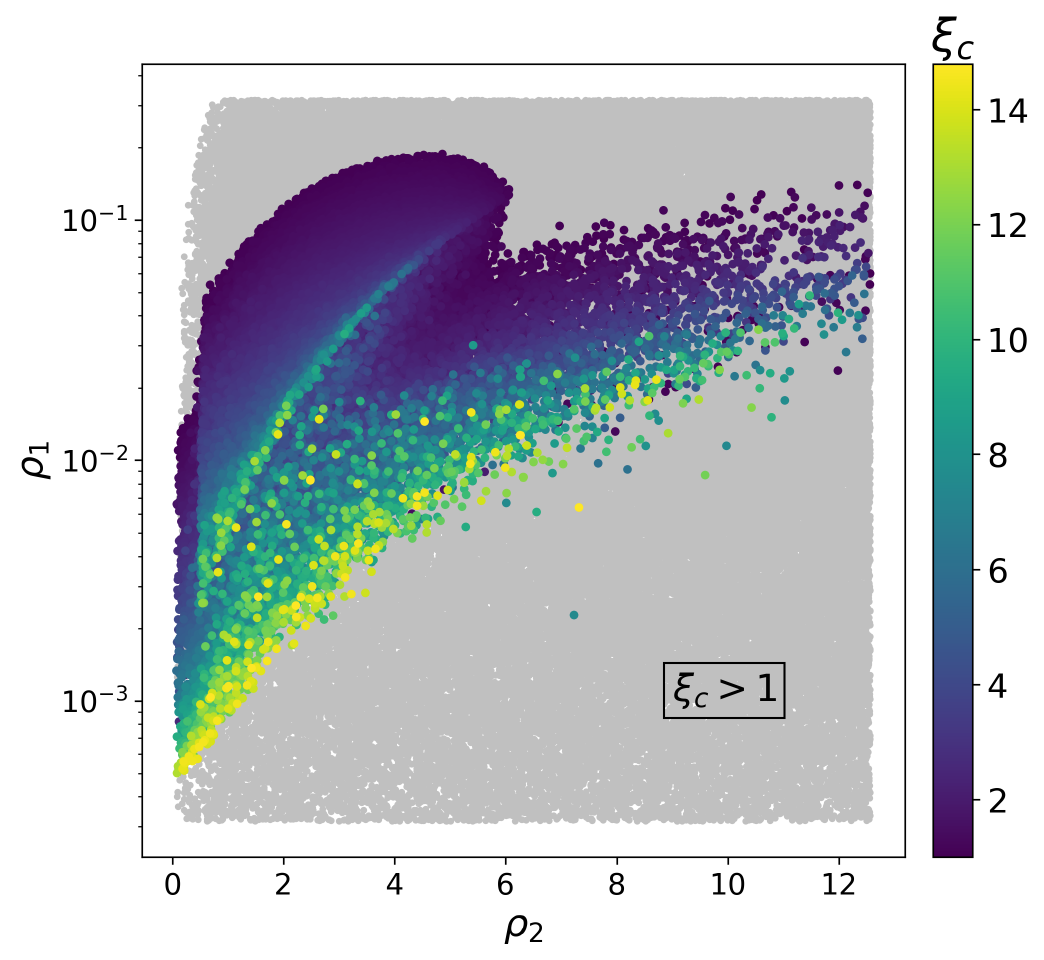}
\caption{\label{fig: rho1rho2} Points with strong FOPT in the $\rho_1-\rho_2$ plane, for $v_R=30$ TeV. The grey points have passed the theoretical and experimental constraints. The left panel shows all points showing FOPT with $\xi_c>0$, whereas in the right panel, the points satisfy the condition $\xi_c>1$.}
\end{figure}

\section{Parameter scan}
\label{sec: paramscan}
As discussed earlier, DLRSM has a large number of parameters: ten quartic couplings, along with $r,w,\mu_4$, and $v_R$. This is called the \textit{generic basis}. To reduce the number of parameters for our analysis, we work in the \textit{simple basis}, introduced in Ref.\,\cite{Karmakar:2022iip}. The condition of boundedness from below, discussed in Sec.\,\ref{subsec: theory}, requires that the ratio $x = \l_2/\l_4$ is restricted to the range $x \in [0.25, 0.85]$. Therefore, we keep $\l_2$ as a separate parameter, while we equate $\l_1=\l_3=\l_4\equiv \l_0$. Similarly, guided by the approximate mass relation, $m_{H_1}\approx \frac{1}{2}(\a_3-\a_4)$, we allow for the possibility of having $\a_3\neq \a_4$ by keeping them independent, while setting $\a_1=\a_2=\a_4\equiv\a_0$. Thus the {\it simple basis} contains six quartic couplings
\bea 
&&\{ \l_0,~\l_2,~\a_0,~\a_3,~\rho_1,~\rho_2 \}\,.
\eea
Along with these quartic couplings, we also scan over the \vev ratios $r$, $w$, and take $v_R = 20,~30,$ 50 TeV. As the mass parameter $\mu_4$ plays an insignificant role in the effective potential, we set $\mu_4=0$ in our analysis. Using the {\it simple basis} allows us to capture the key features of GW phenomenology of DLRSM while retaining the interplay of the existing theoretical and collider constraints.

In preliminary scans, we find that promising scenarios of strong first-order phase transition occur for small values of $\rho_1$. For points with relatively large couplings, the daisy potential, $V_{\rm{D}}$, given in Eq.\,\eqref{eq: Vdaisy} starts dominating over the contribution from the thermal potential, $V_{1T}$, given in Eq.\,\eqref{eq: 1_V_1T}. When this happens, the symmetry-restoring property of the finite temperature effective potential is lost, and instead, symmetry non-restoration is observed. Then the minimum at the non-zero field value becomes deeper at high temperatures, implying the absence of a phase transition, as discussed in Refs.\,\cite{Weinberg:1974hy,Kilic:2015joa,Meade:2018saz}. Based on these observations, we choose the following parameter ranges: 
\bea
    &\log\a_0 \in [-3,0],~\log\a_3 \in [-3,0],~\log\rho_1 \in [-3.5,-0.5],~\rho_2 \in [0,4\pi],&\nn\\
    &~x \in [0.25, 0.85],~\log r \in [-3,0],~\log w \in [-6,1],~v_R = 20, 30, 50\,\text{TeV}.&
\eea
Each parameter is selected randomly from a uniform distribution in the respective range. The parameter $\lambda_0$ is chosen in the following manner:
\begin{itemize}
\setlength\itemsep{0.1 em}
\item To increase the number of points satisfying the bound on SM-like Higgs mass ($m_h$), we solve the equation, $m_{h, \text{analytic}}\,(\l_0 = \Lambda_0) = 125.38$\,GeV, for a fixed set of values $\{\a_0, \a_3, \rho_1, \rho_2, x\}$. 
\item Using the solution $\Lambda_0$, we choose a random value of $\l_0$ as, $\lambda_{0} = (1+y)\,\Lambda_0$, with $y \in [-0.1, 0.1]$. 
\item Finally, each parameter point is defined by the set: $$\big\{\l_0,~\l_2 = x\l_0,~\a_0,~\a_3,~\rho_1,~\rho_2,~r,~w,~v_R\big\}$$
\end{itemize}
Given a parameter point, we first check if it satisfies the theoretical constraints: boundedness from below, perturbativity, and unitarity, discussed in Sec.\,\ref{subsec: theory}. Next, the Higgs constraints described in Sec.\,\ref{subsec: higgs} are checked. Furthermore, the constraint from meson mixing $m_{H_1} > 15$\,TeV is imposed.

If the parameter point passes all the aforementioned theoretical and experimental constraints, we construct the effective potential using the Arnold-Espinosa method. We satisfy the Linde-Weinberg bound\,\cite{Linde:1975sw,Weinberg:1976pe} by numerically checking that the minimum of the zero-temperature effective potential at $R=v_R$ is the absolute minimum. We reject the point if symmetry non-restoration persists at high temperatures. Next, we check for a possible first-order phase transition, using the Python-based package \texttt{CosmoTransitions}\,\cite{Wainwright:2011kj}. The strength of FOPT can be quantified by the ratio
\beq
    \xi_c = \frac{v_c}{T_c},
\eeq
where $T_c$ is the critical temperature at which the two minima become degenerate and $v_c$ is the \vev at $T_c$. The FOPT is considered to be strong if the following criterion is met\footnote{It is known that the field value, $v_c$, and $T_c$ are gauge dependent, therefore, so is the ratio $\frac{v_c}{T_c}$. However changing the gauge-fixing parameter has a subleading effect on $\frac{v_c}{T_c}$\,\cite{Patel:2011th,Chatterjee:2022pxf}. In the subsequent analysis, we work in the Landau gauge, for which the gauge dependence is numerically minimized. }\,\cite{Quiros:1994dr},
\beq
    \xi_c >1.
\eeq

\begin{figure}[tbp]
\centering 
\includegraphics[width=.98\textwidth]{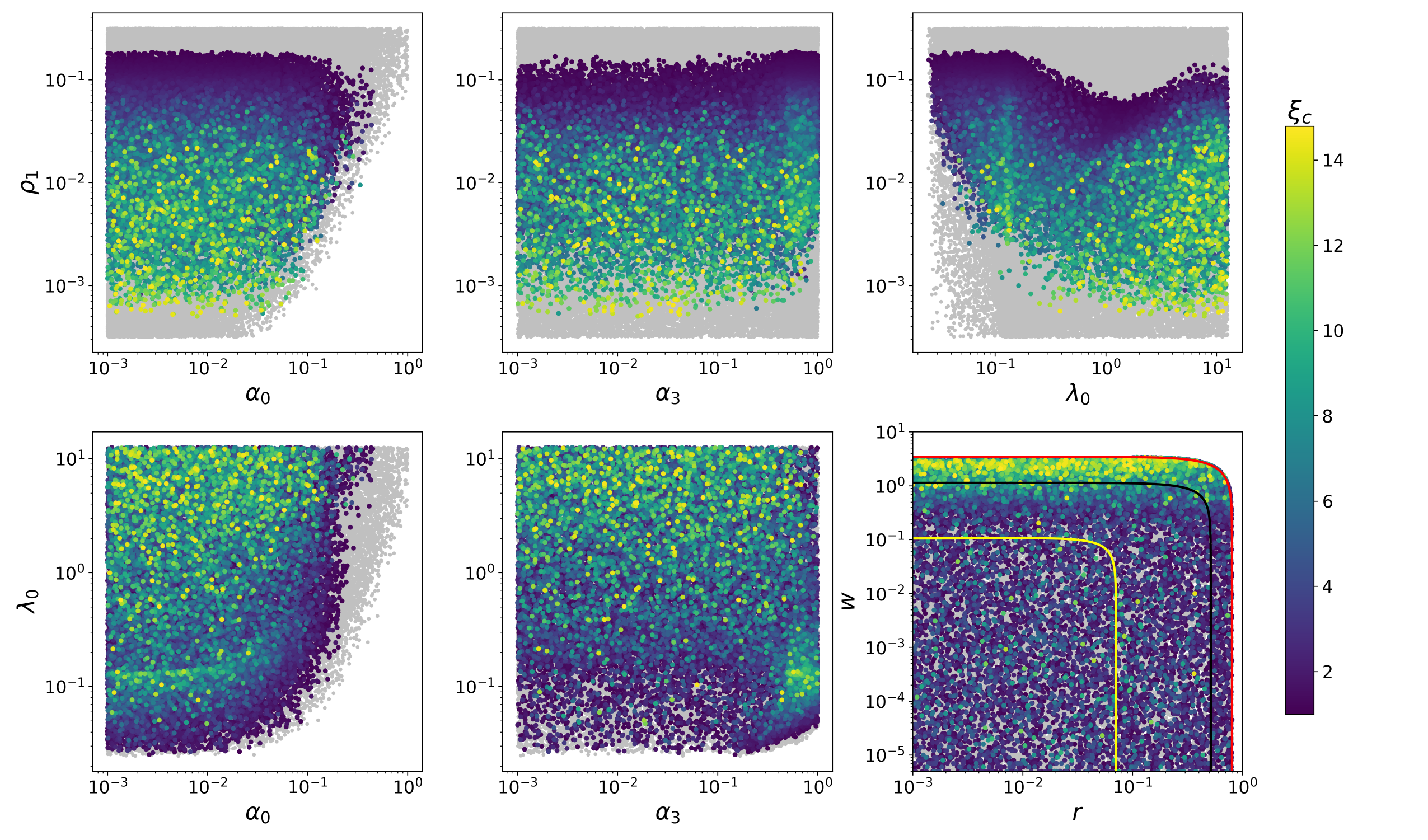}
\caption{\label{fig: projections} Projections showing the points with SFOPT on different parameter planes, for $v_R=30$ TeV. The grey points show all points passing the theoretical and experimental constraints. The points satisfy the condition $\xi_c>1$.}
\end{figure}

In Fig.\,\ref{fig: rho1rho2}, we show the points with FOPT projected onto the $\rho_1-\rho_2$ plane for $v_R=30$ TeV, color-coded according to the value of $\xi_c$. The left panel shows all points with $\xi_c>0$, while the right panel only shows points satisfying the SFOPT criterion $\xi_c>1$. The grey dots depict parameter points passing the existing theoretical and experimental bounds. As suggested by the preliminary scans, SFOPT prefers $\rho_1 \lesssim \mathcal{O}(0.1)$. Points with $\rho_1\lesssim\mathcal{O}(10^{-2})$ and $\rho_2\gtrsim \ordone$ violate the Linde-Weinberg bound. Therefore, there are no points showing SFOPT in this region. A large number of points with $\rho_2 \gtrsim 6$ also exhibit symmetry non-restoration at high temperatures. 

Fig.\,\ref{fig: projections} shows various two-dimensional projections of the DLRSM parameter space for $v_R=30$\,TeV, depicting points with SFOPT. The parameter $\a_0$ is always smaller than 1, as indicated by the left panels in the top and the bottom row. We also restrict ourselves to $\a_3<1$ to avoid points showing symmetry non-restoration. Along the $\a_3$ direction, there is a sharp change in the density of points around $\a_3\approx 0.5$, coming from the bound $m_{H_1} > 15$\,TeV. The value of $\a_3$ where the density changes is different for $v_R = 20$, 30, and 50 TeV. Since the couplings are small for a large number of parameter points, the approximate relation given in Eq.\,\eqref{eq: mH_approx} tells us that $\l_1$ can take values close to $\l_{\rm{SM}}\approx 0.13$. In the top right and bottom left panels, we indeed observe an over-density of points clustered around $\l_0\approx 0.13$. In the $\rho_1-\l_0$ plane, a majority of points with large $\xi_c$ occur for small $\rho_1$, and large $\l_0$. In the $r-w$ plane, points with large $\xi_c$ occur mostly at higher values of $w$ ($\gtrsim \mathcal{O}(0.1)$) and are less frequent for smaller values of $w$. So this parameter region can lead to a detectable GW background. There is no preference along the $r$ direction. The points with large $\xi_c$ also have relatively large values of $y_{33}$, as indicated by the yellow, black, and red contours corresponding to $y_{33}=1,~1.5$, and $\sqrt{4\pi}$ respectively.

The strength of FOPT is more rigorously characterized by the parameters, $\a,~\beta/H_*,$ and $T_n$, defined in Sec.\,\ref{1_FOPT_params}.
 \begin{figure}[tbp]
\centering 
\includegraphics[width=.99\textwidth]{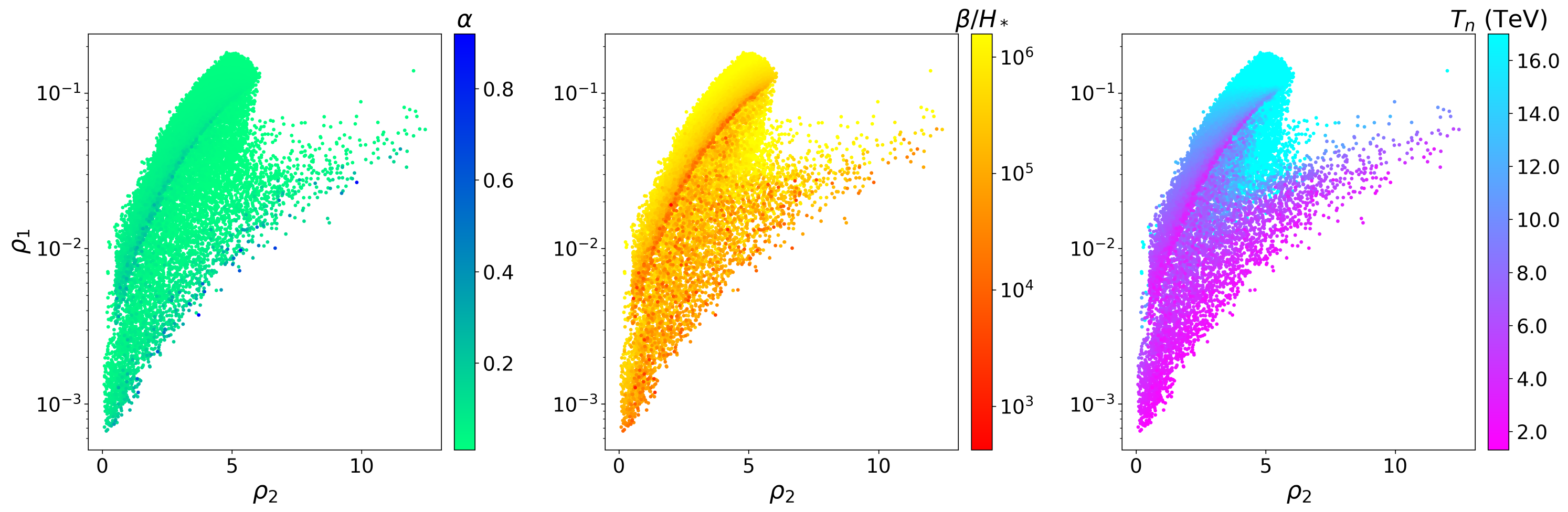}
\caption{\label{fig: abt} Variation of PT parameters in the $\rho_1-\rho_2$ plane. Color code shows the variation of $\alpha$ (left panel), $\beta/H_*$ (middle panel), and $T_n$ (right panel). Here, $v_R=30$ TeV.}
\end{figure}
For points satisfying $\xi_c>1$, we compute the nucleation temperature $T_n$. We find the solution of Eq.\,\eqref{eq: 1_nucl_criterion2} using the secant method, where the tunneling action is calculated by \texttt{CosmoTransitions}. We remove any points with $T_n<0$, as this indicates that the PT is not completed till the present time. Moreover, we set a lower bound of $T_n > 500\,$GeV to ensure that the PT is completed before the EW epoch. Once $T_n$ is obtained, $\a$ and $\beta/H_*$ can be computed using Eqs.\,\eqref{eq: 1_alpha} and \eqref{eq: 1_beta} respectively. Fig.\,\ref{fig: abt} shows the variation of the PT parameters $\a$ (left panel), $\beta/H_*$ (middle panel), and $T_n$ (right panel), in the $\rho_1-\rho_2$ plane. The evaluated ranges roughly are, $\a\in[0,0.8]$, $\beta/H_*\in[10^2,10^6]$, and $T_n\in [2,16]$ TeV. $T_n$ is observed to take smaller values in regions where the strength of SFOPT is high. 

\section{Gravitational wave background}\label{sec: 3_GW}
The GW spectrum is defined as \cite{Caprini:2015zlo}
\beq
\Omega_{\rm{GW}}(f)\equiv \frac{1}{\rho_c}\frac{d\rho_{\rm {GW}}}{d\ln f},
\eeq
where $f$ is the frequency, $\rho_{\rm {GW}}$ is GW energy density, and $\rho_c$ is the critical energy density of the universe, given by,
\beq
\rho_c = \frac{3H_0^2}{8\pi G}.
\eeq
\noindent Here, $H_0 = 100\,h~{\rm{km \,s^{-1} Mpc^{-1}}}$ is the Hubble constant with the current value of $h=0.6736\pm 0.0054$\,\cite{Planck:2018vyg} and $G$ is  Newton's gravitational constant. 

A strong FOPT proceeds by nucleation of bubbles of the stable phase, which expand rapidly in the sea of the metastable phase. GWs are produced when the expanding bubbles collide and coalesce with each other. If sufficient friction exists in the plasma, the bubble walls may reach a terminal velocity $v_w$. We take $v_w = 1$ in our analysis. GW production happens via three main processes: bubble wall collisions ($\Omega_{\rm{col}}$), sound waves produced in the thermal plasma ($\Omega_{\rm{sw}}$), and the resulting MHD turbulence ($\Omega_{\rm{turb}}$). For a recent review of the different GW production mechanisms, please refer to \cite{Athron:2023xlk}. In the non-runaway scenario \cite{Caprini:2015zlo}, GW production happens primarily through sound waves and turbulence, i.e.,
\beq
h^2 \Omega_{\rm{GW}} \simeq h^2 \Omega_{\rm{sw}}+h^2 \Omega_{\rm{turb}}\, ,
\eeq
where\,\cite{Guo:2020grp,Caprini:2015zlo},
\bea
h^2\Omega_{\rm{sw}}(f) &=& 2.65\times 10^{-6} \left(\frac{100}{g_*}\right)^{1/3}\left(\frac{H_*}{\beta}\right)^2 \left(\frac{\k_{\text{sw}}\a}{1+\a}\right)^2v_w ~S_{\rm{sw}}(f)~\Upsilon(\tau_{\text{sw}}),\label{eq: 3_soundwaves}\\
h^2\Omega_{\rm{turb}}(f) &= & 3.35\times 10^{-4} \left(\frac{100}{g_*}\right)^{1/3}\left(\frac{H_*}{\beta}\right)^2 \left(\frac{\k_{\rm{turb}}\a}{1+\a}\right)^{3/2}v_w ~S_{\rm{turb}}(f)\,. \label{eq: 3_turbulence}
\eea

\noindent Here, $\k_{\text{sw}}$ and $\k_{\text{turb}}$ are the efficiency factors for the respective processes. The efficiency factor $\k_{\text{sw}}$ is given by
\beq
\k_{\text{sw}} = \frac{\a}{0.73+0.083\sqrt{\a}+\a},
\eeq
and $\k_{\text{turb}}$ is known to be at most $5-10\%$ of $\k_{\text{sw}}$. Here we take $\k_{\text{turb}} = 0.05 ~\k_{\text{sw}}$. We have included the suppression factor $\Upsilon(\tau_{\text{sw}})$ that arises due to the finite lifetime $\tau_{\text{sw}}$ of sound waves \cite{Guo:2020grp},
\beq
\Upsilon(\tau_{\text{sw}}) = 1 - \frac{1}{1+2\tau_{\text{sw}}H_*}\,,
\eeq
with
\beq
\tau_{\text{sw}} = \frac{R_*}{\overline{U_f}}\,,
\eeq
where the mean bubble separation $R_*\simeq (8\pi)^{1/3} v_w/\beta$ and the mean square velocity is
\beq
\overline{U_f}^2 = \frac{3}{4}\frac{\a}{1+\a}\k_{\rm{sw}}\,.
\eeq

The spectral shape functions, $S_{\rm{sw}}$ and $S_{\rm{turb}}$ determine the behavior of each contribution at low and high frequencies. These are
\bea
S_{\rm{sw}}(f) &=& \left(\frac{f}{f_{\rm{sw}}}\right)^3 \left(\frac{7}{4+3(f/f_{\rm{sw}})^2}\right)^{7/2},\nn\\
S_{\rm{turb}}(f) &=& \left(\frac{f}{f_{\rm{turb}}}\right)^3\frac{1}{[1+(f/f_{\rm{turb}})]^{11/3}(1+8\pi f/h_*)}\,.
\eea

Here, $h_*$ is the Hubble rate at $T=T_*$, 
\begin{equation}
h_* = 1.65\times 10^{-7}~{\rm{Hz}}\left(\frac{T_*}{100~{\rm{GeV}}}\right)\left(\frac{g_*}{100}\right)^{1/6}.
\end{equation} 
The red-shifted peak frequencies, after taking into account the expansion of the universe, are,
\begin{eqnarray}
f_{\rm{sw}} &=& 1.9\times 10^{-5}{\rm{Hz}}\left(\frac{g_*}{100}\right)^{1/6}~\frac{1}{v_w}\left(\frac{\beta}{H_*}\right)\left(\frac{T_*}{100~{\rm{GeV}}}\right),\label{eq: fsw}\label{eq: 3_f_sw}\\
f_{\rm{turb}} &=& 2.7\times 10^{-5}{\rm{Hz}}\left(\frac{g_*}{100}\right)^{1/6}~\frac{1}{v_w}\left(\frac{\beta}{H_*}\right)\left(\frac{T_*}{100~{\rm{GeV}}}\right).\label{eq: 3_fturb}
\end{eqnarray}

\begin{figure}[tbp]
\centering
\includegraphics[width=.5\textwidth]{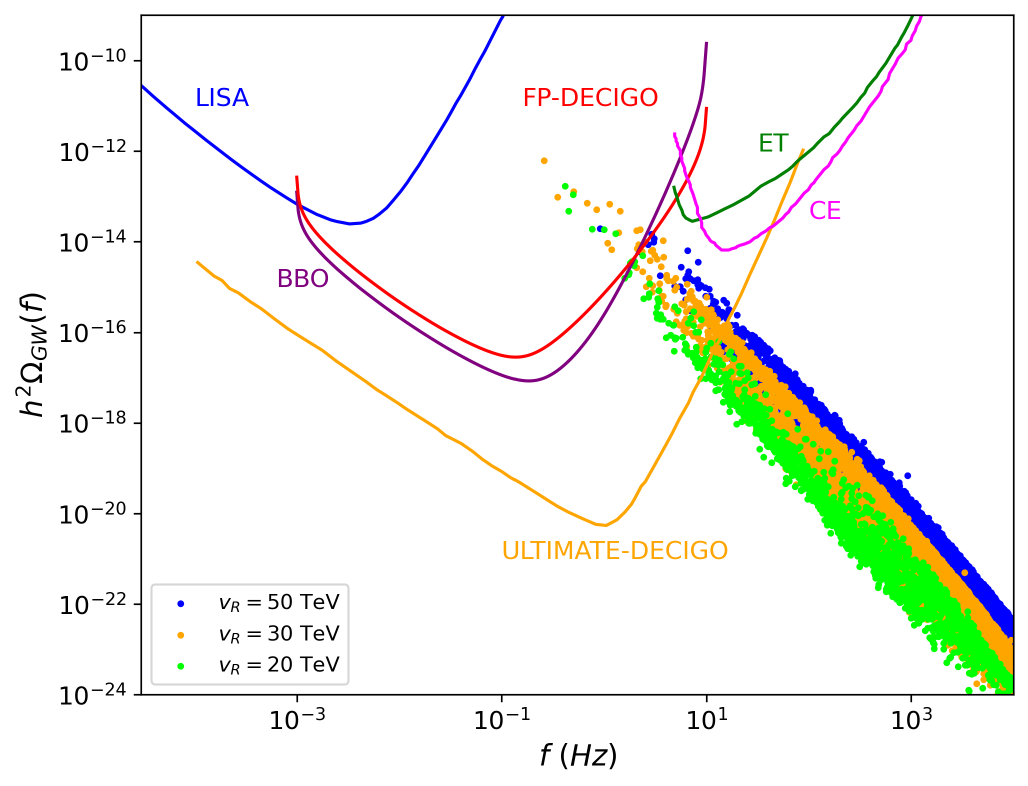}
\hfill
\includegraphics[width=.49\textwidth]{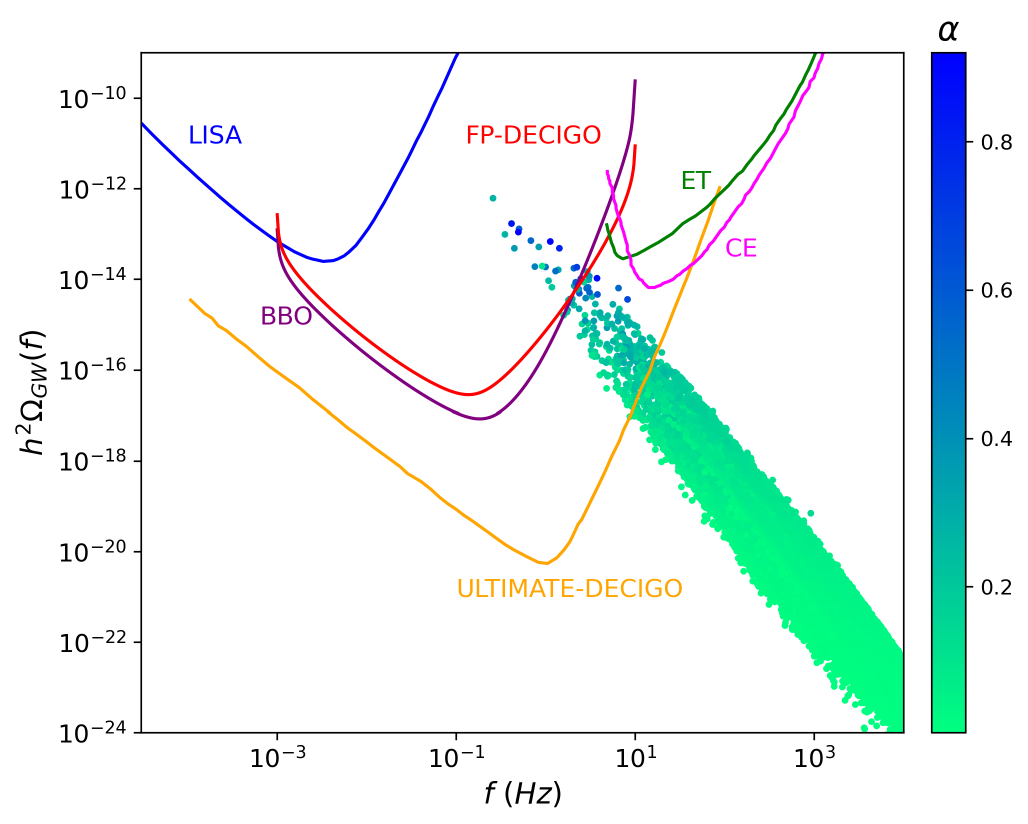}
\caption{\label{fig: gw_scatter} The peak of the GW spectrum $\Omega_{\rm{GW}}$ for points with SFOPT, along with the power-law integrated sensitivity curves of various upcoming GW detectors. Left panel: points corresponding to $v_R=20,~30$, and 50 TeV are shown. Right panel: Points are color-coded according to the value of $\a$, for $v_R=20,~30$, and 50 TeV combined.}
\end{figure}

From the expressions of $\Omega_{\rm{sw}}$ and $\Omega_{\rm{turb}}$, it is clear that large $\a$ and small $\beta/H_*$ lead to a strong GW spectrum. The peak frequency is proportional to $T_n\sim v_R$ and hence, the peak shifts to the right for larger $v_R$. This is illustrated in Fig.\,\ref{fig: gw_scatter}, where we show scatter plots of the parameter points for which $\a,~ \beta/H_*$, and $T_n$ have been computed. Each point represents the peak value corresponding to the GW spectrum, $h^2\Omega_\text{GW}$. The left panel shows that these points shift to the right as $v_R$ is progressively increased between $v_R=20,~30$ and $50$ TeV. The strength of the GW signature is not affected by varying $v_R$. The right panel shows the variation of $\a$ for the points corresponding to $v_R = 20,~30$, and $50$ TeV combined. There is clearly a positive correlation between large $\a$ and the strength of GW. The solid lines represent the power-law integrated sensitivity curves\,\cite{Thrane:2013oya} corresponding to various planned detectors calculated for an observation time of $\tau = 1$ year, and a threshold SNR=1 (see Eq.\,(\ref{eq: SNR})). The curve for Ultimate-DECIGO is obtained following the prescription of Ref.\,\cite{Kuroyanagi:2014qza}, while the other curves are taken from\,\cite{schmitz_2020_3689582}. Points lying above the sensitivity curve of a detector feature SNR$>1$, and have strong detection prospects. The DLRSM phase transition has good detection prospects for the detectors FP-DECIGO, BBO, and Ultimate-DECIGO for the chosen set of $v_R$ values. The GW spectrum is too weak to be detected at ET and CE for the chosen range of $v_R$. If the scale $v_R$ is increased by a factor of $\sim 10-100$, these two detectors may be able to detect them, but we ignore this region as the complementary collider constraints would be too weak. 

\begin{figure}[tbp]
\centering 
\includegraphics[width=.99\textwidth]{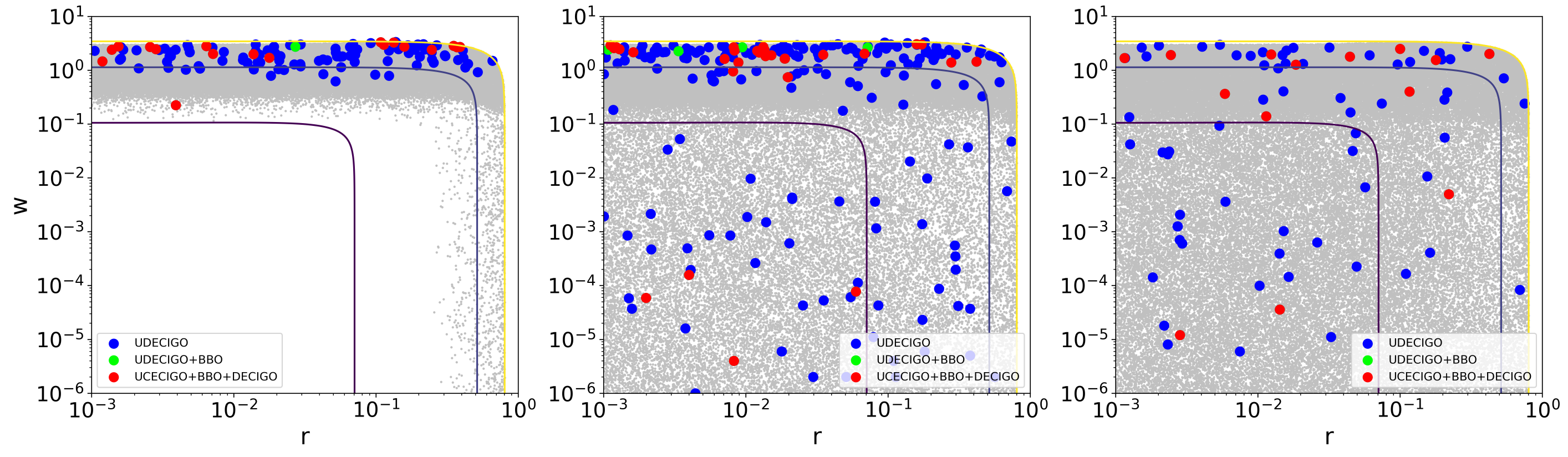}
\caption{\label{fig: rw_203050} Points with detectable GW signature at upcoming observatories: Ultimate-DECIGO (UDECIGO), BBO, and FP-DECIGO. The scale is chosen to be $v_R = 20$ TeV (left panel), $v_R=30$ TeV (middle panel), and $v_R = 50$ TeV (right panel). The purple, blue, and yellow contours represent the upper limits on $y_{33} = 1, 1.5, \sqrt{4\pi}$ respectively based on Eq.\,\eqref{eq: y_33}.}
\label{fig:rwplane}
\end{figure}

In Fig.\,\ref{fig: rw_203050} we illustrate the distribution of the points with detectable GW signal in the $r-w$ plane. The grey points pass all the theoretical and experimental constraints. The blue points are only detectable at Ultimate-DECIGO, the green points are detectable by Ultimate-DECIGO as well as BBO, and the red points can be detected at all three detectors. Interestingly, for $v_R=20$ TeV, the red, green, and blue points are densely clustered around $w\sim\mathcal{O}(1)$. For most of these points, $y_{33}$ is also large, $y_{33} \sim 1.5 - \sqrt{4\pi}$. In the middle panel, i.e. $v_R=30$ TeV, the majority of points still prefer $w\sim\ordone$, but now there are also points at lower values of $w$. In the case of $v_R=50$ TeV, we see that the clustering of points around $\ordone$ values of $w$ is even more diffuse. In all three cases, i.e. $v_R = 20,~30,$ and 50 TeV, there is no particular preference in the $r$ direction, as also seen from the SFOPT plots given in Fig.\,\ref{fig: projections}. 

\begin{table}[tbp] 
\begin{center}
\begin{tabular}{c | c c c c c c}
\hline
 & BP1   & BP2   & BP3    & BP4   & BP5  & BP6\\ 
\hline
$v_R$ (TeV) & 30        & 30       & 30       & 30       & 20         & 50       \\
$\lambda_0$ & 0.126796  & 0.466090 & 0.308396 & 0.324564 & 1.982649   & 0.799371 \\  
$\lambda_2$ & 0.097015  & 0.253725 & 0.141320 & 0.267655 & 1.670007   & 0.413236 \\ 
$\alpha_0$  & 0.004789  & 0.003504 & 0.007640 & 0.012450 & 0.012042   & 0.021020 \\  
$\alpha_3$  & 0.957421  & 0.005786 & 0.006466 & 0.004839 & 0.001015   & 0.003094 \\ 
$\rho_1$    & 0.019071  & 0.001274 & 0.001929 & 0.005930 & 0.009976   & 0.003445 \\  
$\rho_2$    & 2.003479  & 0.627225 & 1.166146 & 1.674371 & 5.574184   & 2.275937 \\
$r$         & 0.008261  & 0.008136 & 0.418869 & 0.020970 & 0.390416   & 0.424048 \\  
$w$         & $4\times 10^{-6}$  & 0.950364 & 1.439902 & 0.766492 & 2.702912   & 2.018973\\
\hline
$m_{W_R^{\pm}}$ (TeV) & 9.81  & 9.81   & 9.81     & 9.81     & 6.54      &  16.35  \\
$m_{Z_R}$ (TeV)       & 11.58 & 11.58  & 11.58    & 11.58    & 7.72      &   19.30  \\
$m_{H_1}$ (TeV)       & 20.72 & 15.97  & 32.79    & 20.89    & 81.06     & 99.90    \\
$m_{H_2}$ (TeV)       & 29.74 & 23.13  & 45.58    & 34.46    & 116.99    & 144.97 \\  
$m_{H_3}$ (TeV)       & 5.86  & 1.51   & 1.86     & 3.27     & 2.82      & 4.15 \\ 
\hline
$\alpha$    & 0.280     & 0.274    & 0.243    & 0.122    & 0.428      & 0.273 \\ 
$\beta/H_*$ & 422       & 1050     & 2648     & 8267     & 975        & 3204    \\  
$T_c$ (TeV) & 5.78      & 3.26     & 3.46     & 4.83     & 2.82       & 5.87  \\
$T_n$ (TeV) & 3.08      & 1.68     & 1.86     & 2.91     & 1.37       & 3.26  \\
\hline
\end{tabular}
\end{center}
\caption{\label{table: bp} Benchmark points for DLRSM in the simple basis.
}
\end{table}

\section{Detection prospects}\label{sec: detection prospects}
The prospect of detecting a GW signal in a given GW observatory can be quantified using the signal-to-noise ratio (SNR), defined as \cite{Athron:2023xlk,Thrane:2013oya}
\beq\label{eq: SNR}
    {\rm{SNR}} = \sqrt{n_{\rm{det}}\tau\int_{f_{\rm{min}}}^{f_{\rm{max}}} df \left[\frac{\Omega_{\rm{GW}}(f) h^2}{\Omega_{\rm{sens}}(f) h^2}\right]^2} ,
\eeq
where $\tau$ is the time period (in seconds) over which the detector is active, and the integration is carried out over the entire frequency range $[f_{\rm{min}},f_{\rm{max}}]$ of the detector. For calculations, we take $\tau = 1$ year. The factor $n_{\rm{det}}$ is two for experiments aimed at GW detection via cross-correlation measurement, or one for experiments aimed at detection via auto-correlation measurement (eg. LISA). $\Omega_{\rm{sens}}(f)$ is the noise energy density power spectrum for the chosen detector. A signal is detectable if the observed SNR value exceeds a threshold SNR, denoted as $\rm{SNR_{thres}}$. The value of $\rm{SNR_{thres}}$ varies from detector to detector. We take $\rm{SNR_{thres}} = 1$ for the purpose of discussion.

Table\,\ref{table: bp} presents six benchmark points\,(BP) with high SNR values for FP-DECIGO, BBO, and Ultimate-DECIGO, obtained using Eq.\,\eqref{eq: SNR}. BP1, BP2, BP3, and BP4 have been chosen at the $SU(2)_R$ breaking scale $v_R=30$ TeV, while for BP5 and BP6 the chosen scales are $v_R=20$ TeV and $50$ TeV respectively. The top segment of the table shows the values of the quartic couplings, while the middle segment gives the mass spectrum corresponding to each BP. The bottom segment gives the values of PT parameters $\a$, $\beta/H_*$, $T_c$ and $T_n$. Barring BP1, all other BPs have $w\sim\mathcal{O}(1)$. All BPs have $\rho_1\lesssim \mathcal{O}(10^{-1})$ and hence smaller values of $m_{H_3}$ are preferred. 

\begin{figure}[tbp]
\centering 
\includegraphics[width=.8\textwidth]{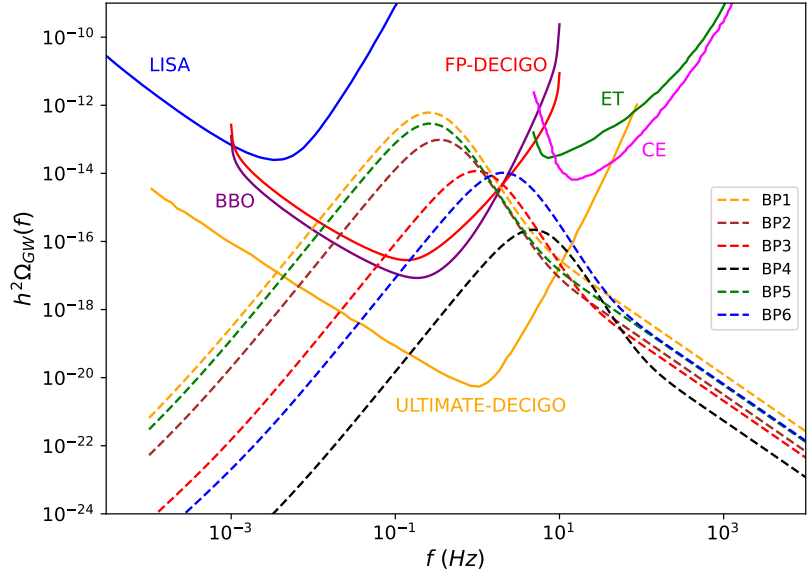}
\caption{\label{fig: benchmarks} GW spectra for the benchmark points listed in Table\,\ref{table: bp}.} 
\end{figure}

The full GW spectra for the BPs are shown in Fig.\,\ref{fig: benchmarks}. The peak of the spectrum corresponds to the frequency $f_{\rm{sw}}$ defined in Eq.\,\eqref{eq: 3_f_sw} since $\Omega_{\rm{sw}}$ gives the dominant contribution. The peak of BP4 lies only above Ultimate-DECIGO and below BBO and FP-DECIGO, while all other BPs have GW peaks above the sensitivity curves of Ultimate-DECIGO, BBO, and FP-DECIGO. The low- and high-frequency tails are dominated by the power law behavior of $\Omega_{\rm{turb}}$.

The SNRs of the BPs are listed in Table\,\ref{table: SNR}. As proclaimed in the previous section, the BPs generally yield high SNR values for FP-DECIGO, BBO, and Ultimate-DECIGO. The SNR values for BP1, BP2, BP3, BP5, and BP6 are higher than $1$ for FP-DECIGO, BBO, and Ultimate-DECIGO, and hence have good detection prospects. Ultimate-DECIGO, being the most sensitive, can detect all the BPs listed in Table\,\ref{table: SNR} with large SNR values $>10^4$. The point BP4 is not detectable at FP-DECIGO and BBO, but can be detected by Ultimate-DECIGO.

\vfill

\begin{table}[tbp] 
\begin{center}
\vspace{0.5 cm}
\setlength{\tabcolsep}{7pt}
\begin{tabular}{|c|c c c c c c|}
\hline
 SNR     & BP1             & BP2              & BP3                & BP4                 & BP5              & BP6                \\ 
\hline
FP-DECIGO   & $6.5\times10^3$ & $736.0$  & $14.4$    &   $6.5\times 10^{-3}$     & $3.0\times10^3$  & $ 2.4$      \\  
BBO      & $5.4\times 10^4$ &  $7.0\times 10^3$  &  $174.2$  & $6.5\times 10^{-2}$    &  $2.5\times 10^4$ &  $28.2$  \\ 
Ultimate-DECIGO  & $1.2\times 10^9$ & $2.6\times 10^8$ & $2.9\times 10^7$ & $2.2\times 10^4$   & $6.0\times 10^8$  & $8.0\times 10^6$ \\
\hline
\end{tabular}
\end{center}
\caption{\label{table: SNR} SNR values corresponding to different detectors for the benchmark points.
}
\end{table}

\section{Complementary collider probes}
\label{Sec: collider}
Now we describe the collider probes that complement the GW signatures discussed in the previous sections. We discuss two important collider implications, namely the precision of $\k_h$ and detection of $H_3$.

\begin{table}[tbp] 
\begin{center}
\vspace{0.5 cm}
\begin{tabular}{|c|c|c|c|c|}
\hline
$ \delta \k_{h}$ & $20$ TeV & $30$ TeV & $50$ TeV  & Combined      \\ 
\hline
$> 5\%$ & $52\%$  & $58\%$ & $50\%$ & $ 54\%$  \\  
$>10\%$ & $21\%$  & $34\%$ & $33\%$ & $ 30\%$  \\  
$>20\%$ & $8\%$   & $20\%$ & $25\%$ & $ 17\%$  \\ 
$>50\%$ & $1.3\%$ & $12\%$ & $15\%$ & $ 9\%$   \\
\hline
\end{tabular}
\end{center}
\caption{\label{table: khhh} 
Percentage of points detectable at Ultimate-DECIGO to be ruled out when the sensitivity of $\k_h$ reaches $5\%, 10\%, 20\%,$ and $50\%$, for $v_R = 20, 30,$ and $50\,$TeV. 
}
\end{table}

\begin{itemize}
\item As argued in Sec.\,\ref{subsec: higgs}, in DLRSM the trilinear Higgs coupling can deviate significantly from its SM value. In Table\,\ref{table: khhh}, we present the percentage of points leading to detectable GW signal at Ultimate-DECIGO, which also shows deviation of $\k_h$ at $5\%,~10\%,~20\%,$ and $50\%$. The current ATLAS measurement allows for a rather large range of $\k_h\in[-2.3,10.3]$. However, future colliders will significantly tighten the bound. Here we quote the projected sensitivities of $\k_h$ from Ref.\,\cite{deBlas:2019rxi}. HL-LHC will achieve a sensitivity of $50\%$ from the di-Higgs production channel. The proposed colliders, such as HE-LHC, CLIC$_{3000}$, and FCC-hh are expected to improve the sensitivity of $\k_h$ to $\sim 20\%, 10\%,$ and $5\%$ respectively. These colliders, therefore will rule out a considerable number of points showing a strong GW signal.

\begin{figure}[tbp]
\centering 
\includegraphics[width=.8\textwidth]{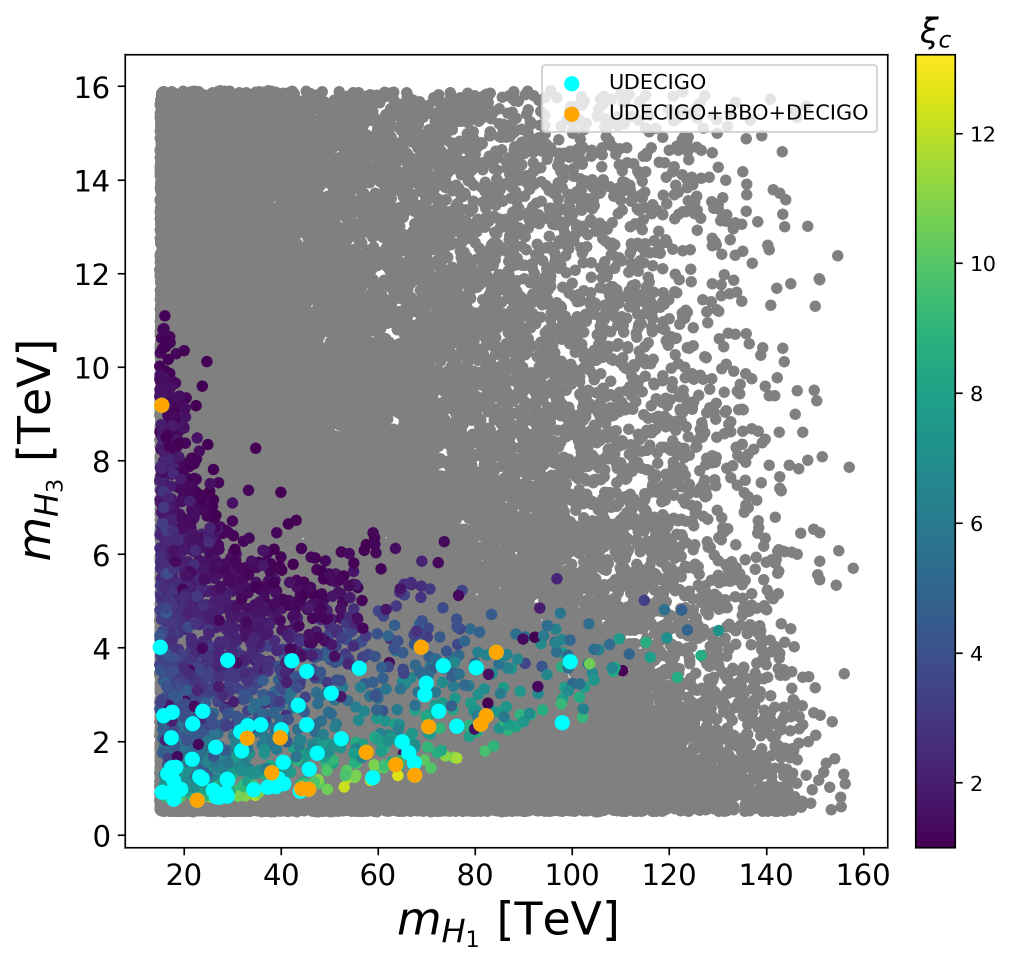}
\caption{\label{fig: mH1_mH3} The mass spectrum of DLRSM for $v_R = 20$\,TeV depicting points with $\xi_c > 1$. The cyan and orange points lead to a GW signal detectable at Ultimate-DECIGO and Ultimate-DECIGO+BBO+FP-DECIGO respectively.}
\label{fig:rwplane}
\end{figure}

\item The scalar $H_3$ can be produced at $pp$ colliders through several channels, for example\,\cite{Dev:2016dja}, 
\begin{enumerate}[(i)]
    \item $H_1$-decay, $p p \rightarrow H_1 \rightarrow h H_3$,
    \item decay of boosted $h$, $p p \rightarrow h^* \rightarrow h H_3, H_3 H_3$,
    \item Higgsstrahlung, $p p \rightarrow V_R^{*} \rightarrow V_R H_3$,
    \item $V_R V_R$ fusion, $p p \rightarrow H_3 jj$\,\,.
\end{enumerate}
The relative strength of these processes depends on the mass spectrum of DLRSM. In Fig.\,\ref{fig: mH1_mH3}, we show the distribution of SFOPT points in the $m_{H_1} - m_{H_3}$ plane for $v_R = 20$\,TeV, overlaid with points which are detectable at Ultimate-DECIGO, BBO, and FP-DECIGO. The detectable points mostly occur for small $m_{H_3}$, with the minimum value of $m_{H_3} = 741 $\,GeV. For the range $m_{H_3} = 740$\,GeV $-$ 1.2\,TeV, the production cross-section of $H_3$ at FCC-hh with $\sqrt{s} = 100$\,TeV can be $\sim \mathcal{O}(\text{fb})$\,\cite{Dev:2016dja}.

For $v_R = 20$\,TeV, $m_{H_3} \lesssim 500$\,GeV can be ruled out from the observations of the channel\,(iv) at FCC-hh with a luminosity of $30\,\text{ab}^{-1}$. For large values of quartic couplings, the decay width $h^{*} \rightarrow h H_3$ and $h^{*} \rightarrow H_3 H_3$ can be large and subsequently, channel (ii) can rule out $m_{H_3} \lesssim 700$\,GeV. For channel (i), $H_1$ with mass $15$\,TeV can be produced with a cross-section $\sim 0.5$\,fb and have sizable branching ratios of $H_1 \rightarrow h H_3,\, H_3 H_3$. As a result, channel (i) can rule out masses up to $m_{H_3} \sim 2$\,TeV. Thus, these searches are capable of ruling out a large number of points with low-$m_{H_3}$, thus low-$\rho_1$, providing a complementarity to the GW probe of DLRSM. 

\end{itemize}
\section{Summary and conclusions}\label{sec: summary}
In this chapter, we studied the possibility of an observable stochastic GW background resulting from SFOPT associated with the spontaneous breaking of $SU(2)_R\times U(1)_{B-L}$ in DLRSM. The gauge symmetry of DLRSM breaks in the following pattern:
$$SU(2)_L\times SU(2)_R \times U(1)_{B-L}\xrightarrow{\mathit{~~v_R~~}} SU(2)_L\times U(1)_{Y} \xrightarrow{\mathit{\k_1,\k_2,v_L}} U(1)_Y . $$
The non-observation of a right-handed current at colliders puts a lower bound on the scale $v_R$ to be around 20 TeV. Due to the hierarchy $v_R \gg v$, the $SU(2)_R\times U(1)_{B-L}$-breaking dynamics is decoupled from the EWPT.  We chose the scale $v_R=20,~30,$ and $50$ TeV to study the possible detection of GW background at the planned observatories. For these values of $v_R$, complementary searches for new scalars of DLRSM are feasible at future colliders. 

Our analysis was carried out using the {\textit{simple basis}} defined in Ref.\,\cite{Karmakar:2022iip}, to reduce the number of independent parameters. It should be noted that analysis with the full set of parameters also gives similar patterns of SFOPT in the $\rho_1-\rho_2$ and $r-w$ planes. The parameters in the {\textit{simple basis}} include the quartic couplings: $\l_0, \l_2, \a_0, \a_3, \rho_1, \rho_2$. In addition, we defined EW \vevs through the ratios $r$ and $w$. Most studies on LRSM take the simplified limit $r,w\rightarrow 0$. However, it was pointed out in Refs.\,\cite{Bernard:2020cyi,Karmakar:2022iip} that the DLRSM phenomenology allows for significant deviation from this limit. Therefore, we also scanned over $r$ and $w$. 

We constructed the one-loop finite temperature effective potential for each parameter point and analyzed the nature of PT using the package \texttt{CosmoTransitions}. Due to the large separation between $v_R$ and the EW scale, the effective potential depends solely on the background field value of the neutral CP-even scalar, $\chi^0_{Rr}$. The condition for SFOPT, $\xi_c>1$ was used to identify viable regions of the parameter space. SFOPT favors small values of the quartic coupling $\rho_1\lesssim \mathcal{O}(10^{-1})$, which leads to $m_{H_3} \ll v_R$. This feature has also been observed in other variants of LRSM, discussed in refs.\,\cite{Brdar:2019fur, Graf:2021xku, Li:2020eun}. 

We find that for very small values, $\rho_1\lesssim 10^{-3}$, however, the zero temperature minimum of the one-loop effective potential at $R = v_R$ becomes metastable, violating the Linde-Weinberg bound. Hence, there is a lower bound on $\rho_1$ below which FOPT is not observed. Most points with SFOPT also feature $w\sim \mathcal{O}(1)$, while for smaller values of $w$, very few points show SFOPT. Out of the chosen set of parameters, the SFOPT region is most sensitive to the parameters $\rho_1$ and $w$ and to some extent $\l_0$. However, we see no particular preference for the \vev ratio $r$ and the quartic couplings relating the bidoublet and the doublet fields, i.e., $\a_0$ and $\a_3$, as illustrated by the projections given in Fig.\,\ref{fig: projections}. 

For parameter points showing SFOPT, we computed the PT parameters, $\a$, $\beta/H_*$, and $T_n$, needed for the calculation of the GW spectrum. In the non-runaway scenario, the stochastic GW background resulting from SFOPT comes primarily from sound waves and turbulence, while the contribution from bubble wall collisions remains sub-dominant. Fig.\,\ref{fig: gw_scatter} shows the position of the peak of the GW spectrum for points satisfying the SFOPT criterion. While for a large number of points, the GW spectrum is too weak to be detected, there is a significant number of points lying above the sensitivity curves for Ultimate-DECIGO, BBO, and FP-DECIGO. Such points will be accessible to these detectors in the coming years. The detectable points also prefer $w \sim \mathcal{O}(1)$, which in turn, correspond to a large value of $y_{33}$ as seen in Fig.\,\ref{fig: rw_203050}. 

The strength of the GW spectrum does not depend on the scale $v_R$. On the other hand, since the peak frequency is proportional to $T_n\sim v_R$, the points shift to the right as $v_R$ changes from $20$ to $30$ to $50$ TeV. To quantify the detection prospects, we computed the signal-to-noise ratio at these detectors for the detectable points. Six benchmark points are given in Table\,\ref{table: SNR}, featuring SNR values higher than $10^5$. We see that for all the BPs, $m_{H_3}\lesssim 5$ TeV. 

There are primarily two complementary collider probes for the points with detectable GW signals. It was found that a significant fraction of points leads to $50, 20, 10, \text{and}\, 5\%$ deviation of $k_h$ from unity, which can be ruled out at HL-LHC, HE-LHC, CLIC$_{3000}$, and FCC-hh, respectively. Due to a relatively low mass of $H_3$, it can be produced at future colliders through various channels. In particular, FCC-hh can rule out up to $m_{H_3} \sim 2$\,TeV.  

Here we make a note of some subtleties involved in computing the GW spectrum that contribute to theoretical uncertainty: \textbf{(i)} The suppression factor $\Upsilon$, introduced in \eqref{eq: 3_soundwaves} was recently proposed, to take the finite lifetime of sound waves into account. For the chosen benchmark points, this suppression factor takes $\mathcal{O}(0.1)$ values. \textbf{(ii)} As pointed out earlier, the value of $v_c/T_c$ depends on the particular choice of gauge since the effective potential is gauge-dependent. The effect of gauge dependence is minimized in the Landau gauge, which we use for our calculations. \textbf{(iii)} In principle, the bubble wall velocity can be computed from the model parameters, as seen in Refs.\,\cite{Moore:1995ua,Moore:1995si,Bodeker:2009qy}. We use $v_w=1$, which is valid when the friction on the walls is low. Thus a particular choice of $v_w$ can cause small shifts in the GW spectrum. We checked that the uncertainties mentioned above contribute to roughly $\mathcal{O}(0.1-1)$ deviations in the GW spectrum of the BPs. However, the BPs would still be detectable at respective detectors, BBO and/or DECIGO.

The spontaneous breaking of the discrete LR symmetry $\mathcal{P}$ can lead to the formation of domain walls. The GW imprint from the domain wall network peaks at much lower frequencies, as compared to that from FOPT \cite{Borah:2022wdy, Borboruah:2022eex}. Since there is no overlap of the GW signals, we have focused our discussion on FOPT. 

Although DLRSM does not account for neutrino masses, it is interesting to ask if incorporating them by adding extra fields to the model could modify the strength of FOPT. In Appendix\,\ref{sec: neutrino mass}, we have shown that it is possible to include neutrino masses without impacting the results of our analysis.
 %FOPT in DLRSM
\newpage
\null
\newpage
\chapter{Domain walls in DLRSM}
\label{chapter4}
\linespread{0.1}
\graphicspath{{Chapter4/Figures}}
\pagestyle{headings}
\hrule height 1mm
\vspace{4mm}

\section{Introduction}
The 15-year dataset (NG15) of the NANOGrav collaboration\,\cite{NANOGrav:2023gor} shows compelling evidence of a stochastic gravitational wave background (SGWB) at nanoHertz frequencies. This evidence has been corroborated at varying significance levels by other pulsar timing arrays (PTAs) such as the European Pulsar Timing Array (EPTA) in collaboration with the Indian Pulsar Timing Array (InPTA)\,\cite{EPTA:2023fyk}, Parkes Pulsar Timing Array (PPTA)\,\cite{Reardon:2023gzh}, which are all part of the International Pulsar Timing Array (IPTA) consortium. While more data is needed to claim a discovery, discussing the possible sources of such a SGWB is interesting. The standard astrophysical interpretation, of the SGWB produced by in-spiralling supermassive black hole binaries (SMBHBs) scattered across the universe, is in slight tension with the data\,\cite{NANOGrav:2023gor,Ellis:2023dgf,Ghoshal:2023fhh,Shen:2023pan,Broadhurst:2023tus,Bi:2023tib,Zhang:2023lzt}. Other explanations of cosmological origin have been discussed in the literature, including the gravitational waves (GWs) from the density perturbations after inflation\,\cite{Franciolini:2023pbf,Vagnozzi:2023lwo,Inomata:2023zup,Ebadi:2023xhq,Liu:2023ymk,Abe:2023yrw,Unal:2023srk,Firouzjahi:2023lzg,Bari:2023rcw,Cheung:2023ihl,Bhaumik:2023wmw,Gorji:2023sil}, first-order phase transitions\,\cite{Fujikura:2023lkn,Addazi:2023jvg,Bai:2023cqj,Megias:2023kiy,Han:2023olf,Zu:2023olm,Ghosh:2023aum,DiBari:2023upq,Cruz:2023lnq,Gouttenoire:2023bqy,Ahmadvand:2023lpp,An:2023jxf,Wang:2023bbc}, and topological defects such as domain walls (DWs)\,\cite{Ferreira:2022zzo,Kitajima:2023cek,Guo:2023hyp,Blasi:2023sej,Gouttenoire:2023ftk,Barman:2023fad,Lu:2023mcz,Li:2023tdx,Du:2023qvj,Gelmini:2023kvo,Zhang:2023nrs} and cosmic strings\,\cite{Ellis:2023tsl,Kitajima:2023vre,Wang:2023len,Lazarides:2023ksx,Eichhorn:2023gat,Servant:2023mwt,Antusch:2023zjk,Fu:2024rsm,Yamada:2023thl,Ge:2023rce}. 

Comparative analyses of the possible SGWB sources reveal that many of these models provide a better fit compared to the standard SMBHB interpretation\,\cite{NANOGrav:2023hvm,Ellis:2023oxs,Wu:2023hsa}. In this chapter, we consider DWs as the possible source of the signal, for which the Bayes factor is $\mathcal{O}(10)$ when compared to the fiducial SMBHB model\,\cite{NANOGrav:2023hvm}. From the particle physics perspective, DWs are formed when a discrete symmetry is spontaneously broken. The GW spectrum from DWs depends on the surface tension, $\sigma$, of the walls, and the bias potential, $V_{\rm{bias}}$. The PTA data is compatible with DWs for values roughly, $\sigma\sim (100\,\rm{TeV})^3$, and the bias, $V_{\rm{bias}}\sim (100\,\rm{MeV})^4$\,\cite{Ferreira:2022zzo}. A microscopic model for DWs at the electroweak (EW) scale, $v_{\rm{EW}}\sim 246.02$\,GeV, cannot yield DWs with such a high surface tension, and thus a viable microscopic DW model must incorporate high-scale physics. Given that the EW interactions in the standard model (SM) maximally violate parity, high-scale extensions of SM that respect the parity symmetry, $\mathcal{P}$, provide an interesting way to generate DWs with sufficiently large surface tension.

\begin{comment}
Left-right symmetric models\,(LRSMs)\,\cite{Pati:1974yy,Mohapatra:1974gc,PhysRevD.11.566,Senjanovic:1975rk,Senjanovic:1978ev} are well-motivated extensions of the standard model (SM) where the gauge group is extended from $\mathcal{G}_{\rm{SM}} = SU(3)_c\times SU(2)_L\times U(1)_{Y}$ to $\mathcal{G}_{\rm{LRSM}} = SU(3)_c\times SU(2)_L\times SU(2)_R\times U(1)_{B-L}$. An additional discrete $\mathcal{P}$ symmetry can be easily incorporated into $\mathcal{G}_{\rm{LRSM}}$, allowing for the possibility of DW formation. The various realizations of LRSM differ from each other, depending on the scalars involved in the spontaneous breaking of $\mathcal{G}_{\textrm{LRSM}}$ to $\mathcal{G}_{\rm{SM}}$. They also differ in the mechanism of fermion mass generation. A widely studied realization is the triplet LRSM (TLRSM), where the scalar sector involves two triplets and a bidoublet\,\cite{Maiezza:2016ybz,PhysRevD.44.837,Senjanovic:2016bya}. When the scalar sector contains two doublets and a bidoublet, it is called the doublet LRSM (DLRSM)\,\cite{Senjanovic:1978ev,Mohapatra:1977be}. Other versions of LRSM are also studied in the literature \,\cite{Ma:1986we,Babu:1987kp,Ma:2010us,Frank:2019nid,Frank:2020odd,Graf:2021xku}. It was recently shown that the pattern of electroweak symmetry breaking (EWSB) in DLRSM can be quite different from the other versions of LRSM, with interesting consequences from precision observables\,\cite{Bernard:2020cyi} and Higgs data\,\cite{Karmakar:2022iip}. 
\end{comment}
In this chapter, we study the DWs arising in parity-symmetric DLRSM. Previous discussions on DWs in LRSM\,\cite{Yajnik:1998sw,Borah:2022wdy,Chakrabortty:2019fov,Mishra:2009mk,Borah:2011qq,Banerjee:2020zxw,Borboruah:2022eex,Barman:2023fad,Banerjee:2023hcx}  mainly centered around TLRSM. In refs.\,\cite{Yajnik:1998sw,Borboruah:2022eex}, the kink solutions for `left-right' ($LR$) DWs were presented for a few benchmark points. The GW signature was studied in refs.\,\cite{Borah:2022wdy,Barman:2023fad}, where the benchmarks were chosen based on an approximate dependence of the DW surface tension without explicitly solving the kink equations. In this chapter, we show that two types of DWs are formed in DLRSM, namely, $Z_2$ and $LR$ DWs, with different surface tensions. For both types, we solve the kink equations to obtain the parametric dependence of the DW surface tension and show that the $Z_2$ DWs are unstable. We then obtain the GW signature in terms of the DLRSM parameters. While qualitative similarity is expected in the GW signature of DWs
from DLRSM and TLRSM, we perform a Bayesian analysis on the PTA data to constrain the model parameters. We consider the GW spectrum from DLRSM DWs with and without the contribution from SMBHBs. The discussion of this chapter can be easily carried over to TLRSM.   
  
In Sec.\,\ref{Sec: 4_Model}, we briefly discuss the scalar potential of the $\mathcal{P}$-symmetric DLRSM. In Sec.\,\ref{sec: DW}, we discuss the vacuum structure of the DLRSM effective potential and study the DW solutions. In Sec.\,\ref{sec: GW}, we discuss the GW spectrum resulting from DLRSM DWs. We present our results in Sec.\,\ref{sec: results}, where we perform the Markov chain Monte Carlo (MCMC) analysis, and discuss the detection prospects at upcoming GW observatories. Finally, we summarize our findings and make concluding remarks in Sec.\,\ref{sec: discussion}.

\section{The model}
\label{Sec: 4_Model}
The model of Chapter\,\ref{chapter3} is briefly recapped here. The gauge group of parity-symmetric DLRSM is,
$$\mathcal{G}_\text{LRSM} = \mathcal{P}\times SU(3)_c\times SU(2)_L\times SU(2)_R\times U(1)_{B-L}.$$ 
For an overview of DLRSM, please refer to\,\cite{Senjanovic:1978ev,Mohapatra:1977be,Bernard:2020cyi}. The scalar sector has a complex bi-doublet $\Phi$, and two doublets $\chi_L$ and $\chi_R$. The scalar multiplets are,
\bea
\Phi = \begin{pmatrix}
    \phi_1^0 & \phi_2^+ \\
    \phi_1^- & \phi_2^0
\end{pmatrix} \sim (1,2,2,0), ~ \chi_L = \begin{pmatrix}
    \chi_L^+ \\
    \chi_L^0
\end{pmatrix} \sim (1,2,1,1), ~ \text{and}~\chi_R = \begin{pmatrix}
    \chi_R^+ \\
    \chi_R^0
\end{pmatrix} \sim (1,1,2,1),\nn\\
\eea
where the parentheses indicate the representation of the multiplets under $SU(3)_c$, $SU(3)_L$, $SU(3)_R$, and $U(1)_{B-L}$ respectively. The group $\mathcal{P}$ denotes the discrete parity symmetry under the exchange $L\leftrightarrow R$, with the action given by,
\beq
\mathcal{P}: Q_L\leftrightarrow Q_R,~l_L\leftrightarrow l_R,~\chi_L\leftrightarrow\chi_R,~\Phi\leftrightarrow\Phi^{\dagger},
\eeq
where $Q_L,~Q_R,~l_L,~l_R$ are the left and right-handed quark and lepton doublets respectively\,\cite{Bernard:2020cyi}. The $\mathcal{P}$ symmetry imposes the condition $g_L=g_R$ on the $SU(2)_L$ and $SU(2)_R$ gauge couplings. The scalar potential is given by\,\cite{Bernard:2020cyi},
\bea\label{eq: potential}
V  &=& V_{\chi} + V_{\chi \Phi} + V_{\Phi},\nn\\
V_{\chi} &=& - \m_3^2\ [\chi_L^{\dagger} \chi_L + \chi_R^{\dagger} \chi_R] + \rho_1\  [(\chi_L^{\dagger} \chi_L )^2 + (\chi_R^{\dagger} \chi_R )^2]
 + \rho_2\  \chi_L^{\dagger} \chi_L \chi_R^{\dagger}\chi_R, \nn\\ 
V_{\chi \Phi} &=& \m_4\  [\chi_L^{\dagger} \Phi \chi_R + \chi_R^{\dagger} \Phi^{\dagger} \chi_L] + \m_5\  [\chi_L^{\dagger} \tilde{\Phi} \chi_R + \chi_R^{\dagger}\tilde{\Phi}^{\dagger}\chi_L ]\nn\\
 &&+ \alpha_1\tr(\Phi^{\dagger} \Phi ) [\chi_L^{\dagger}\chi_L + \chi_R^{\dagger}\chi_R ]
 + \Big\{ \frac{\alpha_2}{2} \ [\chi_L^{\dagger} \chi_L  \tr(\tilde{\Phi} \Phi^{\dagger} ) + \chi_R^{\dagger} \chi_R  \tr(\tilde{\Phi}^{\dagger} \Phi )] + {\rm h.c.} \Big\} \nn\\
 &&+ \alpha_3\ [\chi_L^{\dagger}\
 \Phi \Phi^{\dagger}\chi_L + \chi_R^{\dagger} \Phi^{\dagger} \Phi  \chi_R  ] 
 + \alpha_4\ [\chi_L^{\dagger}\
 \tilde{\Phi} \tilde{\Phi}^{\dagger}\chi_L + \chi_R^{\dagger} \tilde{\Phi}^{\dagger} \tilde{\Phi}  \chi_R  ],\nn\\
V_{\Phi}  &=& -\m_1^2\tr(\Phi^{\dagger}\Phi) - \m_2^2\ [\tr(\tilde{\Phi}\Phi^{\dagger}) + \tr(\tilde{\Phi}^{\dagger} \Phi)] + \l_1[\tr(\Phi^{\dagger}\Phi)]^2\nn\\
&&  + \l_2\ [ [\tr(\tilde{\Phi} \Phi^{\dagger})]^2
 + [\tr(\tilde{\Phi}^{\dagger} \Phi)]^2 ] + \l_3\text{Tr}(\tilde{\Phi} \Phi^{\dagger}) \, \tr(\tilde{\Phi}^{\dagger} \Phi)\nn\\
 &&  + \l_4\tr(\Phi^{\dagger}\Phi) \, [\tr(\tilde{\Phi}\Phi^{\dagger})+ \tr(\tilde{\Phi}^{\dagger}\Phi)],
 \label{eq:scalarpotential}
\eea
where $\tilde{\Phi}\equiv \sigma_2\Phi^*\sigma_2$, and we take all parameters to be real. The pattern of symmetry breaking is:
$$\mathcal{P}\times SU(2)_L\times SU(2)_R \times U(1)_{B-L}\xrightarrow{~~\langle\chi_R\rangle~~} SU(2)_L\times U(1)_{Y} \xrightarrow{\langle\Phi\rangle,\langle\chi_L\rangle} U(1)_{\rm{em}} . $$The following charge-preserving and $CP$-preserving \vev structure achieves the desired symmetry-breaking pattern:
\beq
     \langle\Phi \rangle = \frac{1}{\sqrt{2}}\begin{pmatrix}
     \k_1 & 0\\
     0 & \k_2
 \end{pmatrix}, ~ \langle\chi_L\rangle = \frac{1}{\sqrt{2}} \begin{pmatrix}
     0\\
     v_L
 \end{pmatrix}, ~ \langle\chi_R \rangle = \frac{1}{\sqrt{2}}\begin{pmatrix}
     0\\
     v_R
 \end{pmatrix}\,.
 \eeq
The \vevs $\k_1,\k_2$ and $v_L$ follow the relation, $\k_1^2+\k_2^2+v_L^2 = v_{\rm{EW}}^2$, where $v_{\rm{EW}}= 246.02$\,GeV. The absence of a right-handed gauge boson in collider searches\,\cite{Solera:2023kwt} dictates the hierarchy of scales in DLRSM: $v_R\gg \k_1,\k_2,v_L$. Since the potential is parity-symmetric, an alternative hierarchy where $v_L\gg \k_1,\k_2,v_R$, is also possible but is not realized in nature. We denote the $\mathcal{P}$-symmetry breaking scale by $v_0$, such that $v_0\gg v_{\rm{EW}}$. In the next section, we discuss how spontaneous breaking of the discrete $\mathcal{P}$-symmetry gives rise to regions of disconnected vacuua separated by DWs. 

\section{Domain walls in DLRSM}\label{sec: DW}
The existence of DWs can be inferred from the minima of the effective potential in the $v_R-v_L$ plane. The tree-level effective potential of DLRSM is, 
\beq
V_{0}\equiv V_{\chi}(\langle\chi_L\rangle,\langle\chi_R\rangle) + V_{\chi \Phi}(\langle\chi_L\rangle,\langle\chi_R\rangle,\langle\Phi\rangle) + V_{\Phi}(\langle\Phi\rangle).
\eeq

The contributions to the one-loop finite temperature effective potential were discussed in Ref.\,\cite{Karmakar:2023ixo}. A thorough analysis would require numerically calculating the full one-loop effective potential. Here we make some approximations to simplify the discussion. Symbolically,
\beq
V_{\rm{eff}} = V_0 + V_1 + V_{1T},
\eeq
where $V_1$ and $V_{1T}$ are the one-loop zero-temperature and finite temperature corrections respectively. The effective potential obeys the symmetry of the tree-level potential. While we can always fix the zero-temperature minima and masses at their tree-level values by adding a finite counter-term to $V_1$, the role of $V_{1T}$ should be analyzed. $V_{1T}$ has the form,
\beq
V_{1T}(\phi,T) = \sum_{i\in \rm{heavy}} n_i\frac{T^4}{2\pi^2} J_{b/f}\bigg(\frac{m^2_i(\phi)}{T^2}\bigg) + \sum_{i\in \rm{light}} n_i\frac{T^4}{2\pi^2} J_{b/f}\bigg(\frac{m^2_i(\phi)}{T^2}\bigg),
\eeq
where the sum runs over all fields, generically labeled by $\phi$. The term `heavy' denotes $v_0$-scale fields $\chi_L,\chi_R$, and the gauge bosons,\footnote{$W_L^{\pm}$ and $Z_L$ are taken as heavy fields since $v_L$ can take large values in some domains.} $W^{\pm}_{L,R},Z_{L,R}$. Similarly, `light' denotes the EW-scale fields including $\Phi$, the fermions, the photon, and gluons. $m_i$ are the field-dependent masses, and $n_i$ is the number of degrees of freedom for species $i$. The function $J_b$ ($J_f$) is defined for bosons (fermions) and has well-known high-$T$ and low-$T$ expansions\,\cite{Cline_1997}.  Write $x^2 \equiv \frac{m^2_i}{T^2}$, then for $x^2\ll 1$, 
\begin{eqnarray}%\label{eq: highT}
J_f(x^2) \approx &-&\frac{7\pi^4}{360} + \frac{\pi^2}{24} x^2 \label{eq: highTf}\\
J_b(x^2) \approx &-&\frac{\pi^4}{45} + \frac{\pi^2}{12} x^2 - \frac{\pi}{6}\big(x^2\big)^{3/2}.\label{eq: highTb}
\end{eqnarray}
The $x^2$-dependent term in $J_{b/f}$ leads to symmetry-restoration at high-$T$. For $x^2\gg 1$, both fermions and bosons have the same expansion with an exponential suppression due to the Boltzmann factor\,\cite{Cline_1997},\beq
J_{b/f}(x^2) \approx -\exp\bigg(-(x^2)^{1/2}\bigg)\bigg(\frac{\pi}{2}(x^2)^{3/2}\bigg)^{1/2}\label{eq: lowT}.
\eeq

DWs are formed at temperatures below the parity-breaking scale $v_0$, i.e., $T\lesssim v_0$. The network exists for a temperature range $T_{\rm{ann}}\lesssim T\lesssim v_0$, where $T_{\rm{ann}}$ is the DW annihilation temperature (Section\,\ref{sec: GW}). The DWs of DLRSM are topologically stable even after electroweak symmetry breaking (EWSB) and can survive right till the epoch of Big Bang nucleosynthesis (BBN). The evolution of the DW network before and after EWSB should therefore be considered separately.

\begin{itemize}
\item \textbf{Before EWSB}: When $v_{\rm{SM}}\ll T \lesssim v_0$, we can set the \vevs $\k_1=\k_2=0$ as the EW symmetry is restored. On the other hand, the contribution of the `heavy' fields to the one-loop temperature corrections is Boltzmann-suppressed according to Eq.\,(\ref{eq: lowT}). So the effective potential in terms of the background fields $v_L$ and $v_R$ is given by,
\bea\label{eq: Veff}
    V_{\rm{eff}}(v_L,v_R) &=& V_{\chi}(v_L,v_R)\nn\\
    &=& -\frac{\mu_3^2}{2}(v_L^2+v_R^2) + \frac{\rho_1}{4}(v_L^4+v_R^4) + \frac{\rho_2}{4}v_L^2v_R^2.
\eea
The minima and saddle points of $V_{\chi}$ are given in Eq.\,(\ref{eq: min}) and Eq.\,(\ref{eq: saddle}) respectively.

\begin{figure}[tbp]
    \centering
    \includegraphics[width=.7\textwidth]{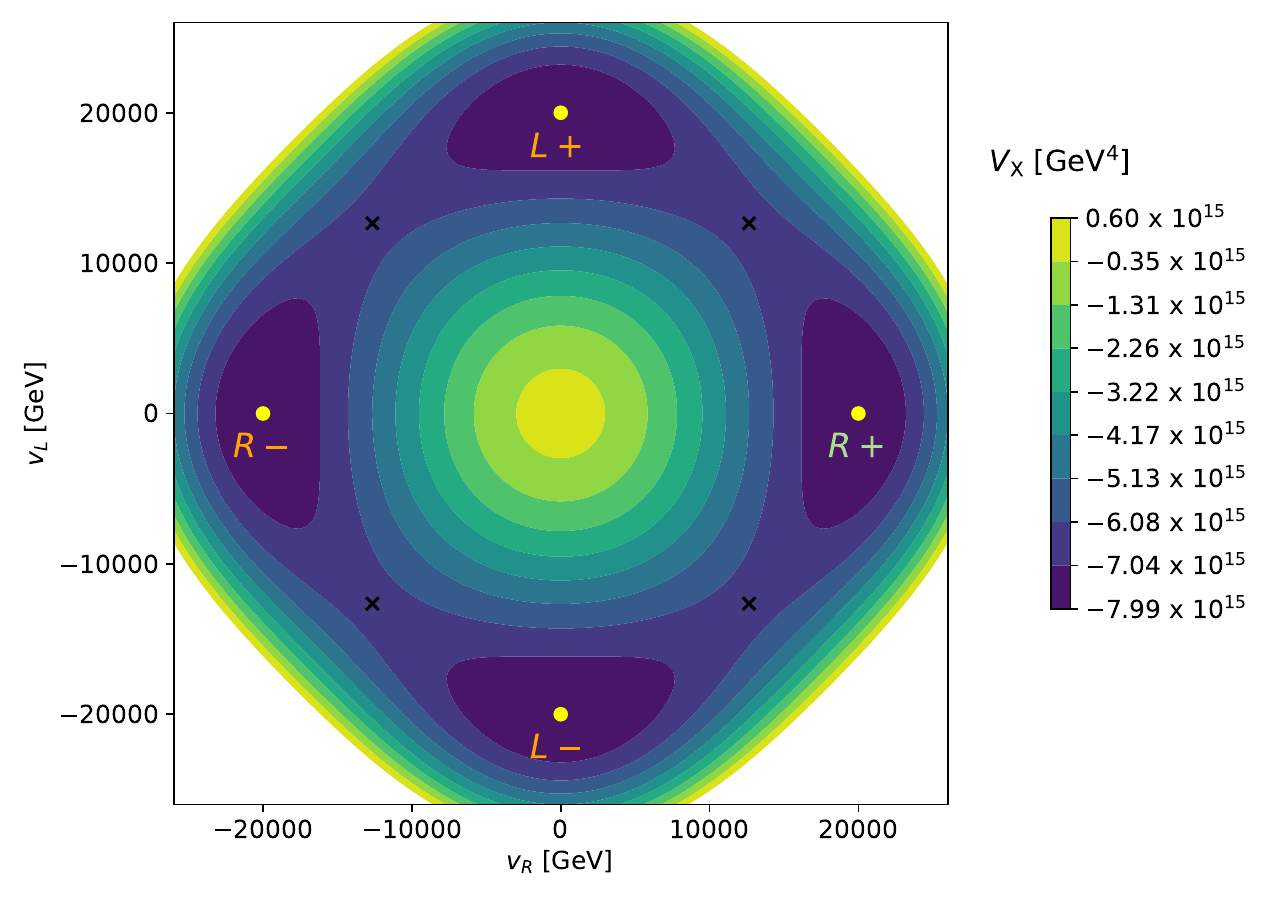}
    \caption{The potential $V_{\chi}$ for $v_0 = 20$\,TeV, $\rho_1 = 0.2$, and $\rho_2 = 0.6$. Yellow dots indicate the minima, while black crosses show the saddle points. The minimum, $R+$, shown in green, is consistent with the observed phenomenology.} 
    \label{fig: Z4}
\end{figure}

%\subsection{After EWSB}
\item \textbf{After EWSB}: When $T_{\rm{ann}}\lesssim T \lesssim v_{\rm{SM}}$, the bidoublet acquires a non-zero \vev. In this regime, the thermal contributions of $v_0$-scale fields as well as EW-scale fields are Boltzmann-suppressed. The effective potential is modified as
\bea
V_{\rm{eff}}(v_L,v_R) &=& V_{\chi}(v_L,v_R) + V_{\chi\Phi}(v_L,v_R;\k_1,\k_2)\nn\\
 &=& c_1 (v_L^2+v_R^2) + c_2 (v_L^4+v_R^4) + c_3 v_L^2v_R^2 + c_4 v_Lv_R.
\eea
The $V_{\Phi}$ term has been dropped since it is independent of $v_L$ and $v_R$. The coefficients, $c_1,~c_2,~c_3,~c_4$, are given by,
\bea
&c_1 = -\frac{\mu_3^2}{2} + \frac{1}{4}[\k_1^2(\a_1+\a_4)+\k_2^2(\a_1+\a_3) + 2\k_1\k_2\a_2],~~c_2 = \frac{\rho_1}{4},&\nn \\
&c_3 = \frac{\rho_2}{4},~~c_4 = \frac{1}{\sqrt{2}}(\k_2\mu_4 + \k_1\mu_5).&\nn
\eea
Non-zero values of $\k_1$ and $\k_2$ slightly change the positions of the minima of the effective potential. We verified that the contribution of $V_{\chi\Phi}$ to the effective potential is numerically insignificant in the parameter space of interest, since $\k_1,\k_2\ll v_0$.
\end{itemize}

Hence, the DW structure is primarily governed by $V_{\chi}$, given in Eq.\,(\ref{eq: Veff}). In polar coordinates, $v_R = v \cos\theta,~v_L = v \sin\theta$, Eq.\,(\ref{eq: Veff}) becomes, 
\beq
V_{\chi}(v,\theta) = -\frac{\mu_3^2}{2} v^2 + \frac{v^4}{4}\Big(\rho_1 + \frac{\rho_{21}}{2}\sin^2 2\theta\Big),
\eeq
where $\rho_{21} = \rho_2/2-\rho_1$. For $\mu_3^2>0,~\rho_{21}>0$, there are four degenerate minima, 
\beq\label{eq: min}
(v,\theta) = \left(v_0, \frac{n\pi}{2}\right),
\eeq
and four saddle points, 
\beq\label{eq: saddle}
(v,\theta) = \left(v_1, \frac{(2n+1)\pi}{4}\right),
\eeq
where $n=0,1,2,3$, and 
\beq
v_0 = \sqrt{\frac{\mu_3^2}{\rho_1}},~\text{and}~v_1 = \sqrt{\frac{\mu_3^2}{\rho_1 + \frac{\rho_{21}}{2}}}.
\eeq
A contour plot of $V_{\chi}$ is shown in Fig.\,\ref{fig: Z4} for a fixed set of parameters. The potential has a $Z_4\simeq\mathcal{P}\times Z_2$ symmetry, with the adjacent minima connected by $\mathcal{P}$, and non-adjacent minima connected by $Z_2$. We denote the four minima $(v_R,v_L)$ as: $R+ \equiv (v_0,0), ~L+ \equiv (0,v_0), ~R- \equiv (-v_0,0), ~L- \equiv (0,-v_0)$. The desired vacuum consistent with phenomenology is the $R+$ vacuum\footnote{The $+$ or $-$ here is just a convention. We could also take $R-$ as the desired vacuum.}. After the $\mathcal{P}$-breaking PT, spatial points separated by distances larger than the correlation length $\xi$ fall into any of the four minima with equal probability. This creates a network of domain walls separating regions of distinct vacuua, each with volume $\sim \xi^3$. Past papers on DWs from LRSM have mostly discussed DWs that separate the $L$-type regions from the $R$-type regions. However, due to the $Z_4$ symmetry, two kinds of DWs are formed: 
 \begin{enumerate}
 \item $LR$ DWs, denoted by $\boxed{L\pm|R\pm}$, which separate adjacent minima, i.e., the $L\pm$ regions from the $R\pm$ regions. 
 \item $Z_2$ DWs, denoted by $\boxed{L+|L-}$ or $\boxed{R+|R-}$, separating non-adjacent minima, i.e. the $L(R)+$ regions from the $L(R)-$ regions. 
 \end{enumerate}
We will later discuss that the $Z_2$ DWs are unstable and decay into pairs of $LR$ DWs. The energy density $\mathcal{E}$ of the DW network is given by the `00' component of the energy-momentum tensor. For a single DW configuration perpendicular to the $x$-axis, separating two vacuua at $x\rightarrow \pm\infty$,
\begin{equation}
\mathcal{E} = \frac{1}{2}\left(\frac{dv_L}{dx}\right)^2 + \frac{1}{2}\left(\frac{dv_R}{dx}\right)^2 + V(v_L,v_R) + C,
\end{equation}

\noindent where $C$ is a constant chosen so that $\mathcal{E}$ vanishes at infinity. 

Due to translational symmetry, we can choose the DW profile to be centered at $x=0$. Integrating $\mathcal{E}$ along the $x$ direction yields the energy per unit area or the DW surface tension, $\sigma$,
\beq
\sigma = \int_{-\infty}^{\infty}\mathcal{E} dx.
\eeq
The kink solution interpolating between the two vacuua minimizes $\sigma$, and therefore obeys,
\beq
\frac{d}{dx}\left(\frac{\partial\mathcal{E}}{\partial (d v_i/dx)}\right) - \frac{\partial\mathcal{E}}{\partial v_i} = 0,~i\in\{L,R\}.
\eeq
We get a pair of coupled ordinary differential equations,
\bea
\frac{\partial^2 v_L}{\partial x^2} = -\mu_3^2 v_L + \rho_1 v_L^3 + \frac{\rho_2}{2} v_L v_R^2,\label{eq: kink1}\\
\frac{\partial^2 v_R}{\partial x^2} = -\mu_3^2 v_R + \rho_1 v_R^3 + \frac{\rho_2}{2} v_L^2v_R\label{eq: kink2},
\eea
which can be solved numerically using relaxation methods (see, for example, Refs.\,\cite{Battye:2011jj,Battye:2020sxy}), with appropriate boundary conditions. To construct DW profiles, it is sufficient to consider $\boxed{L+|R+}$ and $\boxed{R-|R+}$. The boundary conditions are: \\
\noindent\textbf{Case I: $LR$ DWs} 
\beq
    \lim_{x\rightarrow -\infty} (v_R,v_L) = (0,v_0),~\lim_{x\rightarrow +\infty} (v_R,v_L) = (v_0,0). \label{eq: bc1}
\eeq

\noindent\textbf{Case II: $Z_2$ DWs} 
\beq
    \lim_{x\rightarrow -\infty} (v_R,v_L) = (-v_0,0),~ \lim_{x\rightarrow +\infty} (v_R,v_L) = (v_0,0).\label{eq: bc2}
\eeq

\begin{figure}[tbp]
    \centering
    \includegraphics[width=\textwidth]{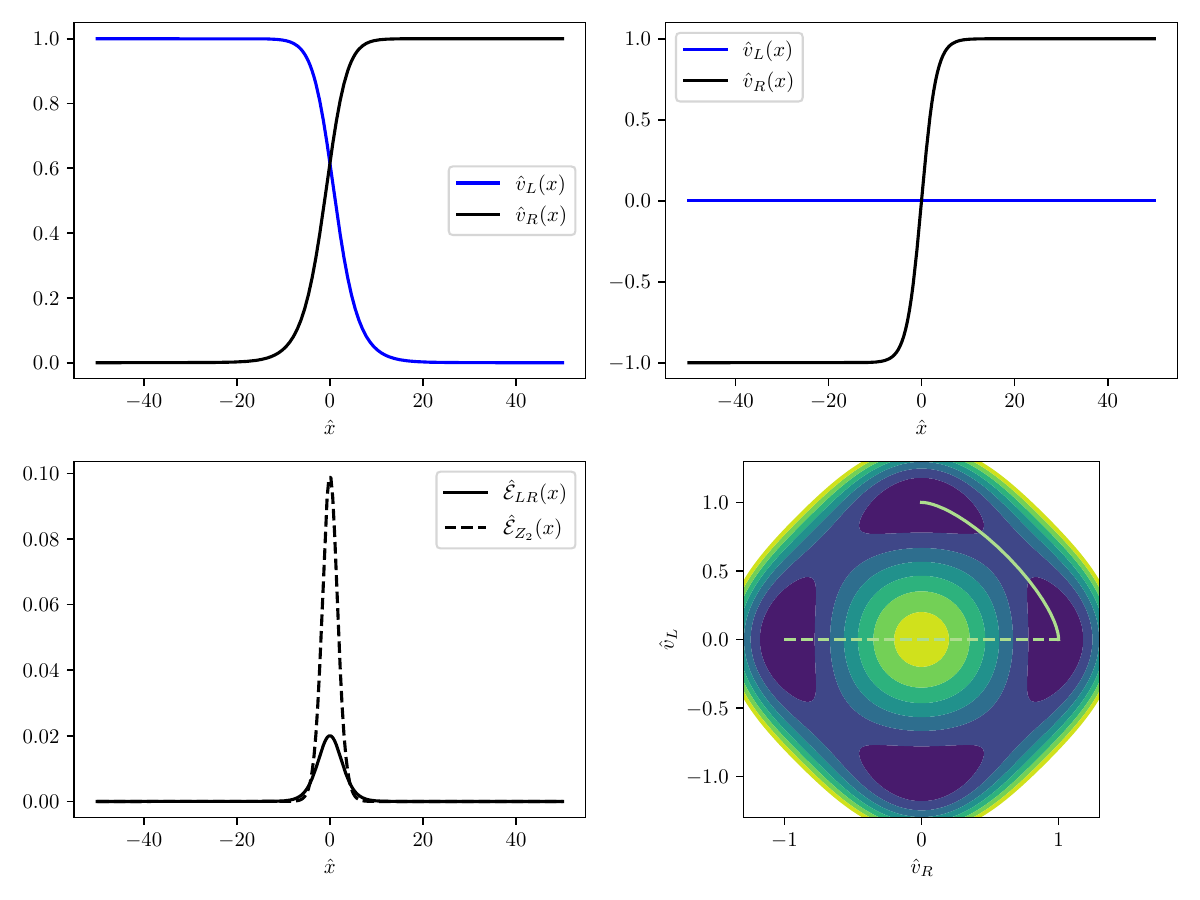}
    \caption{$LR$ (top-left panel) and $Z_2$ (top-right panel) DW profiles, for $\rho_1 = 0.2$, and $\rho_2 = 0.6$. The dimensionless energy density of the DWs is shown in the bottom-left panel. The bottom-right panel shows the $LR$ (solid line) and $Z_2$ (dashed line) DW profiles in field space.}
    \label{fig: DW_profile}
\end{figure}

\begin{figure}[tbp]
    \centering
    \includegraphics[width=\textwidth]{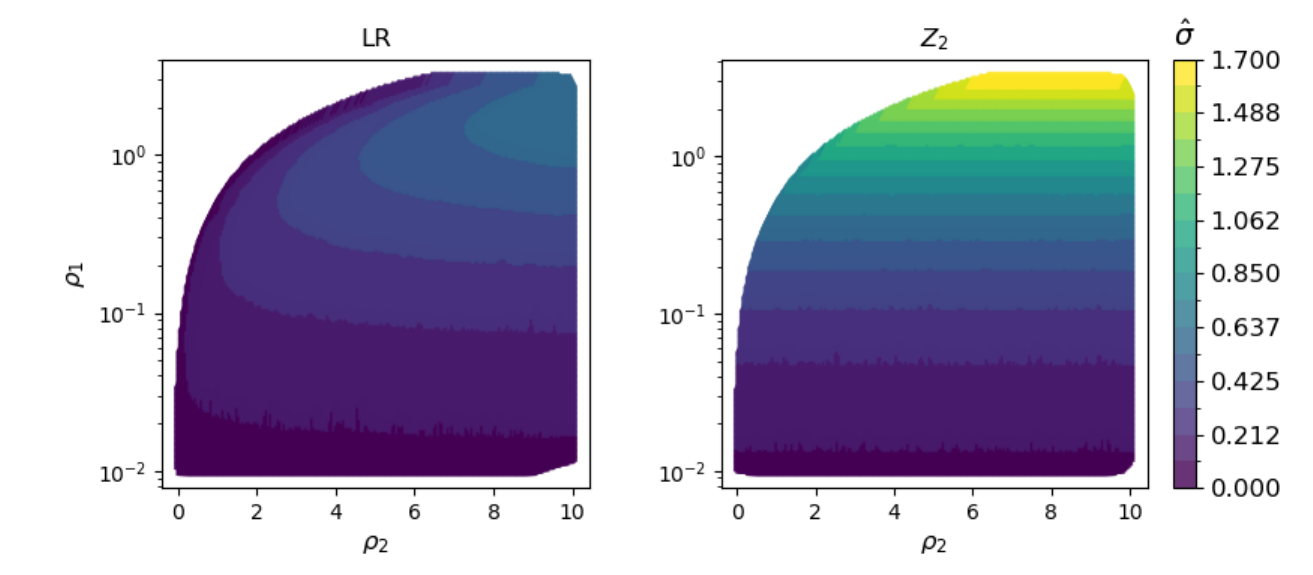}
    \caption{Parametric dependence of $\hat{\sigma}$ on $\rho_1$ and $\rho_2$ for $LR$ (left panel) and $Z_2$ (right panel) DWs. The surface tension of $Z_2$ DWs is greater than that of $LR$ DWs.}
    \label{fig: surf_ten}
\end{figure}

It is convenient to express the equations in terms of dimensionless quantities,
\beq
\hat{v}_L =  \frac{v_L}{v_0},~\hat{v}_R=  \frac{v_R}{v_0},~\hat{\mu}^2_3 =  \frac{\mu_3^2}{v_0^2} = \rho_1, ~\hat{x} = x~v_0,
\eeq
where the hatted variables are dimensionless. Similarly, we can define the dimensionless surface tension,
\beq
\hat{\s} = \frac{\s}{v_0^3} = \int_{-\infty}^{\infty} \hat{\mathcal{E}} d\hat{x},
\eeq
with $\hat{\mathcal{E}} = \frac{\mathcal{E}}{v_0^4}$ as the dimensionless energy density. We describe the procedure to obtain kink solutions in Appendix\,\ref{appendix: kink}.

In Fig.\,\ref{fig: DW_profile} we show the DW profiles for LR and $Z_2$ cases. For the chosen benchmark,
$$\hat{\s}_{LR} = 0.1363,~\hat{\s}_{Z_2} = 0.4216 \implies \frac{\hat{\s}_{Z_2}}{\hat{\s}_{LR}} \approx 3.1 \,.$$ 
The energy density of the two kinds of DWs differs significantly, as seen in the lower left panel. In field space, the $LR$ DW passes through one of the saddle points, while the $Z_2$ DW passes through the origin. Due to the greater energy density, $Z_2$ DWs are unstable and decay into $LR$ DWs, as discussed in Ref.\,\cite{Wu:2022tpe}. By adding a small shift to the initial guess of the $Z_2$ DW along the $v_L$ direction, we checked that the $Z_2$ DW solution converges into a hybrid structure of two $LR$ DWs. 

\begin{figure}[tbp]
    \centering
    \includegraphics[width=0.8\textwidth]{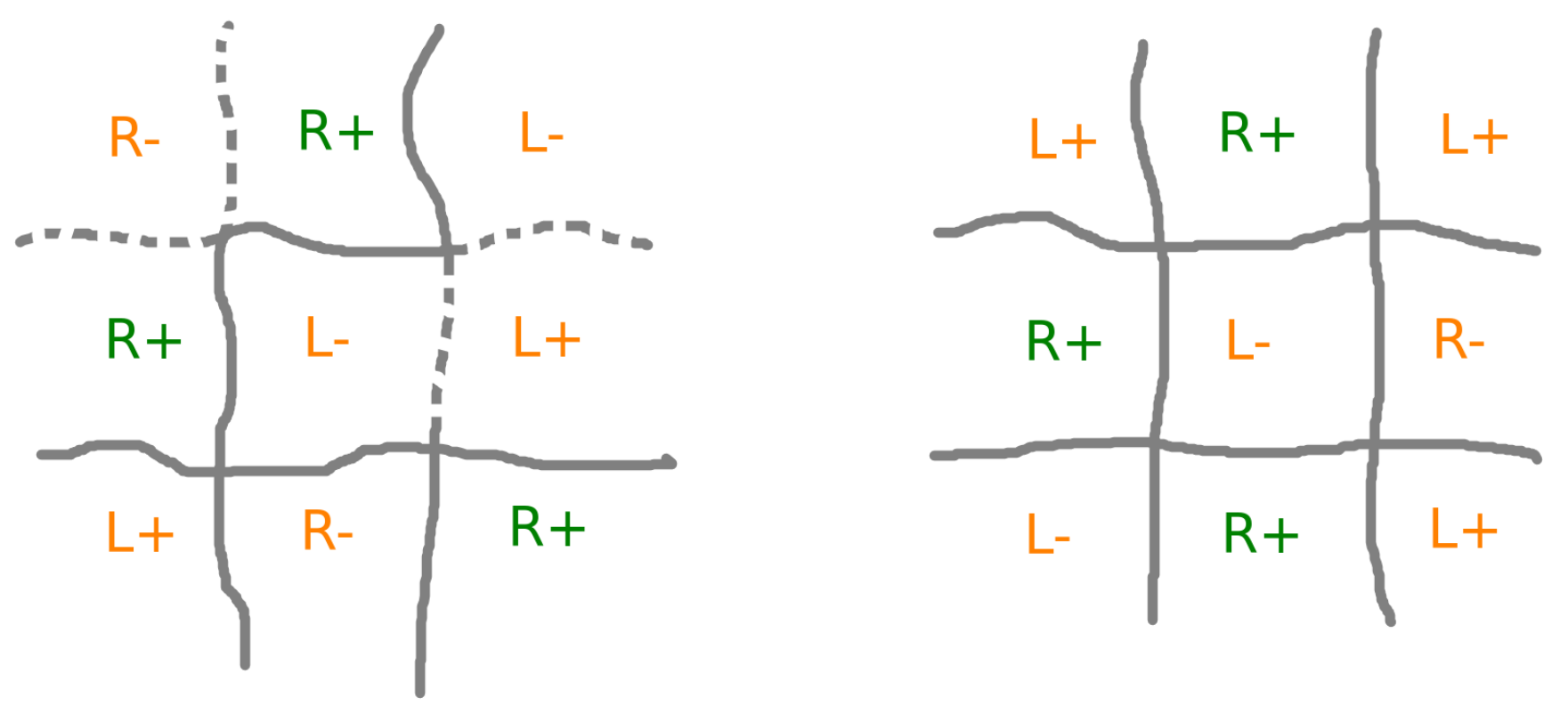}
    \caption{Left: Schematic diagram of a typical DW network showing LR (solid line) and $Z_2$ (dashed line) DWs. Right: In the stable configuration, the $Z_2$ DWs are absent and the network consists entirely of $LR$ DWs. The vacuum consistent with phenomenology, $R+$, is shown in green, while the rest are shown in orange.}
    \label{fig: DW_network}
\end{figure}

Fig.\,\ref{fig: surf_ten} shows the dependence of $\hat{\sigma}$ on the quartic couplings $\{\rho_1,\rho_2\}$, for $LR$ and $Z_2$ DWs. The surface tension of $LR$ DWs is higher for larger values of $\rho_1$ and $\rho_2$, while it is almost independent of $\rho_2$ for $Z_2$ DWs. In both cases, the overall dependence on $\rho_1$ and $\rho_2$ is weak, since the variation in $\hat{\sigma}$ in the $\rho_1-\rho_2$ plane is within an order of magnitude. $Z_2$ DWs have greater energy density than $LR$ DWs in the entire $\rho_1-\rho_2 $ plane. When $\hat{\s}_{Z_2}>2\hat{\s}_{LR}$, a $Z_2$ DW can split into two $LR$ DWs of equal area connecting adjacent minima \,\cite{Wu:2022tpe}. On the other hand if $\hat{\s}_{Z_2}<2\hat{\s}_{LR}$, then a $Z_2$ wall can split into two $LR$ DWs of smaller surface area. The most stable configuration of the DW network consists only of LR DWs after the $Z_2$ DWs have decayed away, as depicted in the right panel of Fig.\,\ref{fig: DW_network}. This explains why it is enough to focus on $LR$ DWs. This discussion applies equally to TLRSM, where the potential also obeys a $Z_4$ symmetry and has a similar vacuum structure.

\section{Gravitational waves from domain walls}\label{sec: GW}
Once formed, the DWs quickly reach a scaling regime in the absence of friction, with $\mathcal{O}(1)$ DWs per Hubble volume moving at relativistic speeds. The energy density in the scaling regime is given by\,\cite{Saikawa:2017hiv},
\beq\label{eq: rho_DW}
\rho_{\rm{DW}} = \mathcal{A} \sigma H,
\eeq
where $\mathcal{A}\sim 0.8$ is a numerical factor obtained from simulations and $H$ is the Hubble parameter. 

At time $t$, $\rho_{\rm{DW}}\propto H(t) \propto 1/t$, while the energy density of matter and radiation falls faster, implying that DWs dominate the energy density of the universe at late times. By introducing a small bias term in the potential, $V_{\rm{bias}}$, via explicit $\mathcal{P}$-breaking operators, we can lift the degeneracy of the four vacuua in such a way that the $R+$ vacuum is favored. This creates a pressure difference across the DWs, causing the domains with the preferred vacuum to grow in size. Due to the bias, the DWs eventually begin to annihilate at a temperature $T_{\rm{ann}}$, defined by the condition, $\rho_{\rm{DW}}(T_{\rm{ann}})\sim V_{\rm{bias}}$. Assuming radiation-domination\,\cite{Ferreira:2022zzo},
\beq\label{eq: Tstar}
T_{\rm{ann}} \simeq \frac{5~{\rm{MeV}}}{\sqrt{\mathcal{A}}}\left(\frac{10.75}{g_*(T_{\rm{ann}})}\right)^{\frac{1}{4}}\left(\frac{V_{\rm{bias}}^{1/4}}{10~{\rm{MeV}}}\right)^2\left(\frac{10^5~{\rm{GeV}}}{\sigma^{1/3}}\right)^{\frac{3}{2}}.
\eeq
where $g_*(T_{\rm{ann}})$ is the number of relativistic degrees of freedom at $T_{\rm{ann}}$. To accommodate cosmological constraints, DWs must annihilate before the BBN epoch, i.e. $T_{\rm{ann}}>T_{\rm{BBN}}\sim 1$\,MeV. This yields a lower bound on $V_{\rm{bias}}$,
\beq
V_{\rm{bias}} > \frac{\mathcal{A}}{25}(10~{\rm{MeV}})^4 \left(\frac{\sigma^{1/3}}{10^5~{\rm{GeV}}}\right)^3.
\eeq

Similarly, an upper bound on $V_{\rm{bias}}$ is obtained by requiring that domains of the preferred vacuum must not percolate\,\cite{Saikawa:2017hiv},
\beq\label{eq: percolation}
\frac{V_{\rm{bias}}}{V_0} > \ln\left(\frac{1-p_c}{p_c}\right),
\eeq
where $p_c=0.311$ is the critical value above which the favored vacuum percolates and $V_0$ is the height of the barrier separating the minima. For $LR$ DWs, 
\beq
V_0 = \frac{1}{4}v_0^4\left(\frac{\rho_2-2\rho_1}{\rho_2+2\rho_1}\right)\approx \frac{1}{4}v_0^4,
\eeq
where the last approximation holds when $\rho_2\gg\rho_1$. The condition of Eq.\,\eqref{eq: percolation} is easily satisfied since we are interested in $V_{\rm{bias}}\ll v_0^4$. 

The bias term can be generated by introducing operators that explicitly break the $\mathcal{P}$ symmetry. Since quantum gravity effects destroy global symmetries, Planck scale-suppressed higher-dimensional operators provide an elegant way to generate the bias term. Indeed, this possibility has been considered in the literature\,\cite{Lew:1993yt,Hiramatsu:2010yz,Borah:2022wdy,Barman:2023fad}. In this chapter, we do not assume any particular origin of the bias term and keep $V_{\rm{bias}}$ as a free parameter.

A dimensionless parameter, $\a_*$, captures the DW energy density at annihilation, 
\bea
\a_* &=& \frac{\rho_{\rm{DW}}(T_{\rm{ann}})}{\rho_{\rm{rad}}(T_{\rm{ann}})},\nn\\
\rho_{\rm{rad}}(T) &=& \frac{\pi^2}{30}g_*(T)T^4,
\eea
%where $g_*(T)$ is the number of relativistic degrees of freedom at temperature $T$. 
Using Eq.\,(\ref{eq: rho_DW}), we get\,\cite{Ferreira:2022zzo},
\beq\label{eq: alpha_star}
\alpha_* \simeq \mathcal{A}\sqrt{\frac{g_*(T_{\rm{ann}})}{10.75}}\left(\frac{\sigma^{1/3}}{10^5~{\rm{GeV}}}\right)^3\left(\frac{{10~{\rm{MeV}}}}{T_{\rm{ann}}}\right)^2 . 
\eeq
We consider the scenario where the DWs decay into standard model particles, in which case, BBN restricts $T_{\rm{ann}}\gtrsim 2.7$\,MeV \cite{Jedamzik:2006xz,PhysRevD.105.095015}. We also impose $\alpha_*<0.3$ to ensure no deviation from radiation domination\,\cite{Ferreira:2022zzo}. 
 
The relic GW spectrum is defined as, 
\beq
h^2\Omega_{\rm{GW}} = \frac{h^2}{\rho_c}\frac{\partial \rho_{\rm{GW}}}{\partial \ln f},
\eeq
where $\rho_c$ is the critical density, given by
\beq
\rho_c = \frac{3H_0^2}{8\pi G},
\eeq
and $H_0 = 100\,h~{\rm{km \,s^{-1} Mpc^{-1}}}$ is the present-day Hubble constant with $h=0.6736\pm 0.0054$\,\cite{Planck:2018vyg}, and $G$ is  Newton's gravitational constant.

GWs are produced due to DW surface oscillations\,\cite{Vilenkin:1981zs,Preskill:1991kd,PhysRevD.59.023505}, with dominant emission happening at $T=T_{\rm{ann}}$. Assuming all the GWs are produced at $T_{\rm{ann}}$, the GW spectrum is given by\,\cite{Ferreira:2022zzo,Hiramatsu:2013qaa}, 
\beq
h^2\Omega^{\rm{DW}}_{\rm{GW}}(f) \simeq 10^{-10}~ \tilde{\epsilon}_{\rm{GW}}\left(\frac{10.75}{g_*(T_{\rm{ann}})}\right)^{\frac{1}{3}}\left(\frac{\a_*}{0.01}\right)^2 S\left(\frac{f}{f_p^0}\right),
\eeq
where $\tilde{\epsilon}_{\rm{GW}} = 0.7\pm 0.4$. The peak frequency $f_p^0$ is given by,
\beq
f_p^0 \simeq 10^{-9}~{\rm{Hz}}~\left(\frac{g_*(T_{\rm{ann}})}{10.75}\right)^{\frac{1}{6}}\left(\frac{T_{\rm{ann}}}{10~{\rm{MeV}}}\right),
\eeq
and the shape function of the GW spectrum $S$ has the form,
\beq
S(x) = \frac{(\gamma+\b)^{\delta}}{\big(\b x^{-\frac{\gamma}{\delta}} + \gamma x^{\frac{\b}{\delta}}\big)^{\delta}}.
\eeq
We set $\gamma = 3$ from causality\,\cite{NANOGrav:2023hvm}, while numerical analyses determine $\delta,\beta\simeq 1$. Following\,\cite{Hiramatsu:2013qaa}, we set $\delta,\beta = 1.$ Note that, $$h^2\Omega^{\rm{DW}}_{\rm{GW}}(f_p^0)\propto \a_*^2\propto\frac{\sigma^2}{V_{\rm{bias}}}\propto \frac{v_0^6}{V_{\rm{bias}}},$$ indicating a strong dependence of the amplitude on the scale $v_0$. 

After the annihilation of DWs, the GW production stops, and $\rho_{\rm{GW}}$ redshifts like SM radiation, contributing to the number of relativistic degrees of freedom, $g_*(T)$. Around BBN temperatures, $T\simeq \mathcal{O}$(MeV), this extra contribution from GWs can be restricted by considering the limits on $\Delta N_{\rm{eff}}$ from CMB and BBN, where $\Delta N_{\rm{eff}} = N_{\nu}-3$, and $N_{\nu}$ is the effective number of light neutrino species at BBN. The upper bound on the GW amplitude is\,\cite{Maggiore:1999vm,Caprini:2018mtu}, 
\beq
h^2\Omega_{\rm{GW}}\lesssim 5.6\times 10^{-6}\Delta N_{\rm{eff}}.
\eeq
The existing limit on $\Delta N_{\rm{eff}}$ from Planck is, $\Delta N_{\rm{eff}}\lesssim 0.28$ at $95\%$ confidence level. Upcoming CMB experiments will be able to probe smaller values of $\Delta N_{\rm{eff}}$\,\cite{Planck:2018vyg}.

\section{Results}\label{sec: results}
\subsection{MCMC analysis}
We use the NG15 data to carry out the Bayesian analysis. Given PTA data $\mathcal{D}$, a hypothesis $\mathcal{H}$, and parameters $\Theta$, we use the posterior distribution $P (\Theta|\mathcal{D},\mathcal{H})$, reconstructed from MCMC analysis, to identify best-fit parameter ranges and set upper limits on them. We consider two hypotheses: (i) $\mathcal{H}_1$: the DLRSM DW model assuming the SMBH background is negligible, and (ii) $\mathcal{H}_2$: the DLRSM DW$+$SMBHB model, where the GW contribution from DLRSM DWs is combined with the contribution from SMBHBs. The DLRSM parameters of interest are,
$$\Theta_1\equiv\{v_0,\rho_1,\rho_2,V_{\rm{bias}}\}.$$

For $\mathcal{H}_1$, we first obtain the functional dependence of $\hat{\sigma}(\rho_1,\rho_2)$ by interpolating the numerical values shown in Fig.\,\ref{fig: surf_ten}, so that the surface tension is obtained as, $\sigma(v_0,\rho_1,\rho_2) = \hat{\sigma}(\rho_1,\rho_2) v_0^3$. Next, we calculate $T_{\rm{ann}}$  according to Eq.\,\eqref{eq: Tstar}, while $\a_*$ is calculated using Eq.\,\eqref{eq: alpha_star}. The constraints $T_{\rm{ann}}\gtrsim 2.7$\,MeV and $\a_*<0.3$,  are imposed as discussed in the previous section. Thus, we obtain the GW spectrum in terms of DLRSM parameters
$$\mathcal{H}_1:~~~ h^2\Omega^{\rm{DW}}_{\rm{GW}}(f;v_0,\rho_1,\rho_2,V_{\rm{bias}}).$$

For $\mathcal{H}_2$, we superimpose the contribution of SMBHBs with the DLRSM DW contribution. For low frequencies, $f\ll 1\,\rm{year}^{-1}$, the SMBHB spectrum is given by a simple power law\,\cite{1995ApJ...446..543R,Jaffe_2003,Wyithe_2003,Sesana_2004,Burke-Spolaor2019}, 
\beq\label{eq: bhb_gw}
h^2\Omega^{\rm{BHB}}_{\rm{GW}}(f) = \frac{2\pi^2h^2 A^2_{\rm{BHB}}}{3H_0^2} \left(\frac{f}{f_{\rm{yr}}}\right)^{5-\gamma_{\rm{BHB}}} f_{\rm{yr}}^2,
\eeq
where $f_{\rm{yr}} = 1\, \rm{year}^{-1} = 3.17\times 10^{-8}$\,Hz. The spectrum falls off rapidly at larger frequencies, $f\gg 1\,\rm{year}^{-1}$. If the orbital evolution of the binaries is purely driven by GW emission, the parameter $\gamma_{\rm{BHB}}$ takes the value $ \gamma_{\rm{BHB}} = 13/3$. To account for environmental effects $\gamma_{\rm{BHB}}$ is taken as a free parameter, along with $A_{\rm{BHB}}$, so that the parameter set is,
$$\Theta_2\equiv\{v_0,\rho_1,\rho_2,V_{\rm{bias}},A_{\rm{BHB}},\gamma_{\rm{BHB}}\}.$$
The GW signal is hypothesised as,
$$\mathcal{H}_2:~~~ h^2\Omega^{\rm{DW}}_{\rm{GW}}(f;v_0,\rho_1,\rho_2,V_{\rm{bias}}) + h^2\Omega^{\rm{BHB}}_{\rm{GW}}(f;A_{\rm{BHB}},\gamma_{\rm{bhb}}).$$

\begin{figure}[tbp]
    \centering
    \includegraphics[width=.8\textwidth]{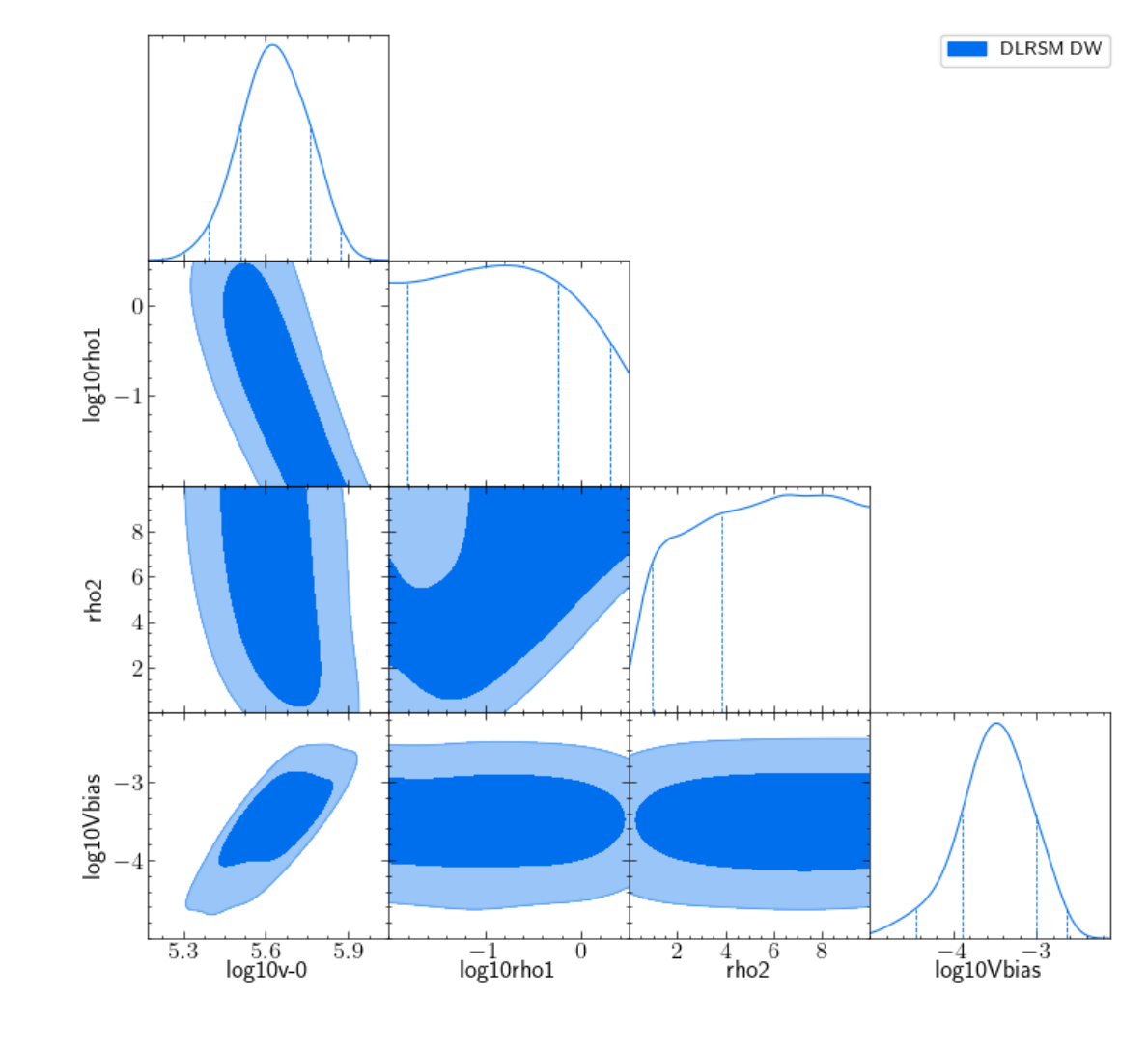}
    \caption{The posterior probability distribution of the DLRSM DW fit to the NG15 data.\label{fig: MCMC}}
\end{figure}

\begin{figure}[tbp]
    \centering
    \includegraphics[width=\textwidth]{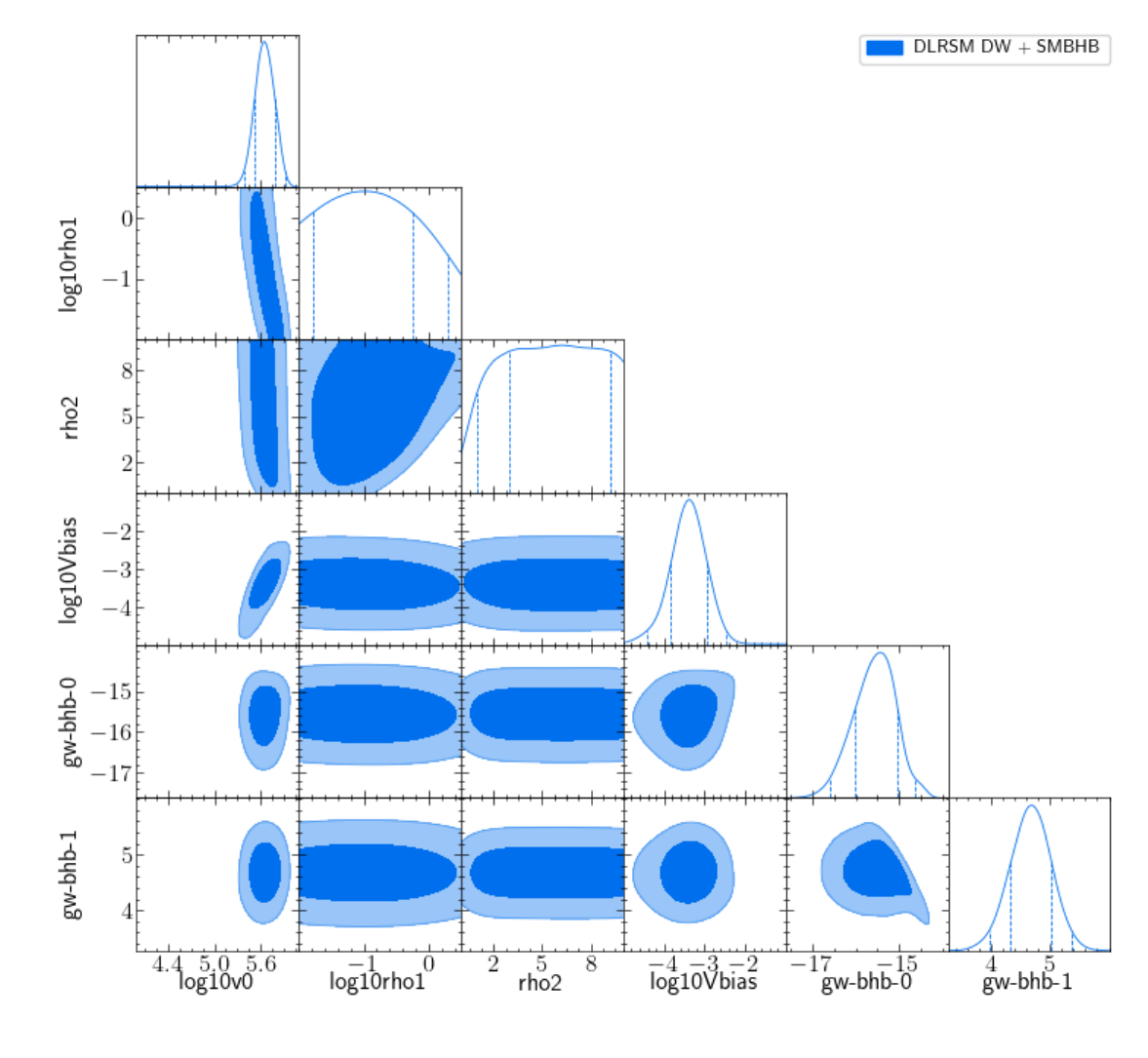}
    \caption{Same as fig.\,\ref{fig: MCMC}, but also including the contribution from SMBHB.}
    \label{fig: MCMC_SMBHB}
\end{figure}

The Bayesian analysis is implemented using the \texttt{PTArcade} package\,\cite{andrea:mitridate_2023,Mitridate:2023oar}, which is a wrapper for the \texttt{ENTERPRISE} code\,\cite{ellis_2020_4059815}, and incorporates PTA data. For the DLRSM DW model, we sample the parameters $v_0,~V_{\rm{bias}}$ and $\rho_1$ from a log-uniform distribution, and $\rho_2$ from a uniform distribution. The parameter ranges for the priors are given in the second column of Table\,\ref{table: priors}. The SMBHB parameters ($\log_{10}A_{\rm{BHB}},\gamma_{\rm{BHB}}$) are sampled from a normal bivariate distribution taken from Ref.\,\cite{NANOGrav:2023hvm}.

The posteriors for DLRSM DWs are shown in Fig.\,\ref{fig: MCMC}. The $\mathcal{P}$-breaking scale $v_0$ and the bias term $V_{\rm{bias}}$ are constrained in a narrow range, as indicated by the closed contours for the $68\%$ and $95\%$ credible intervals. On the other hand, there is a weak dependence on the quartic couplings $\rho_1$ and $\rho_2$. Note also that $\rho_1\ll\rho_2$ in the $68\%$ and $95\%$ credible regions. A positive correlation is observed between $v_0$ and $V_{\rm{bias}}$, which is because, for a given scale $v_0$, we can always find an appropriate value of $V_{\rm{bias}}$, which best explains the data. In Fig.\,\ref{fig: MCMC_SMBHB}, we show the posterior distribution for the DLRSM DW$+$SMBHB model, with the labels gw-bhb-0 and gw-bhb-1 denoting the SMBHB parameters $\log_{10}A_{\rm{BHB}}$ and $\gamma_{\rm{BHB}}$ respectively. The DLRSM parameters follow a similar distribution as in the previous case, and the other two parameters take the maximum posterior values $(\log_{10}A_{\rm{BHB}}, \gamma_{\rm{BHB}}) = (-15.44,4.69)$. The predicted spectral index, $\gamma_{\rm{BHB}} = 13/3$, lies just outside the $68\%$ credible interval. The likelihood ratio of the DLRSM DW relative to the the DLRSM DW$+$SMBHB model corresponds to, $-2 \Delta\ln l_{\rm{max}} = -1.2$, indicating that the pure DLRSM DW model is slightly favored.

Table\,\ref{table: priors} summarizes the prior ranges, maximum posteriors, and $68\%$ credible intervals for all the parameters in the two scenarios. In particular, the maximum posterior values of $v_0$ and $V_{\rm{bias}}$ for DLRSM DW model are,
\beq
v_0 = 4.36\times 10^5\,\rm{GeV}, ~ V_{\rm{bias}} = 3.31\times 10^{-4}\,\rm{GeV}^4.
\eeq 

\begin{figure}[tbp]
    \centering
    \includegraphics[width=\textwidth]{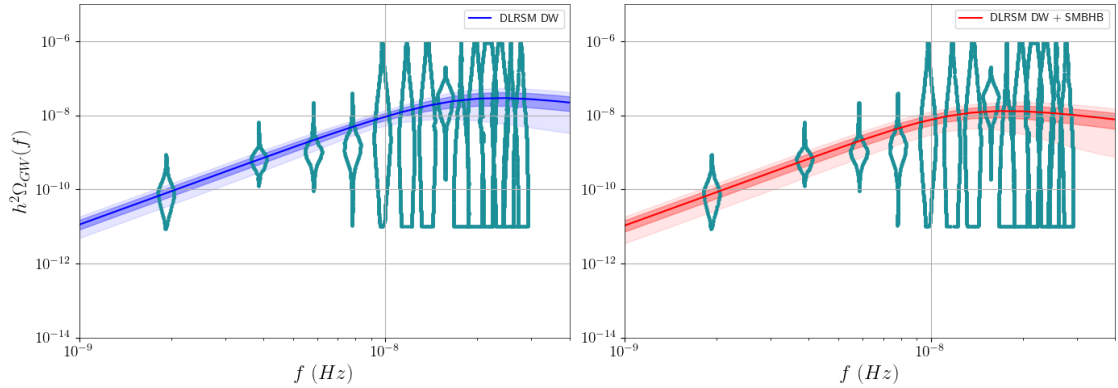}
    \caption{Median GW spectra for DLRSM DWs, along with their $68\%$ and $95\%$ posterior envelopes.}
    \label{fig: GW spectrum}
\end{figure}

\begin{table}[tbp]
\begin{center}
%\vspace{0.5 cm}
\def\arraystretch{1.5}
\setlength{\tabcolsep}{8pt}
\resizebox{\textwidth}{!}{
\begin{tabular}{c c c c c c}
\hline
\multicolumn{1}{c}{\multirow{2}{*}{\textbf{Parameter}}} & \multicolumn{1}{c}{\multirow{2}{*}{\textbf{Uniform prior range}}} & \multicolumn{2}{c}{\multirow{2}{*}{\textbf{Maximum Posterior}}} & \multicolumn{2}{c}{\multirow{2}{*}{\textbf{$68\%$ credible interval}}}  \\ 
\multicolumn{1}{c}{}  & & $\mathcal{H}_1$ & $\mathcal{H}_2$  & $\mathcal{H}_1$ & $\mathcal{H}_2$    \\ \hline
%\hline
%\hline
$\log_{10}~v_0/\rm{GeV}$ &  $[4,8]$ & $5.63$ & $5.66$&  $[5.52,5.77]$ & $[5.52,5.79]$ \\
$\log_{10}~V_{\rm{bias}}/\rm{GeV}^4$ & $[-5,-1]$ & $-3.47$ & $-3.37$ & $[-3.90,-3.05]$ & $[-3.82,-2.91]$\\  
$\log_{10}\rho_1 $    & $[-2.5,0.5]$ & $-0.77$ & $-1.00$ & $[-1.90,-0.35]$ & $[-1.76,-0.23]$\\  
$\rho_2 $    & $[0,10]$ & $6.55$ & $6.50$ & $[2.88,9.16]$ & $[3.03,9.23]$\\   
$\log_{10}A_{\rm{BHB}}$ & - & - & $-15.44$ & - & $[-15.97,-15.02]$\\
$\gamma_{\rm{BHB}}$     & - & - & $4.69$  & - & $[4.35,5.04]$\\
\hline
\end{tabular}}
\end{center}
\caption{\label{table: priors} Priors, along with the maximum posterior values and $68\%$ credible intervals for the parameters, for the two hypotheses considered.
}
\end{table}

\subsection{Detection prospects}
We now discuss the prospects of detecting the GW signal using the best-fit parameter values at upcoming GW observatories. In Fig.\,\ref{fig: GW spectrum} we show the median GW spectra fitted to the NG15 data along with $1\sigma$ and $2\sigma$ confidence intervals, for the DLRSM DW model (left panel), and DLRSM DW combined with SMBHB (right panel). The green violins correspond to the posterior distribution of $\Omega_{\rm{GW}}$ in 14 frequency bins, predicted by the NG15 data\,\cite{NANOGrav:2023gor}. With more data, the posteriors would narrow down, enabling a more precise parameter estimation in the future. 

Fig.\,\ref{fig: spectra}  shows the median GW spectra for the two models for a wider range of frequencies, with the power-law integrated sensitivity curves (PLISCs)\,\cite{Thrane:2013oya} of upcoming GW detectors, SKA\,\cite{Weltman:2018zrl}, $\mu$Ares\,\cite{Sesana:2019vho}, LISA\,\cite{LISA:2017pwj}, BBO\,\cite{Corbin:2005ny}, FP-DECIGO\,\cite{Seto:2001qf}, CE\,\cite{LIGOScientific:2016wof}, and ET\,\cite{Punturo:2010zz}. The shaded grey region shows the region excluded by the $\Delta N_{\rm{eff}}$ bound coming from Planck data. Except $\mu$Ares, the PLISCs of all upcoming detectors are taken from Ref.\,\cite{schmitz_2020_3689582}, using the threshold signal-to-noise ratio (SNR) of 1, and time of observation, $\tau=1$\,year, while $\tau=20$\, years for SKA. The PLISC for $\mu$Ares is calculated for a threshold SNR of 10, and $\tau=7$\,years. If the GW spectrum results from the DLRSM DW model, considered with or without SMBHBs, it would be observed by future observatories, particularly $\mu$Ares, LISA, FP-DECIGO, and BBO, with a high SNR. The signal will also be confirmed by upcoming PTAs such as SKA. If the $\mathcal{P}$-breaking PT is first-order, one would observe a double-peaked GW spectrum, with the additional higher frequency peak coming from FOPT, as discussed in Ref.\,\cite{Karmakar:2023ixo}. For $v_0\gtrsim 10^5$\,GeV, the peak from FOPT would be detected by CE and ET. 

In DLRSM, the extra heavy degrees of freedom including neutral $CP$-even scalars $H_{1,2,3}$, $CP$-odd scalars $A_{1,2}$, charged scalars $H_{1,2}^{\pm}$ and gauge bosons $Z_2, W_2^{\pm}$, all have $\mathcal{O}(v_0)$ masses. Due to the high parity-breaking scale favored by the NG15, $v_0\gtrsim 10^5$\,GeV, the prospects of detecting these heavy particles at upcoming collider experiments are weak.

\section{Discussion}\label{sec: discussion}
After the 15-year NANOGrav dataset analysis reported convincing evidence of a low-frequency GW background, several works have compared the possible models to explain it. In addition to the standard interpretation of GWs produced by SMBHBs, many cosmological models have been proposed. One such model is the DW model, in which the GW background is due to a network of DWs in the early universe.  The formation of DWs requires the existence of a discrete symmetry that is spontaneously broken in a phase transition. In LRSMs, the discrete parity symmetry can be elegantly incorporated to explain the parity violation observed in SM via spontaneous symmetry breaking. The parity-breaking PT in LRSM can give rise to DWs. Since LRSMs typically require the scale of parity-breaking, $v_0$, to be large compared to the EW scale, the DW surface tension, $\sigma\propto v_0^3$, can be made large enough to explain the NG15 result. In this work, we considered the DWs of the parity symmetric DLRSM as the source of the NG15 signal.

Earlier discussions on DWs in LRSM mostly focused on TLRSM and reported the scale of $\mathcal{P}$-breaking required to explain the PTA signal using benchmarks obtained from order of magnitude estimates. In this chapter, for DLRSM, we found the explicit parameter dependence of the surface tension and carried out a Bayesian analysis to constrain model parameters. Due to the $Z_4$ symmetry of the DLRSM potential, two kinds of DWs, i.e., $LR$ and $Z_2$ DWs, are formed. The surface tension of the $Z_2$ walls is higher than that of $LR$ DWs. Earlier works on DWs in LRSM mainly focused on $LR$ DWs and did not discuss the fate of $Z_2$ DWs in detail. We argued that the $Z_2$ DWs are unstable and decay into $LR$ DWs, thus providing a rationale for considering only $LR$ DWs. The arguments presented here can also be applied to TLRSM. The DW surface tension depends weakly on the quartic couplings $\rho_1,\rho_2$, and the primary dependence is on $v_0$. The DW network must annihilate before the epoch of BBN to respect cosmological constraints, which can be achieved via explicit parity-breaking terms in the potential, resulting in a bias, $V_{\rm{bias}}$. While such operators can be motivated by quantum gravity and grand unified theories, we considered $V_{\rm{bias}}$ as a free parameter. 

\begin{figure}[tbp]
    \centering
    \includegraphics[width=0.8\textwidth]{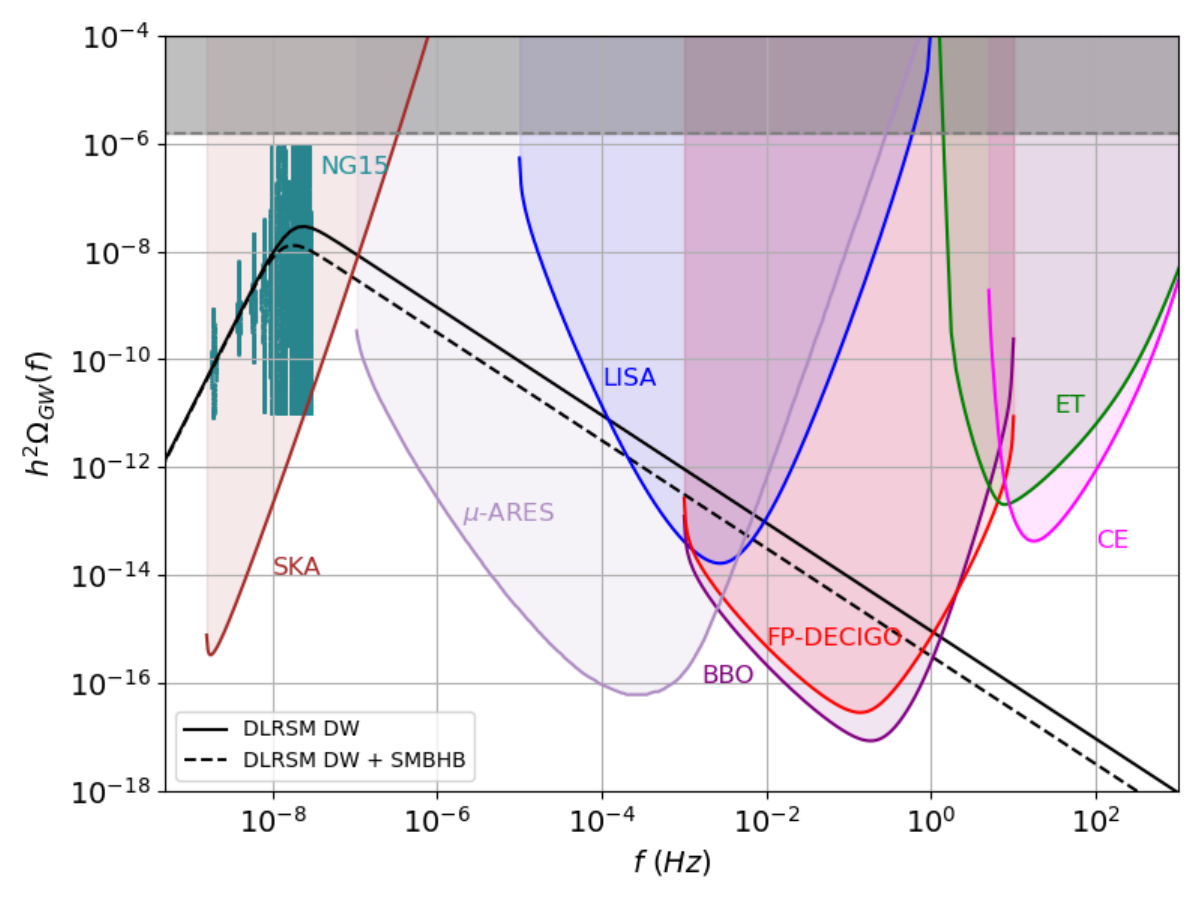}
    \caption{Median GW spectrum for DLRSM DW (solid black) and DLRSM DWs$+$SMBHB model (dashed black) along with sensitivity curves for upcoming GW detectors. The shaded grey region is excluded by the Planck $\Delta N_{\rm{eff}}$ bound.}
    \label{fig: spectra}
\end{figure}

For the Bayesian analysis, we considered the case where the NG15 signal is entirely explained by DLRSM DWs and the case where the SMBHB contribution is also included. The maximum posterior values of the parameters and the $68\%$ credible intervals are summarized in Table\,\ref{table: priors}. The strong dependence of the peak amplitude of the GW spectrum on $\sigma$, and $V_{\rm{bias}}$, i.e. $\Omega^{DW}_{\rm{GW}}(f_{\rm{peak}})\propto\sigma^2/V_{\rm{bias}}$, results in a tight constraint on $v_0$ and $V_{\rm{bias}}$, as reflected by the respective $68\%$ credible intervals. On the other hand, the quartic couplings $\rho_1$ and $\rho_2$ are less constrained. For the DLRSM DW$+$DW case, the spectral index $\gamma_{\rm{BHB}}$ lies outside the $68\%$ credible interval, indicating a tension with the NG15 results. 

The median GW spectra are presented in Fig.\,\ref{fig: GW spectrum} and Fig.\,\ref{fig: spectra}, which show a good agreement with the violins of NG15 data. If the signal is due to DWs, future GW observatories such as $\mu$Ares, LISA, BBO, and FP-DECIGO would also observe a GW background at higher frequencies. Moreover, if the parity-breaking PT is first-order, it would give rise to a double-peaked spectrum, with the higher frequency peak observable by ET and CE. 

In this chapter, we have neglected the effect of friction on the DWs from the thermal plasma, which could dampen the GW signal. Since the couplings of the fields constituting the wall with the SM fields are small, the friction is expected to be negligible. In addition, the GW spectrum from SMBHBs is modeled assuming the binaries lose their energy entirely from GW production. Since current observations and numerical simulations have a large uncertainty in the spectral shape, the power-law given in Eq.\,\eqref{eq: bhb_gw} serves as a reasonable approximation\,\cite{NANOGrav:2023hvm}.
 %DLRSM DWs
\chapter{Summary and conclusions}
\label{chap:Conclusions}
\linespread{0.1}
\pagestyle{headings}
\noindent\hrule height 0.5mm 
\vspace{5mm}

The SM is one of the most successful theories in physics, describing the nature and interactions of elementary particles. 
Adding to its long list of achievements, the discovery of the Higgs boson at the LHC in 2012 was the final piece in the establishment of the SM, and was predicted about fifty years earlier. Despite its tremendous success, the SM is faced with many challenges that hint at possible new physics, including the existence of small neutrino masses, the particle nature of dark matter, the origin of a non-zero baryon asymmetry, the EW hierarchy problem, the SM flavor puzzle, etc. Many of the BSM scenarios that address these issues involve high energy scales not accessible to present-day colliders such as the LHC, which operates at $\sqrt{s}=14$\,TeV. Although the next generation of colliders, such as the ILC and the FCC, will be able to achieve relatively higher energies, they may not be able to test BSM theories beyond a scale of $\mathcal{O}(10)\,$TeV. Furthermore, the timeline for these colliders is uncertain, and they may take several decades to become operational. 

The revolutionary detection of GWs from a binary black hole merger in 2015 by LIGO marked the beginning of GW astronomy. Unlike the CMB, which screens the photons before recombination, GWs can reach us from the earliest epochs of the universe. Planned GW ground and space-based observatories, and PTAs will significantly enhance our capability to detect feeble GW signals in the coming decades. In addition to momentary foreground signals such as those from binary black hole and binary neutron star mergers, these planned observatories present good prospects for detecting a stochastic GW background across a wide range of frequencies, from $\sim 10^{-9}$ Hz to $\sim 10^4$ Hz. Among various possibilities, special types of cosmological PTs are important sources of a stochastic GW background. In particular, FOPTs and topological defects like DWs and CSs produce a GW imprint that carries information about the PT. The future GW observatories will be able to probe the scale of the PT ranging from $\sim 100\,$MeV to $\sim 10^7\,$GeV, enabling us to constrain the underlying particle physics model far beyond the reach of colliders. 

In this thesis, we analyzed the nature of PTs for a few well-motivated BSM scenarios using the one-loop finite temperature effective potential and determined the conditions under which they can lead to strong FOPTs or the formation of DWs. We found the appropriate parameter space where the resulting GW imprint is strong enough to be observed at future GW observatories. To benefit from the ability of GWs to probe high-scale physics, the BSM models were chosen to be at scales of $\mathcal{O}(10)\,$TeV or higher. Whenever possible, we used collider and cosmological constraints to complement the GW constraints on the model parameter space. 

In Chapter\,\ref{chapter2}, we studied FOPTs from $U(1)_{\fn}$-breaking in two UV completions of the FN mechanism in the quark sector. Collider constraints put a weak lower bound of a few TeVs on the cutoff scale or the VLQ mass scale, but GWs from FOPT would allow us to probe the cutoff from $10^4\,$GeV to $10^7\,$GeV at upcoming GW observatories. We found that the parameter space can be constrained by BBO and DECIGO for $v_s=10^5$\,GeV, while CE and ET can constrain the parameter space for $v_s = 10^7$\,GeV. Although the parameter space with a detectable GW background requires $\mathcal{O}(1)$ values of the Higgs-flavon quartic coupling, it was shown to be compatible with the observed Higgs mass and mixing constraints. We also found that when the two models are considered at the same scale, the GW spectrum in the two models is similar, and does not discriminate between them. It would be interesting to find complementary methods so that two models at the same scale can be discriminated. 

In Chapter\,\ref{chapter3}, we explored the possibility of 
FOPT in DLRSM during $SU(2)_R\times U(1)_{B-L}$ breaking. The breaking scale was taken to be $v_R=20,\,30,\,50$\,TeV, and the parameter space was scanned to determine the GW strength from FOPT. For these values of $v_R$, the GW background can be observed at BBO, FP-DECIGO, and Ultimate DECIGO, in certain regions of parameter space. We found that a strong GW background prefers a light CP-even scalar $H_3$, which can be probed by 
upcoming colliders such as ILC and FCC. Interestingly, in contrast to the more popular version of LRSM called TLRSM, where the ratio $w=v_L/\k_1 \approx 0$, in DLRSM, the detectable points show a preference for $w\sim\mathcal{O}(1)$. Finally, it was seen that a large number of points with detectable GW signals can be ruled out from the precise measurement of the trilinear Higgs coupling at future colliders.

In Chapter\,\ref{chapter4}, we studied the formation and evolution of DWs due to the breaking of the discrete $\mathcal{P}$-symmetry imposed on DLRSM. The scale of $\mathcal{P}$-breaking coincides with the scale of $SU(2)_R\times U(1)_{B-L}$-breaking considered in Chapter\,\ref{chapter3}. We saw that there are two types of DWs, namely $LR$ and $Z_2$ DWs. By explicitly solving the kinks solutions for the two types of DWs, we showed that the $Z_2$ DWs are unstable due to higher surface tension and decay into $LR$ DWs, thus providing a rationale for considering only $LR$ DWs. Considering the GW signal from the DLRSM DW model with and without the contribution from SMBHBs, we performed a Bayesian analysis using the NG15 data and found the best-fit value of the surface tension and the bias potential. The model with only DLRSM DWs is slightly favored over the model where additional SMBHB contribution is considered.

In conclusion, the era of GW astronomy has opened up a completely new window to explore particle physics. The methodology adopted in this thesis can be used to constrain many more interesting BSM scenarios. Several other possibilities can be explored as an extension of the thesis work. Some of these include, (i) supercooled PTs leading to exciting possibilities such as PBH production, (ii) identifying new methods to discriminate models with GWs, (iii) exploring other possible sources of the PTA signal, and (iv) identifying the possible sources of uncertainty in modeling the GW spectrum from PTs.

\graphicspath{{Appendix/}}
\begin{appendices}
\chapter{Theory and background}
\section{Procedure to obtain field-dependent masses}\label{app: field}
The field-dependent masses for all species are evaluated in the presence of the background field $\phi_c$, as described below
\begin{itemize}
\setlength\itemsep{.1em}
    \item \textbf{Scalars}: The relevant part of the Lagrangian is
    $$\mathcal{L} \supset \frac{1}{2}\partial_{\mu}\phi^a \partial^{\mu}\phi^a - V(\phi)$$
    Write the scalar field as a combination of the background field plus a fluctuation, $\phi = \phi_c+\tilde{\phi}$. The potential can then be Taylor expanded 
    \beq 
    V(\phi_c+\tilde{\phi}) = V(\phi_c) + \left.\frac{\partial V}{\partial\phi^a_{c}}\right\vert_{\tilde{\phi}=0} \tilde{\phi}^a + \frac{1}{2}\left.\frac{\partial^2 V}{\partial\phi^a_{c} \partial\phi^b_{c}} \right\vert_{\tilde{\phi}=0}\tilde{\phi}^a \tilde{\phi}^b + \cdots
    \eeq 
   The term quadratic in $\tilde{\phi}$ is clearly the mass term in the presence of a constant background field $\phi_c$. Hence the field-dependent mass matrix is 
   \beq 
    [\mathcal{M}_s^2(\phi_c)]_{ab} = \left.\frac{\partial^2 V}{\partial\phi^a_{c} \partial\phi^b_{c}} \right\vert_{\tilde{\phi}=0}\, .
    \eeq 
    Note that for spontaneously broken theories, the squared field-dependent mass for scalars can be negative in a certain range of field values, leading to a complex effective potential. The imaginary part is related to the decay rate of the vacuum as shown in Ref.\,\cite{Weinberg:1987vp}. However, phase transitions are governed only by the real part of the effective potential.
    
    \item \textbf{Fermions}: The relevant part of the Lagrangian is
    $$\mathcal{L} \supset - \Gamma^i_{ja}\phi^a~ \overline{\psi}_i~\psi_j$$
    Putting $\phi = \phi_c+\tilde{\phi}$, we can simply read off the field-dependent mass matrix,
   \beq 
    [\mathcal{M}_f^2(\phi_c)]^i_j = \Gamma^i_{ja}\phi_c^a\, .
    \eeq 

    \item \textbf{Gauge bosons}: The relevant part of the lagrangian is
    $$\mathcal{L}\supset\frac{1}{2}\tr[(D_{\mu}\phi)^{\dagger}D^{\mu}\phi]$$
    Again, write $\phi = \phi_c+\tilde{\phi}$, and $D_{\mu} = \partial_{\mu}\mathbb{I} + i \sum_{\alpha}g_{\alpha}\mathbb{T}_{\alpha} A_{\mu}^{\alpha}$, where $\mathbb{T}$ are the generators of the gauge group in the representation of $\phi$. Then the gauge boson mass term is
    \beq 
    \frac{1}{2}[\mathcal{M}^2(\phi_c)]^{\alpha}_{\beta} A_{\mu,\alpha} A^{\mu}_{\beta}\, ,
    \eeq 
    where the field-dependent mass matrix is, 
   \beq 
    [\mathcal{M}_{gb}^2(\phi_c)]^{\alpha}_{\beta} = g_{\alpha}g_{\beta} \tr[(\mathbb{T}^a_{\alpha b}\phi_a)^{\dagger}\mathbb{T}^b_{\beta a}\phi_b]\, .
    \eeq 
    
\end{itemize}
If the mass matrix for scalars or gauge bosons is non-diagonal, then the squared field-dependent masses are given by the eigenvalues of the corresponding mass matrix. For fermions, they are the singular values of the field-dependent mass matrix, i.e. eigenvalues of $\mathcal{M}^{\dagger}\mathcal{M}$. 

\section{Spontaneous $U(1)$ symmetry breaking}\label{app: ssb_toy}
Consider the lagrangian for a complex scalar field $\phi$, obeying a global $U(1)$-symmetry
\bea 
\mathcal{L}(\phi) &=& \partial^{\mu}\phi^*\partial_{\mu}\phi - V(\phi)\, ,\\
V(\phi) &=& \mu^2\phi^{*}\phi + \lambda(\phi^{*}\phi)^2 \, ,\label{eq: Vscalar}
\eea
where $\mu$ is a mass-parameter, and $\l$ is the scalar self-coupling. To ensure the scalar potential $V(\phi)$ is bounded from below, $\l$ must be non-negative, i.e. $\l\geq 0$. Depending on the sign of $\mu^2$, we have two possibilities for the minima of the potential, 
\begin{itemize}
\setlength\itemsep{0.1 em}
    \item $\mu^2\geq 0$: $V(\phi)$ has a minimum at the origin, i.e. $\phi = 0$. In this case, $m$ can be interpreted as the mass of $\phi$, and we can quantize the theory around the origin. 
    \item $\mu^2<0$: $V(\phi)$ has a maximum at the origin, with infinitely many degenerate minima lying on a circle of radius $|\phi| = v\equiv \sqrt{-\mu^2/\l}$, centered at the origin. The field can no longer be quantized around the origin. Rather, we must quantize the theory around a specific minimum of $V(\phi)$, chosen from all the possible minima.  
\end{itemize}
The $U(1)$ symmetry is the rotational symmetry of the potential in field space. In the second scenario, we see that although the potential is $U(1)$-symmetric, the choice of a minimum along a specific direction is not $U(1)$ symmetric. 
%This phenomenon is known as \textit{spontaneous symmetry breaking}. 
The chosen minimum represents the non-zero vacuum expectation value of the scalar field: $\sqrt{\langle 0_{\rm{out}}|\phi^*\phi|0_{\rm{in}}\rangle} \neq 0$. Without loss of generality, we can choose the minimum along the real $\phi$ direction. We can write $\phi =(v+\rho+ i\,\eta)/\sqrt{2}$, where $\rho$ and $\eta$ are real scalar fields corresponding to the real and imaginary components of $\phi$. The potential becomes
\bea 
V(\rho, \eta) &=& \left(\mu^2 v + \l v^3\right)\rho + \left(\frac{\mu^2}{2} v^2 + \frac{3}{4} \l v^2\right) \rho^2 +  \left(\frac{\mu^2}{2} + \frac{\l}{2} v^2\right) \eta^2 \nn\\
&+& \l v \left(\rho^3 + \eta^3\right) + \frac{\l}{4}\left(\rho^4 + \eta^4\right) + \frac{\l}{2}\left(\eta^2\rho^2 + 2v\rho\eta^2\right)\, ,
\eea 
where the term independent of the fields has been removed since it is a constant. Using the relation $v = \sqrt{-\mu^2/\l}$, the above expression simplifies to
\bea 
V(\rho,\eta) &=& \frac{m_{\rho}^2}{2} \rho^2 + \l v \left(\rho^3 + \eta^3\right) + \frac{\l}{4}\left(\rho^4 + \eta^4\right) + \frac{\l}{2}\left(\eta^2\rho^2 + 2v\rho\eta^2\right)\, ,
\eea 
%\left(\frac{m^2}{2} v^2 + \frac{\l}{4} v^4\right) +
where $m_{\rho}^2 = 2\l v^2$, is the mass of the scalar associated with $\rho$. The potential contains terms such as $\rho^3, \eta^3$, etc., indicating that the $U(1)$ symmetry is spontaneously broken\footnote{Without SSB, $v=0$, so the potential is a function of $\rho^2+\eta^2$.}. The mass term of the $\eta$ field vanishes, indicating the presence of a massless boson. Such a massless boson always arises when a continuous symmetry is spontaneously broken and is called a \textit{Goldstone boson}. 

Now let us promote the $U(1)$ symmetry to a gauge symmetry, which introduces a gauge field in the Lagrangian
\bea 
\mathcal{L} &=& -\frac{1}{4}F^{\mu\nu}F_{\mu\nu} + (D^{\mu}\phi)^*D_{\mu}\phi - V(\phi)\, ,\label{eq: gaugedU}\\
D_{\mu} &=& \partial_{\mu} + i g A_{\mu}\, ,~F_{\mu\nu} = \partial_{\mu}A_{\nu} - \partial_{\nu}A_{\mu}\, .
\eea
where $V(\phi)$ is given in Eq.\,\eqref{eq: Vscalar}, and $g$ is the gauge coupling. Expanding the kinetic term for $\phi$
\beq 
(D^{\mu}\phi)^*D_{\mu}\phi = \partial^{\mu}\phi^*\partial_{\mu}\phi + g^2|\phi^2|A^{\mu}A_{\mu} + ig A^{\mu}\left(\phi\partial_{\mu}\phi^* - \phi^*\partial_{\mu}\phi\right) \,,
\eeq 
Due to gauge symmetry, an explicit mass term for the gauge boson is disallowed. However, after SSB, writing $\phi = (v+\rho + i\eta)/\sqrt{2}$, we see the magical appearance of a mass term 
\beq 
\frac{m_A^2}{2} A^{\mu}A_{\mu}\,, ~~m_A^2 = g^2v^2\, .\nn
\eeq 
Hence we can get massive gauge bosons by introducing scalars which spontaneously break the symmetry. The mass of the gauge boson is proportional to the gauge coupling, and the symmetry-breaking scale.

We can go one step further and introduce a massless fermion to the lagrangian of Eq.\,\eqref{eq: gaugedU}
\beq 
\mathcal{L}\supset i\,\overline{\psi} \gamma^{\mu}D_{\mu}\psi - y \overline{\psi}\phi\psi + \rm{h.c.} \, .
\eeq 
Expanding $\phi$ as before, we generate a mass term for the fermion
\beq 
\mathcal{L}\supset -m_f \overline{\psi}\psi + {\rm{h.c.}}, ~~ m_f = \frac{y}{\sqrt{2}} v \, .
\eeq 

\chapter{More on flavor models}
\section{A two-flavon model}\label{app: model3}
Consider the horizontal group: $G_{\fn} = U(1)_{\rm{H_1}}\times U(1)_{\rm{H_2}}$. The flavor symmetry is then broken by two flavons $S_1,~S_2$, one for each $U(1)$ group, which gets a \vev. A possible two-flavon model is given below \cite{Nir-Seiberg1}:

\begin{itemize}
\item \textbf{Model 3}: $\epsilon_1 \sim \lambda^2 = 0.04,~\epsilon_2\sim \lambda^3 = 0.008$,
$$Q_{\fn}(\overline{Q}_L)=((0,1),(1,0),(0,0)),$$
$$Q_{\fn}(u_R)=((0,1),(-1,1),(0,0)),$$
$$Q_{\fn}(d_R)=((0,1),(1,0),(1,0))$$
The Yukawa matrices are:
\begin{gather}
{\bf{Y^u}} \sim \begin{pmatrix}
\epsilon_2^2 & \epsilon_1^{-1}\epsilon_2^2 & \epsilon_2\\
\epsilon_1\epsilon_2 & \epsilon_2 & \epsilon_1\\
\epsilon_2 & \epsilon_1^{-1}\epsilon_2 & 1
\end{pmatrix},~~~
{\bf{Y^d}} \sim \begin{pmatrix}
\epsilon_2^2 & \epsilon_1\epsilon_2 & \epsilon_1\epsilon_2\\
\epsilon_1\epsilon_2 & \epsilon_1^2 & \epsilon_1^2\\
\epsilon_2 & \epsilon_1 & \epsilon_1
\end{pmatrix} .
\end{gather}

The determinants are:
\begin{equation}\label{det3}
\det {\bf{Y^u}} \sim \epsilon_2^3,~~~\det {\bf{Y^d}} \sim \epsilon_2^2\epsilon_1^3
\end{equation}
\end{itemize}
In supersymmetric models, negative powers of $\epsilon$ are prohibited due to holomorphy; this gives rise to texture zeros (see \cite{Nir-Seiberg1}). In this work, we have only considered non-supersymmetric realizations of FN. 

\section{Stabilizing the potential with heavy bosons}\label{app: bosons}
Here we illustrate a scenario under which the effective potential can be stabilized by heavy bosons. Consider $N$ bosons at mass scale $M_b$, with an unbroken $O(N)$ symmetry:
\begin{eqnarray}
\mathcal{L} &\supset & \frac{1}{2}\big(\partial^{\mu}\Phi\big)^{\dagger}\big(\partial_{\mu}\Phi\big) - V(\Phi,S)\nonumber\\
V(\Phi,S) &=& \frac{1}{2}M_b^2\big(\Phi^{\dagger}\Phi\big) + \frac{1}{4}\lambda_b\big(\Phi^{\dagger}\Phi\big)^2 + \frac{1}{2}\lambda_{bS} \big(\Phi^{\dagger}\Phi\big)|S|^2,
\end{eqnarray} 
where $\Phi$ is a column vector of $N$ real scalar fields, and $\lambda_b$ and $\lambda_{bS}$ are the new couplings. Including the Coleman-Weinberg correction from these bosons modifies $V_{\rm{CW}}$ as,

\begin{equation}\label{eq: CWboson}
V_{\rm{CW}} = \frac{1}{64\pi^2}\sum_{i} (-1)^{f_i}n_i m^4_i\bigg[\log\bigg(\frac{m^2_i}{\mu^2}\bigg)- c_i\bigg] + \frac{1}{64\pi^2}\sum_{b} m^4_i\bigg[\log\bigg(\frac{m^2_b}{\mu^2}\bigg)- c_b\bigg],
\end{equation}
where the sum over $i$ includes SM fields, VLQs and the flavon, while sum over $b$ includes the new heavy bosons whose field dependent mass $m^2_b$ is given by,
\begin{equation}
m^2_b(s) = M_b^2+\frac{1}{2}\lambda_{bS} s^2.
\end{equation}

We take $\mu^2=v_s^2$, implying $m_b^2>>\mu^2$, and hence the heavy bosons can be integrated out, with their effect captured by higher dimensional operators (see for example \cite{Blum:2015rpa}). Keeping the leading order EFT operator, \eqref{eq: CWboson} becomes, 

\begin{equation}\label{eq: CWboson}
V_{\rm{CW}} = \frac{1}{64\pi^2}\sum_{i} (-1)^{f_i}n_i m^4_i\bigg[\log\bigg(\frac{m^2_i}{\mu^2}\bigg)- c_i\bigg] + \frac{c_6}{\Lambda_b^2}s^6,
\end{equation}
with $c_6$ as the Wilson coefficient.

\chapter{Details of DLRSM the effective potential}
\section{Minimization at the EW vacua}\label{appendix: min}
The minimization conditions are given by:
\bea \label{eq: minimization}
\mu_1^2 &=& \frac{1}{2(r^2 -1)}\Bigg(\k_1^2 \Big(w^2((r^2-1)\a_1 + r^2 \a_3 - \a_4) + 2 (r^2 -1)((r^2 + 1)\l_1 + 2 r \l_4) \Big) \nn\\
&& + 2 \sqrt{2} r v_R w \m_4 + v_R^2 \Big((r^2 -1)\a_1 + r^2 \a_3 - \a_4  + 2 w^2 \rho_{12}\Big) \Bigg) \,\,, \nn\\
\mu_2^2 &=&  \frac{1}{4(r^2 -1)}\Bigg( \k_1^2 \Big(w^2 (r^2-1) \a_2 - w^2 r \a_{34} + 2 (r^2 -1)(2 r \l_{23} + (r^2+1)\l_4) \Big)\nn\\
&& - \sqrt{2}(r^2+1) v_R w \m_4 + v_R^2 \Big((r^2-1)\a_2 - r \a_{34} - 2 w^2 \rho_{12} \Big) \Bigg)\,\,, \nn\\
\m_3^2 & = & \frac{1}{2} \k_1^2 ((r^2+1)\a_1 + 2 r \a_2 + r^2 \a_3 + \a_4 + 2 w^2 \rho_{1}) + v_R^2 \rho_1\,\,, \nn\\
\m_5 &=& - r \m_4 - \sqrt{2} v_R w \rho_{12},
\label{eq:minimisationcondition}
\eea
 where, $\rho_{12}=\rho_2/2 -\rho_1$, $\a_{34}=\a_3-\a_4$, and $\l_{23} = 2 \l_2 + \l_3$. 
 
\section{Field-dependent masses in DLRSM}\label{appendix: field}
The field-dependent mass matrices are obtained from the tree-level effective potential:
\beq
    m^2_{ij}(R) = \left.\frac{\partial^2}{\partial \varphi_i \partial \varphi_j} V_0\right\vert_{\langle \cdots \rangle}
\eeq
where $\langle \cdots \rangle$ denotes the background field value. This amounts to replacing $v_R\rightarrow R$, and $\k_1,\k_2,v_L\rightarrow 0$ in the usual mass matrices.

For the CP-even sector, in the basis $\{\phi_{1r}^0,\phi_{2r}^0,\chi_{Lr}^0,\chi_{Rr}^0\}$, we obtain,
\beq
    \mathcal{M}^2_{\rm{CPE}} = \begin{pmatrix}
        -\mu_1^2 + \frac{1}{2}(\alpha_1+\alpha_4)R^2 & -2\mu_2^2+\frac{1}{2}\alpha_2 R^2 & \frac{1}{\sqrt{2}}\mu_5 R & 0\\
        -2\mu_2^2+\frac{1}{2}\alpha_2R^2 & -\mu_1^2 + \frac{1}{2}(\alpha_1+\alpha_3)R^2 & \frac{1}{\sqrt{2}}\mu_4 R & 0\\
        \frac{1}{\sqrt{2}}\mu_5 R & \frac{1}{\sqrt{2}}\mu_4 R & -\mu_3^2+\frac{1}{2} \rho_2 R^2 & 0 \\
        0 & 0 & 0 & -\mu_3^2+ 3\rho_1 R^2\\
    \end{pmatrix}.
\eeq
For the CP-odd scalars, in the basis $\{\phi_{1i}^0,\phi_{2i}^0,\chi_{Li}^0,\chi_{Ri}^0\}$,
\beq
    \mathcal{M}^2_{\rm{CP0}} = \begin{pmatrix}
        -\mu_1^2 + \frac{1}{2}(\alpha_1+\alpha_4)R^2 & 2\mu_2^2-\frac{1}{2}\alpha_2R^2 & -\frac{1}{\sqrt{2}}\mu_5 R & 0\\
        2\mu_2^2-\frac{1}{2}\alpha_2R^2 & -\mu_1^2 + \frac{1}{2}(\alpha_1+\alpha_3)R^2 & \frac{1}{\sqrt{2}}\mu_4 R & 0\\
        -\frac{1}{\sqrt{2}}\mu_5 R & \frac{1}{\sqrt{2}}\mu_4 R & -\mu_3^2+\frac{1}{2} \rho_2 R^2 & 0 \\
        0 & 0 & 0 & -\mu_3^2+ \rho_1 R^2\\
    \end{pmatrix},
\eeq
and for the charged scalars, in the basis $\{\phi_{1}^{\pm},\phi_{2}^{\pm},\chi_{L}^{\pm},\chi_{R}^{\pm}\}$ we get,
\beq
    \mathcal{M}^2_{\rm{charged}} = \begin{pmatrix}
        -\mu_1^2 + \frac{1}{2}(\alpha_1+\alpha_4)R^2 & 2\mu_2^2-\frac{1}{2}\alpha_2R^2 & -\frac{1}{\sqrt{2}}\mu_5 R & 0\\
        2\mu_2^2-\frac{1}{2}\alpha_2R^2 & -\mu_1^2 + \frac{1}{2}(\alpha_1+\alpha_3)R^2 & \frac{1}{\sqrt{2}}\mu_4 R & 0\\
        -\frac{1}{\sqrt{2}}\mu_5 R & \frac{1}{\sqrt{2}}\mu_4 R & -\mu_3^2+\frac{1}{2} \rho_2 R^2 & 0 \\
        0 & 0 & 0 & -\mu_3^2+ \rho_1 R^2\\
    \end{pmatrix}.
\eeq

The neutral gauge boson mass matrix, in the basis $\{Z_L^{\mu},Z_R^{\mu},B^{\mu}\}$, is,
\beq
    \mathcal{M}^2_{Z} = \begin{pmatrix}
        0 & 0 & 0\\
        0 & 0 & 0\\
        0 & \frac{1}{4}g_R^2 R^2 & -\frac{1}{4}g_{BL}g_R R^2\\
        0 & -\frac{1}{4}g_{BL}g_R R^2 & \frac{1}{4}g_{BL}^2 R^2\\
    \end{pmatrix}.
\eeq

For the charged bosons, in the basis $\{W_L^{\mu\pm},W_R^{\mu\pm}\}$,
\beq
    \mathcal{M}^2_{W} = \begin{pmatrix}
        0 & 0\\
        0 & \frac{1}{4}g_R^2 R^2\\
    \end{pmatrix}.
\eeq

In addition to the field-dependent masses, we also need thermal self-energies of the fields for daisy resummation. These are obtained from the high-T expansion of the one-loop thermal potential. Substituting eq.\,\eqref{eq: 1_highT}  in eq.\,\eqref{eq: 1_V_1T} gives, to leading order,
\beq
    V_{1T}^{\rm{high}} = \frac{T^2}{24}\left(\sum_b n_b m_b^2 + \frac{1}{2}\sum_f n_f m_f^2\right).
\eeq
Here, index $b$ runs over bosons, while index $f$ runs over fermions. Each sum can be expressed as the trace of the respective matrix. Thermal mass matrices are then expressed as, $\Pi_{ij} = c_{ij} T^2$, where $c_{ij}$ are,
\beq
    c_{ij} = \frac{1}{T^2}\left.\frac{\partial^2}{\partial\varphi_i\partial\varphi_j}V_{1T}^{\rm{high}}\right\vert_{\langle\cdots\rangle}.
\eeq
We define,
\bea
     d_1 &=&\frac{1}{48}(9 g_L^2 + 9 g_R^2 + 8(2\alpha_1+\alpha_3+\alpha_4+5\lambda_1+2\lambda_3)),\\
     d'_1 &=& d_1 + \frac{y_{33}^2}{4} + \frac{\tilde{y}_{33}^2}{4},\\
     d_2 &=& \frac{1}{3}(2\alpha_2+3\lambda_4),\\
     d'_2 &=& d_2 + \frac{y_{33}\tilde{y}_{33}}{4},\\
     d_L &=& \frac{1}{48}(3 g_{BL}^2 + 9 g_L^2 + 8(2\alpha_1+\alpha_3+\alpha_4+3\rho_1+\rho_2)),\\
     d_R &=& \frac{1}{48}(3 g_{BL}^2 + 9 g_R^2 + 8(2\alpha_1+\alpha_3+\alpha_4+3\rho_1+\rho_2))\,.
\eea
We obtain the following thermal mass matrices:
\beq
    \Pi_{\rm{CPE}} = T^2 \begin{pmatrix}
        d'_1 & d'_2 & 0 & 0\\
        d'_2 & d'_1 & 0 & 0\\
        0  & 0 & d_L & 0 \\
        0 & 0 & 0 & d_R\\
    \end{pmatrix},
\eeq
\beq
    \Pi_{\rm{CP0}} = T^2 \begin{pmatrix}
        d'_1 & -d'_2 & 0   & 0\\
      -d'_2 & d'_1   & 0   & 0\\
         0  & 0     & d_L & 0 \\
         0  & 0     & 0   & d_R\\
    \end{pmatrix},
\eeq
\beq
    \Pi_{\rm{charged}} = T^2 \begin{pmatrix}
        d_1 & -d_2 & 0 & 0\\
        -d_2 & d_1 & 0 & 0\\
        0  & 0 & d_L & 0 \\
        0 & 0 & 0 & d_R\\
    \end{pmatrix}.
\eeq

The thermal mass matrices for the longitudinal gauge bosons are, 
\beq
    \Pi_Z = \frac{T^2}{6}\begin{pmatrix}
        13 g_L^2 & 0 & 0 \\
        0 & 13 g_R^2 & 0 \\
        0 & 0 & 6 g_{BL}^2\\
    \end{pmatrix}, 
\eeq
\beq
    \Pi_{W^{\pm}} = \frac{13}{6}T^2\begin{pmatrix}
        g_L^2 & 0 \\
        0 &  g_R^2 \\
    \end{pmatrix}.
\eeq
The mass of each species with the above thermal corrections is obtained as the eigenvalue of the matrix, $m^2(R)+\Pi(T)$. After diagonalization, the longitudinal polarization of the photon becomes massive, while the transverse components remain massless. 

\section{Neutrino masses in DLRSM}\label{sec: neutrino mass}
We have not taken into account a mechanism for generating neutrino mass in our version of DLRSM.  In this section, we argue the minimal way of incorporating neutrino mass in this model do not give any additional contribution to the GW phenomenology of the model. 
%rewrite if necessary

To demonstrate our point, we consider the model discussed in refs.\,\cite{FileviezPerez:2016erl,FileviezPerez:2017zwm}. Small neutrino masses are generated radiatively by the Zee mechanism, by adding a charged singlet scalar $\delta^+\sim(1,1,1,2)$ to DLRSM. In our notation, the Majorana Lagrangian is,
\beq
-\mathcal{L}^M_{LR} = \gamma_L L_L L_L\delta^+ + \gamma_R L_R L_R\delta^+ + \gamma_1 \chi_L^T i\sigma_2\Phi\chi_R\delta^- + \gamma_1 \chi_L^T i\sigma_2\tilde{\Phi}\chi_R\delta^- + \text{h.c.}\,,
\eeq
where, $\gamma_{L,R},~\gamma_1,~\gamma_2$ are the new Yukawa couplings.
As there is no tree-level right-handed neutrino mass, the contribution of the RH neutrinos to the effective potential is zero.  However, the quartic terms involving $\delta^+$ modify the mixing between the charged scalars. In the basis of $\{\phi_1^{\pm},\phi_2^{\pm},\chi_L^{\pm},\chi_R^{\pm},\delta^{\pm}\}$, the additional contribution to the charged mass matrix, $\mathcal{M}^2_{\rm_{charged}}$ is, 
\beq
\delta \mathcal{M}^2_{\rm_{charged}} = v_R^2\begin{pmatrix}
0 & 0 & 0 & 0 & \frac{\gamma_2}{2} \frac{v_L}{v_R}\\
0 & 0 & 0 & 0 & -\frac{\gamma_1}{2} \frac{v_L}{v_R}\\
0 & 0 & 0 & 0 & \frac{(\gamma_1\k_2+\gamma_2\k_1)}{2 v_R} \\
0 & 0 & 0 & 0 & -\frac{v_L(\gamma_1\k_1+\gamma_2\k_2)}{2v_R^2}\\
\frac{\gamma_2}{2} \frac{v_L}{v_R} & -\frac{\gamma_1}{2} \frac{v_L}{v_R} & \frac{(\gamma_1\k_2+\gamma_2\k_1)}{2 v_R}  & -\frac{v_L(\gamma_1\k_1+\gamma_2\k_2)}{2v_R^2}  & 0
\end{pmatrix}.
\eeq

Each of the non-zero entries is suppressed by a factor $v_L/v_R$ or $\k_{1,2}/v_R$ compared to $v_R^2$. Therefore the mixing of the charged scalars of DLRSM with $\delta^+$ is negligible, while their mixing among themselves remains unchanged. In the field-dependent mass matrix, we put $v_L\rightarrow 0,~\k_{1,2}\rightarrow0$, and $v_R\rightarrow R$, by which the additional mixing matrix, $\delta \mathcal{M}^2_{\rm_{charged}}(R)$, vanishes entirely. Hence the presence of $\delta^+$ does not alter the field-dependent mass matrices and therefore does not contribute to the effective potential. 

\section{DLRSM kink equations}\label{appendix: kink}
The dimensionless energy density is given by,
\bea
\hat{\mathcal{E}} = \frac{\mathcal{E}}{v_0^4} = \frac{1}{2}\left(\frac{d\hat{v}_L}{dx}\right)^2 + \frac{1}{2}\left(\frac{d\hat{v}_R}{dx}\right)^2 + \hat{V}(\hat{v}_L,\hat{v}_R) + \hat{C}.
\eea
We can rescale eq.\,(\ref{eq: kink1}) and eq.\,(\ref{eq: kink2}) as,
\bea
\frac{\partial^2 \hat{v}_L}{\partial \hat{x}^2} &=&  - \hat{\mu}_3^2 \hat{v}_L +\rho_1 \hat{v}_L^3 + \frac{\rho_2}{2} \hat{v}_L\hat{v}_R^2\label{eq: kink1_app}\\
\frac{\partial^2 \hat{v}_R}{\partial \hat{x}^2} &=&  - \hat{\mu}_3^2 \hat{v}_R +\rho_1 \hat{v}_R^3 + \frac{\rho_2}{2} \hat{v}_L^2\hat{v}_R.\label{eq: kink2_app}
\eea
In relaxation methods, a fictitious `time' variable, $\hat{t}$ is introduced, and the above equations are written as, 
\bea
\frac{\partial \hat{v}_L}{\partial \hat{t}} &=& \frac{\partial^2 \hat{v}_L}{\partial \hat{x}^2} + \hat{\mu}_3^2 \hat{v}_L -\rho_1 \hat{v}_L^3 - \frac{\rho_2}{2} \hat{v}_L\hat{v}_R^2\\
\frac{\partial \hat{v}_R}{\partial \hat{t}} &=& \frac{\partial^2 \hat{v}_R}{\partial \hat{x}^2} + \hat{\mu}_3^2 \hat{v}_R -\rho_1 \hat{v}_R^3 - \frac{\rho_2}{2} \hat{v}_L^2\hat{v}_R.
\eea
Eq.\,(\ref{eq: kink1_app}) and  eq.\,(\ref{eq: kink2_app}) are recovered in the limit $\frac{\partial }{\partial \hat{t}}\hat{v}_{L,R}\rightarrow 0$. 
We discretize the spatial and temporal coordinates with step size $\Delta \hat{x}$ and $\Delta \hat{t}$ respectively and express the derivatives in terms of second-order finite differences. The solution converges if $\Delta \hat{t} \leq \Delta\hat{x}^2/2$.  

For numerical purposes, we approximate spatial infinity by a value $R=50$ and linearly interpolate the boundary conditions eq.\,(\ref{eq: bc1}) and eq.\,(\ref{eq: bc2}), for the initial guess of $LR$ and $Z_2$ solutions. 

\end{appendices}

%%%%%%%%%%%%%%%%%%%%%%%%%%%%%%%%%%%%%%%%%%%%%%%%%%%%%%%%%%
\bibliographystyle{unsrtnat}
%\bibliography{bibliographychapterwise}
\bibliography{reference}
\phantomsection
\addcontentsline{toc}{chapter}{Bibliography}
\printindex

%%%%%%%%%%%%%%%%%%%%%%%%%%%%%%%%%%%%%%%%%%%%%%%%%%%%%%%%%%
\end{document}